unifying cloud
and carrier networks

# Deliverable D4.2

Proposal for SP-DevOps network capabilities and tools

| | |
|---|---|
| Dissemination level | PU |
| Version | 1.0 |
| Due date | 30.06.2015 |
| Version date | 30.06.2015 |

This project is co-funded
by the European Union

# Document information

## Editors and Authors:

Editors: Dr. Rebecca Steinert (SICS) and Dr. Wolfgang John (EAB)

Contributing Partners and Authors:

| | |
|---|---|
| ACREO | Pontus Sköldström, Bertrand Pechenot |
| BME | András Gulyás, István Pelle, Tamás Lévai, Felicián Németh |
| DTAG | Juhoon Kim |
| EAB | Wolfgang John, Catalin Meirosu, Xuejun Cai, Chunyan Fu |
| EICT | Kostas Pentikousis |
| iMinds | Sachin Sharma |
| OTE | Ioanna Papafili |
| POLITO | Guido Marchetto, Riccardo Sisto, Fulvio Risso |
| SICS | Rebecca Steinert, Per Kreuger, Jan Ekman, Shaoteng Liu |
| TI | Antonio Manzalini |
| TUB | Apoorv Shukla, Stefan Schmid |

## Project Coordinator

Dr. András Császár

Ericsson Magyarország Kommunikációs Rendszerek Kft. (ETH) AB

KONYVES KALMAN KORUT 11 B EP

1097 BUDAPEST, HUNGARY

Fax: +36 (1) 437-7467

Email: andras.csaszar@ericsson.com

## Project funding

7th Framework Programme

FP7-ICT-2013-11

Collaborative project

Grant Agreement No. 619609

## Legal Disclaimer

# Table of contents

## List of abbreviations and acronyms

| Abbreviation | Meaning |
|---|---|
| CP | Control Plane |
| DP | Data Plane |
| ESCAPE | Extensible Service ChAin Prototyping Environment |
| LCP | Local Control Plane |
| LDP | Local Data Plane |
| MF | Monitoring Function |
| NF-FG | Network Function Forwarding Graph |
| OF | OpenFlow |
| OP | Observability Point |
| UN | Universal Node |
| URM | Universal Resource Manager |
| VNF EE | Virtual Network Function Execution Environment |
| VSE | Virtual Switching Environment |
| NFV | Network Function Virtualization |
| OAM | Operations, Administration and Maintenance |
| QoS | Quality of Service |
| SAP | Service Access Point |
| SDN | Software Defined Network |
| SG | Service-graph |
| SLA | Service Level Agreement |
| VNF | Virtual Network Function |

# Summary

This deliverable summarizes relevant updates of the SP-DevOps concept, with focus on architectural integration within UNIFY and technical descriptions of SP-DevOps functions and tools that implement the observability, troubleshooting, verification, and VNF development processes as first defined in D4.1. The role of SP-DevOps processes in relation to programmability support and orchestration is described, along with SP-DevOps specific metrics that add further support in examining the lifecycle of a service.

The integration and interaction of SP-DevOps tools rely on a number of important support functions that are necessary for Orchestration and for composing SP-DevOps processes. The support functions enable communication, configuration, and information exchange between SP-DevOps tools and other functional blocks of the UNIFY architecture. They also implement functionality across all the architectural layers down to the infrastructure and the Universal Node. The main observability and verification processes, supported by the majority of the SP-DevOps tools, have been further established across the functional architecture and refined for orchestration support. The concept of local observability points has contributed to an extension of the Universal Node architecture by a monitoring management plugin component, enabling node-local analytics and access to monitoring information. Detailed specification of such a plugin remains as further work for the remainder of the project.

The deliverable, provides a greater understanding of how the technically maturing SP-DevOps concept can be applied and integrated with the UNIFY architecture as a whole. As part of the technical development in WP4, several tools and functions are planned to be released with individual licenses but packaged in a SP-DevOps Toolkit. Until now, the focus has mainly been on developing stand-alone monitoring functions and verification tools addressing specific problems of scientific interest, which may also be used for troubleshooting. Initial integration efforts between stand-alone tools and support functions are reported. Support for increased automation will be further developed during the remainder of the UNIFY project with focus on automated observability, troubleshooting and verification workflows.

A case study in this deliverable exemplifies how SP-DevOps functions and tools can support fulfillment and assurance phases of an elastic virtual network function. The example demonstrates how SP-DevOps supports both service deployment and dynamic scaling of virtual network functions on programmable nodes in an automated fashion. Automated processes will be absolutely necessary for managing future networks and we believe that the SP-DevOps way is a promising approach. However, the case study is an illustrative reminder of the substantial changes of current IT management systems that are required towards the adoption of SP-DevOps practices.

Further aspects of the SP-DevOps concept will be demonstrated in a project-wide integrated prototype. Until now, the work towards such a prototype encompassed development of smaller integrated and stand-alone prototypes developed within Wp4, which have been demonstrated during the first year of the project. Currently, work is carried out within WP4 to migrate parts of the already existing implementations to a testbed that is more compliant with the UNIFY architecture.

# 1 Introduction

## 1.1 Work Package Progress

The UNIFY D4.1 [D41] deliverable provided a first outline of the SP-DevOps concept, identifying the differences of DevOps applied in a telecommunications provider compared to a data centre environment. Moreover, the deliverable defined the three SP-DevOps roles, namely Operators, VNF Developer and Service Developer along with necessary processes for supporting the needs in the respective role.

Overall, the work in WP4 revolves around four key processes: a) observability, b) troubleshooting, c) verification, and d) VNF development support. D4.1 presented the state-of- the art literature with respect to existing tools and approaches relative to the aforementioned processes. Based on the identified limitations of the available approaches with respect to observability and scalability, a set of main challenges along with the necessary requirements for addressing various aspects were described. Throughout the project, WP4 aims to develop and evaluate solutions to many of these challenges, and as part of this work, relevant sets of requirements have been provided to other WPs together with an initial description of each process relative to the functional architecture. Moreover, D4.1 provides a first definition of the monitoring function concept and the components of an observability point.

The subsequent M4.1 [M41] milestone report presented the progress of WP4 towards the present deliverable D4.2. The integration of the SP-DevOps concept, processes, and tools into the UNIFY Architecture were further detailed in terms of initial APIs in collaboration with WP2. Moreover, a first description of the SP-DevOps Toolkit was provided along with initial technical descriptions of available tools and capabilities. Based on the refined concept of Monitoring Functions (MF) (first introduced in D4.1) and the observability levels depending on SP-DevOps role, each tool was initially mapped onto the UN architecture for obtaining a better understanding of the practical aspects of future prototyping work and for the purpose of closer integration of the work between WP4 and WP5. Finally, a first report on prototyping activities and integration plans with WP4 were presented.

Milestone M4.2 [M42] is an internal status report on the evaluation and prototyping activities in WP4. This milestone summarizes the progress of the work carried out in T4.2 and T4.3, as well as progress and results on integrated prototyping efforts in WP4. Since the previous milestone report M4.1, WP4 presented one integrated demonstration and one stand-alone demonstration (Section 7) at the first year review of the UNIFY project, created and organized within T4.4.The development process gave valuable insights for finding an efficient workflow of integrating individual prototypes, which was partially simplified by the availability of the messaging system that greatly facilitated integration. The work towards an integrated prototype in WP4 is developed in parallel to the project-wide prototype IntPro led by WP2. M4.2 summarizes the maturity level of the individual technical contributions based on WP2 questionnaires, for the purpose of defining common scenarios, identifying software and hardware requirements, dependencies, etc. The results from the questionnaires have been used as input to WP2 for further planning of the IntPro and have also been useful for finding synergies within WP4. The work in WP2 will be further supported by WP4 in terms of continuous discussions on practical integration aspects as well as a more refined mapping to the functional architecture as part of the D4.2.

## 1.2 Scope of this Deliverable

This deliverable provides a detailed description of the SP-DevOps Concept and the SP-DevOps Toolkit, including:

- an updated integration of SP-DevOps processes in the UNIFY overall functional architecture and the progress towards an integrated prototype in WP2;
- a description of how SP-DevOps tools and processes support orchestration and programmability in WP3 ;
- a detailed view of how SP-DevOps tools can be integrated with the UN architecture developed in WP5.

Support functions and concepts enabling programmability and information exchange in the UNIFY architecture are presented in this deliverable, including SP-DevOps metrics, interfaces for monitoring configuration and legacy support, as well as functions for dissemination and retrieval of monitoring information and observed network state.

Moreover, this deliverable provides detailed technical descriptions of the tools and capabilities included in the SP-DevOps Toolkit together with initial evaluation results, motivating the design choices compared to existing solutions and challenges identified in D4.1. The deliverable includes an example on how the SP-DevOps Concept as a whole can support different processes based on a case study on the Elastic Firewall [D21b] (as one of the use cases first outlined in D2.1[D21a]). Further, current integrated prototyping activities and plans are reported based on M4.2 and the progress since its publication. Finally, we highlight the main conclusions from the WP4 work so far and outline the roadmap towards next milestones and deliverables.

The remainder of this document is organized as follows: in Section 2 we provide a summary of latest developments in WP2, WP3, and WP5, where we specifically highlight progress and updates relevant to WP4. Section 3 provides an overview of the updated SP-DevOps concept, and details architectural integration aspects with respect to orchestration and programmability support. Moreover, the mapping of monitoring functions and observability points to the Universal Node architecture is described. Section 4 summarizes support functions and SP-DevOps tools and presents them in the context of a first version of the SP-DevOps Toolkit, followed by technical descriptions of each tool in Section 5. A case study exemplifying the interaction between SP-DevOps tools and functions is presented in Section 6. In Section 7, prototyping activities within WP4 and the progress towards project-wide integrated prototypes are reported, followed by conclusions and gap analysis in Section 8.

# 2 Overview of WP4-relevant Developments in Other WPs

## 2.1 Summary of Development in WP2

The main aims of WP2 are multi-fold. First, WP2 collects and categorizes requirements for modern service creation from the perspective of service providers and consumers. Second, it designs the architecture of the UNIFY framework based on the discovered requirements. Third, it defines various use cases that sketch examples of creating network services or handling potential issues (e.g., scalability). Finally, WP2 integrates prototypes implemented by other partners to a united test-bed, including the outcomes of WP4 such as monitoring and troubleshooting tools. Therefore, WP2 embraces all technical WPs (WP3-5) and supervises their activities and development.

In this regard, WP2 has released Deliverable 2.1 [D21a] and Deliverable 2.2 [D22]. D2.1 defines the key design principles of the UNIFY framework and provides the initial proposal of the UNIFY architecture based on the review of the state of the art with over 30 aspects that support the development of UNIFY, identification of 46 requirements collected from the technical work packages, and three wider use-case groups (including 11 detailed use-cases). Furthermore, D2.1 identifies six fundamental processes that describe the behavior of the UNIFY architecture, i.e., boot-strapping, programmability, verification, observability, troubleshooting, and VNF development. As the result of putting these identifications and definitions together, the initial proposal of the UNIFY architecture was described in two steps in D2.1, i.e., overarching architecture and a functional architecture. An addendum regarding the Elastic Firewall use case was released in [D21b].

Deliverable 2.2 [D22] mainly addresses the detailed interface definitions of the architecture. The deliverable describes use cases and presents the reference architecture by illustrating the main components, abstract interfaces and primitives, which reflect the priorities of design, functional and business requirements (an overview is given in Figure 3.2 in Section 3.1). In order to create a resilient, trustworthy and energy-efficient architecture supporting the complete range of services and applications, the UNIFY architecture supports the full flexibility of NFV, isolation among service requests (see Section 3.2.7 in D2.2), policy enforcement per service consumer (see Section 3.2.8 in D2.2), embedded monitoring services (see Section 3.2.9 in D2.2) to provide enablers for Quality of Service (QoS) support, and service resilience at any level of the virtualization hierarchy.

WP2 has developed and demonstrated a prototype integration platform that is built on a set of existing operating system and development tools such as Mininet, Click, and NETCONF and POX components, namely, the Extensible Service ChAin Prototyping Environment (ESCAPE) [D22]. Some of the WP4 partners plan to or have already integrated their tools into the platform. Currently, WP2 discusses the integration method of the remaining SP-DevOps tools.

## 2.2 Summary of Development in WP3

WP3 works on the UNIFY programmability framework, and defines the necessary functionality and primitives required. Furthermore, WP3 designs a generic optimization framework, which supports a variety of services and service chains, infrastructures, and objective functions. The derived solution will jointly optimize the network and node resources from the data centre over the network core to the access network. The optimization framework in WP3 and the approaches developed in WP4, are jointly aimed to reduce the complexity of service creation processes, as well as management and maintenance processes. Altogether, this will contribute to making the overall network management less intensive (e.g. in terms of labour costs, time, etc.), thereby minimizing OPEX and CAPEX.

Task 3.1 was concluded in 2014 and its outcome was documented in Deliverable D3.1 [D31]. D3.1 presented the major components of the UNIFY Programmability Framework, which is of high relevance to WP4 and included: i) the core primitives of this model, i.e. endpoints, network functions (NF), network elements and monitoring parameters, and ii) the monitoring parameters that could be measured through either standard-defined counters and OAM tools, or using SP-DevOps tools developed within WP4. D3.1 introduced a representation of these primitives in terms of the Network Function Forwarding Graph (NF-FG), which is it is a data structure representing the ordered interconnection of VNFs as well as their resources in order to fulfil a service. The SP-DevOps concept has so far been developed based on the D3.1 definition of a forwarding graph, and was first used for illustrating the VPN case-study in D4.1 [D41].

Furthermore, D3.1 discussed the integration of monitoring in the Programmability Framework as annotation to the NF-FG through joint work in WP3-WP4, and gave an overview of the "MEAsurements, States, and REactions" (MEASURE) language described in M 4.1 [M41]. Deliverable D3.1 also presented initial definitions of interfaces and operations between the separate layers of the UNIFY architecture. A set of such operations is related to data that could be obtained through SP-DevOps observability processes. The implementation of such processes, in compliance with the recursion property of the UNIFY Architecture, was detailed in M4.1. Finally, D3.1 approached the problem of defining the differences between mapping functionality that is an inherent part of the Resource Orchestrator and verification functions aligned with the SP-DevOps Verification process.

Both WP3 and WP4 aim to provide the refined descriptions of the UNIFY programmability framework and the SP-DevOps framework in D3.2 and D4.2, respectively, accompanied by an updated architecture, associated APIs and information model definition. D3.2 [D32] is prepared in parallel with this document, and addresses the following problems: i) programmability interfaces and data models, ii) NF-FG data model, iii) resource orchestration, including scalable orchestration algorithms, and scalable orchestration architectures and traffic steering, iv) service decomposition, and v) elastic network services, and vi) architectural components of the service programming and orchestration framework. Section 3 of D4.2 describes how SP-DevOps supports orchestration and programmability, which are two major subjects of investigation throughout WP3 as part of refining the UNIFY programmability framework.

## 2.3 Summary of Development in WP5

WP5 activities are focusing on the development of the Universal Node (UN) to enable the design of a flexible and easily programmable network. The UN aims to provide this by designing a software architecture optimized for executing dataplane-oriented workloads on standard high-volume servers, such as Intel-based blades. Depending on the specific use case, applications can either be executed on bare hardware or through lightweight virtualization primitives such as Linux Containers or Docker, as well as in a fully virtualized environment. This technical approach opens up wider possibilities not only for reducing OPEX and CAPEX but also for building up new services or reengineering existing ones with a far more flexible approach.

The first deliverable D5.1 [D51] assessed the requirements for the UN and the impact on the key aspects of the data plane with respect to network virtualization and resource sharing, as well as switching and traffic steering. Requirements relevant to support of SP-DevOps processes were derived in various WP4-WP5 discussions based on the requirements outlined in D4.1 [D41]. In addition, the D5.1 introduced the first functional specification of the UN data plane involving three modules for management, control and execution.

In the D5.2 [D52] deliverable, further development of the high-level UN architecture introduced in D5.1 [D51] were reported, in terms of more concrete definitions of architectural blocks and their interactions. The key element of the UN architecture is the Universal Resource Manager (URM), where compute and networking resources are commonly managed. This solution supports packet switching and processing both inside and outside of a VNF for the purpose of efficient management of dedicated resources and processes. Discussions between WP4 and WP5 have included issues about the architectural support for locally deployed MFs and access to various data sources (e.g. counters) on the UN. This has lead to an extension of the UN architecture in terms of a Monitoring Management Plug-in (MMP). The introduction of the MMP essentially enables the orchestrator to interact with the control plane of a particular MF and observability point (OP) through a uniform interface, which is crucial for achieving increased observability in a scalable manner. These efforts have also resulted in integration guidelines for MFs and a messaging system for ensuring communications between OPs and other components of the architecture. This enables efficient data processing close to the OPs, where the results can be transferred outside the UN to the MF control applications without significantly affecting the external bandwidth, yielding good observability together with limited performance impairments. Section 3 and 5 of this deliverable include an overview of the architectural aspects of the MMP and detailed descriptions of how different SP-DevOps tools may be implemented in a UN.

Finally, the D53 [D53] deliverable is mainly aimed at describing WP5 prototyping implementation details and phases towards a project-wide integrated prototype. In this context, the implementation details of the MMP will be discussed between WP4 and WP5 as part of the work towards a project-wide integrated prototype with the aim of showcasing aspects of the SP-DevOps MF concept.

# 3 UNIFY SP-DevOps

In this section we present the final SP-DevOps concept, refined from the first descriptions provided in D4.1 [D41] and M4.1 [M41]. Recall that SP-DevOps derives from DevOps, a concept born by the fast software development practices of major software companies such as IBM and HP.

Through a number of tools and support functions, generally referred to as the SP-DevOps tools, the UNIFY SP-DevOps framework maintains the deployment and re-deployment cycle of production through four major processes, namely, *VNF Developer support, Verification, Observability,* and *Troubleshooting*. These four processes are mapped on the general service lifecycle of UNIFY as shown in Figure 3.1. They are effectively dealing with three out of four crucial principles of the general DevOps concept (i.e., *monitoring and validation of the operational quality, development and testing against production-like systems*, and *reliable and repeatable deployment*). The four processes of SP-DevOps are detailed D4.1. Note that, although the description of each process in D4.1 is based on an early representation of the functional architecture, the overall mapping remains the same, and is therefore not included in this document. In this deliverable, we focus on the description of the SP-DevOps functions and tools that support one or several of the aforementioned processes.

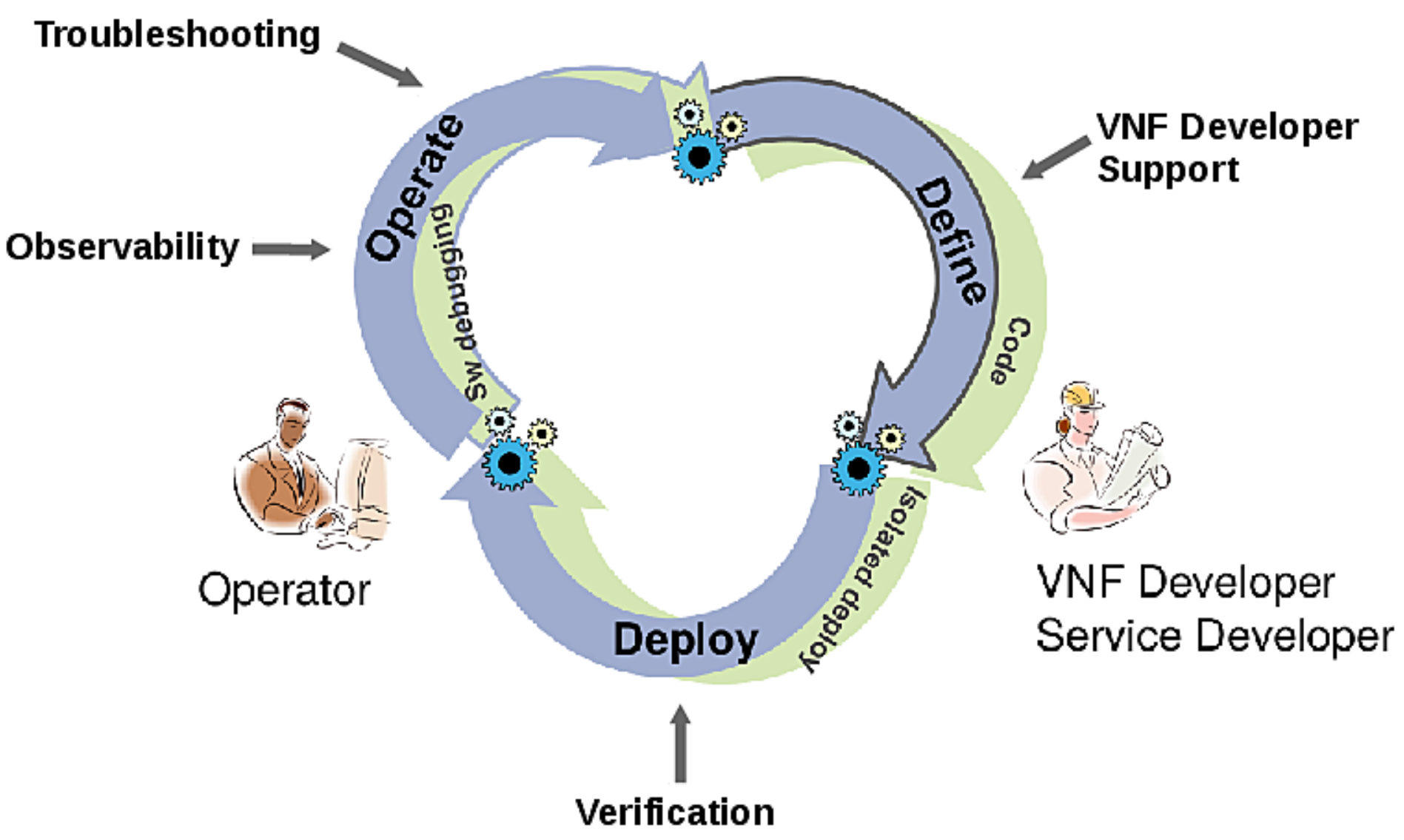

*Figure 3.1: SP-DevOps processes mapped on the UNIFY service lifecycle.*

The main aim of this sectoin is to explain how the SP-DevOps framework operates in the functional architecture presented in D2.2 [D22] and describe the key processes and interactions between the architectural layers and the SP-DevOps components. In Section 3.1, we provide an overview of the UNIFY architecture and highlight the main functional components relevant to the SP-DevOps framework and associated processes. One of the main objectives in this deliverable is to describe how SP-DevOps supports orchestration and programmability in the sense of providing relevant monitoring information and verification feedback upon deployment - this is explained layer by layer in Section 3.2 and is partially based on input from D3.1 [D31] and D3.2 [D32] (which is produced in parallel with this document).

Deployment and operation of MF control apps and observability points in the Infrastructure Layer the UN are detailed in Section 3.3, where we describe the necessary extensions to the UN architecture and outline several deployment options of an observability point and its components. As part of supporting programmability at different levels across the architecture, SP-DevOps includes functions for configuration, information dissemination and retrieval, as well as support for integration of OAM legacy tools, as described in Section 3.4. In Section 3.5, a set of SP-DevOps metrics is presented for the purpose of evaluating the SP-DevOps processes and service and network performance, followed by concluding remarks in Section 3.6.

## 3.1 Overview of SP-DevOps in the UNIFY Architecture

The UNIFY architecture defines the main architectural components and reference points relevant to the UNIFY concept which allows architecture designers to focus on developing orchestration capabilities within the system, introducing SP-DevOps tools and processes, and high performance data plane based on Commercial Off-the-Shelf (COTS) hardware. The overarching view of the UNIFY architecture comprises three layers, namely, the Service Layer (SL), the Orchestration Layer (OL) and the Infrastructure Layer (IL). Each layer contains a set of management components and Network Function Systems (NFS), as well as reference points, which define communication channels between them. Such elements and their interconnections are illustrated in Figure 3.2.

The **Service Layer (SL)** comprises traditional and virtualization-related management and business functions concerned with the service lifecycle. SL management functions should be infrastructure-agnostic and should deal with the management of the offered services. We denote the users of the management systems by operators, enterprise users, and VNF developers.

The **Orchestration Layer (OL)** has two major functional components. The first component is the Resource Orchestrator (RO) in which virtualizers (see Sec. 3.2.1 in [D22]), policy enforcement (see Sec. 3.2.8 in [D22]), and resources orchestration between virtualizers and the underlying resources (see Sec. 3.2.7 in [D22]) play important roles. The second component is the Controller Adapter (CA) which is responsible for the domain-wide resource abstraction and for virtualization of different resource types, technologies, vendors, or administrative domains. The RO and CA are managed by a corresponding management system including, for example, an OSS as shown in the right-hand side of Figure 3.2.

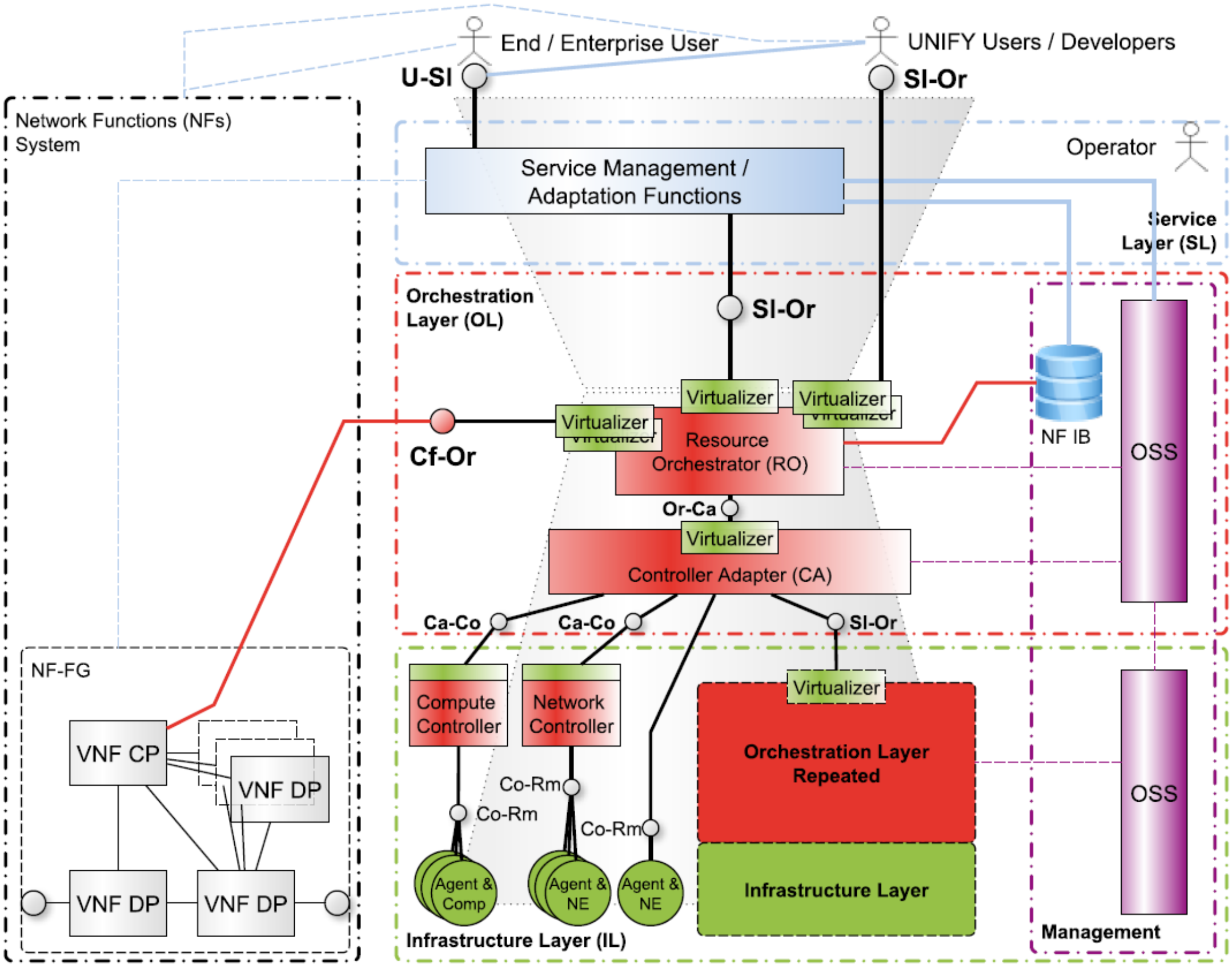

*Figure 3.2: The UNIFY Overarching Architecture.*

The **Infrastructure Layer (IL)** includes resources (i.e., compute, storage and network), local resource agents (e.g., OpenFlow switch agent and Open Stack Nova compute interface), and/or various controllers (e.g., an SDN Controller for a transport network domain and compute controllers for a data center).

On these layers, UNIFY implements a set of tools and methods in order to support dynamic, stable, repeatable, and agile service deployment.

### 3.1.1 Functional components of the SP-DevOps

For applying the SP-DevOps concept to the UNIFY framework, diverse functional components are required to be implemented. Especially, monitoring components are considered essential building blocks in terms of ensuring performance of a service (i.e., verification and observability) and rapid reaction to infrastructure failures (i.e., troubleshooting and debugging). Monitoring components within the UNIFY architecture collect various information from the infrastructure that indicates performance and availability of virtualized and physical system components.

This subsection presents the components and methods of SP-DevOps that build the monitoring/verification system in the UNIFY architecture. Figure 3.3 is a panoptical illustration of how SP-DevOps-related functional components can be mapped on the UNIFY functional architecture. The left box of the figure shows SP-DevOps-related functional components and methods (yellow boxes) and the right box of the figure describes functional components with regard to the information collection/delivery (yellow boxes) and the information flow (blue lines and arrows) between infrastructure layer and users.

#### Monitoring Functions and Observability Points

Monitoring capabilities (e.g., information collection, information collector control, measurement, and/or analysis) and components are collectively referred to as a MF. An MF is typically responsible for probing performance properties (e.g., network latency, or utilization of network and compute resources), or processing and analyzing measurement outcomes (e.g., throughput, bandwidth utilization, energy consumption), and/or notifying concerned components about resource shortages.

As defined in D4.1 [D41], an MF consists of a set of monitoring network functions or components, which are referred to as OPs. An OP is responsible for collecting and processing information, or for relaying the monitoring data between components. OPs consist of a local data plane component (LDP), which is the source of the actual measurement data. Additionally, OPs may consist of a node-local control plane component (LCP), which allows the central control app to distribute certain control functionality in the infrastructure for increased scalability. This offers the operator a tradeoff between monitoring fidelity on one hand and monitoring overhead on the other hand. Examples of scalable MFs utilizing LCP are described in Section 5.5 – 5.9 in form of scalable network performance measurement tools.

#### Messaging system (DoubleDecker)

For exchanging data between distributed monitoring DBs and the recursive query engine, the DoubleDecker (see Section 5.1) messaging system is used. DoubleDecker is developed atop the ZeroMQ [ZeroMQ] messaging library and extends its capabilities to support complex messaging systems such as the UNIFY architecture. By applying DoubleDecker to the UNIFY monitoring system, functional components can exchange information in a rapid and scalable manner. DoubleDecker routes messages between applications connected to it, allowing for point-to-point synchronous and asynchronous communication between applications, via a lightweight publish/subscribe interface.

#### Monitoring description language (MEASURE)

Service developers design a Service Graph (SG) by specifying the requirement based on various Key Quality Indicators (KQIs). On the service layer, the service graph is transformed into network function chains, namely, the network function forwarding graph (NF-FG). The KQIs are translated into the Key Performance Indicators (KPIs) during this transform from SG to NF-FG. The MEASURE language (see Section 5.2) is developed in order to specify the KPIs in a unified form. MEASURE is widely used on interfaces between distinct layers and modules throughout the UNIFY architecture. However, at the infrastructure layer of the system, MEASURE may be translated further into a technology-specific description method, e.g., extensions to OpenFlow or other protocols. NF-FGs can be (recursively) decomposed into smaller NF-FGs at the OL. Along with this decomposition, the MEASURE expression is split into multiple MEASURE expressions and appropriate MFs are added into NF-FGs depending on the availability of monitors (e.g., by invoking GetPerformanceValue() API). Based on the requirement and specification of MFs described in NF-FGs (in the form of MEASURE), MF control apps are instantiated and OPs are configured in suitable infrastructure components.

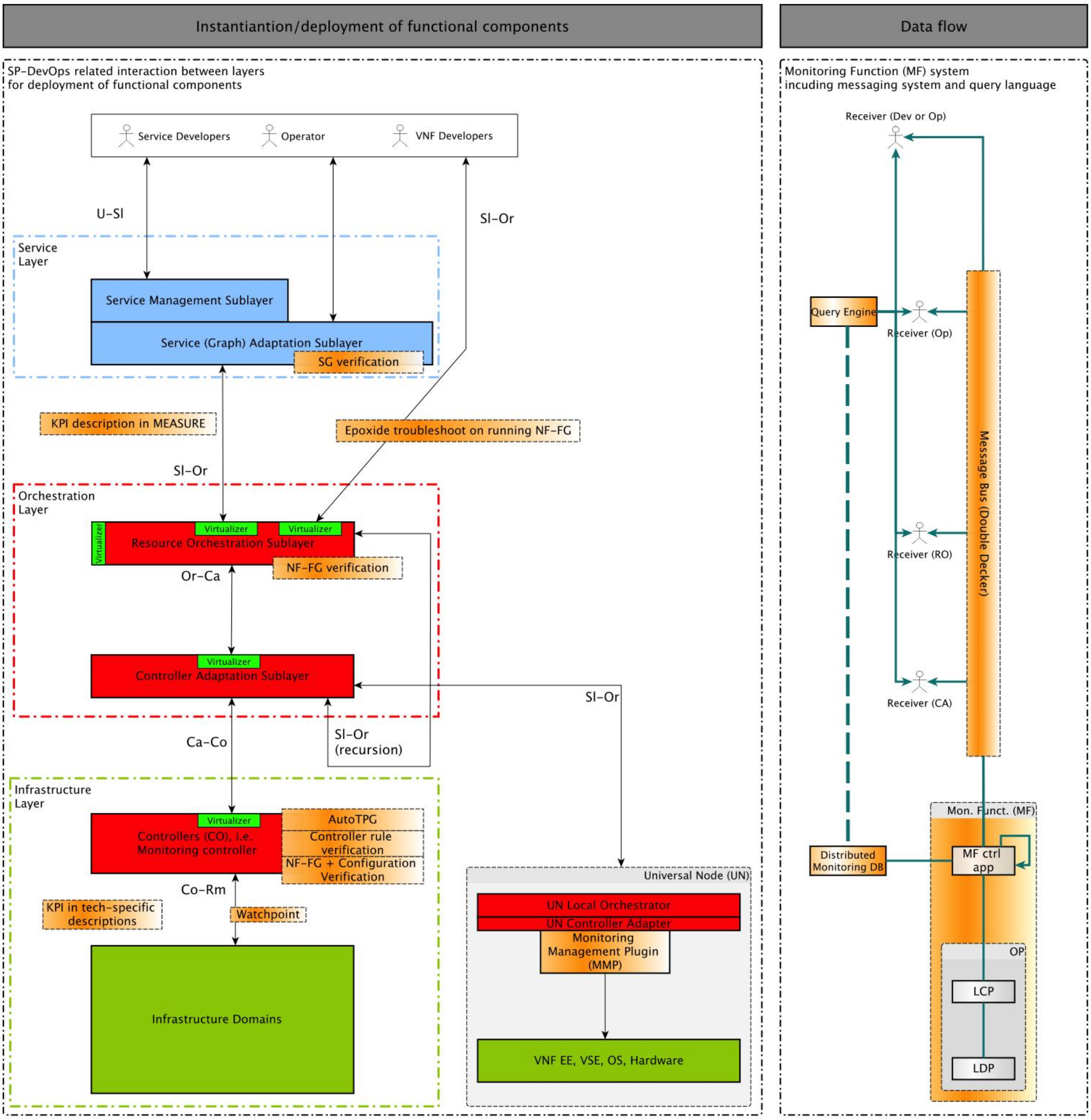

*Figure 3.3: Architectural mapping of functional components of SP-DevOps on the UNIFY architecture.*

### Distributed monitoring data and recursive query language

Developers and operators can collect/receive monitoring data gathered by the MF control app from the OPs by asking the query engine with a specific query language (see Section 5.3). The monitoring results are first stored in distributed monitoring DBs (e.g. OpenTSDB [OpenTSDB]) and the desired information is delivered to the query engine through the messaging system. More specifically, the high-level query from users and developers, described in the recursive query language, can be decomposed into a set of smaller queries in order to gather desired information from distributed monitoring DBs.

### Monitoring function deployment methods: Universal Node vs. legacy devices

Functional components within UNIFY are based on the UN as the IL, which is the main development focus of WP5. The UN monitoring capabilities are extended by the MMP (see Section 3.3.3 for the further details). MMP is operated as a monitoring controller on a level with the VNF and VSE management plugins. However, it is important to note that the legacy devices will be still in use after the introduction of UNIFY. For integrating legacy devices into the UNIFY framework, the unification of the management plane based on a standardized measurement protocol (i.e., Two-Way Active Measurement Protocol - TWAMP) is discussed within UNIFY (see Section 3.4.2 and 5.4).

### Verification before service deployment

Some properties of services such as network reachability, forwarding loops, and undetected route failures (a.k.a. network black holes) can be verified before the service is actually deployed. In this case, verification is solely based on the description of the service. This verification is performed across all UNIFY layers. As a first step of this process, verification is carried out based on the service graph on the service layer. This step typically checks the integrity of the service description provided by the users. The second step of the verification is performed on the orchestration layer based on the NF-FG. This step additionally examines the possible loops within the service chain along with the service policy provided by users. Finally, the third step of the verification is executed on the infrastructure layer. In this step, verification checks the correctness of the technology-specific description such as OpenFlow rules. Further details on this topic is provided in Section 5.10.

### Verification after service deployment

Verification actions after the service deployment are typically performed at the infrastructure layer. For example, the Automatic Test Packet Generation Tool (AutoTPG, see Section 5.8) checks the correctness of the flow table at OpenFlow switches and routers. By verifying the flow table, AutoTPG ensures faultless delivery of data packets within the service. Another example of the verification after the service deployment is the run-time NF-FG verification tool (see Section 5.11) that is currently under development. The target of the tool is to robustly verify SLAs from the data plane traffic based on sampling techniques.

**Troubleshooting components**

Highly distributed physical and logical resources in today's network infrastructure make pinpointing the cause of the problem difficult. To effectively deal with this challenge, WP4 develops tools that support the troubleshooting process of the SP-DevOps in the UNIFY framework. For service troubleshooting, it is crucial to acquire a comprehensive view of the state of the infrastructure. Moreover, developers need to examine the state of the infrastructure via command line (or GUI-based) operations based on developer-friendly software. To this end, the controlling components of the troubleshooting tool are commonly operating on U-Sl and Sl-Or interfaces of the UNIFY architecture, whereas components that collect the required information (e.g., agents) are positioned at the infrastructure layer. Epoxide is illustrated as a representative troubleshooting tool of SP-DevOps in Section 5.13.

**VNF developer support components**

One of the major tasks of VNF developers is debugging the problem of VNFs. For supporting this task, UNIFY implements several debugging tools. Debugging tools are commonly deployed on the infrastructure layer in order to closely observe and fix anomalies at the data plane and the control plane. Since identification of the problem and discovery of its causes are accompanying activities on debugging process, debugging tools often covers other SP-DevOps processes such as the verification process and the troubleshooting process. Watchpoint is a good example of debugging tools which serves multiple processes of SP-DevOps in UNIFY (see Section 5.12 for further detail).

In the next subsection, we will go through in detail how the SP-DevOps concept support orchestration by describing both deployment and operations phases, together with the actions at each layer.

## 3.2 SP-DevOps Orchestration Support

In this section we will go through each of the layers in the architecture and describe the necessary actions and processes in each layer, first during the service *deployment phase* (i.e. the workflow from top to bottom), and then during the *operational phase* (i.e. information flowing from bottom to top). In the service deployment phase we currently do not consider the resource costs of any monitoring functions that we deploy and view them as functionality provided by the infrastructure for free. Depending on the actual operation environment, this may not be a realistic model. Instead, a similar resource orchestration process as to the one performed in WP3 to allocate network, compute, and storage resources could be employed in order to properly model and account for resources consumed by deployed monitoring functions. How to more accurately model resource requirements of monitoring functions and integrate this model into the general resource orchestration process is a question that needs further study.

### 3.2.1 Service Layer SP-DevOps support

The deployment phase involves four main steps and starts when a service request in the form of a SG arrives at the SL. It contains a description of the service that should be realized together with the associated requirements. Specifically, the description contains the NFs that should be instantiated, a description of their required connectivity, and the SLA requirements associated with the service. For monitoring related aspects, the Service layer has to perform three steps when converting the Service Graph into a NF-FG; 1) KQI-KPI translation, 2) expressing KPIs in MEASURE, and 3) inserting resource constraints in the NF-FG that the orchestration layer can utilize to map VNFs (shown on the left side of Figure 3.4). A fourth step is taken by the SP-DevOps Verificaton process by applying graph-theoretic algorithms to the SG to verify user-defined policies (see Section 5.10 for more details).

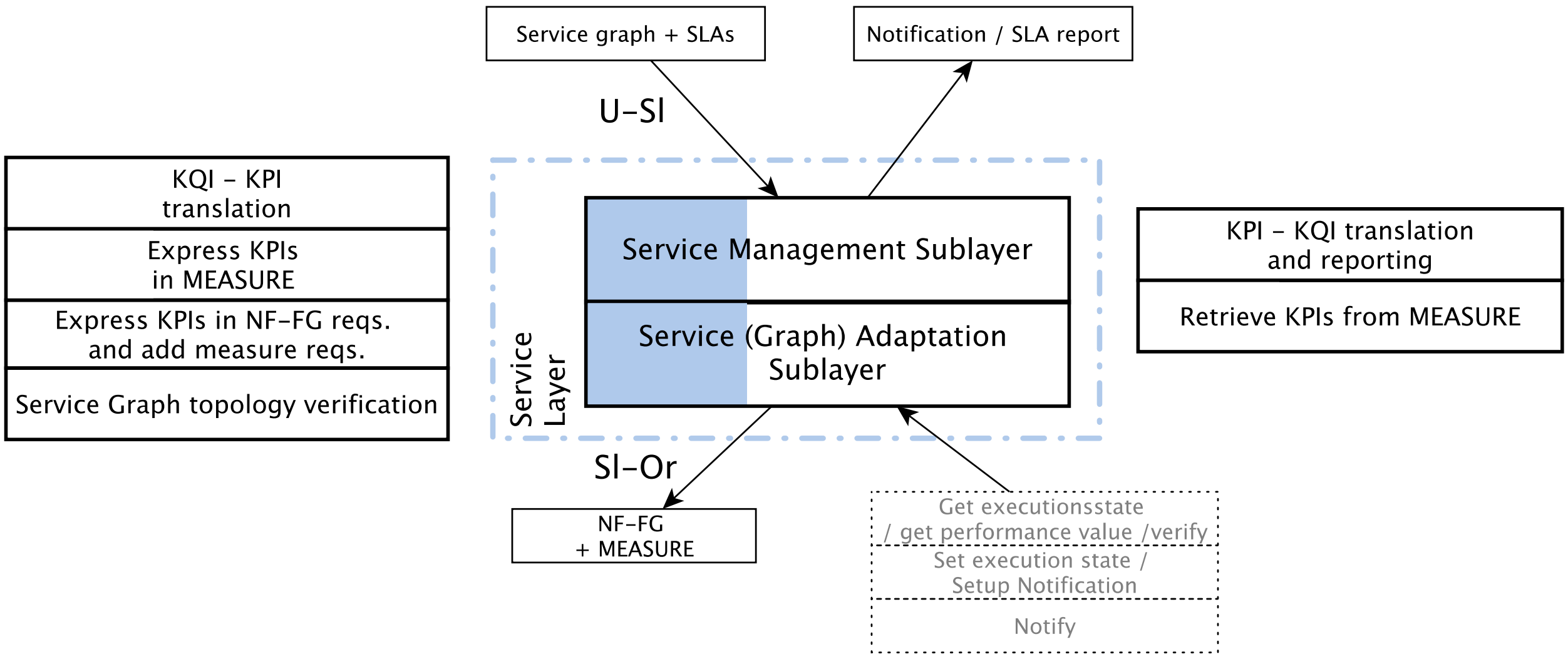

*Figure 3.4: SP-DevOps actions taken at the Service layer.*

The first step is to perform KQI-KPI translation which translates the SLA requirements into: 1) link and node resources constraints that can be used to assist the orchestration layer to fulfil a service request and 2) continuous monitoring requirements. Examples of KPIs that can be directly translated into constraints usable by the RO to find an appropriate mapping for network resources include: minimum bandwidth, maximum latency, maximum jitter, and maximum packet loss. Similarly for compute resource KPIs, such constraints can, for example, include maximum response time, number of CPU units, and the amount of RAM and non-volatile memory. Other types of constraints for example could be placed on privacy and integrity (e.g., traffic encryption or restrictions on co-location with other customer's traffic or data) or regional restrictions (e.g., to avoid crossing national borders) which translate into resource capability constraints. All of these constraints on resources and capabilities can be directly used by the orchestration layers to find a suitable placement for the requested SG and resulting NF-FG (i.e. the virtual network embedding VNE, as discussed in [D32]). A list of relevant KQIs/KPIs can be found in Annex 1.

The second step is to handle other KPIs that cannot be directly mapped to resource requirements but rather require active monitoring to continuously perform SLA assurance. These include KPIs such as mean time between failure for individual components or the service as a whole, service availability, and performance requirements. These are instead translated into MEASURE which specifies which monitoring tools needs to be activated and where to monitor in order to observe these KPIs.

In the third step the defined monitoring tools add additional capability constraints in the NF-FG. For example, the node where a function is placed has to support a particular monitoring tool that can test availability and operability of the function. The same can be true for directly mapped KPIs that may require continuous monitoring as well, such as latency or bandwidth. These SLA translation steps are illustrated in Figure 3.5.

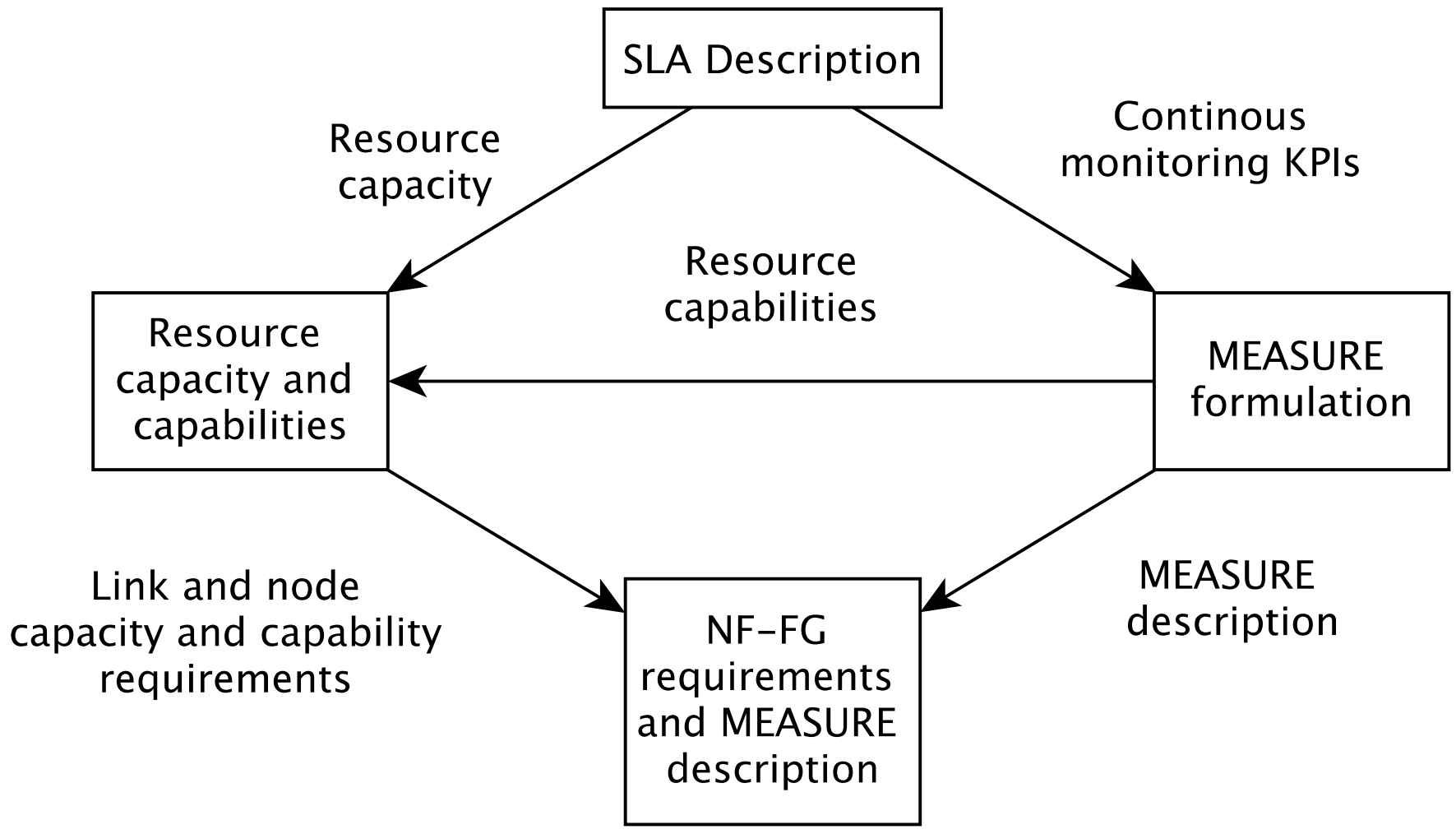

*Figure 3.5: SLA translation into resource requirements and MEASURE description*

When these three translation steps have been performed the resource and capability constraints are inserted in the NF-FG together with the MEASURE code. Before sending the NF-FG to the OL, MEASURE (Section 3.1.1 and 5.2) code for the local monitoring aggregator is prepared, and will be injected in the aggregator once the NF-FG has been successfully placed and relevant monitoring identifiers resolved by lower layers. More details about this process are provided in Section 5.2.

During the operational phase monitoring results from OL starts flowing into the aggregator in the Service layer, which starts evaluating the results based on the MEASURE code that it has been configured with. For simpler KPIs, such as latency, this would typically be monitoring of thresholds and reporting to some entity when they are breached. For some KPIs the calculations can become more complex as for example measuring service availability requires storing of the timestamps when the service failed followed by summarizing them. Storing long term data is better suited for a database than in-memory in the aggregator, so the aggregator may send the timestamps of service available / unavailable events and notify another entity when the service becomes unavailable, which can evaluate the events stored in the database. Periodically storing results in the database allows for both asynchronous

SLA reporting, e.g. by notifying a customer or management system when a service is unavailable or a threshold breached, as well as synchronous reporting, e.g. by compiling weekly status reports.

### 3.2.2 Orchestration Layer SP-DevOps support

The OL is composed of the RO and the CA sublayers. The RO sublayer is responsible for 1) presenting virtual views to the higher layer, 2) enforcing policies, and 3) performing resource orchestration on resources presented by the CA sublayer. The CA sublayer is responsible for 1) constructing a network view of the underlying resources (which may come from different technological and/or administrative domains) in a technology and vendor agnostic fashion, and 2) for adapting the configuration made on the northbound vendor and technology-agnostic interface to the different underlying controller technologies over various southbound interfaces.

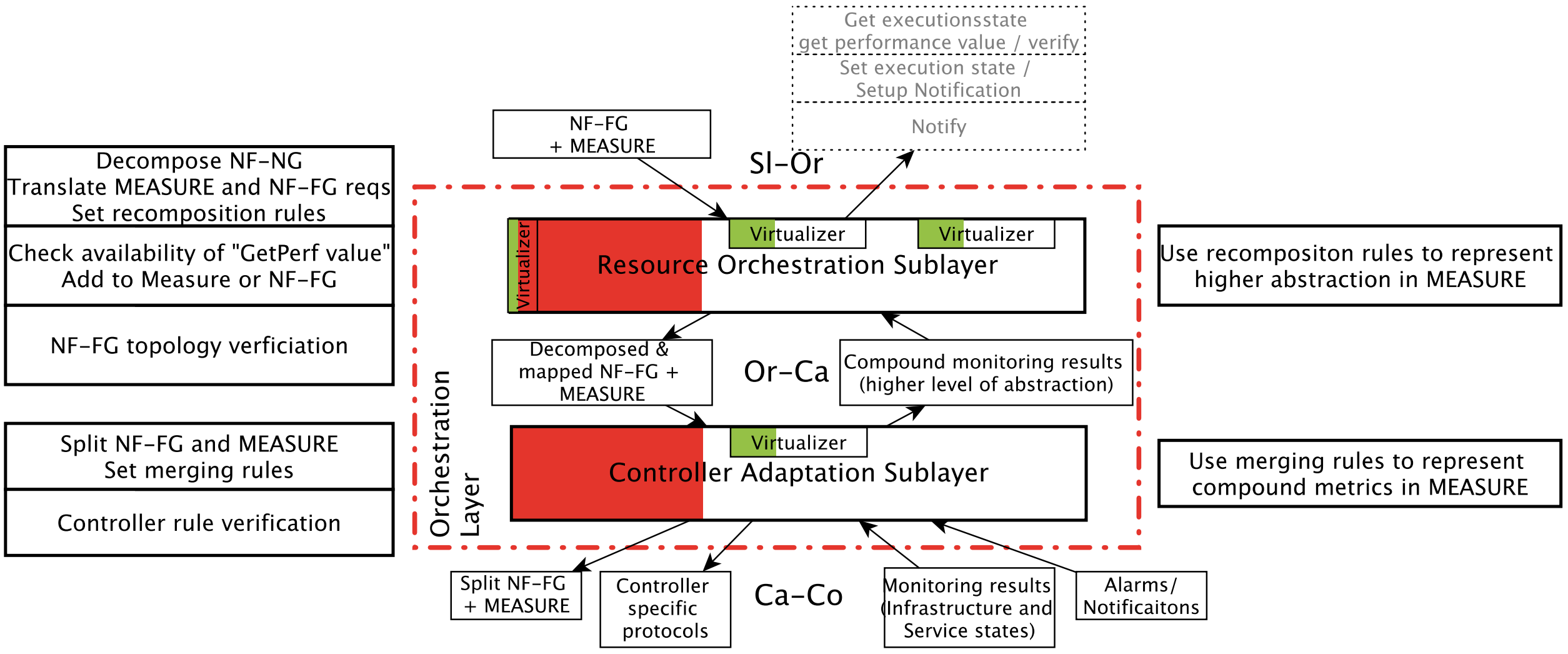

*Figure 3.6: SP-DevOps actions taken at the Orchestration layer*

#### 3.2.2.1 Resource Orchestration sublayer

The RO Sublayer and the SP-DevOps related actions it performs are depicted at the top of Figure 3.6, involving decomposition of the NF-FG, updating of the MEASURE definitions of the decomposed NF-FGs, mapping to the actual underlying resources as well as verification of network topology properties. In the deployment phase, NFs in the incoming NF-FGs get decomposed into atomic blocks suitable for orchestration in the layer below, which can be another OL or the actual IL. During the decomposition one NF can be for example expanded into a set of parallel NFs, a set of serially connected NFs (e.g. a processing pipeline), or into a set of NFs connected in a particular fashion. During this operation both NF-FG resource capacity and capability requirements as well as the MEASURE definitions need to be updated and inserted in the decomposed NF-FG. Both capacity and capability requirements on the incoming NF-FG also have to be taken into account during the decomposition, for example, if an NF in the incoming NF-FG requires the capability for latency monitoring, the decomposed version must support it as well.

Note that translating MEASURE descriptions from the incoming NF-FG to a decomposed version of itself is not a trivial problem that can easily be generalized (see Section 5.2.2 for more discussion). The translation has to be done differently for different decompositions and metrics. If, for instance, an incoming NF has a latency measurement defined over it, and is decomposed into two parallel NFs, a reasonable translation could be to translate it into the maximum latency of two latency measurements over each of the two in parallel decomposed NFs. If however the decomposition is two serially placed NFs, one option of translation could be to measure the latency over both of them with a single measurement. For a packet loss measurement in the parallel case it does not make sense to take the maximum of two loss measurements, but rather to take the sum of the two measurements. Instructions on how to decompose different metrics could be placed in the NF-Information Base (NF-IB) as part of the decomposition model and used by the resource orchestrator to perform the translation.

Translating resource capacity and capabilities constraints during the decomposition is also an issue that has to be handled. If the incoming NF-FG contains for example one NF connected to two end points with an end-to-end latency requirement this has to be taken into account when decomposing that NF so that the end-to-end latency requirement is respected even for new links created by the decomposition.

Next, the decomposed NF-FG is mapped to the available resources, taking the updated resource capacity and capability requirements into account. The mapped NF-FG is then ready to be transmitted to the CA sublayer.

Finally, the RO sublayer will trigger the verification process sending the generated NF-FG to the verification engine for verification of network properties considering only the input topology (more details in Section 5.10). If the verification process does not raise any concerns, the NF-FG is forwarded to the CA sublayer.

During the operational phase, and similarly to what happens in the SL, local aggregation code has to be generated and injected to the aggregator in this layer, in order to “recompose” monitoring values before transmitting them to the higher layer, for example by calculating and transmitting the maximum of two measurements to the higher layer for the parallel NFs described in the previous paragraph.

When presenting the virtual resource view to the higher layer, the Virtualizers must take monitoring capabilities into account, presenting appropriate capabilities to the higher layer.

#### 3.2.2.2 Controller Adaptation sublayer

The CA sublayer and the actions it performs are depicted at the bottom of Figure 3.6, which mainly involves partitioning (splitting) the NF-FG, updating the MEASURE definitions of the partitioned NF-FGs and performing verification on network controller rules (specifically OF rules, as detailed in Section 5.10).

In the deployment phase, the mapped NF-FG is partitioned according to the mapping created by the RO and the underlying resources, which may belong to different technological and/or administrative domains. With regards to monitoring components this means that a single session of a set of monitoring functions may need to be configured in different domains, or be split into multiple measurements. For instance, if an end-to-end latency measurement is defined and the two end-points are in two separate domains, either a single end-to-end measurement may be created or three separate measurements may be necessary, one within each domain and another between the domains. In the first case with a single measurement it is likely that some parameters have to be assigned to the monitoring functions at each end, for example a session identifier, so that probe packets from one domain can be correctly interpreted in the other. In the second case each domain may handle its own session identifiers but instead appropriate MEASURE code has to be generated in order to assess the different latencies and present it as a single value. Which approach to take for measurements that cross domain boundaries may depend on the type of measurement and the technological capabilities of the different domains, these capabilities should be exposed by lower orchestration layers. If one domain is using e.g. MPLS tunnels to isolate customer traffic while the other is using VLANs the OAM tools are likely incompatible and separate monitoring sessions have to be instantiated together with appropriate MEASURE code in the aggregator that combines the results.

If recursive OLs are stacked on top of each other, the lower Orchestrator would receive its split part of the NF-FG with MEASURE annotation and the procedure described in these subsections would be repeated. In other scenarios, the CA sublayer is connected to an actual IL. In case the next layer is a UN the split NF-FG with the MEASURE code is sent directly. However, in the case that the underlying layer is not UN the NF-FG and MEASURE description has to be adapted and modified to what is supported by the IL, with the MEASURE part terminating on a Monitoring controller.

In the operational phase, the CA sublayer receives regular monitoring results from the MF deployed in the infrastructure, as well as asynchronous alarms and notification. Monitoring results can come from both infrastructure monitoring (i.e. physical links, actual CPU consumptions, etc.) as well as service monitoring (virtual compute and resource consumption) and can be merged to compound metrics based on the rules defined and stored during the splitting of the NF-FG and MEASURE at deploy-time. Infrastructure results could be received by the RO sublayer to updated resource and topology databases, supporting the orchestration algorithms. Service results would typically be forwarded upwards toward the service layer for logging and reporting purposes. Alarms and notification can either be directed to the RO sublayer and used as triggers for NF-FG optimization and redeployment mechanisms, or be forwarded upwards as notifications to Operators or Developers on top of the SL.

In the operational phase, like in the RO sublayer, monitoring capabilities have to be taken into account when multiple resources views coming from the different Infrastructure layers below are constructed into a single resource view by the Virtualizer. This could be problematic in the case above where the different domains support different technologies (e.g. MPLS or IP) for the same type of measurement metrics (e.g. delay or loss), depending on the type of abstraction desired in the resource view.

### 3.2.3 Infrastructure Layer SP-DevOps

The Monitoring controller (analogous to the MMP in the UN) in the Infrastructure layer is responsible for instantiating and/or configuring the MF control applications residing on an infrastructure domain controller (e.g. a dedicated monitoring controller, or co-located on an SDN network controller). The instantiation is based on the description provided by MEASURE, as well as further configuring the OPs. This may require that the Monitoring Controller interacts with the Network and Compute controllers to e.g. find the correct port or VM instance that should be monitored. While it functionally is a separate controller, it could be implemented as part of a Network or Compute controller, for example as a plugin.

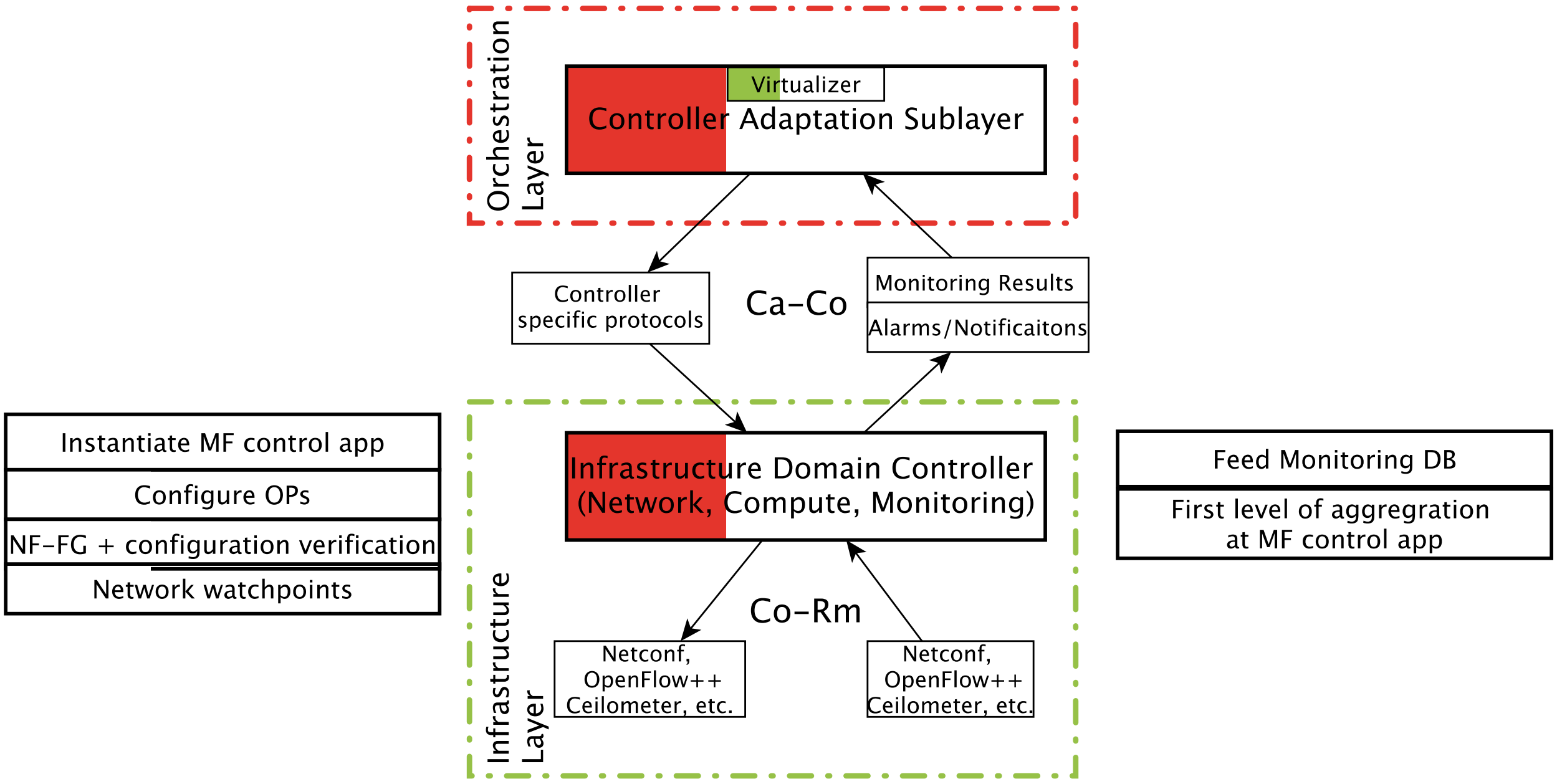

*Figure 3.7: SP-DevOps interactions with a Monitoring Controller.*

The domain controller acting as the Monitoring Controller obtains the description of the MFs that should be instantiated, and returns an identifier for the different MFs, used to tag the monitoring results going into the higher layer aggregation point (Figure 3.7). In case these identifiers cannot be generated in the Infrastructure layer the Controller adaptation has to supply them. The MF control application instantiated on top of the Monitoring Controller is also responsible for configuring its OPs for operation, e.g., which services/links/flows to measure, monitoring rates, thresholds, etc. Additionally, the OPs need to be configured to transmit their measurement results using the appropriate identifiers to the aggregation points, which may be in the IL (e.g. on an UN) or in the Monitoring Controller itself - depending on the capabilities of the technology of the particular infrastructure domain.

MFs can have various purposes. A prime task of MFs is infrastructure and service monitoring with respect to faults and performance metrics. For instance, in this deliverable we describe novel efficient and scalable network performance monitoring functions in Section 5. Another tasks that can be fulfilled by MFs is functional verification of service chains (i.e. service activation testing), for which the AutoTPG tools (Section 5.8) or the NF-FG run-time verification tool (Section 5.11) would be examples of.

Besides the deployment of monitoring functions in the actual infrastructure (which will be discussed on the example of a UN in the following section), domain controllers (e.g. a monitoring controller co-located with the network controller) are also hosting functionality for deploy-time verification of NF-FGs and network configurations. This NF-FG verification differs from the one in the upper-layers (i.e., OLs) as it considers both the NF-FG topology as well as the actual configuration of the Network Functions contained within the graph (see Section 5.10). Policy violations by specific OF rules can additionally be detected and mitigated by the Network Watchpoint tool (Section 5.12), which is a standalone debugging tool on the southbound interface of OF controllers, that helps to define network watchpoints for certain datapath or control traffic events.

## 3.3 SP-DevOps on Universal Nodes

Above, we described how MFs are deployed in the IL by instantiating an MF control app on a domain controller (i.e. Monitoring Controller). In D4.1 [D41] we introduced our idea of MFs that besides a control app consist of one or several OPs to increase the observability of network state and behavior.

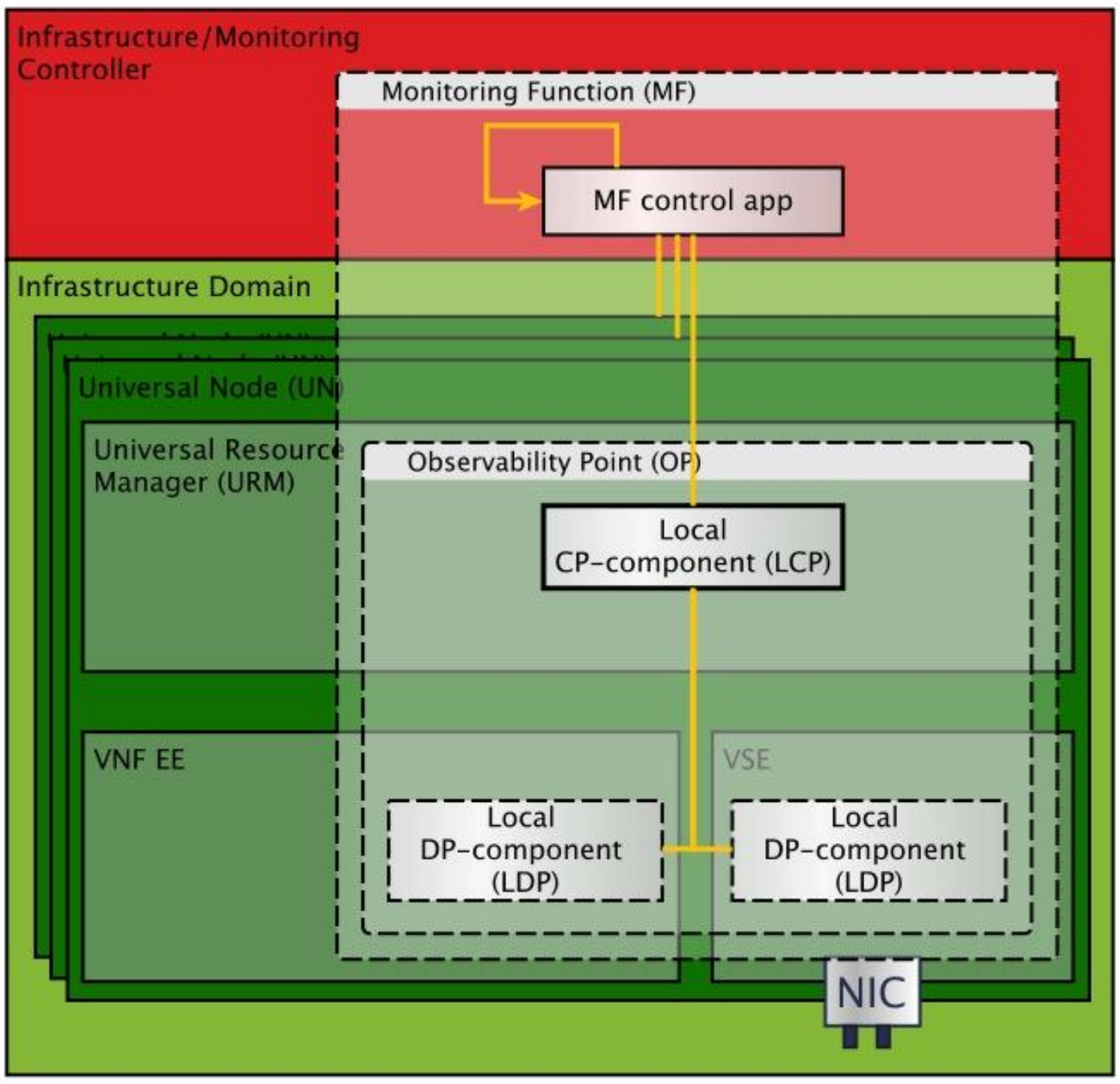

*Figure 3.8: An overview of the scalable software defined monitoring concept.*

The split of OPs into node-local control and data plane components (LCP and LDP, detailed in the following subsection) provides the flexibility to trade-off between scalability (i.e. monitoring overhead) and monitoring accuracy. In M4.1 [M41], we extended the concept by introducing the DoubleDecker as messaging system that allows MFs and their OPs to transfer data between them and make it available to the rest of the UNIFY architecture in an efficient and flexible manner. Furthermore, M4.1 provided a first idea of how MFs would be realized within the UNIFY architecture, and in the IL on UNs in particular. In the following subsections we will update the concept, which we now collectively label *Scalable Software Defined Monitoring*. Furthermore, we will follow up on the advancements in the definition of the UN in WP5 [D53] and further detail the mapping of our software defined monitoring components on UN architecture.

### 3.3.1 Overview of the Monitoring Function concept

Scalable Software Defined MFs typically implement functionality for collecting IT resource (e.g., CPU, memory, and storage) and network performance metrics (e.g., bandwidth, latency, jitter, and packet loss), but can also implement management task beyond pure monitoring, such as verification of configurations, etc. MFs may not only collect data, but also pre-process monitoring information (e.g., aggregate, filter, etc.) from other MFs across one or several UNs. An MF is implemented as one or several OPs and an MF control app. The following components form one MF (Figure 3.8):

- An MF control app is the software defined monitoring equivalent to SDN control apps, i.e., a logically centralized control application taking care of the configuration of one or multiple OPs and parts of the MF-related processing operations.
- An OP is a MF component that runs locally on the UNs. In general, the implementation of OP capabilities encompasses measurement or verification mechanisms, node-local aggregation and analytics, depending on the type of the MF (for instance, node-local operations are not necessarily performed by all types of MFs). An OP operates in terms of the LCP and LDP, and is managed by an MF control app.
    - An LCP is splitting certain control functionality from the MF control app for scalability and resource consumption purposes. It reflects essentially a local monitoring controller which provides functions for retrieving data from the LDP; processes obtained data; and controls the monitoring behavior (e.g., measurement intensity).
    - An LDP is basically any kind of data source in the UN (e.g., statistics from logical switches, resource metrics from the VNF executing environment, hardware counters, meters, injected packets, log data, etc.), retrievable from the virtualized environments (VNF execution environment (EE) or virtual switches environment (VSE)), the UN operating system, or specific monitoring VNFs deployed on the node as part of service chains/NF-FGs.
- A monitoring channel enables information exchange between LCPs within one UN and between LCPs and their MF control apps. In UNIFY, we propose the DoubleDecker messaging system for this purpose. The messaging system is composed of lightweight generic nodes (Brokers) able to forward any message following DoubleDecker formatting, and route them through the messaging system. Brokers have an arbitrary number of clients attached to them, e.g., LCPs in the UN, MF control apps, databases or any other module in higher orchestration layers. The clients only need to have a minimal knowledge of the actual architecture, but need to follow a standardized protocol to communicate with the messaging system and have to define their own sub-protocol to communicate with other clients. More details about the technology choices behind the DoubleDecker messaging system developed in UNIFY can be found in Section 5.1.

### 3.3.2 Overview of the Universal Node architecture

In parallel with WP4 detailing the scalable software defined monitoring concept, WP5 was finalizing its proposal on the architecture for the UN, shown in Figure 3.9.

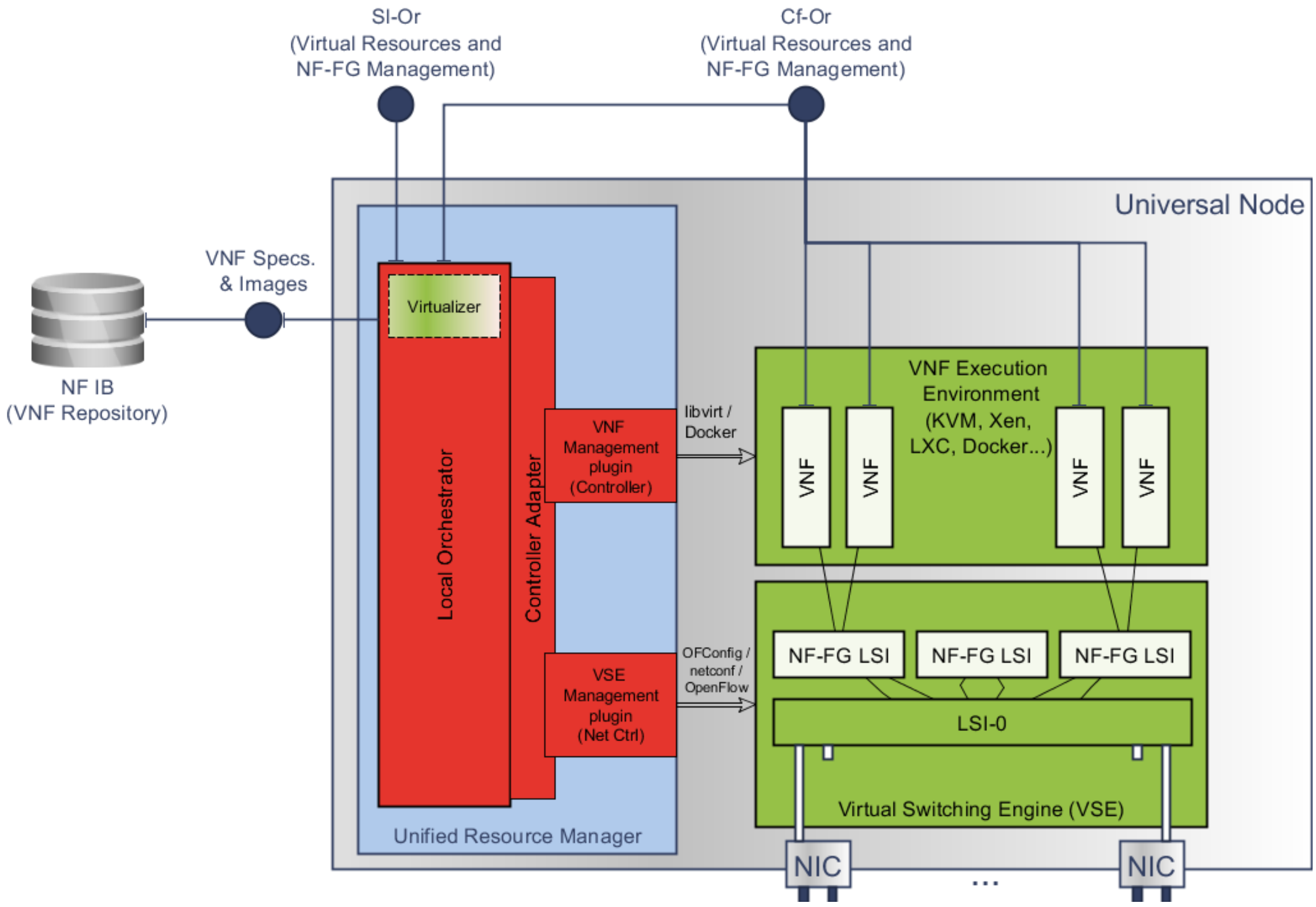

*Figure 3.9: UNIFY Universal Node Architecture.*

The UN architecture follows the strictly hierarchical model targeted by the UNIFY architecture (see [D22] and Section 3.1 in this deliverable). The goal of the hierarchical model is to allow recursive stacking of orchestration layers across domains, technologies, and vendors. Recursions in turn require the northbound and southbound interface of each orchestration layer to use the same semantics in order to allow the orchestrator of level n to control all the orchestrators at level (n-1) for every N (i.e. an arbitrary number of hierarchical levels). For an UN, which is providing the Infrastructure layer and is thus assumed to be the lowest component of the orchestration chain, this architectural principles translates into two consequences:

- First, this means that every UN needs to implement the Sl-Or reference point as a northbound interface (in Figure 3.9 called NF-FG management interface). This interface accepts resource requests in the form of NF-FGs, i.e. a graph that describes the service that has to be deployed in terms of network functions and interconnections between them. Additional information annotated in an NF-FG include QoS-related parameters as well as monitoring primitives that need to be instantiated for that graph, e.g. in the form of MEASURE (see Section 5.2 for more details).
- Second, this requires every UN to implement a complete UNIFY orchestration layer with the scope of this UN. To accommodate this requirement, the Universal Resource Management (URM) plays the role of a local orchestrator that has complete and detailed view on the resources available on the node, their topology and related usage constraints and limitations. It provides the main UN interfaces and controls the actual execution environments of the infrastructure (i.e. the VNF EE and VSE) to fulfil the NF-FG deployment and management requests. Control adapters guarantee the independence of the orchestrator logic (e.g., optimized network embedding of the NF-FG) from the VNF EE and VSE actually being used, which are controlled by the proper plug-ins. The VNF and VSE Management plugins are thus acting as compute- and network controller respectively. Particularly, the VSE management plugin can include OpenFlow-based network controllers.

Besides the NF-FG management interface (one part of the Sl-Or interface) and the URM (including a local orchestrator) mentioned above, the UN shown in Figure 3.9 also contains:

- The Resource management interface (part of the Sl-Or interface), which covers the discovery of resources exposed by the node as well as possible updates to those resources (such an update may be the result of actions performed through the other interfaces, or from a reconfiguration of the node) and also reporting to the upper layers the current availability of resources due to the NF-FGs already deployed in the UN.
- The Cf-Or interface allows the Orchestration layer to provide resource control functions directly to NFs deployed within the UN. For instance, the Cf-Or will take over the communication of scaling requests toward the Resource Orchestration layer.
- VNF Template and Images repository interface: When the UN is instructed to deploy a NF-FG, it needs to fetch the detailed specification and the related binaries of the involved VNFs. This constitutes an outbound interface of the UN towards a central VNF repository.
- The VNF Execution Environment (VNF EE) consists of one or more compute platform virtualization solutions, including hypervisors or simpler container based approaches.
- The Virtual Switching Engine (VSE) implements packet switching on the UN, managing the physical network interfaces (NICs) and the inbound and outbound traffic steering of the deployed NF-FGs, as well as the internal traffic steering between the different NFs deployed as part of the NF-FG. Each NF-FG encompasses a distinct logical switch (Logical Switch Instance, or LSI) that provides the traffic steering between the components of the graph itself, while the traffic between the network and each single NF-FG (and between different NF-FGs) is handled by the base switch, called LSI-0. The creation of different logical LSIs makes easier to control the traffic steering as well as obtaining the statistics (e.g., traffic counters) associated to each NF-FG.

### 3.3.3 SP-DevOps updates to the Universal Node

Given the additional requirements originating from the WP4 monitoring function concept, the architecture of the UN has been extended by a Monitoring Management Plugin and the DoubleDecker messaging system, as described in Figure 3.10.

**Monitoring Management Plugin**: The main extension of the UN for monitoring purposes is the MMP, which operates as a monitoring controller on the same level as the VNF and VSE Management plugins. When the local orchestrator receives an NF-FG via the NF-FG management interface (part of Sl-Or), it also receives monitoring requests expressed in MEASURE. These requests are then translated into concrete OPs, and the corresponding LCPs are instantiated within the MMP. The MMP provides the local orchestrator with a unified and platform-independent interface to the different components that are involved in the monitoring process, hence enabling the orchestrator to interact with all the LCPs through a uniform interface. In fact, different information (e.g., counters) need to be extracted through different commands/interfaces, which requires the definition of multiple LCPs, each one hiding the platform-specific details for getting access to that specific data (Figure 3.10). The data sources for this data resemble the LDPs defined as part of an OP. Some examples of LCPs and corresponding LDPs are:

- An LCPs can be defined to read the CPU load in */proc/stat*
- An LCP can obtain CPU and memory statistics using the *ps* command
- An LCP can reads basic traffic statistics from a OpenvSwitch softswitch through *ovs-ofctl dump-flows*
- An LCP can retrieve resource usage and performance data from a Docker realization of the VNF EE via *CAdvisor*; .
- An LCP can retrieve OpenFlow port- and flow statistics from logical switch instances (LSI) by either requesting the statistics from the VSE management plugin (which acts as a network controller) or by acting as a secondary OpenFlow controller using a separate OpenFlow channel to the LSI.

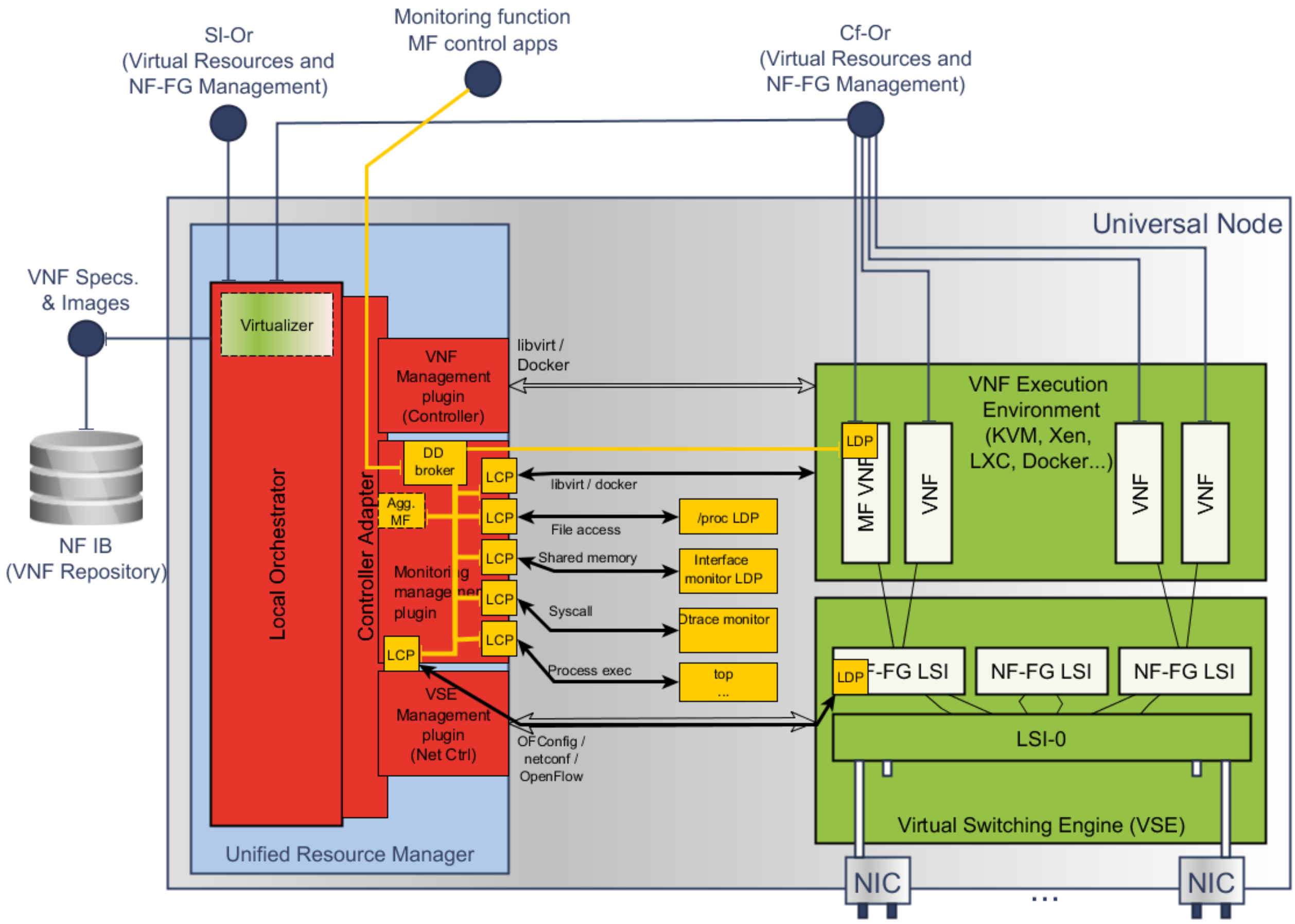

*Figure 3.10: UN architecture with extensions for software defined monitoring.*

From these examples it can be seen that the implementation of the LCP depends on the operating system or the specific component in use. Although LCPs are used mainly to read information, LCPs can also pre-process the obtained data to offload the central MF control app (e.g. with data aggregation or filtering steps). Furthermore, LCP can control certain LDPs monitoring behavior and thus also configure them. To give an example, an LCP can reconfigure a given OpenFlow switch by replacing the single flow entry by multiple entries, providing the same forwarding behavior but allowing statistics on different sub-flows of the original flow definition (e.g. different counters for TCP and UDP).

**DoubleDecker:** In addition to LCPs, the Monitoring Management Plugin also contains a function for message aggregation and filtering (e.g. a MEASURE aggregation point) and a DoubleDecker messaging system Broker. The Broker terminates the intra node messaging system that provides the LCPs with an efficient communication channel towards their MF control apps (the DoubleDecker messaging system is depicted in yellow in Figure 3.10). Since the MF control apps in turn can notify modules in the Orchestration layer with monitoring results of their OPs, the Cf-Or interface for scaling request originating from MFs is provided by the DoubleDecker messaging system. The DoubleDecker broker also enables communication with observability points that are implemented as VNFs, which may be needed when advanced counters or non-standard functions are needed, such as application-layer counters, active monitoring probes, or deep packet inspections.

## 3.4 Programmability Support and Information Dissemination

The SP-DevOps concept encompasses several components supporting programmability and information dissemination across the UNIFY architecture and between SP-DevOps functions and tools, including

- support functions for configuration of SP-DevOps tools and dissemination of monitoring information and reports to existing databases and other receivers in the architecture;
- a standardized data-model for common interfacing of monitoring configuration and information between the architecture and monitoring tools, including support for legacy OAM tools;
- distributed information retrieval via a query system capable of gathering data within the scope of a certain service and user role.

The components cover all the layers of the functional architecture, allowing for information being disseminated and exchanged both vertically and horizontally between resource management functions, controllers, and monitoring entities of the infrastructure, supporting programmability and management of software-defined infrastructures.

### 3.4.1 Communication and configuration for programmability and automated workflows

To be able to programmatically start, control, configure, and retrieve information from Monitoring Functions a communication channel is necessary. This channel should allow MFs to communicate not only with pre-configured components in the architecture (such as databases) but allow flexible communication patterns with components that may be started dynamically, for example other MFs. The DoubleDecker messaging system (Section 3.1.1 and 5.1) provides scalable and easy to use communication between MFs and other components.

While primarily intended to run in the lower layers of the architecture, connecting monitoring components in the IL, it can be seen as a transport middleware, providing message transport wherever needed. This could for example be to connect aggregation points in multiple OLs, VNF Control Applications, or other components such as automated troubleshooting applications that needs to control various SP-DevOps tools.

DoubleDecker consists of two components, brokers and clients. Brokers are responsible for routing messages between the clients, and can only be connected to each other in a hierarchical, tree-like, fashion, in order to keep the routing mechanism simple. Messages between clients connected to different brokers will take the shortest path through the tree structure, off-loading more central brokers when possible. A client connects to the closest broker (e.g. a broker running on the same UN as the client) using a simple protocol and one of the supported transport protocols, and registers itself with a unique name. Using the unique names, messages are routed between clients. The use of names allows a split between address and location, which simplifies for example migration of components as well as initial configuration of the control network as e.g. locally unique IP addresses can be used within the UN. Technical details about implementation and usage are found in Section 5.1.

The MEASURE configuration and aggregation language (Section 3.1.1 and 5.2) provides a higher level programming interface to MFs and a way to define where and how monitoring should be configured, how the resulting measurements should be aggregated, and where triggers should be sent. This allows for automated workflows to be implemented using for example a combination of MEASURE and a simple scripting language where a script may be waiting for certain monitoring triggers to start a troubleshooting process.

In the typical case measurements results are sent to a local aggregation client via the DoubleDecker messaging system. The aggregation client performs the first step of data aggregation by executing the MEASURE code provided during the NF-FG deployment, which could for example combine multiple measurements into single metric or propagate results to other components when a threshold is breached. Multiple aggregation points for a metric may have been configured during the deployment in order to merge metrics from different domains or to "recompose" metrics when NFs have been decomposed, this means that measurement results may flow upwards through the architecture either directly to the SL or via these merging and/or re-composition points in different orchestration layers. Keeping the amount of messages going to higher layers low is a key scalability goal. By defining MEASURE rules that are triggered on state changes, i.e. that only send data when breaching thresholds, the amount of data propagated to higher layers should be lowered by orders of magnitude compared to sending results directly to a centralized database, assuming relatively stable results. This process is described with more technical details in Section 5.2.

### 3.4.2 Standardized data-model supporting programmability

The environment in which the UNIFY SP-DevOps concept will be deployed may not be a fully programmable infrastructure from Day 1. At least for certain domains we could anticipate a period of migration to the new modus operandi until the majority and most critical parts of the infrastructure can support the new operations. That said, a certain part of the infrastructure could take a very long time until it is replaced with fully SP-DevOps compatible elements. Although Task 2.4 will look into the strategic analysis for Telcos, it is reasonable to expect that intermediate standards will be advanced to allow for more programmability within specific component and protocol scopes. An excellent example in this case is traffic measurements, which are extensively used today to determine SLA adherence as well debug and troubleshoot pain points in service delivery. For these reasons, we propose a model for common interfacing of monitoring configuration and information between the architecture and monitoring tools, including support for legacy OAM tools. This will enable supporting a higher degree of programmability as well as information exchange to and from the architecture.

In IP networks, the Two-way Active Measurement Protocol, TWAMP, is both widely implemented by all established vendors and deployed by most global operators. However, TWAMP management and control today relies solely on diverse and proprietary tools provided by the respective vendors of the equipment. For large, virtualized, and dynamically instantiated infrastructures where network functions are placed according to orchestration algorithms as discussed in [RSZ15] proprietary mechanisms for managing TWAMP measurements have severe limitations. For example, today's TWAMP implementations are managed by vendor-specific, typically command-line interfaces (CLI), which can be scripted on a platform-by-platform basis. As a result, although the control and test measurement protocols are standardized, their respective management is not. This hinders dramatically the possibility to integrate such deployed functionality in the UNIFY SP-DevOps concept.

A key part that is missing in this case is a standardized data model which would permit the integration of such measurement components in deployed equipment with the rest of the WP4 work, especially in the likelihood of the aforementioned migration phase. A way forward therefore is standardization action, in this case at the IETF, where a TWAMP data model can be defined and published as a Standards Track RFC.

In short, for the case of software-defined and virtualized nature of network infrastructures where we apply the SP-DevOps concept based on dynamic service chains [WJO13] and programmable control and management planes [EHA15] we identify the need for a well-defined data models for legacy but widely deployed implementations. Specifically, within the IPPM WG, UNIFY is contributing to the development of a TWAMP data model and the associated YANG module that will enable full, cross-vendor programmability of TWAMP implementations that are in use and therefore can be included in the UNIFY SP-DevOps. This work is a joint collaboration across partners from WP4 and WP6. Interested readers are referred to the latest version of [RCI15] for further details on the YANG module.

### 3.4.3 Retrieving distributed information – a recursive monitoring query language

In NFV environments, operators or developers sometimes need query the performance of virtualized Network Functions (VNF), for example, the delay between two VNFs. In existing systems, this is usually done by mapping the performance metrics of VNFs to primitive physical network functions or elements, statically and manually when the virtualized service is deployed. However, in UNIFY a multi-layer hierarchical architecture is adopted, and the VNF and associated resources, expressed NF-FGs, may be composed recursively in different layers of the architecture. This will put greater challenge on performance queries for a specific service, as the mapping of performance metrics from the service layer (highest layer) to the infrastructure (lowest layer) is more complex when compared to cloud infrastructure with single layer orchestration. We argue that it is important to have an automatic and dynamic way for decomposition of the performance query in a recursive way, following the different abstraction levels expressed in the NF-NFs at hierarchical architecture layers. Hence, we propose to use a declarative language such as Datalog to perform recursive queries based on input in form of the resource graph depicted as NF-FG. By reusing the NF-FG models and monitoring database already existing in the UNIFY architecture, the language can hide the complexity of the multilayer network architecture with limited extra resources and efforts. However, the expression of performance queries is significantly simplified for any potential receivers (i.e. operator and developer roles).

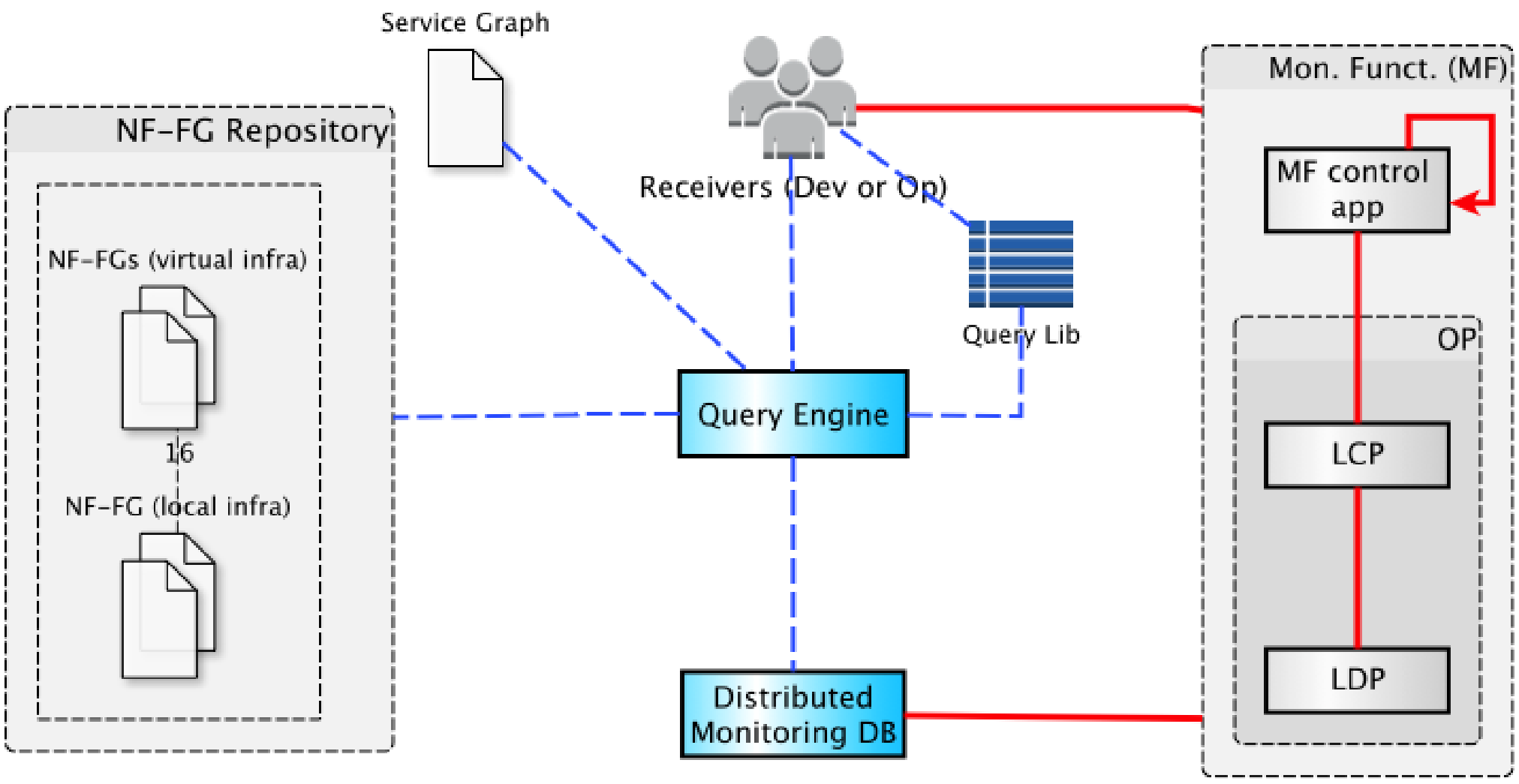

*Figure 3.11: Recursive Monitoring Query.*

As shown in Figure 3.11, the Query Engine is responsible to receive query scripts written in a variant of declarative language (i.e., Datalog) from receivers (e.g. VNF and service developers or operators) and automatically translate these query requests into corresponding queries on the distributed monitoring database in the infrastructure control layer. The translation of the database queries is based on resource graphs described in NF-FG repository, which can contain multiple nested NF-FGs for different orchestration layers. Based on the instruction of the query scripts, the Query Engine will parse the NF-FGs and decompose the performance metrics in a recursive way until some primitive infrastructure metrics available in the monitoring DB (such as CPU of VM/PM, delay between two physical network elements) are identified or terminated according to instruction. Then the query engine will query the distributed Monitoring DB for these decomposed performance metrics and perform aggregation according to query scripts. Further details of the language are described in Section 5.3 and in the Annex 3.

The monitoring DB contains the measurement results collected by monitoring functions via the DoubleDecker messaging system, as described in Section 3.4.1 and 5.1. The OpenTSDB, a distributed time-series database, is an example of a DB used for storing the monitoring data (e.g. in [FNE15]. The query engine can use the provided HTTP API of OpenTSDB to query it and re-use the schemas defined in monitoring functions. The Query Library is used to store some pre-defined query templates or libraries that are developed by the receivers or predefined by the service provider. The templates or libraries can then be called by receivers to simplify the query request or create more complex query requests. Some example templates for monitoring functions will be described in Section 5.3.

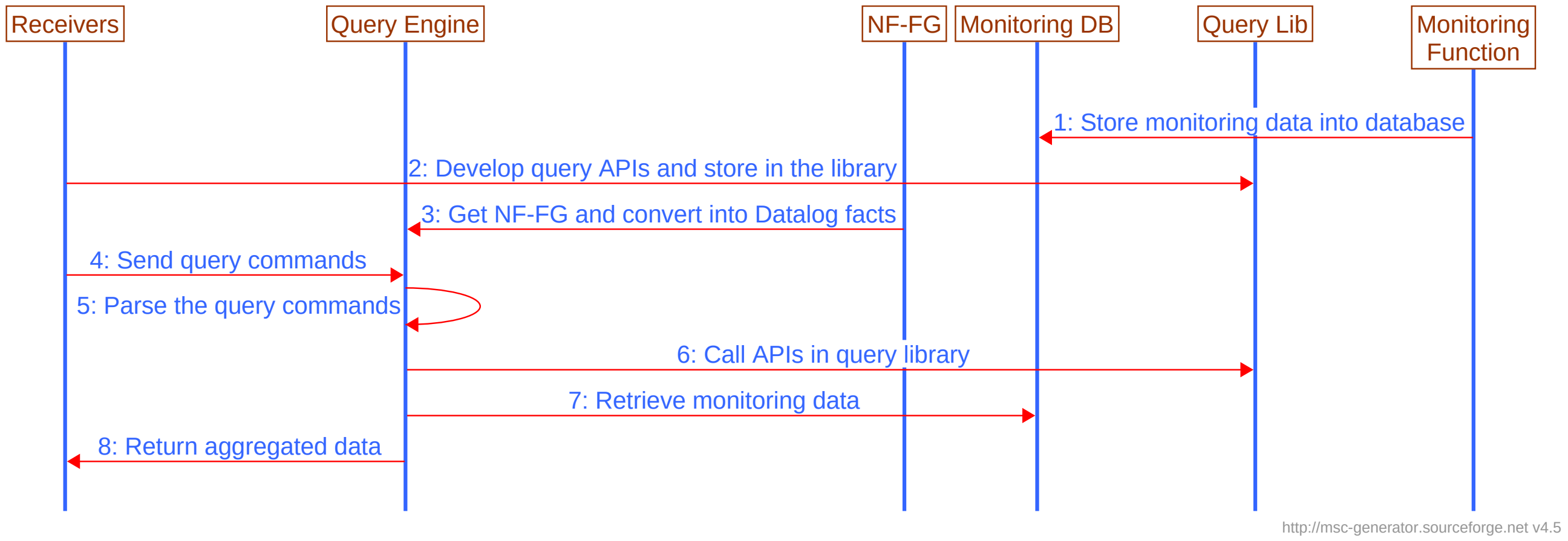

*Figure 3.12: The flow of the query procedure.*

Below is a sequence flow of the query procedure and is illustrated in Figure 3.12:

1. The Monitoring Functions collect the data from the infrastructure and store them into the monitoring database;
2. The receivers develop the query template or APIs and store into the query library;
3. The query engine gets the NF-FG and converts it into Datalog based facts;
4. The receivers send query commands to the query engine;
5. The query engine parses the query commands;
6. The query engine calls corresponding APIs in the query library if needed;
7. The query engine generates primitive query and retrieve corresponding monitoring data from the monitoring database;
8. The aggregated monitoring data is returned to the receivers by the query engine.

Typically, the recursive query language can be used to request end-to-end delay from NF1 to NF2 according to a Service Graph as in e.g. Figure 3.13, using Datalog-based query scripts. The detailed explanation of the Datalog rules is described in Section 5.3. After running such a Datalog program, the delay between NF1 and NF2 in service layer is mapped recursively to the delay between vm7 and vm10 in infrastructure layer. The Datalog rules can be stored into the Query Library as an API so that the receivers could just send a simple request, e.g., e2e_delay(NF1, NF2) to the Query Engine in order to get the end-to-end delay between NF1 and NF2.

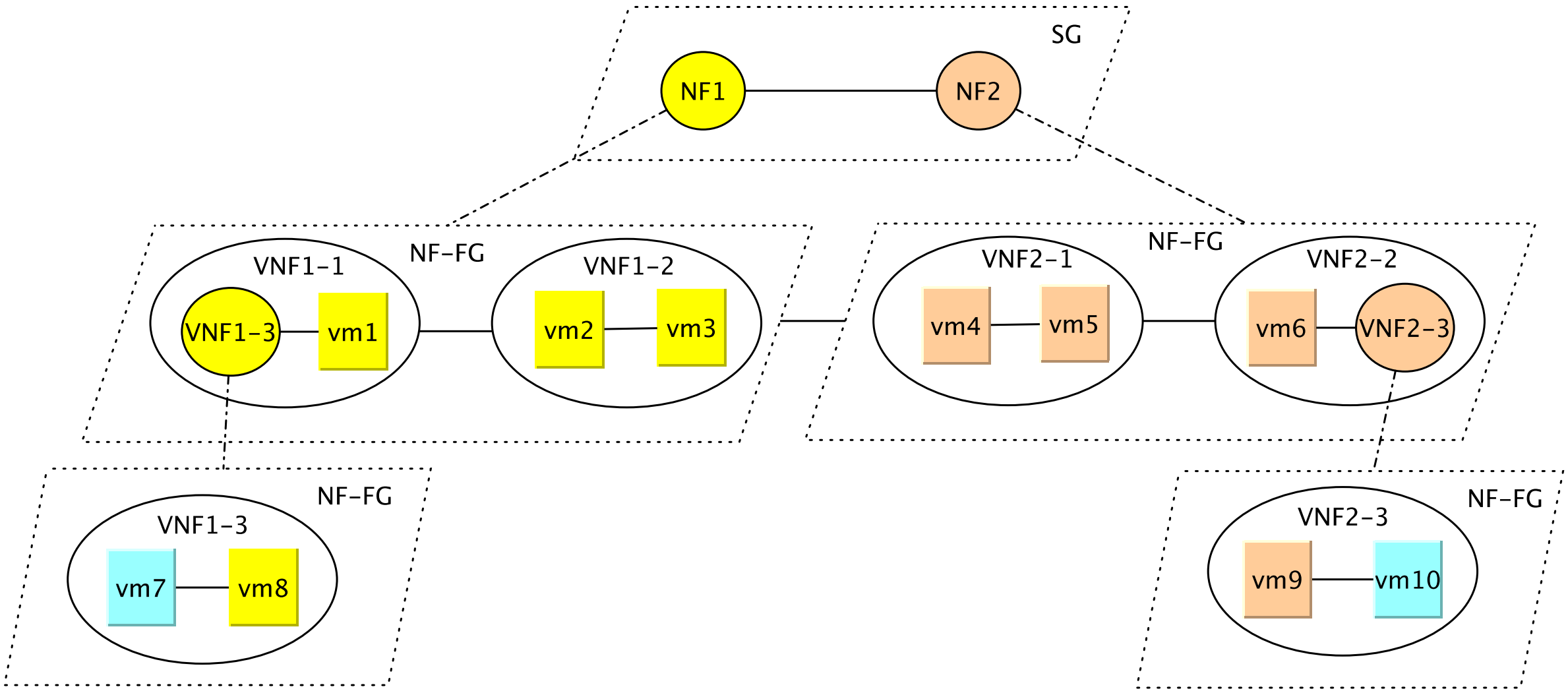

*Figure 3.13: A sample Service Graph and NF-FG*

This example in Figure 3.13 shows how the proposed query engine allows automatic mapping of detailed performance metrics from the infrastructure to abstracted, higher level views by following the abstraction presented by NF-NFs at various intermediate architecture layers. This provides receivers (i.e. users of the query engine) with a very flexible way of receiving monitoring data without the need to know details beyond the level of abstraction present in their layer of operation.

The query engine can support many SP-DevOps processes, most notably observability and troubleshooting tasks relevant for both operators and developer roles, e.g. for high-level troubleshooting where various information from different sources need to be retrieved. Additionally, the query engine might be used by specific modules located in the control and orchestration layers, e.g. a module realizing infrastructure embedding of NF-FGs might query monitoring data for an up-to-date picture of current resource usage. Also scaling modules of specific network functions might take advantage of the flexible querying engine for on-demand "pulling" of monitoring information (e.g. related to resource usage, traffic trends, etc.), as complement to notification that "passively" push this information based on pre-defined thresholds.

## 3.5 Measuring Performance Using the SP-DevOps Metrics

For the purpose of expressing, measuring and evaluating different aspects of network and service performances, as well as different development and operations processes, we have defined a set of key performance metrics that are specific to the SP-DevOps concept (Annex 1, Table A1.1-A1.3). In particular, we identify the key performance metrics of highest importance to the key stakeholders, namely the Operator, VNF developer and Service Developer, as defined in [D41]. Some of these metrics are already employed in practice by some of these roles, such as performance metrics employed by the Operator to evaluate the quality-of-service (QoS) of a specific service or application. The metrcis proposed are aimed to meet SP-DevOps specific needs for each one of the defined roles, meaning that the metrics:

- should provide support for evaluating the efficiency of the typical actions of a VNF Developer, Service Developer, or Operator, carried out in a SDN/NFV environment,
- are selected to be useful indicators of the efficiency and performance of software-defined infrastructures and virtual components,
- can be used to indicate and measure management efficiency, e.g., time and costs, at various levels relevant to a VNF Developer, Service Developer, or Operator.

In the process of selecting relevant SP-DevOps metrics, we have qualitatively estimated the importance of each metric from the viewpoint of a given role, while we take into account the expected impact on service creation, deployment, provision and maintenance In addition, concerning the evaluation of the SP-DevOps tools, we estimate the importance, relevance and easiness of measurement for each of these metrics based SP-DevOps processes and tools developed within WP4. Note that a subset of SP-DevOps metrics will be used for evaluating different aspects of the UNIFY prototype.

#### 3.4.1.1 Categories of SP-DevOps metrics

We identify three types of metrics that are particularly relevant for the SP-DevOps framework: 1) technical metrics directly related to the service provided, 2) process-related metrics concerning the deployment, maintenance and troubleshooting of the service, i.e. concerning the operation of VNFs, and 3) cost-related metrics associated to the benefits and costs from using a software-defined telecom infrastructure.

First, technical performance metrics shall be service-dependent/-oriented and will address service performance in terms of *latency*, *throughput*, *congestion*, *energy consumption*, *availability*, etc. Acceptable performance levels should be mapped to SLAs, which are appropriately designed to address the requirements of the service users. Metrics in this category have been defined in standardization documents such IETF RFCs, e.g. [JFA14], [GAL14], [JMA99], and [GF011], addressing particular services, protocols and types of measurements.

Second, process-related metrics shall serve a wider perspective in the sense that they shall be applicable for multiple types of services. These metrics include: *number of probes for end-to-end QoS monitoring, number of on-site interventions, number of unused alarms, number of configuration mistakes, incident/trouble delay resolution, and delay between service order and deliver..*

Third, cost-related metrics shall be used to monitor and assess the benefit of employing software-defined telecom infrastructure compared to the usage of legacy hardware infrastructure with respect to operational costs, e.g. *possible man-hours reductions, elimination of deployment and configuration mistakes*, etc.

Identifying metrics for SP-DevOps and measuring the performance in these identified metrics are highly challenging tasks because of the amount and availability of data sources, e.g., calculation of human intervention, or secret aspects of costs. Below, we define and briefly discuss such a list of key metrics for the three roles: Operator, VNF Developer and Service Developer - the complete lists can be found in Annex 1, Table A1.1-A1.3.

#### 3.4.1.2 SP-DevOps specific metrics relevant to role and prototype evaluation

**Operator**: In the context of UNIFY, we consider the technical metric "reaction time to increasing/decreasing load" (see Annex 1) or the process-related metric "trouble reaction time" as highly important metrics both for the Operator role in general and the evaluation of the UNIFY prototype. Such metrics consist of the combination of other simpler metrics such as measurement of traffic/load on specific links or network nodes, time to react to potential events, etc., and may thus be more complex in some cases. These are considered highly important, due to the fact that they enable the service developer or operator to measure the (expectedly) significantly faster reaction and restore times in case of outages, e.g. so as to design suitable SLAs for their customers. Moreover, they also indicate costs in terms of e.g., SLA violation for Operators in an actual production environment, practically constituting elements for the quantification of other cost-related metrics such as "cost of outages". Finally, only for UNIFY purposes, they will allow us to demonstrate the significant improvement in reaction time against legacy techniques.

**VNF Developer:** We assess the "time to fix", which could be decomposed into smaller constituents, i.e., "time to discover bug" and "time to deploy solution", as highly important for the VNF Developer role, due to the fact that it may imply service outage for a specific time duration and thus, cost in terms of SLA violation. Moreover, the "time to fix" metric may contribute to more complex metrics associated with the VNF Developer role, e.g., the so-called "length of compile-deploy-debug-modify cycle". This metric is considered to be different than the already-known software deployment cycle, and may have significant impact on the service management, update and evolution.

For instance, such a metric could be very helpful in cases where a novel service development and maintenance procedure is followed, such as the "error budget" approach introduced in [BTR14] by Google's Service Reliability Engineering Group Head. In such a scenario, the longer the "time to fix", the lower the "error budget" for a specific VNF Developer, thus the VNF Developer has clear incentive to assure that the VNF code is reliable and stable before launching in a production network.

**Service Developer:** From Service Developer's point of view, the "ratio of the number of automated operations to the number of total operations" to develop, update, maintain, monitor and troubleshoot a service is significantly important, as it translates to operating cost by means of person hours, loss of time due to transportations or even cost due to violation of SLA requirements. The aforementioned metrics are associated to cost either indirectly by means of performance, or directly by means of person hours, penalties for SLAs, etc. depending on the stakeholder/role they refer to. Moreover they are meaningful in the specific DevOps concept, where mechanisms that fall under the SP-DevOps framework and vision are expected to demonstrate significant and noticeable improvements of such metrics.

## 3.6 Concluding Remarks

In this section the updated SP-DevOps concept was presented in terms of the architectural integration of support functions and tools into the functional architecture and the UN. Moreover, we have described how the SP-DevOps processes and functions support the ideas of UNIFY programmability and orchestration, by outlining how SP-DevOps functions and tools facilitate deployment and operations of service graphs. For evaluating service performance and management processes, a set of SP-DevOps relevant metrics have been identified with respect to the needs of an Operator, Service Provider and VNF developer. In the following, Section 4 and 5 will present the SP-DevOps technical contributions in more detail, including how each function or tool integrates with the UNIFY architecture. A case study in Section 6 provides examples of how SP-DevOps components support deployment and scaling processes for the Elastic Firewall scenario [D21b]. Relevant steps and discussion points for a stronger integration of the SP-DevOps concept with respect to orchestration and operations in UNs are outlined in further detail in Section 8.

# 4 The SP-DevOps Toolkit

In this section we provide a short summary of SP-DevOps support functions and tools developed in WP4. Detailed descriptions of each item and the architectural mapping can be found in Section 5. Further, we refine the SP-DevOps Toolkit introduced in M4.1 and list the subset of tools and functions which will be included and are aimed to be released as open source. Making parts of the SP-DevOps tools publically available is, together with publication efforts, an important channel to disseminate the results of WP4 and concretely demonstrate several aspects of the SP-DevOps concept as defined in the UNIFY project.

## 4.1 Summary of SP-DevOps Support Functions and Tools

A number of methods implementing various aspects of the SP-DevOps observability, troubleshooting, verification and VNF development processes have been developed within WP4. Due to the varying nature of developed approaches, we categorize them in terms of support functions and tools. Support functions can here be viewed as generic mechanisms, capabilities or tools that support programmability and provide means for connecting functional components across one or several layers of the UNIFY architecture. The set of SP-DevOps tools encompasses monitoring functions, troubleshooting methods, verification mechanisms, and VNF development tools, embedded into the UNIFY architecture with help of  several generic support functions. Table 4.1 summarizes all SP-DevOps tools and support functions developed in the scope of WP4, and shows how they relate to the SP-DevOps processes.

The next subsections contain  short summaries of the SP-DevOps tools ordered in line with the main process supported. The troubleshooting process will be supported by several observability, verification and VNF-development tools. In order to avoid overlap, SP-DevOps tools covering the troubleshooting process are therefore not reported in a separate  troubleshooting category. Note, that automated troubleshooting methods are part of the ongoing UNIFY Task 4.3 and are a main focus area for the remainder of the project and will be discussed in future WP4 deliverables.

### 4.1.1 Support functions

Support Functions facilitate the integration of other tools. They provide infrastructure for exchanging data, a common way of describing parameters that need to be measured and an abstract way of accessing from the Service Layer information collected by tools monitoring a particular metric.

**Support Function 1**: DoubleDecker Messaging system

DoubleDecker provides scalable communication services for Monitoring Functions. It is a lightweight implementation based on  ZeroMQ [ZeroMQ].  It allows exploiting Inter-Process Communication (IPC) low-latency high throughput communications within a host while also supporting TCP-based or reliable multicast-based transport in a heterogeneous software environment built on virtual machines, containers and even plugins. DoubleDecker extends ZeroMQ by implementing a client and broker hierarchy together with a simple message routing algorithm that supports the capability of sending opaque data between virtual network and monitoring functions, as well as to higher layer orchestration modules.

| Tool/Function name | Section | Observability | Verification | VNF Development | Troubleshooting |
|---|---|---|---|---|---|
| Support functions | | | | | |
| DoubleDecker | 5.1 | x | x | x | x |
| MEASURE | 5.2 | x | | | x |
| Rec. Query Lang. | 5.3 | x | x | x | x |
| TWAMP data-model | 5.4 | x | x | x | x |
| SP-DevOps Tools | | | | | |
| Rate monitoring | 5.5 | x | | x | x |
| E2E link mon. | 5.6 | x | | x | x |
| EPLE | 5.7 | x | | x | x |
| AutoTPG | 5.8 | x | x | | x |
| IPTV Quality monitor | 5.9 | x | | | |
| Deploy-time verification | 5.10 | | x | | |
| Run-time verification | 5.11 | x | x | | x |
| Network watchpoint | 5.12 | | x | x | x |
| EPOXIDE | 5.13 | | | x | x |

*Table 4.1: Summary of the SP-DevOps tools and support functions in relation to SP-DevOps processes.*

**Support Function 2**: MEASURE description language

Additional to the definition of NF-FGs itself, we need a way to express and annotate monitoring intents and requirements to the NF-NG. In collaboration with WP3, we are developing the MEASURE language that describes MFs and OPs. MEASURE provides machine-readable configurations of monitoring functions, including thresholds for various monitoring conditions, as well as instructions for aggregating and disseminating monitoring data.

**Support Function 3**: Recursive query language

This tool allows accessing low level monitoring information from the infrastructure from descriptions situated at higher levels of abstraction aligned with the UNIFY architecture. Since UNIFY resources are represented in a hierarchical and recursive structure, we also need to present monitoring results (fault, performance, etc) at higher abstraction levels. The query language, inspired from Datalog and based on a Python implementation, will support querying a distributed DB located in the infrastructure domains (UNs, controllers) and representing the results at an abstracted level following the local SG or NF-FG definition.

**Support Function 4**: Standardized TWAMP data-model

As the adoption of programmable nodes (UNs, OF switches, etc.) will increase over time, the transition from current infrastructures to software-defined infrastructures and networking requires interfaces capable of bridging the gap between legacy equipment and programmable network entities. In this respect, most legacy devices (router, switches, etc.) have some OAM capabilities like BFD, LSP Ping, TWAMP, etc., that are usually configured rather statically via proprietary interfaces, which limits flexible and dynamic instantiation across diverse infrastructures. The TWAMP data model is an instance of a standardized data-model supporting increased programmability for legacy OAM tools .

### 4.1.2 Observability tools and troubleshooting support

The observability tools provide enhanced monitoring abilities and troubleshooting support in software-defined infrastructures. A major focus is on scalable observability: reducing the amount of data transmitted towards a management system or controller while maintaining good accuracy of the metric estimates. The SP-DevOps observability tools also offer enhanced programmability capabilities for integration with orchestration and controllers in software-defined infrastructure.

**Observability Tool 1**: Rate Monitoring

This tool determines, locally on the node instead of in a centralized setting, a high-quality estimate of the link utilization. The method employed is a statistical analysis algorithm that makes use of two counters for storing the first and second statistical moments of each monitored metric, as compared to the current practice of using a single counter for each of the byte and packet rates. The monitoring function is implemented in terms of an observability point operating locally in the UN in the infrastructure layer.

**Observability Tool 2**: E2E link delay and loss

This tool determines, locally at the node instead of in a centralized setting, a high-quality estimate of the delay and loss on a link. A first approach to end-to-end link monitoring of delay and loss has been evaluated. It will be extended with a mechanism for automatically controlling the measurement behavior relative to specified precision requirements.

**Observability Tool 3**: Efficient Packet Loss Estimate (EPLE)

EPLE is a monitoring function that estimates packet loss for aggregated flows in OpenFlow switches in a manner that is efficient from both dataplane (no probe packets) and control plane (piggybacked on OpenFlow messages) perspectives.

**Observability Tool 4**: AutoTPG

The existing automatic test packet generation (ATPG) tool is able to verify (or rather test) a productive network for error conditions (e.g., incorrect firewall rules, software, hardware, and performance errors). However, ATPG is not able to verify the matching header part of the forwarding entries for software or hardware failures. In AutoTPG, the existing tool is complimented by a tool able to verify/test the matching header part of the forwarding entries, even in the presence of aggregated (i.e. wildcarded) flow rules.

**Observability Tool 5:** IPTV Quality Monitor

The IPTV service quality monitor classifies IPTV traffic from the overall traffic (i.e., MPEG-2 transport streams) by observing packet-level network traffic. The tool extracts required parameters such as video/audio codec, bitrates, timestamps, and GoP (Group of Pictures) structures for the purpose of estimating MOS of the multimedia stream, which can be used to indicate the quality of the service in terms of packet losses, startup delay, and audio/video/audiovisual quality.

### 4.1.3 Verification mechanisms and tools

Verification tools improve the capabilities to check the correctness of NF-FG deployments in a software-defined infrastructure. As in any development process, service development can greatly benefit from identification of problems early in the service or product lifecycle to significantly reduce time and costs spent later on debugging and troubleshooting processes. We argue that this is especially true for telecommunication service definitions and configurations (like UNIFY SGs and NF-FGs) due to the high spatial distribution and the lower levels of redundancy in operator environments. Aspects of both formal verification at deploy time and operational verification during runtime are thus covered by a set of SP-DevOps tools.

**Verification Tool 1**: Deploy-time verification of SG, NF-FG and forwarding rules

The objective is to verify SGs, NF-FGs and forwarding rules at deployment time (or before instantiation). Verification should include predefined properties (routing loops, black holes, etc.) as well as user-specified properties in the form of policy definitions. This tool set encompasses a number of modules, located at the service layer for SG verification, the resource orchestration layer for NF-FG verification, and the infrastructure controller layer for forwarding (i.e. OpenFlow) rule verification.

**Verification Tool 2**: Run-time verification of NF-FGs

A run-time verification tool for NF-FGs, based on sampling and tagging of packets for the purpose of verifying if traffic is actually fulfilling the defined service graph, as observed in the infrastructure layer. The tool supports both verification and troubleshooting processes.

**Verification Tool 3**: Network watchpoint tool

A watchpoint module intercepts the OpenFlow control channel between the controller and switches looking for packet_in messages. If a packet matches user-defined criteria, then different actions can be triggered, e.g., dropping the message, sending alarms, or creating switch-state snapshots..

### 4.1.4 VNF development support

While the observability and verification tools are universal in addressing all three roles defined for SP-DevOps actors, we identified the need for at least one tool that would prioritize addressing the needs specific to the least traditional SP-DevOps actor, the virtual network function developer.

**VNF Development Support Tool**: Multi-component debugging tool (EPOXIDE)

We define a framework capable of assembling general and SDN specific debugging and troubleshooting tools in a single platform and make it possible to flexibly combine them. Our framework facilitates the work of SDN developers and network operators by orchestrating available troubleshooting tools in a flexible, automated, reproducible and re-usable manner.

## 4.2 SP-DevOps Toolkit

The SP-DevOps Toolkit groups together a set of tools developed in WP4, targeted to be made publicly-available under open source licenses, to solve particular problems on network performance monitoring, troubleshooting and service graph verification. The purpose of the Toolkit is to create a vehicle that facilitates integration between tools developed by the partners as well as to create a pull effect whereby potential popularity of one tool may lead to the discovery of some of the other components. Note that not all tools developed in WP4 are "candidates" to be made available publicly for reasons ranging from intellectual property rights to concerns about the overhead of supporting the tool beyond the end of UNIFY.

The Toolkit covers all three types of roles defined in SP-DevOps, namely Service and VNF Developers and Operators, with emphasis placed on performance monitoring and service graph verification. The way the information generated by the tools is accessed through the use of the DoubleDecker messaging system makes the tools accessible to both developers that might want to include them in scripts as well as operators that would employ the tools either individually or embed them within workflows. Such an integrated approach allows VNF Developers to publish performance and debugging data in a format that could be easily accessible by scripts, or other VNFs, as well as the infrastructure orchestration and controller functions. This is in contrast with many of the current Application Performance Management instrumentation solutions (for example, Java Management Extensions - JMX) that are aimed at providing the data in log files to be consumed by an offline management tool such as Splunk or Loggly. It also creates better opportunities for communication between Developer and Operator roles, as individuals in these roles may access the same tools through similar interfaces.

Each of the tools released through the SP-DevOps Toolkit may be accompanied by specific license terms. The Toolkit as such does not force a particular type of license on any of its components. The candidate components of the SP-DevOps Toolkit are presented in Table 4.2 The table also shows an *indicative* allocation of candidate tools to two releases of the Toolkit. The first release is expected in November 2015, and the second release is scheduled at the end of the project.

| SP-DevOps Tool / Function | First Toolkit Release | Second Toolkit Release |
|---|---|---|
| DoubleDecker | x | |
| MEASURE language | | x |
| Rate Monitoring | x | |
| AutoTPG | x | |
| Deploy-time Verification | | x |
| EPOXIDE | x | |

*Table 4.2: Overview of candidate tools and support functions included in the SP-DevOps Toolkit.*

## 4.3 Concluding Remarks

In addition to summarizing a number of support functions and tools developed under the SP-DevOps concept, this section presents the SP-DevOps Toolkit, which is the result of an integrated set of tools developed in WP4 that are candidates to be made publicly available. The integration aspects are represented by the use of the common DoubleDecker messaging system to exchange data as well as expressing parameters, where applicable, via the MEASURE language. Licensing aspects related to each of the tools will be discussed within the project and agreed in time for the scheduled release of the tool in question.

# 5 Technical Description of SP-DevOps Tools and Functions

In this section, we present the technical descriptions and examples for each support function or tool included in the SP-DevOps Toolkit, together with reports of the resulting individual evaluation of the methods performed in stand-alone prototypes. We carefully motivate certain design choices made in order to achieve the WP4 objectives of increased automation, observability and scalability. Moreover, we show how the support function or tool integrates across the FA and in the UN, and indicate which research challenge (RC), WP4 objectives and requirements as specified in D4.1 [D41]. Individual tools or support functions may implement more than one SP-DevOps process at different levels of the functional architecture and UN - how and where parts of the observability, troubleshooting, verification, or VNF developer support processes are deployed and operate are described per tool.

## 5.1 DoubleDecker Messaging System

To provide a scalable system for easily integrating different monitoring functions, and transporting their results, the DoubleDecker messaging system has been designed. As introduced in Sections 3.4.1 and 4.1.1, the system consists of two protocols (client-broker and broker-broker) which allow point to point messaging as well as supporting a publish/subscribe mechanism. Messages are routed between connected clients through a local broker or a hierarchy of brokers if necessary, keeping messages as local as possible in order to avoid network overhead. The hierarchical messaging topology created by connecting brokers fits naturally to the hierarchical, recursive, UNIFY architecture.

As it is built on top of the ZeroMQ messaging library it supports multiple underlying transport protocols, from shared memory within a process, to IPC, TCP, TIPC, and PGM multicast. Client endpoints are identified not by a network address but by a separate identifier, by separating address from identity components can more easily be migrated in the infrastructure without requiring that for example the network routing is updated or tunnels moved between endpoints.

### 5.1.1 Technical description

The main problem the DoubleDecker messaging system is trying to solve is to provide a simple and easy to use transportation mechanism for control and management data for the various monitoring functions developed in the project. As these monitoring functions can be quite different in their complexity and how they are implemented, the solution should be easy to apply in a range of scenarios. For example going from a set of counters in a switch that are read by a SDN Controller app, to daemons running on a UN, to dynamically instantiated VMs. Additionally, it should be easy to use in the dynamic and changing environment envisioned in UNIFY, where functions may migrate and scale-in or out depending on various factors such as incoming traffic load. Finally, it should be easy to implement support for the messaging system in the clients, and to integrate with existing systems.

For these reasons we have chosen to implement the DoubleDecker messaging system as a protocol running on top of the ZeroMQ [ZeroMQ] messaging library. ZeroMQ implements the ZeroMQ Message Transport Protocol in a C library with bindings and native implementations available for many languages (currently more than 20 are supported). While ZeroMQ does not support many features of more complex messaging system out of the box, it can be tailored for many scenarios. ZeroMQ also focuses on being low-latency and scalable which also makes it an attractive solution. The core of ZeroMQ is a set of socket abstractions, similar to traditional TCP sockets, which provides the building blocks for creating various messaging patterns. Patterns range from simple synchronous Request-Reply patterns, to asynchronous point-to-multipoint messaging, automatic message load-balancing and more. The ZeroMQ library supports multiple underlying transport protocols transparently to the messaging layer, this allows us to be flexible when connecting monitoring functions. We can e.g. connect monitoring functions using low-latency IPC mechanisms such as UNIX sockets when possible, or connect monitoring functions from a remote node using TCP/IP if appropriate, without modification to the code.

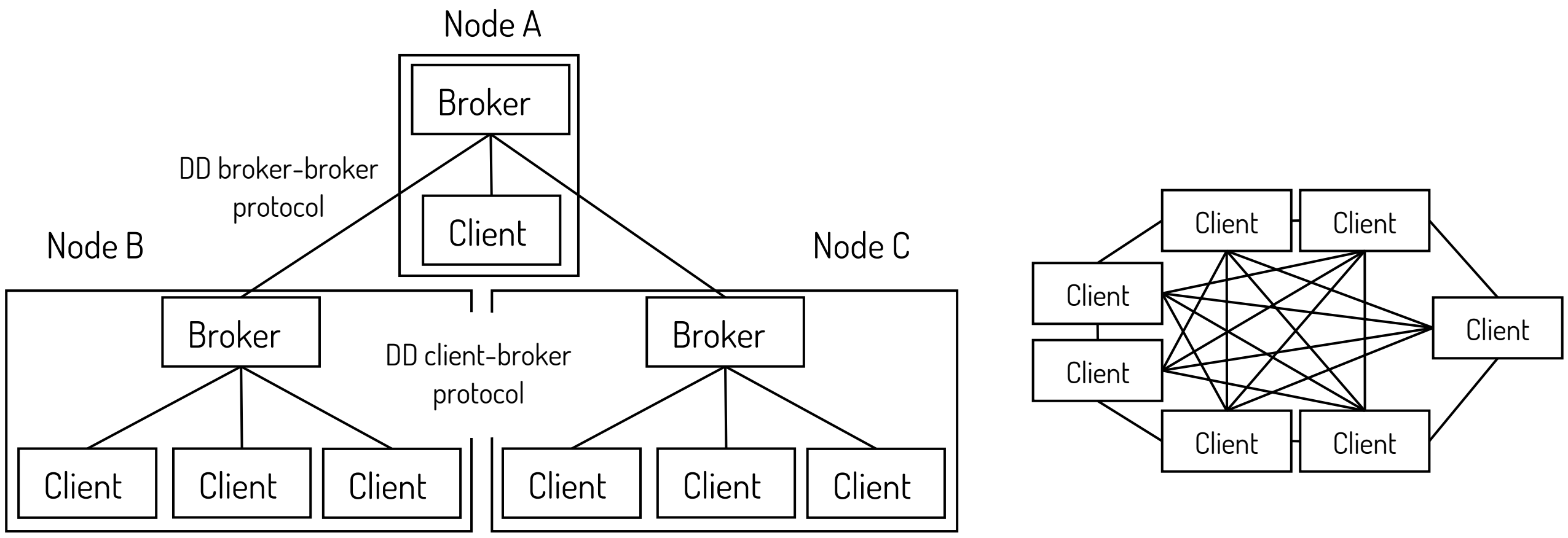

*Figure 5.1: Clients connect to brokers in a hierarchical fashion (left). Logically, all clients are peers that can transmit messages to each other (right).*

The DoubleDecker protocols on top of ZeroMQ combines several of the low-level functions provided by ZeroMQ, creating a more advanced messaging system allowing hierarchical message routing over multiple brokers, where ZeroMQ alone only supports a single broker. The DoubleDecker protocols also include other mechanisms, for example heart beating to detect communication failures, provides authentication, isolation between customers, and topological grouping (e.g. ability to define different scopes for a pub/sub topic). The two components of the messaging system are the brokers and the clients, at the heart there is a set of brokers linked to each other. These brokers are identical and define a hierarchical architecture (Figure 5.1). They are able to route messages from one client to another and keep track of the connected clients, keeping forwarding tables up-to-date. If a client disconnects or unregisters, the brokers are updated, same if a part of the broker topology is partitioned from the rest. As long as there is at least one broker available then the system will provide a minimal set of capabilities. For instance if a UN is disconnected, local messaging is still possible.

Brokers supporting the DoubleDecker protocol have been implemented in both Python3 and C, with client modules primarily in Python3. With the client side of the protocol being relatively simple, it is fairly easy to implement in any of the supported languages. Python3 is particularly useful because of the simplicity of the language itself and the large amount of libraries available. This makes it easy to extend the client modules to act as a proxy to existing tools, e.g. integration with a REST API can be done in just a few lines of code.

The clients are quite different, as they have to follow a simple protocol to register with a broker and maintain a heartbeat. When connected they can use the four simple functions listed in Table 5.1 to communicate with each other. Any other functionality is up to the operator to implement. For instance in the case of the Rate Monitoring tool (described in Section 5.5), the rate monitoring algorithm has been embedded in a client able to transmitting the calculated values to other components.

| Function | Purpose |
|---|---|
| `send(identity, data)` | Client-to-client transmission of *data* to client *identity*. |
| `[identity, data] = receive()` | Received *data* from *identity*, client-to-client. |
| `[topic, data] = subscribe(topic)` | Subscribe to messages beginning with the topic prefix *topic*. E.g. if the subscription is for topic "A", messages published on topics "A" and "ABCD" will both be received. |
| `publish(topic, data)` | Publish message with *data* in the topic starting with prefix *topic*. |

*Table 5.1: List of DoubleDecker client functions and their purpose.*

DoubleDecker solves the issue of assigning and updating network addresses during instantiation and migrations, as clients only use names to communicate with other clients. Brokers, however, do need to know the address of the broker one step higher in the hierarchy, however, since brokers are typically not deployed dynamically nor migrate this is less cumbersome to handle. Moreover, brokers can forward messages no matter the format. A client could monitor the network and send data in a JSON format to an aggregator while in the same node another client could send binary files.

### Design choices and practical considerations

The implementation of the messaging system itself can be compared with messaging systems like RabbitMQ [RabbitMQ] and ActiveMQ [ActiveMQ], both having a centralized broker design. The choice to implement a new system on top of ZeroMQ was motivated by 1) the ability to implement a distributed architecture, where messages are kept as locally as possible, and 2) the flexibility of transport mechanisms (in particular the support for IPC through UNIX sockets). The centralized messages brokers of RabbitMQ and ActiveMQ requires messages between components to always be transmitted to a central location before being send to its destination, adding unnecessary latency e.g. in the case where the destination is actually running on the same node as the sender. The distributed nature of DoubleDecker also makes it more resilient, as a single node or link failure may leave the messaging system partitioned but still operational within the partitions, compared to a crash of a centralized broker where the whole system fails. Centralized brokers like the ones provided by ActiveMQ and RabbitMQ additionally requires IP connectivity from the clients, this is not the case for DoubleDecker which can use local IP addressing (i.e. not routable outside a UN) or even UNIX sockets to connect clients.

An alternative solution to DoubleDecker, which also be scalable in the sense that messages would not need to traverse a central node, could be to have direct connection between clients over IP and using e.g. TCP or STCP to transport data. This however requires IP connectivity between all clients that should communicate, IP routing, and a name lookup service like DNS (one or more of which need to be updated in case of service/VM migrations, e.g. either the DNS or the routing is updated). In addition to this, typically ports for different services have to be defined and client/server relationships clearly defined in order to establish connections in an organized manner (i.e. the server has to open and listen to a port before the client connects). Messaging systems provide a more flexible way to develop distributed systems by providing an easy to use communication interface where communication patterns between various components can easily be created, without having to define client/server relations, port numbers, and low-level protocols.

To illustrate how DoubleDecker can be used, an example focused on deployment is available in Annex 2.

### 5.1.2 Architectural mapping

**Functional architecture:** The DoubleDecker messaging system operates at several levels, primarily to allow clients such as MFs , OPs, and VNFs to communicate with each other and other clients such as control applications and monitoring data aggregation points (see Section 3.2 and 5.2). Monitoring data aggregation points in different layers could be connected to each other using DoubleDecker, however, since these are unlikely to migrate or be dynamically deployed the advantages are fewer in this scenario.

**UN-mapping**: A DoubleDecker broker can be running in the UN, e.g. as part of the Monitoring Management Plugin as described in Section 3.3. Here it can be used to connect various monitoring tools, specifically their LCP component to the Monitoring Management Plugin, and a local aggregation point. It can be connected to higher layer brokers in order to provide connectivity for 1) MFs spanning multiple nodes, and 2) aggregation points to aggregation points and other components, for sending aggregated measurements and notifications. In the UN an important feature of DoubleDecker can be observed as local clients can connect to the messaging system using inter-process UNIX sockets or locally assigned IP addresses, circumventing the complexity of dynamically (re-)configuring IP routing.

### 5.1.3 Support to SP-DevOps

DoubleDocker supports the SP-DevOps flow primarily by providing a simple and scalable way of connecting monitoring components and transporting data between them, thereby supporting observability and troubleshooting. Message patterns can be easily extended in case new components are deployed as part of the SP-DevOps feedback cycle between the different teams. It is a generic tool which needs to be connected to relevant MF to be of any use. The design of DoubleDecker makes it easy for operators to develop their own troubleshooting tool and deploy it in a real situation.

For the Verification workflows and processes, DoubleDecker can be used to communicate with components to obtain e.g. internal state or traces, which could be requested by higher layers and be the input for verification tasks. It is also possible to do local verification where e.g. log files are analysed locally and higher layers informed only in case an issue is detected. The main research challenge addressed is RC2 (scalable observability data transport and processing), while assisting objectives O4.2 and O4.3 by providing an easy and lightweight way to connect various components of the architecture [D41]. The approach relates to several requirements in D4.1: NL1 (NL1.2-1.5), NL2, NL3, and NL4 (NL4.2-4.8).

### 5.1.4 Evaluation on the current prototype

The important performance indicators of the messaging brokers are throughput both in terms of number of messages and amount of data as well as the latency to deliver messages, especially under load. Here we evaluate our two broker implementations (Python3 and C) in regard to those indicators, for the point-to-point messaging primitives. This is a stand-alone prototype meaning DoubleDecker is the only prototype tested. Several demonstrations of integration of DoubleDecker with other prototypes have been realized and DoubleDecker was never a limiting factor. In this case we want to exhaust the resources in the broker to discover its limits.

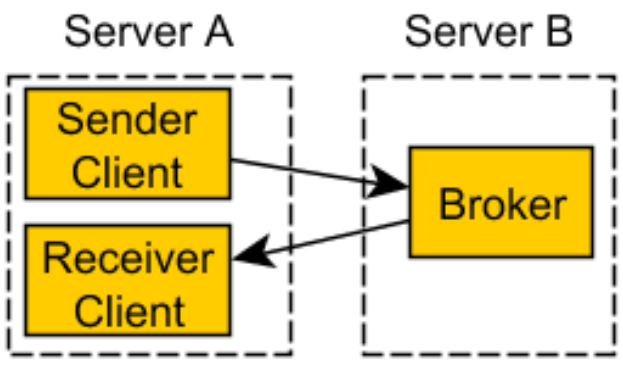

*Figure 5.2: Test setup for measuring messaging throughput and latency*

#### 5.1.4.1 Test conditions

We connected a Sender and a Receiver running on Server A using two gigabit links to Server B, where a broker is running (Figure 5.2). The Sender simply sends messages either with a set size, or with timestamps. Since the Sender and Receiver are running on the same node the received timestamps can be directly compared to the system clock to calculate the one way delay. We measured three aspects of the message forwarding performance, 1) message one-way delay depending on load, 2) maximum messages per second depending on message size, and 3) maximum goodput depending on message size. Both clients and the brokers are running directly on the host system.

#### 5.1.4.2 Message forwarding one-way delay

When measuring the one-way delay over the Python3 broker implementation, we observe two phases. The first phase is when the broker is not saturated (Figure 5.3), the second phase starts when the broker becomes saturated (Figure 5.4). While the broker is not yet saturated the increasing message rate unintuitively decrease the latency per message, this is likely due to optimization strategies in ZeroMQ such as message batching.

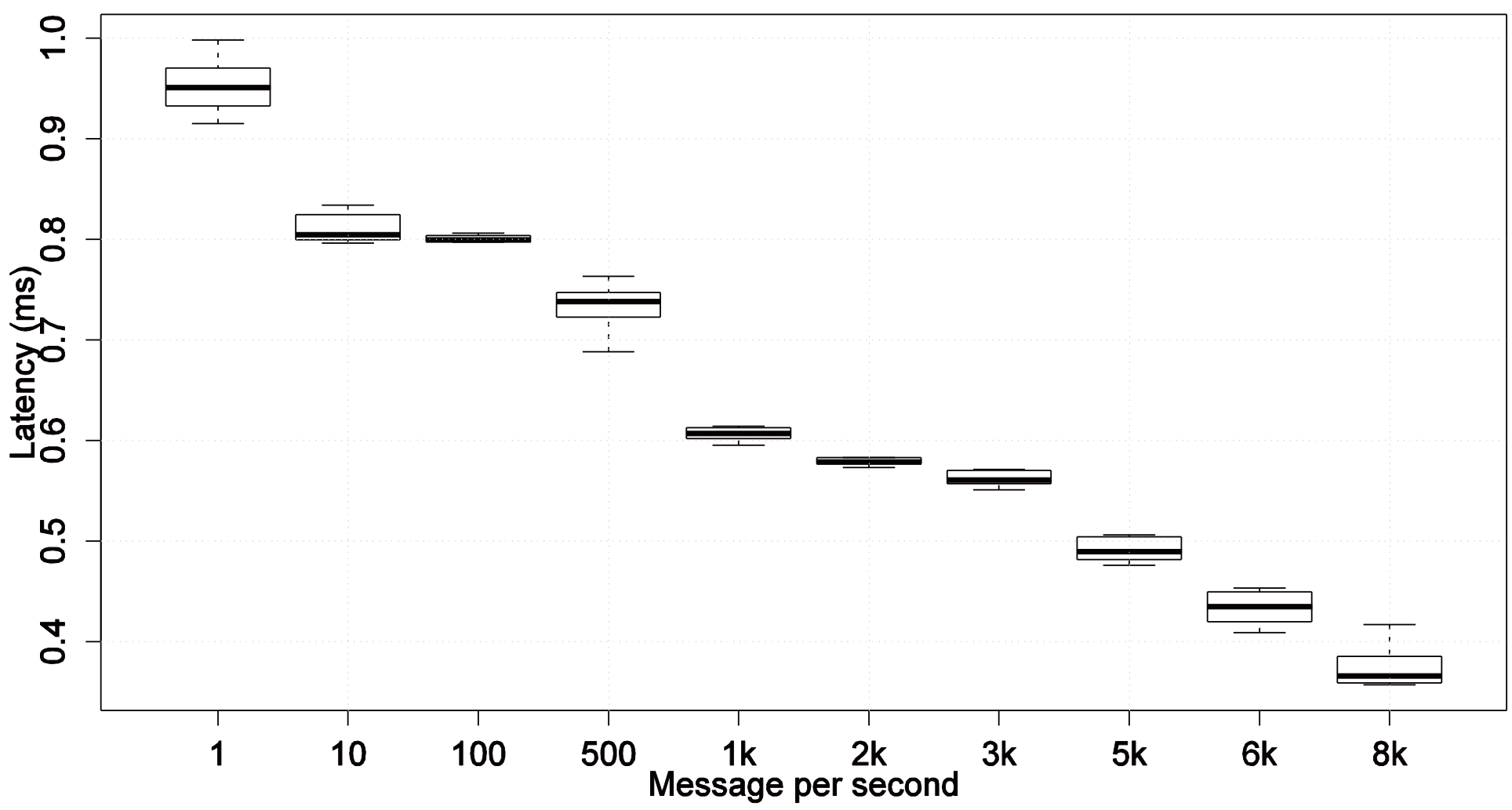

*Figure 5.3: Message latency for messages under increasing load, before link saturation (Python3 implementation).*

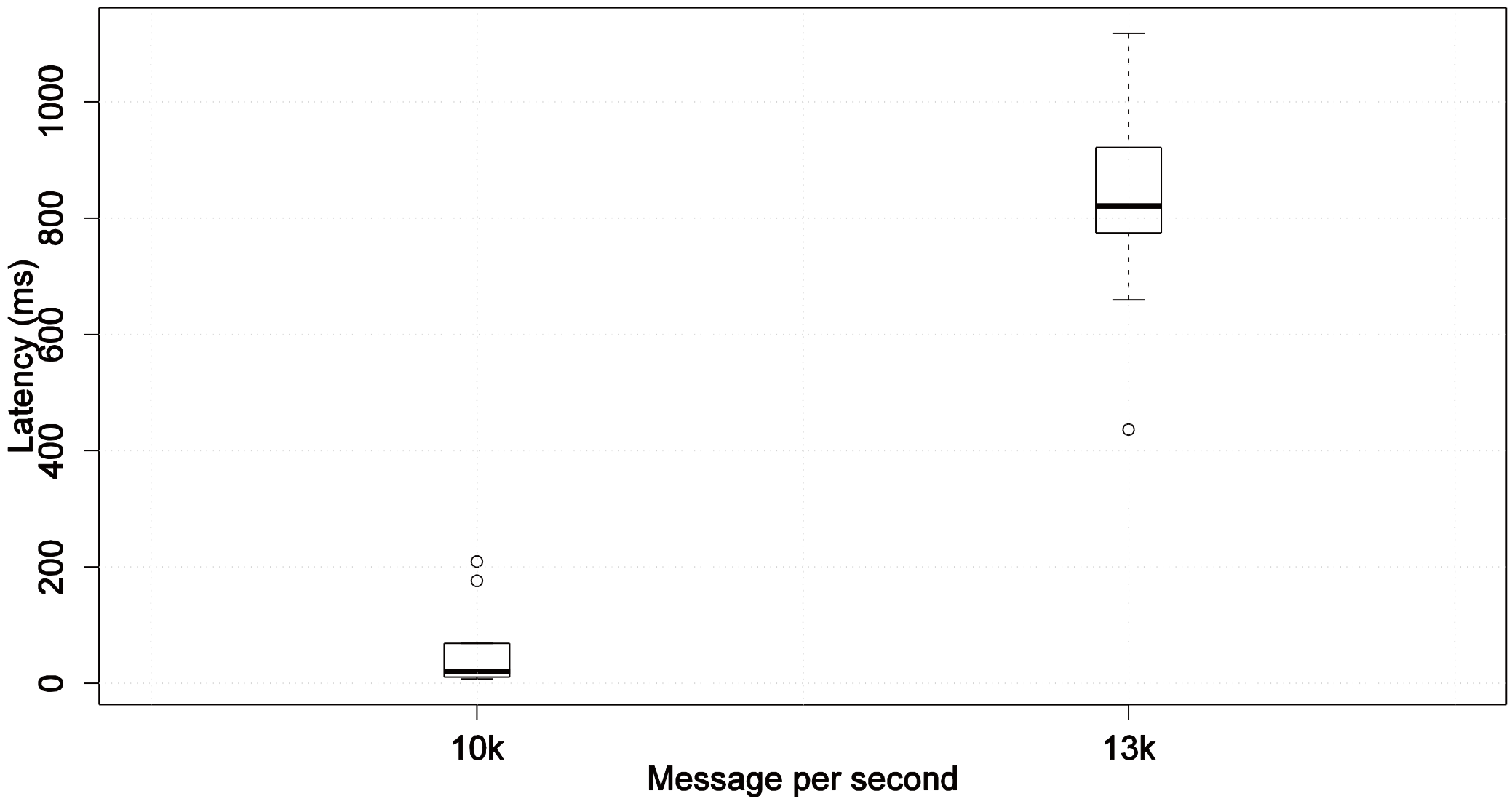

*Figure 5.4: Message latency once the link is saturated (Python3 implementation).*

Once the link is saturated messages starts to be queued, either in the Sender or in the Broker, and the delay starts to increase, as can be seen in Figure 5.4. This figure also shows the highest message rate we were able to reach with the current Python3 implementation, at about 13,000 messages per second. The low performance of the Python3 implementation was expected, this implementation is primarily done as a quick proof-of-concept to assist the development of the protocol itself. The C version is currently experimental as it is currently being implemented, as the protocol design has now stabilized.

### 5.1.4.3 Message throughput

Throughput in terms of messages per second and goodput, i.e. the actual data transferred in the messages, has been measured for different message sizes. Results for the two broker implementations are depicted in Figure 5.5.

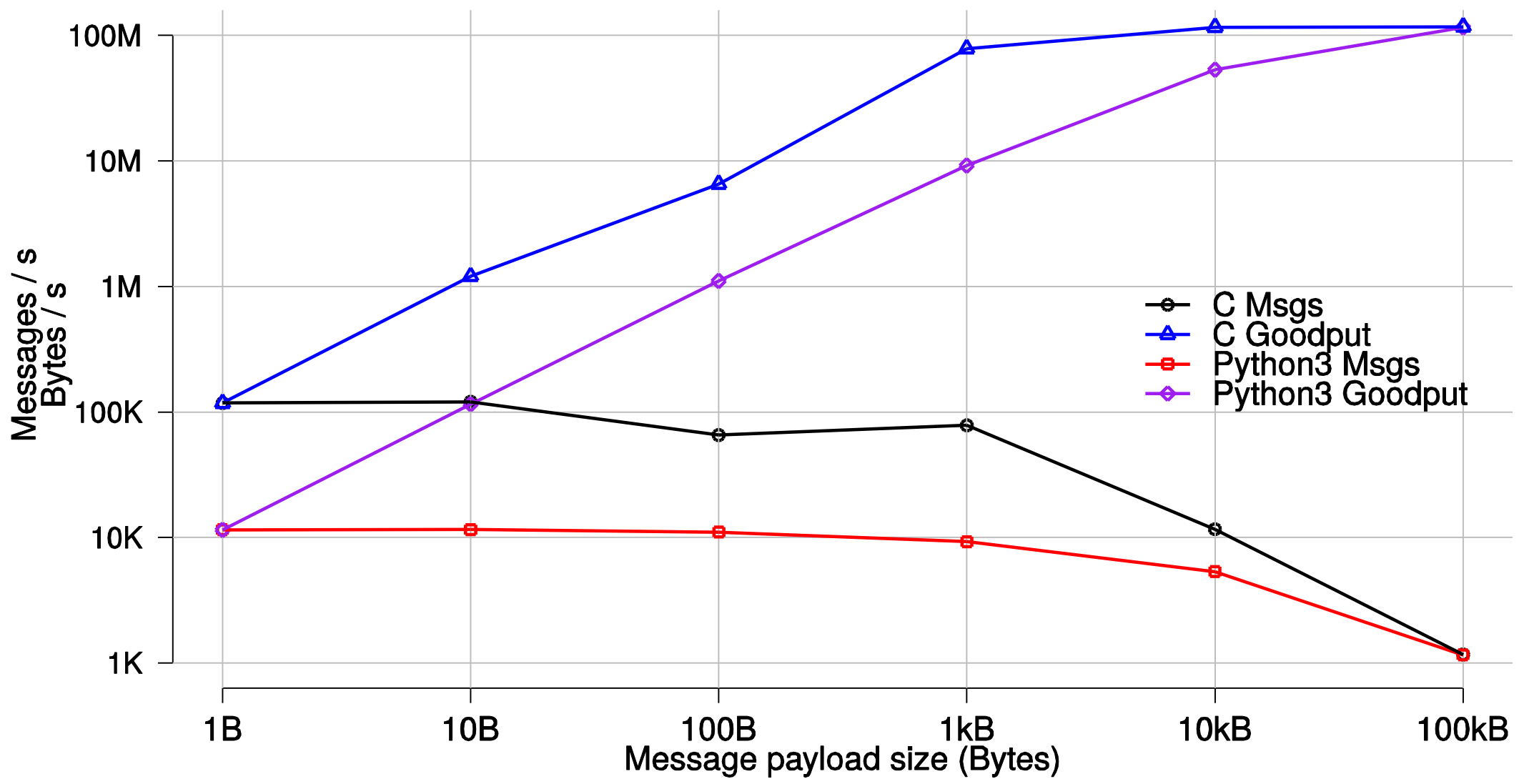

*Figure 5.5: Throughput performance for the C and Python3 broker implementations.*

As can be seen in the figure the maximum number of messages per second for the C and Python versions is around 120k vs. 12k, reachable with message payload sizes between 1 and 1000 bytes. For the C version, the 1 Gbit/s link is saturated at 1000 bytes of message payload. With larger payloads the number of messages per seconds starts to decrease while goodput remains saturated. The behavior of the Python3 implementation is similar; however it is not able to fully saturate the link until message sizes reach 100kb.

### 5.1.5 Next steps

As the current prototype has reached a usable state we are now working on integrating it with components from different partners' prototypes and demonstrators. This will help us have a better understanding of if and how the messaging system should be extended to support additional features needed by other UNIFY components.

Work to simplify deployment of the brokers and clients in different VM environments is ongoing; in Docker environments it is already easily integrated - an example of this is provided in Annex 2. While the performance of the current experimental (single-threaded) C implementation should be adequate for most UNIFY scenarios, we believe it is an order of magnitude or more below what should be possible with ZeroMQ. The main bottleneck for the number of messages per second we can handle appears to be the forwarding lookup task. For improved forwarding performance the initial target would be to divide this task over multiple CPU cores.

## 5.2 MEAsurements, States, and REactions - MEASURE

The MEASURE language was introduced in [M41] as a way of providing:

> "Machine-readable descriptions of the capabilities and their configuration parameters need to be developed. They specify which measurement functions should be activated, what and where they should measure, how they should be configured, how the measurement results should be aggregated, and what the reactions to the measurements should be."

To perform these actions, in a way that is not only machine but also human-readable, a grammar for a C-like monitoring description language has been defined. The language contains three main sections:

1. **Measurement definitions** describe *which* measurement function should be activated, *where* the particular measurement should be taken, and *how* the measurement should be configured. These are defined like functions in a programming language, with parameters for measurement locations (e.g. a particular port or VNF), and MF specific parameters such as measurement frequency, identifiers etc.
2. **Zone definitions** that specify how measurement results should be *aggregated* and define thresholds for a combination of aggregated results. Zones definitions results in variables that are either true of false.
3. **Reactions** that specify what *actions* should be taken both when moving between zones, to and from zones, and while within a particular zone. Generally, actions consist of sending notifications, aggregated measurement results, and configuration to other components in the UNIFY architecture.

MEASURE typically annotates a SG / NF-FG, specifying how monitoring for this particular service should be performed. However it could also be defined for infrastructure monitoring by replacing the virtual components of a SG (such as virtual ports and links) with their infrastructure counterparts.

The work on MEASURE was inspired by the work done in OpenStack Telemetry with Ceilometer [Ceilometer]. Ceilometer is a centralized system targeted to 1) collecting measurement data from distributed nodes, 2) persistently store the data, and 3) analyse the data and trigger actions based on defined criteria's. MEASURE on the other hand relies on distributed aggregation functionality and does not define how to collect and store the data, rather it 1) describes the configuration of the measurement functions, together with where they should be placed and where they should send results, 2) describes how results should be aggregated, and 3) triggers actions based on defined criteria's. When combined with DoubleDecker, measurement results and triggers can be sent to aggregation points for aggregation and analysis, to databases for persistent storage, and/or to higher layer components such as Resource Orchestrators.

The *Measurement definitions* section of the language defines the how a tool should be configured, in that sense it is similar to e.g. the GMPLS RSVP-TE Extensions for Operations, Administration, and Maintenance (OAM) Configuration [JHE14] which allows dynamic configuration of OAM tools, such as packet loss or latency monitoring tools, for certain LSPs .

It could also be compared with the Simple Network Management Protocol (SNMP) [DHA02] or NETCONF [REN11], in the sense that all can be used to configure components. MEASURE could also be combined with SNMP and NETCONF; at the lowest layer where MEASURE terminates, these protocols may be used to finally configure a component. Neither NETCONF or SNMP by themselves contain aggregation logic or the ability to specify reactions, however particular SNMP MIBs or YANG models may provide similar functionality at the level of the configured component.

### 5.2.1 MEASURE example

To illustrate the idea behind MEASURE, we give an example of how it could be used for the simple Service Graph depicted in Figure 5.6.

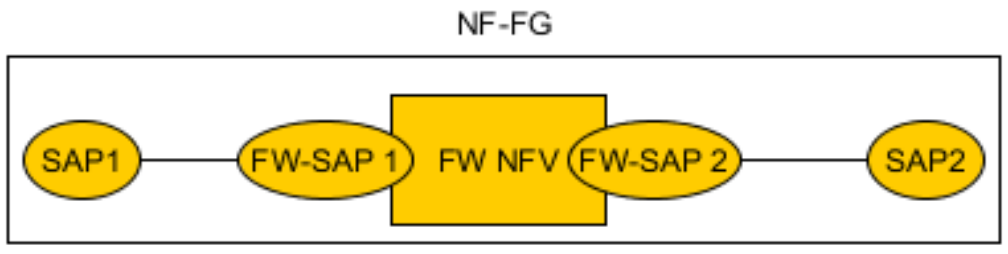

*Figure 5.6: Simple Service Graph depicting a Firewall connected to SAP1 and SAP2.*

In order to measure end-to-end delay in this Service Graph as well as the delay over only the Firewall NFV, with 10 measurements per second, the definition in MEASURE would be:

```
measurements {
  m1 = delay(FW-SAP1, FW-SAP2, 10hz);
  m2 = delay(SAP1, SAP2, 10hz);
}
```

Here `m1` is defined as the output of a delay function between FW-SAP1 and FW-SAP2, with the parameter"10hz". Similarly `m2` is defined as the output of a delay function between SAP1 and SAP2.

With the two measurements defined we can define zones based on them. Below we first define zone `z1` as True when the mean value of `m1` over the 50 last measurements is larger than 10 milliseconds. Zone `z2` is the complement of `z1`, being True when the average over the 50 last measurements is equal or below 10 milliseconds. Finally zone `z3` is True when the average of the total latency minus the Firewall specific latency over the last 50 measurements is over 5 milliseconds.

```
zones {
  z1 = mean(50, m1) > 10ms;
  z2 = mean(50, m1) <= 10ms;
  z3 = mean(50, m2 - m1) > 5ms;
}
```

With zones defined we can define transitions based on them. Below we define that when we go from zone `z2` to `z1` (from a good to a bad state) we send a notification to the Controller alerting it that `m1`is out of threshold. Similarly we define another notification when we transition back to `z2` from `z1`, here we send a message that `m1` is now ok. Finally, whenever we enter `z3`, we send a notification that the link latency is high. Who defines to whom notifications should be sent, and what the notification should be, could depend on the role of the originator of the MEASURE code. For example a VNF Developer may be allowed to specify it freely whereas a typical customer maybe is presented only with a restricted template.

```
actions {
  z2 -> z1 = Notify(Controller, ["Alert", "m1", m1]);
  z1 -> z2 = Notify(Controller, ["OK", "m1"]);
     -> z3 = Notify(Controller, ["Alert", "m2-m1", m2-m1]);
}
```

### 5.2.2 MEASURE process through the layers

The intention is for each layer to parse the MEASURE definition in the SG/NF-FG, update it based on any NF decomposition or NF-FG split performed in the layer (described in Section 3.2), potentially configure a local Aggregation point if a decomposition or split necessitates it, and finally update/merge the NF-FG before transmitting it to lower layers. The updated MEASURE code takes any split or decomposition into account, these typically generate new capability requirements that have to be annotated into the NF-FG(s) so that the next orchestration layer is aware of all monitoring requirements. The typical process flow is shown in Figure 5.7, optional steps are depicted with a dashed box frame.

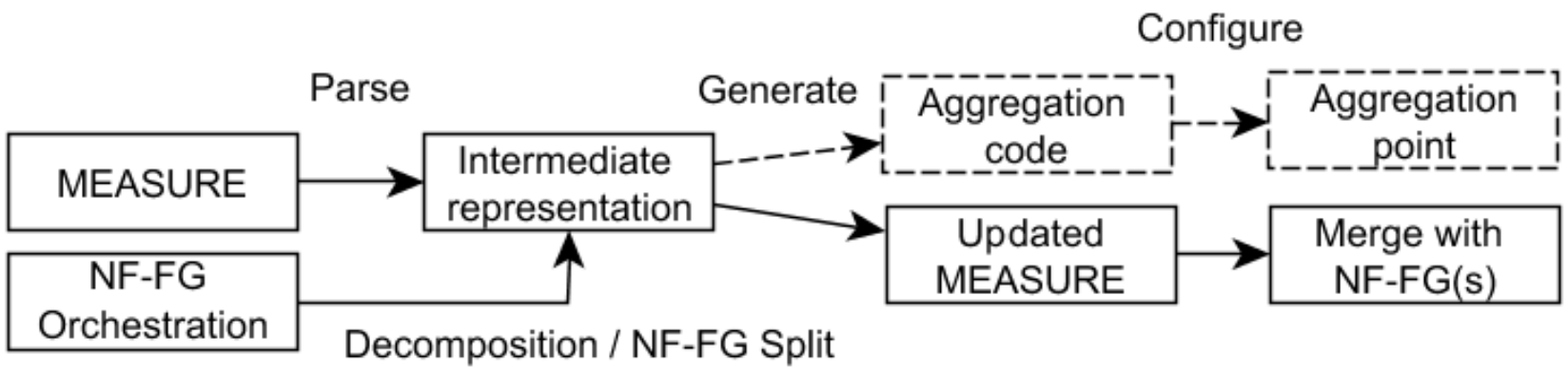

*Figure 5.7: Typical MEASURE behavior in an Orchestration layer*

The expansion of the MEASURE code based on decomposition or NF-FG split is not always easy, in some cases simple substitutions may be enough, without the need for local aggregation code. This would be ideal for scalability reasons, as computations can be performed closer to the measurement source with fewer messages reaching the more central components. In other cases, in particular with NF-FG splits it may be possible, but difficult, to further push aggregation. In these cases, and other cases where measurement aggregation causes complicated dependencies between measurements, it may be necessary to configure local aggregation at the particular orchestration layer in order to "compose and un-split" measurements. In other cases decomposition may be solved by e.g. substituting one measurement with new ones and substituting the use of the measurement variable in the MEASURE code with e.g. the maximum value of the newly defined measurements.  Such an example is shown in Figure 5.8, based on the simple example shown above. However, how to handle decomposition and splitting is a complicated issue that needs further study.

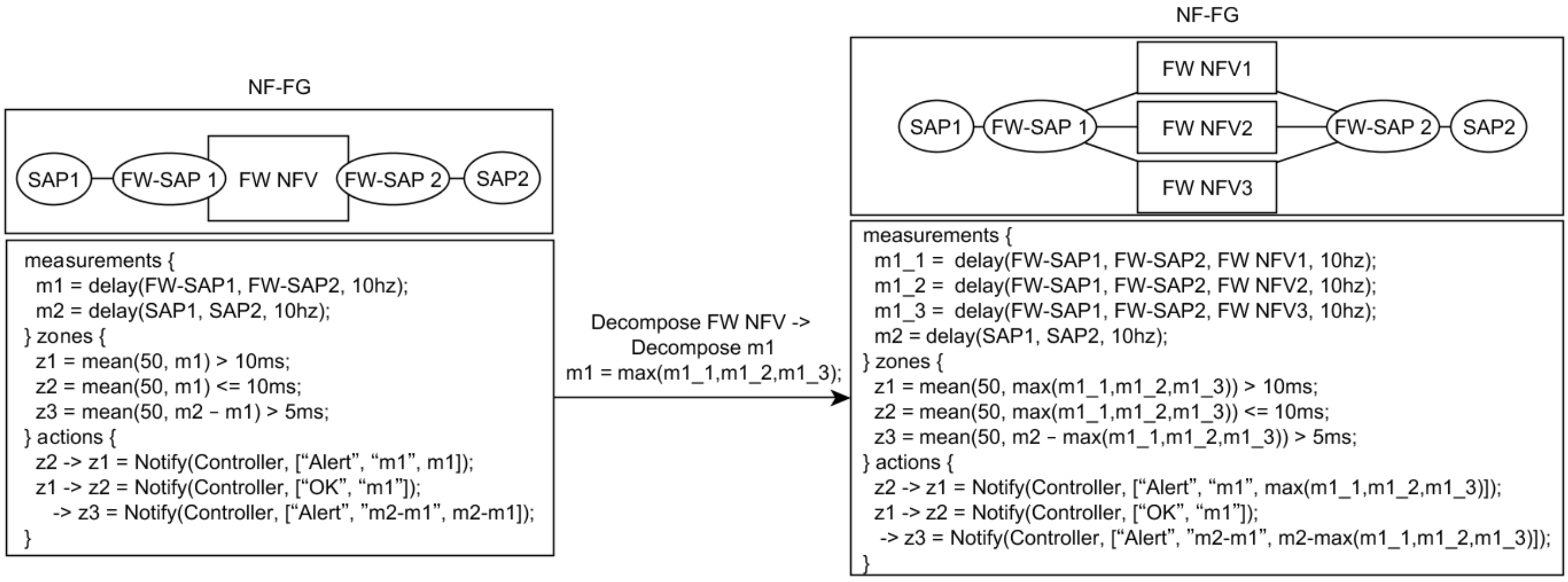

*Figure 5.8: MEASURE translation for decomposition of a Firewall into three Firewalls.*

At the lowest layer the typical process changes, as we are no longer generating new NF-FGs but rather generating configuration of MFs and OPs, and "compiling" the aggregation code that is necessary for processing the measurement results. A simple view of this process, which at this level of detail is similar to the one in an Orchestration layer, is shown in Figure 5.9.

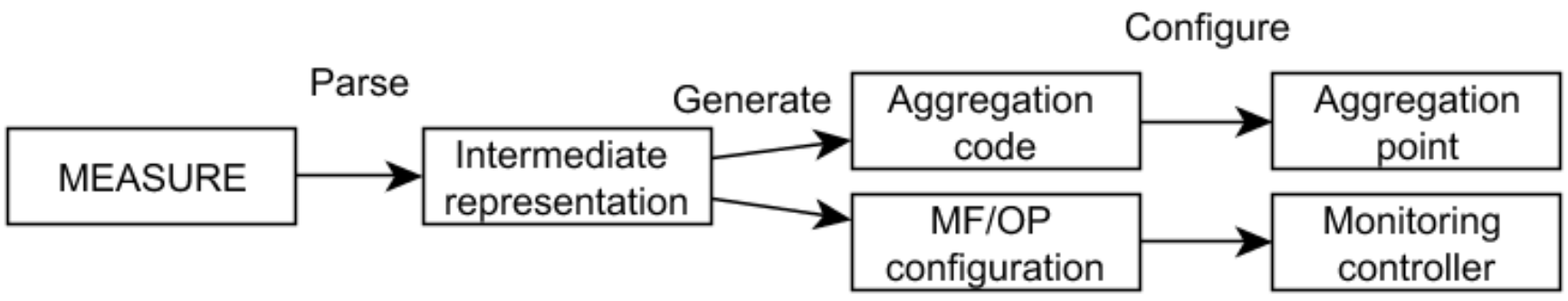

*Figure 5.9: Typical MEASURE process at lowest layer.*

Splitting this high level view of the process into more detailed blocks renders the more complex Figure 5.10. As the figure illustrates, the MEASURE description is first parsed and separated into MF configuration (based on the *measurement* section) and generation of aggregation point code (based on the *zones* and *action* sections). The generated aggregation code is however incomplete until feedback from the MF configuration and instantiation step is supplied, as measurements coming from different tools have to be identified by a MF-ID in the code. Once MFs have been configured, the updated aggregation code can be sent to the aggregation point. Finally, MF-IDs for the different measurements have to be returned to higher layers, which may or may not need them for configuring their own local aggregation points.

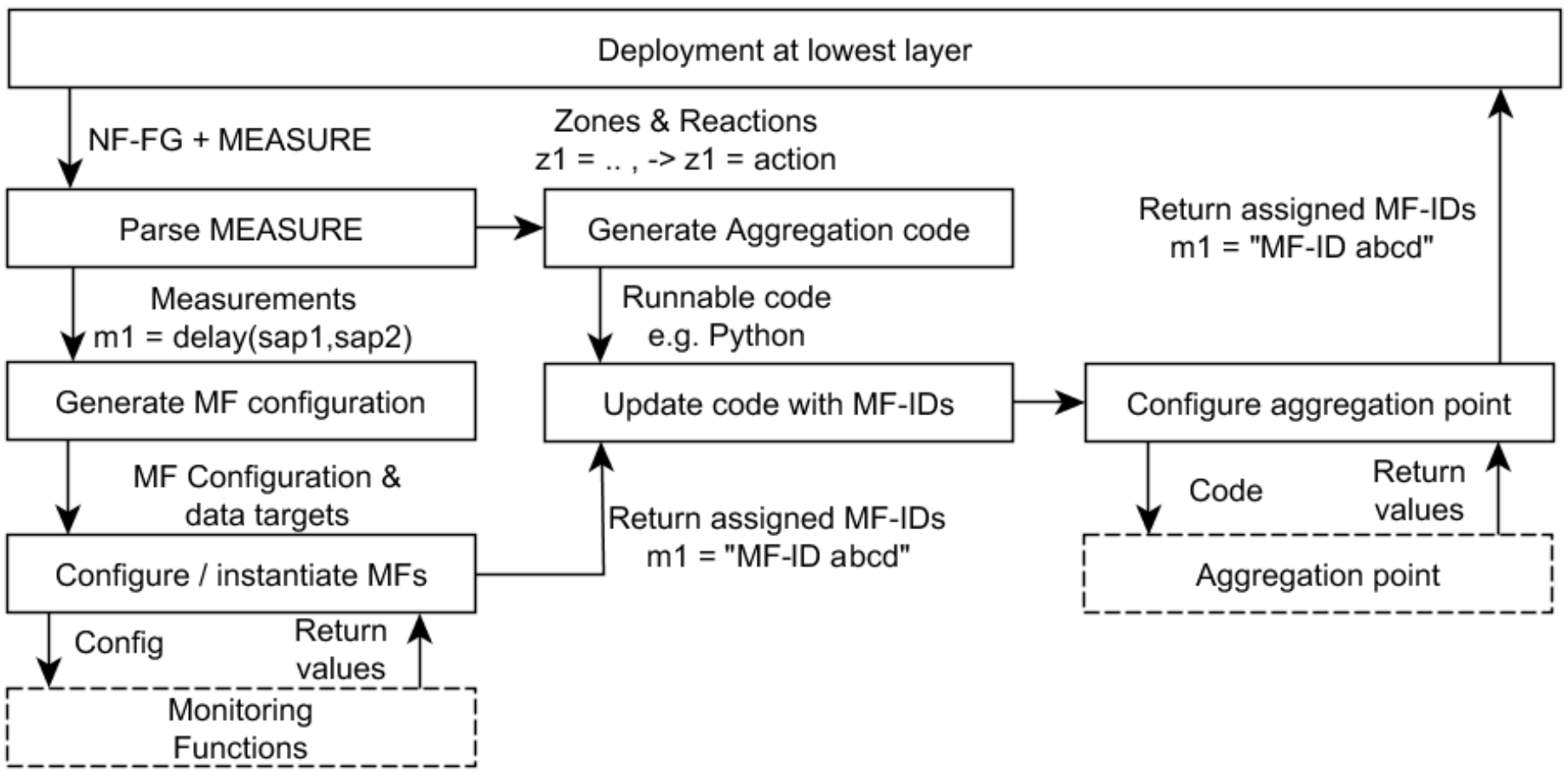

*Figure 5.10: Detailed flow chart of monitoring deployment at the lowest layer.*

Finally, the process once everything has been configured and measurements are running is shown in Figure 5.11. At the bottom a monitoring function is performing a measurement and transmitting the results to the local aggregation point. The aggregation point executes the generated code which may result in notifications being sent to various controlling components, or data being sent to higher layer aggregation points, or databases.

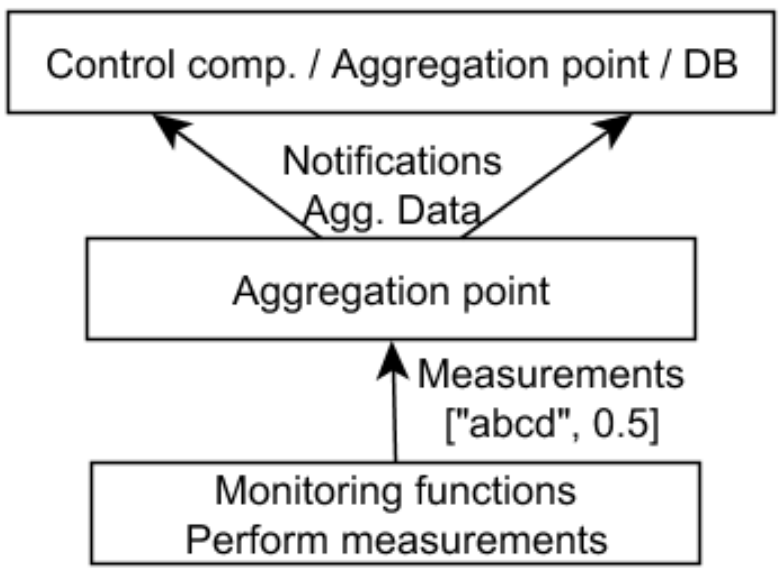

*Figure 5.11: Flow of measurement data up through the layers.*

### Design choices and practical considerations

To perform the processes outlined above we started by defining a grammar for the language itself, which we implemented using PyParsing[PyParsing], an easy to use alternative to traditional parsing tools such as a combination of Flex and Bison [Flex,Bison].Both could be used in a production environment as MEASURE is a language definition more than the software itself. Choosing PyParsing is a suitable for demonstrating a prototype.

This parser is able to parse the MEASURE definitions and generate a parsing tree for them. Conveniently it is also able to output the parsing tree as JSON and XML, which may better fit in the NF-FG model defined by WP3. The JSON and XML models contain the equivalent information as the original MEASURE definitions, but are less readable for a human; perhaps one of these representations could be used through the architecture once the initial MEASURE description has been parsed at the Service Layer.

Currently the implementation parses and compiles the MEASURE description into a Python module that fulfils an abstract class definition by providing a set of functions for evaluating a measurement and registering the module. The module code is then transmitted using DoubleDecker to an aggregation point, implemented as a Python3 daemon. The daemon loads the module, using the same mechanisms as any other Python module, and associates it to one or more MF-IDs. When a measurement result arrives to the aggregation point, the evaluation functions of the modules associated with the MF-ID that originated the measurement are executed (Figure 5.12).

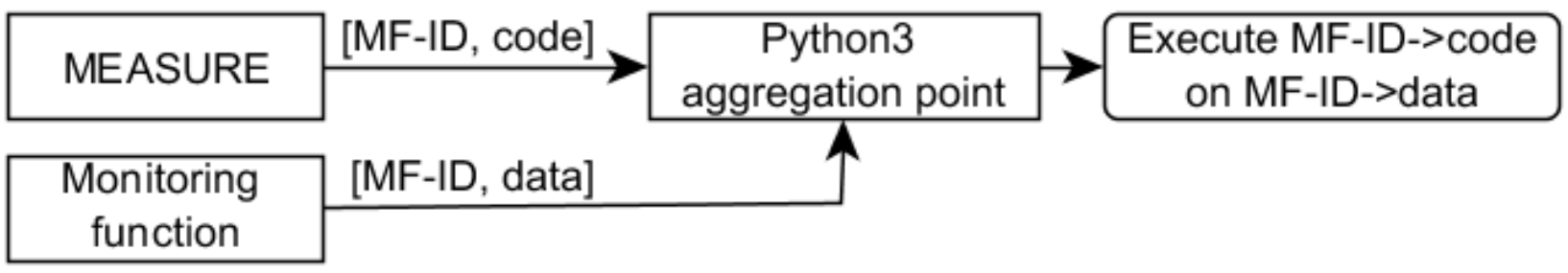

*Figure 5.12: Current complier implementation, generating and dynamically executing Python code.*

### 5.2.3 Changes compared to initial design

Regarding the Zone definitions in the initial design, some changes have been made compared to [M4.1]. The initial idea was to use Zones to model a finite state machine (FSM), where each State definition translated to a state in the FSM. Changing from one state to another could trigger an action, such as sending a notification. However, defining a FSM for this purpose quickly gets complicated to do manually, as the number of possible state transitions grows by the square of the number of states. If the definition is not correct, the FSM will not behave as expected with unwanted or undefined behavior as a result. For example, if the current input doesn't match any defined states, or it matches several states, determining state transitions is impossible. These cases could be solved by having for example an error state, or defining multiple FSMs in cases where we would like to have multiple transitions at once. However, the complexity of this solution seems to be impractical for real-world deployments.

An alternative solution that probably feels more natural for most programmers is to have a list of if - elseif - else statements based on the Zones. The states transitions are then naturally prioritized according to the order of if - elseif - else statements. This approach allows enough flexibility as there is no limit on the number of states or their complexity. If a state is not reachable there are no side effects on the other states. The else state also serves as the default state, which can be used to e.g. for sending error notifications to simplify debugging of the MEASURE description.

### 5.2.4 Architectural mapping

**Functional architecture**: MEASURE descriptions are embedded in the Service Graph/NF-FG and thus exist in all the layers of the architecture. The Service layer includes the initial MEASURE definition, defined either manually or as a template that can be automatically associated to a service. As the SG/NF-FG passes down through the layers this description is updated and partitioned depending on how Resource Orchestration is performed. Finally the MEASURE descriptions reaching the lowest layer are compiled into aggregation code and MF configuration to be instantiated.

### 5.2.5 Support to SP-Devops processes

MEASURE is a key support function for the SP-DevOps observability and troubleshooting processes, as it describes how and where measurements should be taken, and how they should be reported. While initially targeting primarily service monitoring, the same concept can be applied also to infrastructure monitoring. MEASURE is an important tool for the WP3 line of work - comments about the language can be found in D3.1. In WP4, the language partially addresses O4.2 and O4.3. From a research challenge perspective, this function supports programmability which is highly relevant for several of the research challenges related to software-defined monitoring (e.g. RC1, RC3, RC9 and RC10). Moreover, MEASURE relates to the requirements described in D4.1 for the Orchestration Level : OL1, OL2,OL3 (OL3.1-3.5) and OL4(OL4.1-4.3)

### 5.2.6 Next steps

The first next step to perform is to design a generic aggregation module to better understand what the target of the aggregation code compilation is; the current prototype seems to be a promising direction but needs refinement. The initial top-down design, starting with a language definition and proceeding from there, has reached it limits and we now switch to a bottom-up approach to better understand how to proceed.

Another necessary step is to better understand how the decomposition and NF-FG split steps should affect the MEASURE description for a particular NF-FG, and if the necessary actions for affected measurements can be generalized.

Due to the complexity of MEASURE we plan to continue development in a restricted scope, an existing demo. In the demo monitoring is currently hardcoded; our goal is to implement the monitoring configuration and reactions using MEASURE, for this particular demo scenario, using MEASURE.

## 5.3 Recursive Query Language

The concept and of the recursive monitoring language has been introduced in Section 3.4.3. The formal definition of the query language is to be described in M3.4, an internal document of WP3. To make reading easier, the formal syntax defined in WP3 is also included in Annex 3. In this section, we provide further details about the design of the query engine and the query language in the form of concrete examples, querying network delay and CPU/memory consumption.

### 5.3.1 Technical description

The query engine is the key element to support the query based on the language. The main functions of the query engine include:

- receiving a monitoring query from the receivers and sending back query results;
- parsing and compiling the query scripts;
- communicating with NF-FG repository and translating NF-FG graphs into Datalog facts;
- traversing NF-FG graphs according to query scripts;
- querying the distributed monitoring database which contains the measurement results from monitoring functions.

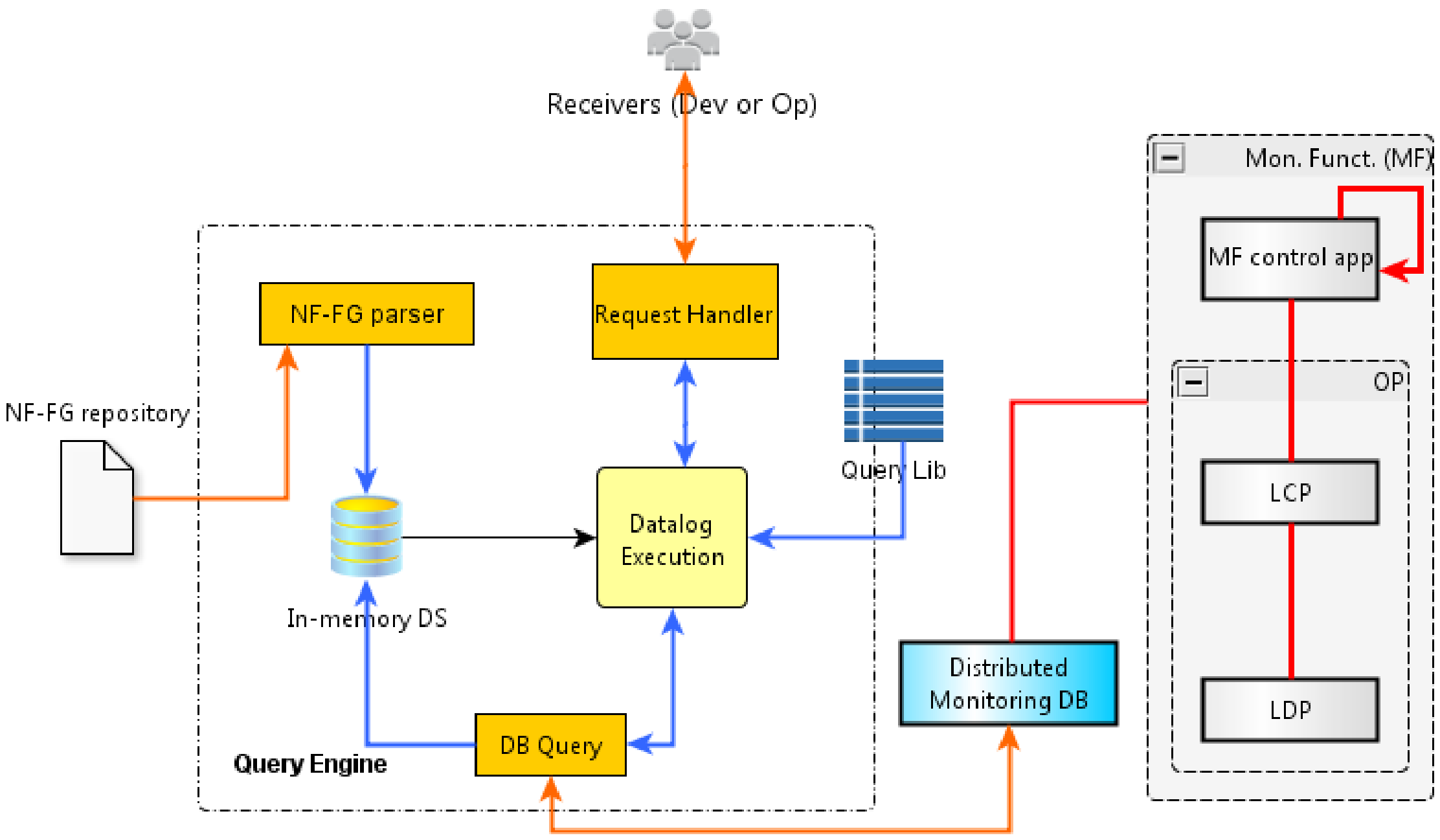

*Figure 5.13: The design of the Query Engine.*

Figure 5.13 shows the design of the query engine implementation. The Request Handler receives and parses the query commands from developers or operators. The query command could be in the format of "query_command [args]", for example, "e2e_delay 'src' 'dst'" is to query the end-to-end delay between the source node and destination node in service graph. The query command is then translated into the python based query API which is stored in Query Library. The Request Handler is also responsible for sending query results to the receivers. The NF-FG parser is used to check the NF-FG repository and translate the graphs into Datalog ground facts and stored into the In-memory Data Store (DS).

The key-value store of FoundationDB is used for the In-memory DS. For example: the facts: "link(x, y)" is represented as "kvp(link, x, y)"; "sub(x, $sub_1$, ..., $sub_n$)" is represented as "kvp(sub, x, $sub_1$, ..., $sub_n$); and "node(x, ..., z)" is represented as "kvp(node, x, ..., z)". In addition, the in-memory DS can also be used to store the intermediate results obtained by querying the monitoring database. The intermediate results can be used as cached results or to aggregate the data to be returned to the receivers.

The Datalog Execution will call the Query API according to received query command, and decompose the high level query request into primitive queries towards the monitoring data stored in the distributed monitoring database. It is also responsible to aggregate the results retrieved from monitoring database and pass the aggregated result to Request Handler.

The DB Query retrieves low level monitoring data from the monitoring database and store intermediate results into the in-memory data store. Because only intermediate results are stored in the query engine and the original monitoring data are still stored in Monitoring DB, the fault of the in-memory data store will not lose the monitoring data.

The Query Library contains the query APIs provided by developers or operators. The APIs could be written by Python scripts with the PyDatalog extension and encapsulated as Python classes. For each monitoring function, an API could be developed. Below in Section 5.3.2 there are some sample usages that utilize the query language defined in WP3 to implement the recursive monitoring query.

The query engine provides a tool for the receivers (e.g., developers and operators) to query the various performance metrics in service layer. It could be created as an application in service layer. The query engine doesn't collect monitoring data directly from infrastructure layer. Instead it relies on the data collected by the monitoring functions developed in WP4.It will have little impact on running network functions except it needs access to NF-FG.

### Design choices and practical considerations

There are several network monitoring languages available today. For example, Frenetic/Pyretic [NF011], Akamai query system [JC010], OGF NM-WG schema [MSW09], and etc. But none of these languages or schemas do not provide the capability to recursively query the performance in the hierarchical and recursive architecture which is adopted by UNIFY. Therefore we propose to use Datalog, a declarative logic programming language with recursive query capability, as the base for the monitoring language [TGR13].

In terms of the compiler of Datalog program, we have decided to use PyDatalog as the basis of the query system. PyDatalog adds Datalog support to Python. One main advantage of using PyDatalog is that the extensive library of Python can be re-used. Furthermore, the query functions can be written by developers or users in Python and provided as python classes/libraries, being able to access multiple databases including noSQL, an open source project.

### 5.3.2 Usage examples of the query language

To recursively query infrastructure or NF performance metrics, the developers or operators need to write query scripts according to the defined language. In this section we illustrate using two general examples how queries can be expressed related to querying network delay as well as CPU and memory metrics.

#### 5.3.2.1 Query the delay between two network functions in service graph

Two kinds of delay between network functions are discussed here: end-to-end delay and hop-by-hop delay. Here end-to-end delay is defined as the delay between the ingress node in the lowest layer of the source network function and the egress node in the lowest layer of the destination network function. And the hop-by-hop delay is defined as the aggregation of the delay of each segment, which consists of the path from the source to the destination network function.

Below are the query scripts according to the language defined in WP3 deliverables:

```
# R1-R2 are used to recursively retrieve all child nodes of network function ‘X’
R1: child(X,Y) <= sub(X,Z), child(Z,Y)
R2: child(X,Y) <= sub(X,Y)


# R3 gets all leaf children ‘Y’ of network function ‘X’. Leaf node is thought as the node in the
lowest layer
R3: all_leaf(X,Y) <= child(X,Y), is_leaf(Y)

# R4 gets the ingress leaf node (Y) of network function ‘X’
R4: leaf_src(X,Y) <= all_leaf(X,Y), is_source(Y)
# R5 gets the egress leaf node (Y) of network function ‘X’
R5: leaf_dst(X,Y) <= all_leaf(X,Y), is_dst (Y)

# R6 and R7 get the end-to-end and hop by hop delay between network function ‘S’ and ‘D’
respectively. fn_e2e_delay() is used to retrieve the actual measurement results from monitoring
database. Fn_h2h_delay() is used to get the delay of each hop and return the aggregated delay.
R6: e2e_delay(S,D,P) <= link(S,D), P == fn_e2e_delay(leaf_src(S,Y), leaf_dst(D,Z))
R7: h2h_delay(S,D,H) <= link(S,D), H == fn_h2h_delay(all_leaf(X,Y))
```

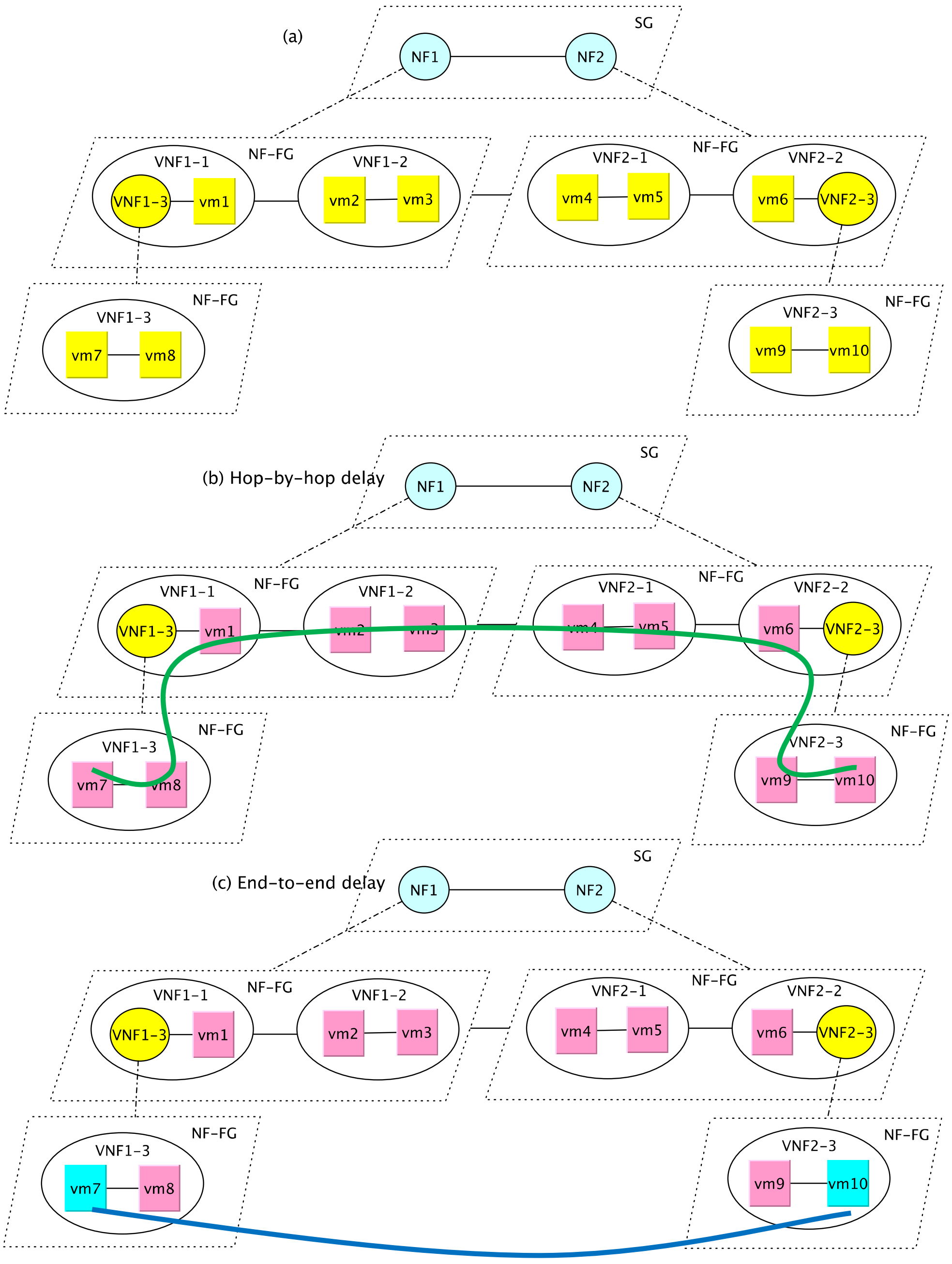

*Figure 5.14: Recursive retrieval of information across a sample SG and the corresponding NF-FG.*

In the scripts, R1-R5 is used to traverse the NF-FG to get the ingress node of NF1 and egress node of NF2. Figure 5.14 depicts a sample SG and corresponding NF-FGs. R1-R2 can recursively traverse the graphs and figure out all child nodes as shown in Figure 5.14(a) (i.e., VNF1-1, VNF1-3, VNF1-2, vm1, vm2, vm3, vm7, vm8, VNF2-1, VNF2-2, VNF2-3, vm4, vm5, vm6, vm9, vm10). R3 is used to figure out all leaf nodes (i.e., virtual machines or physical machines). In the example, they include all virtual machines as shown in Figure 5.14(b). R4 and R5 are used to get the ingress and egress nodes of NF1 and NF2 respectively, i.e., vm7 and vm10 in Figure 5.14(c). Then R6 is used to query the measured end-to-end delay between vm7 and vm10 from Monitoring DB. R7 is used to query the measure delay for each hop as shown in Figure 5.14 (b) and return the aggregated delay results.

Below shows the python scripts when PyDatalog compiler is used. PyDatalog does not support all syntax of Datalog, e.g., negate operation, the logic of the implementation is a bit different from the above.

```
# Get all children of given node
child(X, Y) <= sub(X, Z) & child(Z, Y)
child(X, Y) <= sub(X, Y)

# Get the node which is not leaf
non_leaf(X) <= sub(X, Y)

# a function to remove non-leaf nodes from a list
def fn_list_remove(original_list, element_to_remove):
    for i in element_to_remove:
        if i in original_list:
            original_list.remove(i)
    return original_list

# get the leaf children by removing non-leaf children
all_leaf(X,Y) <= fn_list_remove(child(X,Y), non_leaf(Y))
# get the leaf nodes of the source and destination network functions for delay function
ingress_leaf_node(X,Y) <= all_leaf(X,Y), is_ingress(Y)
egress_leaf_node(X,Y) <= all_leaf(X,Y), is_egress(Y)

h2h_path(X,Y,P) <= all_leaf(X,Y), all_leaf(X,Z), link(Y,Z)

# Get the actual delay between two nodes from monitoring database
def fn_e2e_delay(src, dst):
    return query_db_delay(src, dst)

e2e_delay = fn_e2e_delay(ingress_leaf_node(src_NF,Y), dst_leaf_node(dst_NF,Y))
seg_delay = f_seg_delay(src_leaf_node, dst_leaf_node)
```

#### 5.3.2.2 Query the CPU/memory usage of network functions

Because one network function usually consists of multiple compute nodes, the receivers may be interested in different aggregation ways on CPU or memory usage of network functions. One possible usage is to query the average CPU/memory usage over all compute nodes belonging to the given network function. Another possible usage is to query the maximum CPU/memory usage among the compute nodes of the given network function.

Below are the query scripts for average CPU based on defined language. The scripts for querying the maximum CPU are similar. Take the SG and NF-FGs in Figure 5.14 as example and the receivers want to query the average CPU of network function NF1 , R1-R3 are used to get all leaf children (VNF1-3, vm1, vm2, vm3, vm7, vm8) of Network Function NF1 as shown in Figure 5.14(b). R4 then gets a list of all compute nodes in infrastructure layer (i.e., vm1, vm2, vm3, vm7, vm8). R5 retrieve the measured CPU for each node in this list and return the average CPU usage.

```
# R1-R2 are used to recursively retrieve all child nodes of network function ‘X’
```

```
R1: child(X,Y) <= sub(X,Z), child(Z,Y)
R2: child(X,Y) <= sub(X,Y)


# R3 gets all leaf children ‘Y’ of network function ‘X’. Leaf node is thought as the node in the
lowest layer.
R3: all_leaf(X,Y) <= child(X,Y), is_leaf(Y)

# R4 get all compute nodes in lowest layer
R4: all_compute(X,Y) <= all_leaf(X,Y), is_compute(Y)


# calculate the avarege/maximum cpu usage over all compute nodes in lowest layer
R5: average_cpu(X) <= fn_average_cpu(all_compute(nf, Y))
```

The query scripts writing in PyDatalog is similar to the one regarding delay, and is not shown here.

Here only the query for delay and CPU usage are discussed above. But the idea can be applied to other monitoring functions, where parts of these scripts can be re-used.

The query scripts can be provided as Python classes in query library. So the developers or operators can use simple commands with arguments to query these monitoring metrics.

### 5.3.3 Support to SP-DevOps processes and objectives addressed

The recursive query language can be regarded as a support function to the SP-DevOps tools for the purpose of disseminating monitoring information and observed states at service and network levels. Hence, all processes and workflows relying on actively acquiring monitoring information and states (service and infrastructure level) are supported:

- Observability processes where hierarchical high-level monitoring relies on retrieval of information and states provided by low-level monitoring components, for the purpose of e.g. automated workflows and decision making.
- Troubleshooting processes where information (such as monitoring information and metrics, resource consumption and availability, service and infrastructure states, etc) is needed for processing and diagnosing anomalous states and performance degradations, as well as for e.g. automated triggering of additional observability points as part of identifying a root cause.
- VNF development processes, where monitoring (compute, storage, and network metrics) and state information are needed as part of debugging and troubleshooting activities.
- Verification processes, where monitoring and state information are needed in order for verify the infrastructure requirements for deployment of the actual service graph.

This work is related to the scopes of both WP3 and WP4. The objectives and definition of the recursive query language itself is covered in WP3, while the use cases and samples are covered in WP4.

### 5.3.4 Next steps

The implementation of an individual prototype is ongoing at the time of writing. Once a prototype if finalized, a functional and performance evaluation will be carried out, and results are expected to be included in future deliverables in WP3 and WP4. We are also exploring the possibility to integrate with declarative policy engine like OpenStack Congress.

## 5.4 Legacy OAM Protocol Support – TWAMP

In Section 3.4.2, we motivated the need for legacy OAM protocol support in the context of UNIFY SP-DevOps. In particular we introduced the effort to specify and standardize a data model for the Two-Way Active Measurement Protocol (TWAMP). In short, the TWAMP data model enables programmable active network measurements while at the same time enhances its capacity to be used in a much more scalable and automated manner in software defined infrastructures as targeted by UNIFY as a whole. As we explain below, legacy OAM protocol support can complement other support tools and mechanisms presented in this deliverable including the rate monitoring function (Section 5.5) and successive E2E monitoring of link delay and loss (Section 5.6). Further, TWAMP active measurement results could be stored in the distributed monitoring DB and retrieved using the recursive query language introduced in Section 5.3.

### 5.4.1 Technical description

TWAMP (RFC 5357 [KHE08]) can be used to actively measure path performance both on an end-to-end and a hop-by-hop. RFC 5357 defines four functional entities namely the Control-Client and Server, which communicate to coordinate measurement sessions using the TWAMP-Control protocol; and Session-Sender and Session-Reflector that implement the specified active measurement session using the TWAMP-Test protocol. For an overview see Section 2 of [RCI15]. However, to date, there is no standard way to program measurement sessions, through APIs, let alone in a distributed manner. Typical vendor support is aimed at the network operator level only, which is not suitable for the environment SP-DevOps targets.

For the purposes of UNIFY SP-DevOps such legacy TWAMP session management tools are not sufficient as they do not allow for full programmability for the Operator, and provide no APIs for the VNF Developer and Service Provider users. The TWAMP data model currently under specification, and the associated programmability through the YANG module, allow for more distributed approaches than what is possible today with off-the-shelf TWAMP implementations which typically rely on a centralized management platform. For example, the model supports programmable configuration and scheduling of one-off and repeated active measurements. Such TWAMP sessions can be conditionally fired off based on measurement results, for example, obtained by the rate monitoring function described in Section 5.5. Service Providers can use this programmability to determine if certain SLA parameters are indeed respected by the infrastructure provider, as discussed in Section 5.9. VNF developers can evaluate whether their applications can achieve the required level of network performance in a programmatic manner as well.

### Design choices and practical considerations

Earlier efforts to add standardized programmable management to TWAMP implementations based on the definition of a Management Information Base (MIB) and use of SNMP did not gather enough support in the IPPM working group. On the other hand, given the increased interest in enhancing manageability and programmability as fostered by the ascendancy of SDN fabrics, a data model specified in YANG appeared very attractive and in line with the SP-DevOps work in WP4. As discussed below in the next steps to be taken, the working group participants appear very perceptive to this effort, recognizing the significant benefits of the approach advanced by [RCI15]. Finally, in the meantime, commentary in the working group mailing list indicates that earlier efforts based on SNMP are unlikely to continue.

### 5.4.2 Architectural mapping

Figure 5.15 illustrates the mapping of the TWAMP monitoring function in the functional architecture. In particular, the TWAMP controller app defines the TWAMP tests to be made and configures the Control-Client and Server using the TWAMP data model defined in [RCI15]. The Control-Client and Server coordinate using the TWAMP-Control protocol the active measurements to be undertaken between the Session-Sender and Session-Reflector, including the model to be used, when the measurement should take place and whether it should be repeated and how often. These four functional entities form an OP. As discussed in Section 2 of [RC115], implementers can locate the Control-Client and Session-Sender functional entities in the same device or network function in NFV terms; the same applies to the Server and Session-Reflector functional entities.

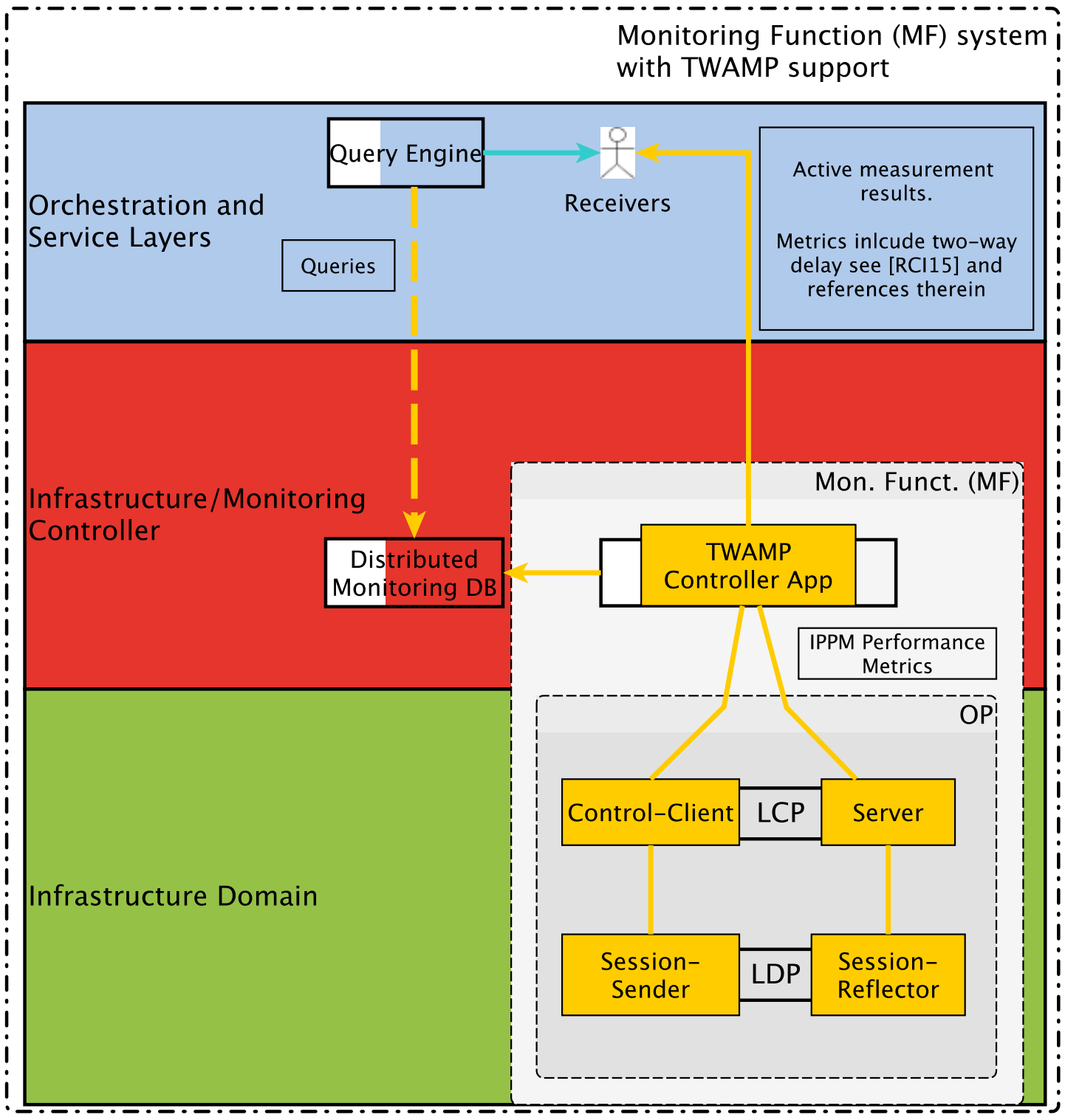

*Figure 5.15: Mapping of the TWAMP monitoring function in the infrastructure layer of the functional architecture.*

As illustrated, TWAMP active measurement results can be stored in the distributed monitoring DB, although this is beyond the scope of the standardization work in IETF (as it relates to implementation aspects). From the distributed monitoring DB, such measurement results can become available to the query engine, although this not planned at this stage for prototyping in the context of the WP4 effort.

### 5.4.3 Support to SP-DevOps processes and objectives addressed

Legacy OAM protocol support in the case of TWAMP provides for the observability, troubleshooting and verification processes. In particular, through the programmability added via the data model in YANG, RC3 and RC8 are addressed, while VNF development support (RC11) is also in scope [D41]. Further, in-network troubleshooting (RC9) and troubleshooting with active measurement methods (RC10) are also addressed. Addressing these research challenges is part of achieving OL1, OL2, and OL3 [D41]. The approach relates to following D4.1 requirements: NL1.3, NL1.5, NL2.1, and NL4.5 [D41].

### 5.4.4 Evaluation and implementation

The TWAMP data model is defined through Unified Modeling Language (UML) class diagrams and formally specified using YANG in [RC115]. The YANG data module is verified for use by three vendors and one operator, as co-authors of the document, both in terms of correctness and compliance with the YANG modelling standards of IETF. More importantly, the draft data model reflects their current implementations in the field, as well as the possibilities outlined in this deliverable for use in a software-defined infrastructure deployment based on the UNIFY architecture and its associated technologies.

### 5.4.5 Next steps

The first version of the TWAMP data model [RC115] was presented at IETF 92 as the result of joint work of the ad-hoc design team. The draft presentation led to active discussion during the IETF meeting and later on the mailing list. So far all comments are constructive and supportive. It appears that there are good chances of adopting this draft as a Standards Track document of the IPPM working group charter, although this will require continuous effort till the end of the UNIFY project. As of this writing, an updated revision of the draft is under development and planned for submission and presentation in the upcoming IPPM meeting at IETF 93.

## 5.5 Rate Monitoring

Dynamic service provisioning and performance management rely on the ability to perform scalable and resource-efficient measurements that can serve as input to richer statistical models of the observed network aspect, so that such information can be used in a proactive manner for resource management and quality assurance.

### 5.5.1 Technical description

The link utilization and rate monitoring MF implements a scalable congestion detector [PKR15] based on the analysis of the traffic rate distribution on individual links at different time scales. For these purposes, we employ a statistical method for node-local analytics based on the use of two *byte* counters for storing the first and second statistical moment ($s_1 = \sum x_i / n$, $s_2 = \sum x_i^2 / n$) observed under $n$ time intervals $\Delta t$. Assuming a log-normal distribution $f(\mu, \sigma)$ for the observed rates, the parameters $\mu$ and $\sigma$ of the distribution can be estimated from the statistical moments:

$$\begin{cases} \hat{\mu} = \ln M - \frac{1}{2}\hat{\sigma}^2 \\ \hat{\sigma}^2 = \ln\left(1 + \frac{V}{M^2}\right) \end{cases} \quad (1)$$

where $M = s_1, V = s_2 - M^2$ are the sample mean and variance. Once the estimates are obtained, the cumulative density function (CDF) $F(\hat{\mu}, \hat{\sigma})$ can be used for detecting increased risks of link overload by inspecting the percentiles. The use of the CDF allows for assessing the probability of observing a traffic rate that overshoots the link capacity, and can therefore be used for identifying shorter or more persistent congestion episodes (e.g. when packets are lost) at various time scales. The estimated parameters of the lognormal distribution can be used as input to a resource manager supporting (re)allocation decisions during deployment and operations of software-defined infrastructures. The counters used are associated with the defined flow space (e.g. OpenFlow rules), meaning that the rate monitoring can be applied to more or less arbitrary aggregation levels, except for individual flows that exhibit sparse traffic characteristics with short bursts.

#### Design choices and practical considerations

Existing approaches (e.g. SNMP [RPR02] polling, sFlow [PPH01]) usually involve forwarding of measurement information to dedicated monitoring equipment for further processing, which impacts the scale at which monitoring can be efficiently performed and thereby the overall network observability. Querying of counter statistics locally at high rates enables high accuracy in the captured aspects of the traffic behavior, with significantly lower overhead forwarding raw measurements to a dedicated management station for further analysis. By varying the querying intervals $\Delta t$, high-granular monitoring information can be obtained in a flexible and scalable manner, without the cost of constant high rate sampling of the counters. Local congestion warnings and estimates can then either be forwarded automatically at user-defined intervals towards relevant receivers in the architecture (e.g. to the distributed storage), and/or be retrieved via direct requests or via the recursive query language. Stored estimates can be used for further processing to generate trends or high-level views of the global performance at the service or infrastructure level.

### 5.5.2 Architectural mapping

**Functional architecture:** The rate monitoring MF operates via a locally deployed OP connected to the MMP of a UN in the infrastructure layer (Figure 5.16). A MF control app is running in the controller layer of the infrastructure, for the purpose of forwarding incoming monitoring information and configuration requests via the Ca-Co reference point down to the active OP of the functional architecture. Communication between the MF control app and the OP involves the Co-Rm reference point (Figure 5.16). Placement and instantiation of one or several rate MFs is part of the decomposition and mapping operations in the Orchestration layer based on resource information (e.g. which node has capability of running the MF, and current resource availability for running such a function). Monitoring information can be forwarded via DoubleDecker  to specified recipients, e.g. a DB accessible from the query language engine.

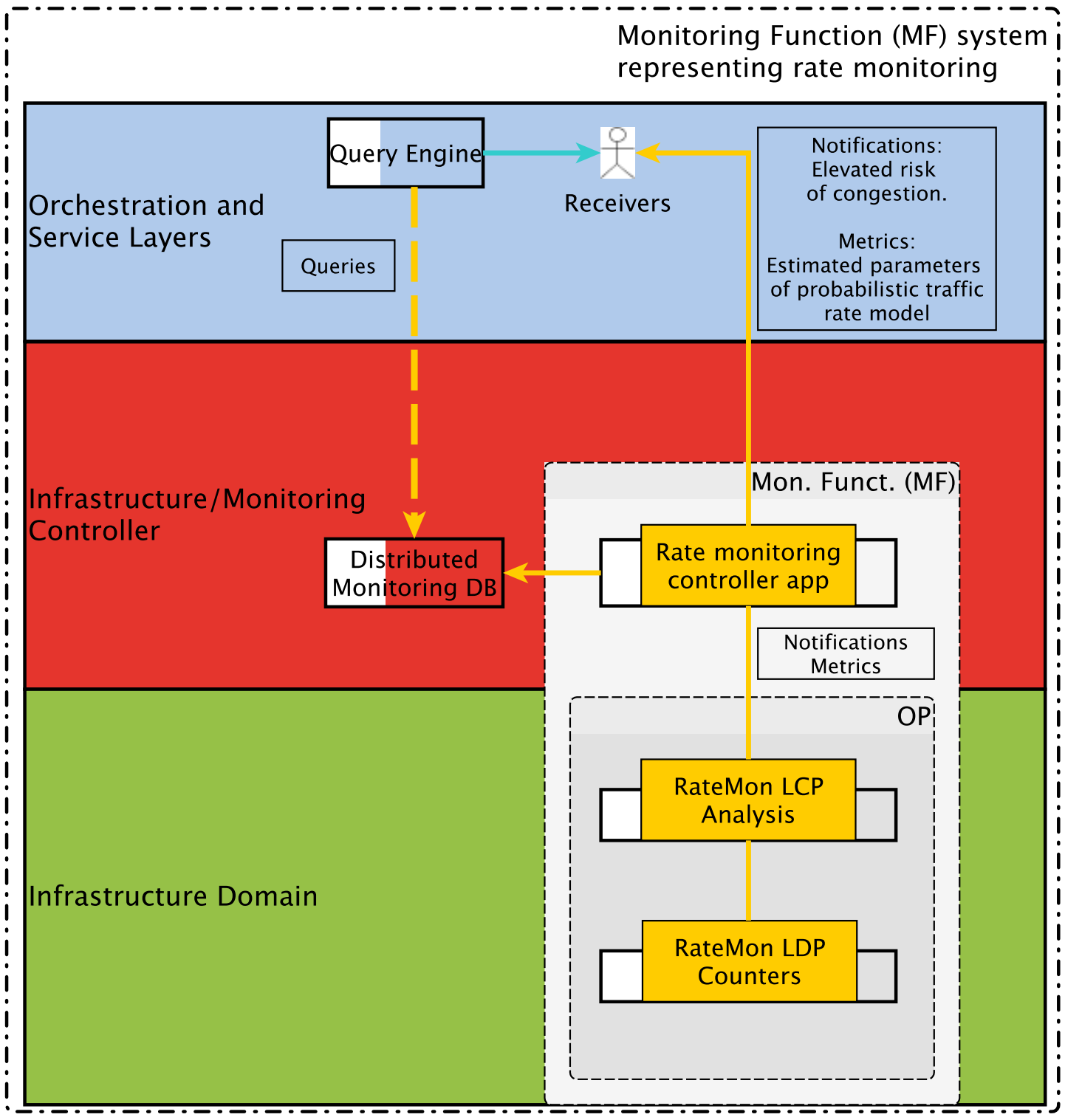

*Figure 5.16: Mapping of the rate monitoring function in the infrastructure layer of the functional architecture.*

**UN-mapping:** A rate monitoring MF may consist of one or several OPs depending on the monitoring needs. For one OP, the node-local statistical modeling takes place in the LCP of the residing in the MMP, and is based on the reading of two counters in the LDP for storing the first and second statistical moments of the observed flow(s) (Figure 5.17).

This deviates from the current practice of using a single counter for the byte rates, and hence the available set of counters in the VSE may need to be extended with a squared-sum byte counter. Additional counters may cause a slight increase of the processing time in the LDP, but, enable light-weight node local modelling via method-of-moments estimates which can be reported to the management layers in a compact form as described above, thereby increasing the level of network observability in a scalable manner. The LCP also implements functions for reading LDP counters. Depending on the scope of monitoring, this can be done in three ways:

- The MF models traffic rates on logical entities running in the VSE which requires accesses to counters in one or several LSI, the LDP consists of an extended variant of OpenFlow (i.e. with one additional counter), whereas the LCP implements a miniature OpenFlow-controller for retrieving counter values similar to a micro-OpenFlow controller (however, in the current design of the UN only one OF channel is available, which may limit the rate with which counter values can be requested).
- The MF is used for modeling traffic rates on the physical NICs, file access (such as /proc) or access to shared memory are two ways of implementing counter access.

- The MF may be instantiated as a MF VNF, implementing the monitoring behavior and rate modelling inside the VNF, while using an LCP running in the Generic Monitoring Manager for information exchange with the rest of the architecture. In all three cases, it is assumed that monitoring and configuration is done via DoubleDecker.

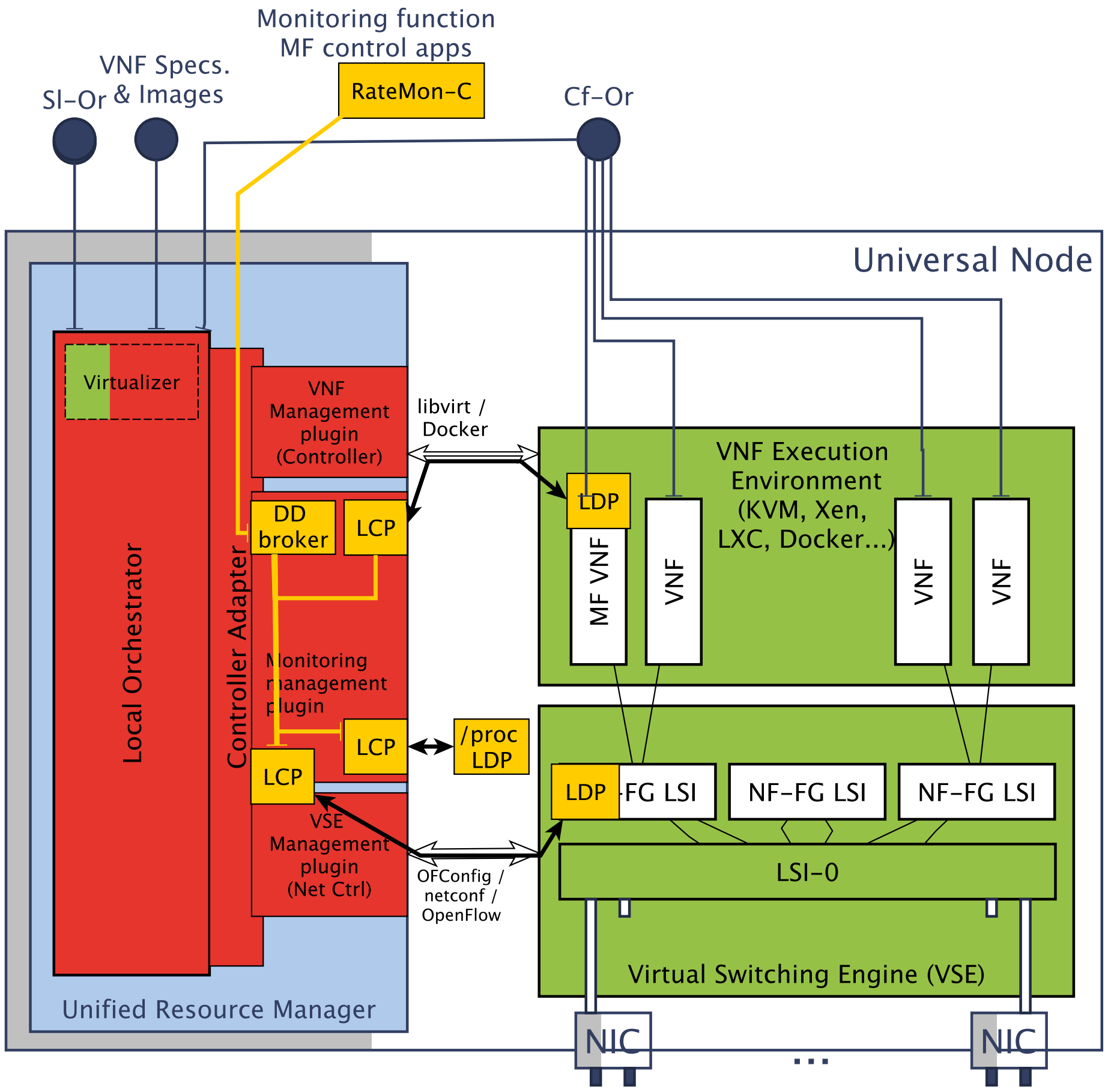

*Figure 5.17: Three examples of how the rate monitoring OP can be mapped in the UN. For prototyping, reading interface statistics using system calls or reading from files would be candidate approaches.*

### 5.5.3 Support to SP-DevOps processes and objectives addressed

The rate monitoring tool is an instance of probabilistic in-network monitoring methods addressing RC1, and as it can be used for detecting congestion it also addresses RC9. Addressing both these research challenges is part of achieving O4.2 and O4.3. The approach relates to several requirements in D4.1: NL1 (NL1.2-1.5), NL2, NL4 (NL4.2-4.7), OL2, OL3 (OL3.1-3.2) and OL4(4.1-4.3)

The tool supports both observability and troubleshooting processes. The observability process is supported in terms of:

- statistics on the observed traffic rates (or link utilization) represented in compact parameter form at different time scales, supporting resource management and deployment processes;
- reports on the risk of congestion supporting dynamic resource management and troubleshooting.

In the case of troubleshooting, in-network congestion detection can be used for triggering:

- local actions for mitigating persistent congestion episodes at critical time scales,
- localization and root-cause analysis at higher levels for correlating input from several MFs across the network in case of more severe failures.

### 5.5.4 Evaluation and individual prototyping results

The approach has been initially evaluated with respect to estimation error and congestion detection rate at varying time scales in a stand-alone simulator framework implemented in Scala [Scala], using data from 1Gb/s and 10Gb/s links. One of the most important points of estimating the rate distributions based on simple statistics produced at high rate, is that we can query those statistics at significantly lower rates, and still obtain reasonable estimates of the underlying behavior. For example, from our experiments performed on the 10Gb/s link, we have observed an error in the order of only a few hundred Mb/s in the value of the 99th percentile, when estimating the lognormal distribution from observations made every $\Delta t = 0.3s$ during a time period of 5 minutes; see [PKR15] for more details.

Standard practice for identifying increased bandwidth consumption is based on low-frequency counter inspections and reporting when the average exceeds a fixed threshold, or by manual inspection. An example of traffic rate averages from the 1Gb/s link over consecutive 5 minute intervals over the last 9h of a single day is shown as the top part of Figure 5.18. Using such low-resolution averages often lead to missed congestion episodes as well as false alarms, as these averages are far below the link capacity and the detection threshold is difficult to set. Instead, by inspecting the CDF of the obtained parameter estimates the risk of congestion can be predicted at varying time scales.

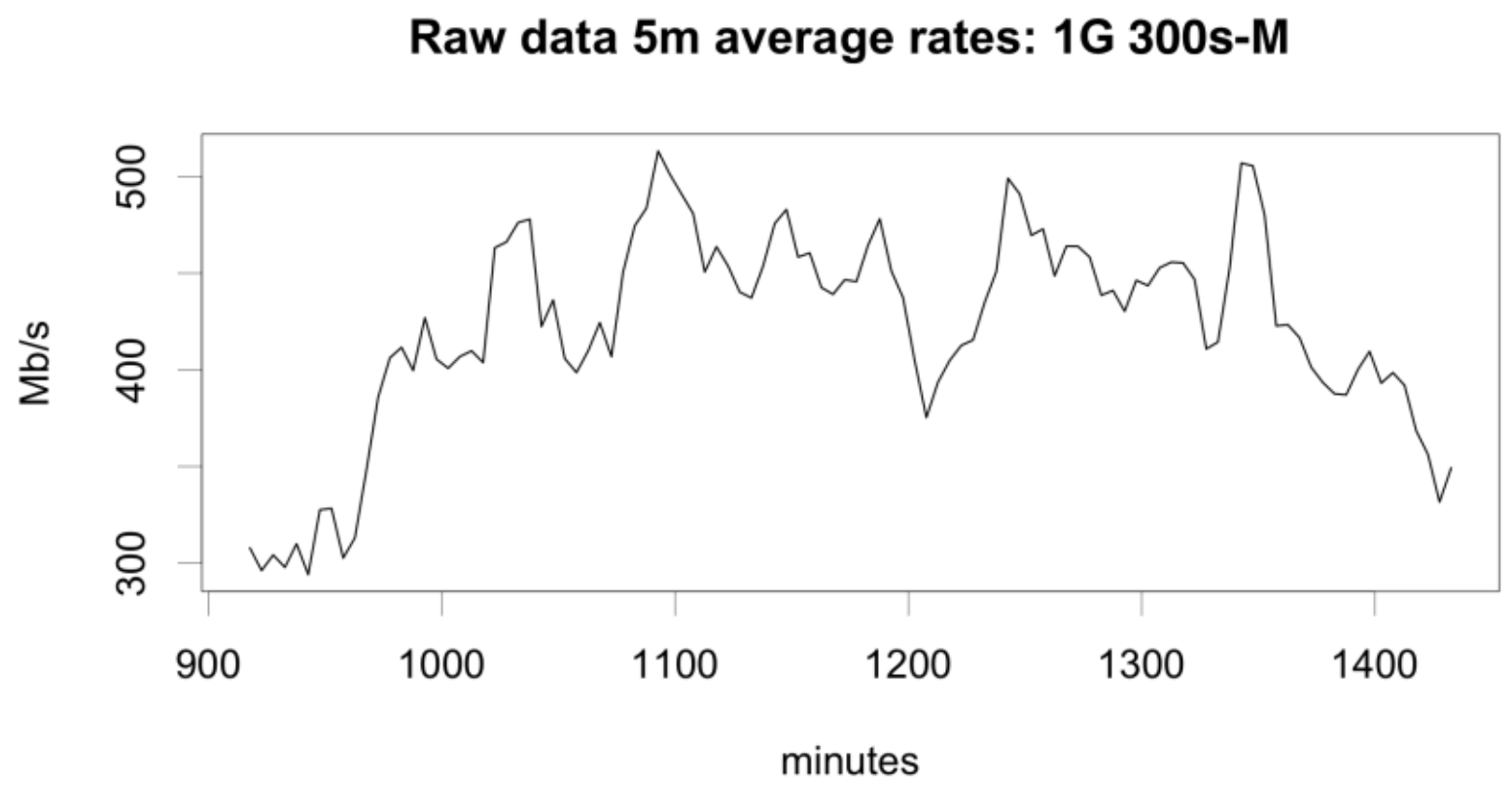

*Figure 5.18: Time series of 5 m averages over one day on a 1G link for assessing congestion risks.*

For comparison, we see in Figures 5.19 and 5.20 that the risk of exceeding the link capacity at estimation intervals of $\Delta t = 300s$ and $\Delta t = 0.3s$, respectively. Compared to Figure 5.18, we can in the clearly observe episodes of increased risk of micro-congestions developing over time at both time scales.

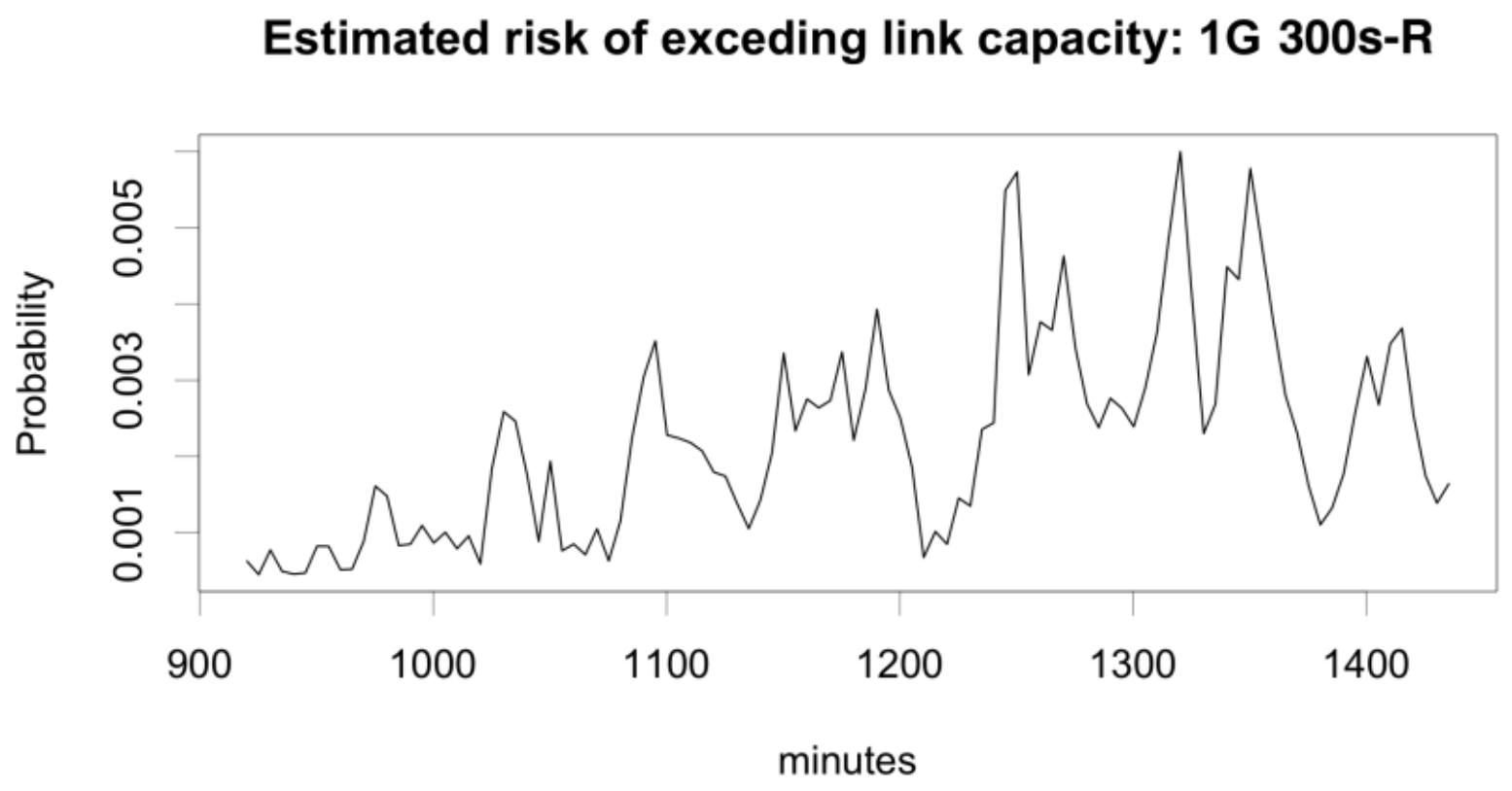

*Figure 5.19: Risk estimates on the same 1G link based on 5 m observation intervals.*

Evaluation using a naive congestion detector based on the CDF, yields a success rate of over 98% when detecting episodes of high congestion risk when updating the estimates at $\Delta t = 0.3s$ intervals.

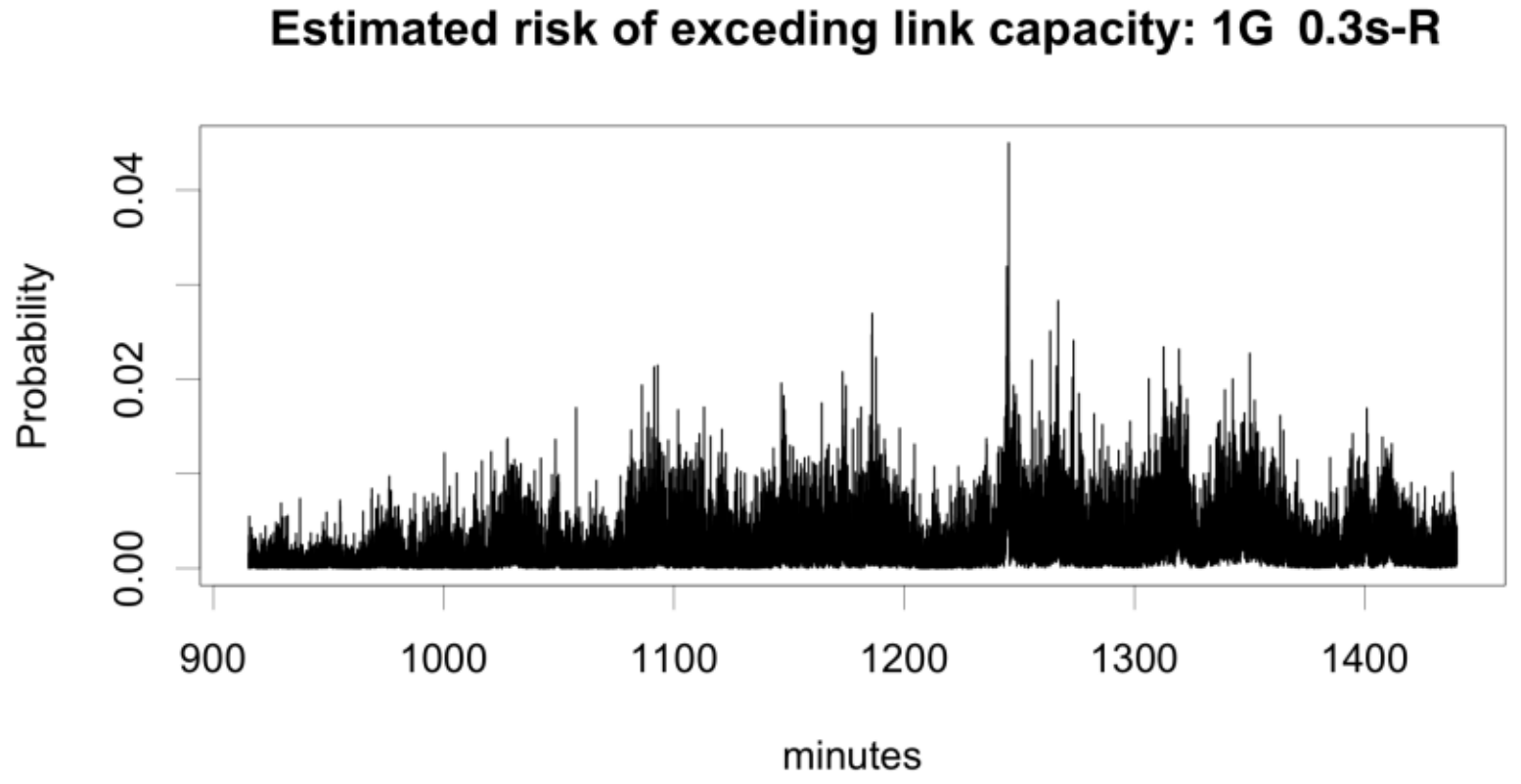

*Figure 5.20: Risk estimates on the same 1G link based on 0.3s observation intervals.*

### 5.5.5 Next steps

Next steps involves refinement of the rate monitoring model, involving a more robust detection mechanism for troubleshooting support and self-adaptive behavior for monitoring at various time-scales. Preliminary results obtained from initial tests with adaptive scaling mechanisms can reduce the rate of local measurements with 75% and produce a 100% congestion detection rate based observations made at varying time-scales between 30s to 30ms, compared to constant rate monitoring at 30ms intervals. The method was part of the Y1 WP4 integrated demo, and is subject for further integrated protoyping activities within the ESCAPEv2 [D32] prototyping framework.

## 5.6 Successive E2E Monitoring of Link Delay and Loss

Common limitations of existing methods for measuring delay and loss are in general that (*i*) the packet loss can be accurately derived only at the end of a flow session (e.g., OpenNetMon [NAD14], FlowSense [CYU14]), and (*ii*) packet injection for measuring delay does not reflect the delay perceived by the monitored flow or service, because of fixed packet sizes [CYU14] and different traffic forwarding priorities [LDE07]. Moreover, these approaches are mainly designed for deriving end-to-end link metrics from a central point. Centralized approaches are not suitable for large-scale monitoring since the communication overhead increases substantially with the number of measurements and network size. Instead, we propose a decentralized method for deriving intermediate link metrics based on one-way measurements along a monitored path, which can be applied to flows of varying aggregation levels, and without measuring each individual link explicitly. Here, the link can be either physical or virtual within a service, defined by a flow space. The measurements and counter values used for the statistical model can, compared to many existing methods, be derived at any time during a flow session and in a consistent manner [RST15].

### 5.6.1 Technical description

The decentralized link monitoring tool derives one-way delay and packet loss for each intermediate link $X_i$ on a monitored path from successive end-to-end measurements (Figure 5.21).

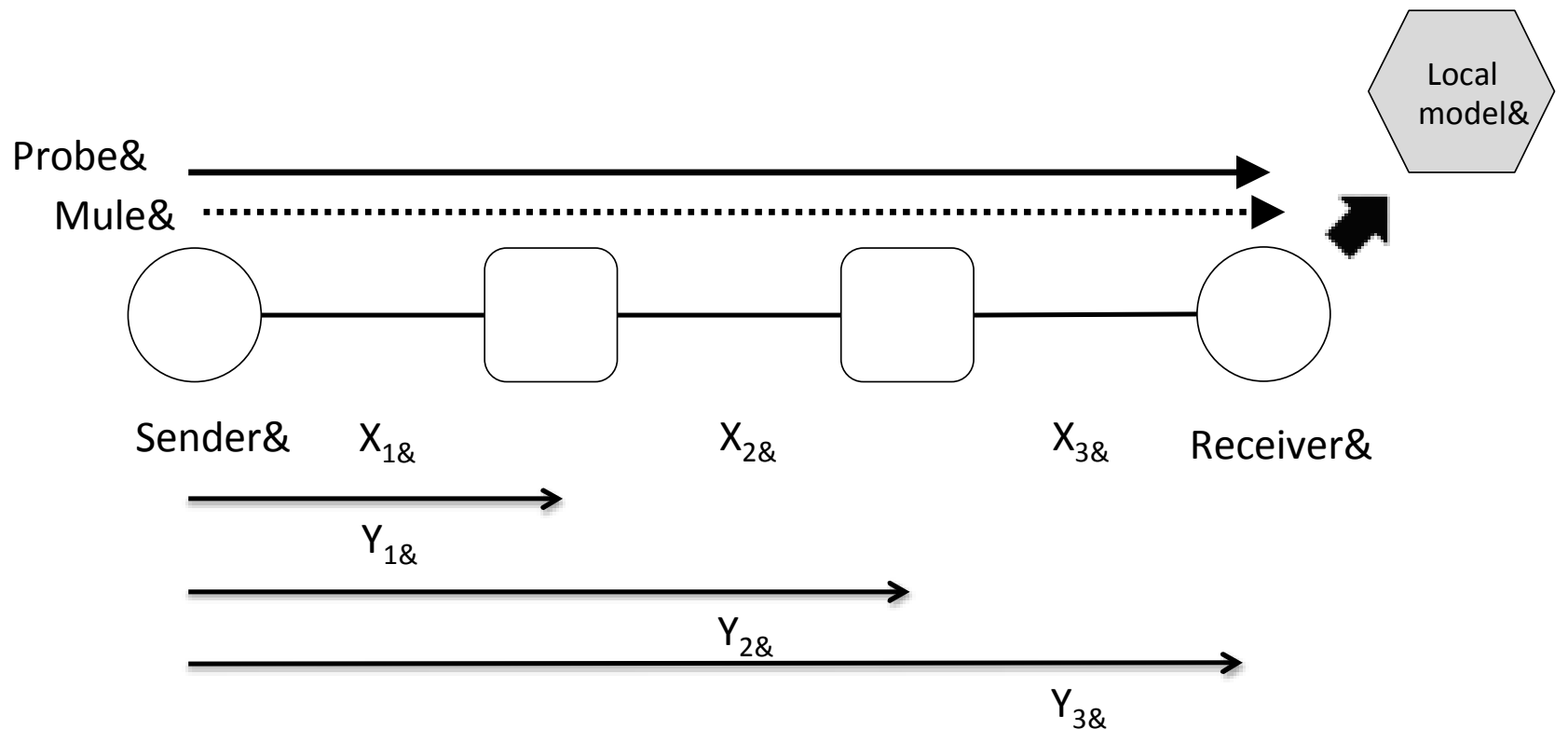

*Figure 5.21: Overview of the link monitoring approach*

The two end-points of the monitored path implement light computational capabilities for performing measurements and node-local analytics, while flow rules in intermediate nodes are exploited for packet forwarding, time stamping, and piggybacking. The measurements $Y_i$ are obtained by observing the trail of information (counter values and timestamps) a randomly selected packet from a flow leaves behind while being forwarded hop-by-hop between the end-points, and collecting the information by piggybacking on another packet (also referred to as a "mule" packet).

The sender node is responsible for randomly indicating when a forwarded packet should be used for measuring delay or loss or act as a mule packet. In addition to packet forwarding, the intermediate nodes update relevant timestamps and counters for measurement packets. Similar to the sender node, the intermediate nodes also perform relevant actions for adding information on to a mule packet before forwarding it towards the receiver. The receiver node performs processing of received information piggybacked by mule packets, and maintains the statistical model used for deriving link delay and loss.

Once collected, the measurements are used as input to a statistical model in the receiver node for estimating link delay and loss on intermediate links without explicitly measuring each link individually. During the measurement session, it is required that the monitored path is static and that the packets are identically forwarded. Link delay estimation is performed by pairwise timestamp comparison of successive link segment measurements $Y_i$ for the purpose of estimating the mean and variance for link $X_i$:

$$E(X_i) = E(Y_i) - E(Y_{i-1}) \quad (2)$$

$$Var(X_i) = Var(Y_i) + Var(Y_{i-1}) - 2Cov(Y_i, Y_{i-1}) \quad (3)$$

For packet loss estimation we considered two approaches: (*i*) information consistency, meaning comparison of piggybacked information, which allows to identify the link where the probe was lost (e.g. if the timestamps obtained along the monitored path are monotonically increasing or not), or (*ii*) letting the mules also read out existing OpenFlow counters for packets received and sent while traversing between the end-points. In either case, the loss model here used is:

$$E(X_i) = \frac{E(Y_i)}{E(Y_{i-1})} \quad (4)$$

Once the statistics for each individual link has been derived from the successive end-to-end observations, the mean and variance can be used for estimating the parameters of probability distributions that are often used for fitting monitoring data, such as Gamma for link delays and Bernoulli distributions for loss [AHE07][CFR05]. Instead of sending raw measurements for further processing and analysis, the estimates can be used for more advanced analysis, such as prediction and anomaly detection.

### Design choices and practical considerations

In general, the one-way delay model requires temporally synchronized nodes on the monitored path which can be done using various protocols (such as NTP) or methods [KR001][RS007]. The temporal granularity that can be captured by the statistical model depends on the clock precision available and the degree of synchronization. Compared to round-trip delays (that do not require temporal synchronization), one-way delay models can provide increased observability of the network (by modelling uplinks and downlinks separately), allowing for a more efficient and effective optimization of resource usage and consumption.

In the case of delay measurements, timestamps could be piggybacked on the measurement packet itself, meaning low packet overhead but with a small perturbation in the observed measurements due to increasing packet size and processing time. Instead, we considered piggybacking the information by using a mule packet sent after the measurement packet, meaning low packet overhead (if the mule is part of the stream and not specifically injected), and less measurement perturbation as the mule is independent from the measurement packet. This is more suitable for deriving not only link delay, but also loss.

The approach has been developed with respect to how flow entries and features from e.g. OpenFlow can be exploited for monitoring the service performance in combination with this particular statistical model. The method has so far been evaluated in NS3 with focus on statistical consistency, model accuracy, and robustness, given a piggybacking measurement strategy exploiting counter values and time stamping [OpenFlow14]. Implementation of necessary OpenFlow extensions is subject to future work and remains to be evaluated, based on the existing NS3 framework [NS3] and Mininet [Mininet].

In general, the necessary mechanisms for storing timestamps (using the IDLE_TIMEOUT [OpenFlow14]) upon a flow entry match and piggybacking (based on tagging) already exist in OpenFlow. However, we see that two types of actions need to be implemented. First, for increased robustness of the method, we suggest an additional action for storing a variable used for identifying a measurement packet. Second, a reference and an action for accessing stored information (i.e. timestamp, loss counter, or identifier variable) are needed. For the mule packet, the header needs to match the same fields as the observed packets of the monitored flow, and has also to be matched against a suitable field in order to read and execute a piggybacking action. Matching or identification of a mule packet can be performed by reusing an unused field of the packet header.

### 5.6.2 Architectural mapping

**Functional architecture:** The monitoring MF operates locally in the infrastructure layer in a series of connected nodes on a monitored path, consisting of OPs of varying complexity. The monitored path consists of a sender and receiver node, with passive forwarding intermediate nodes in the infrastructure layer (Figure 5.22).

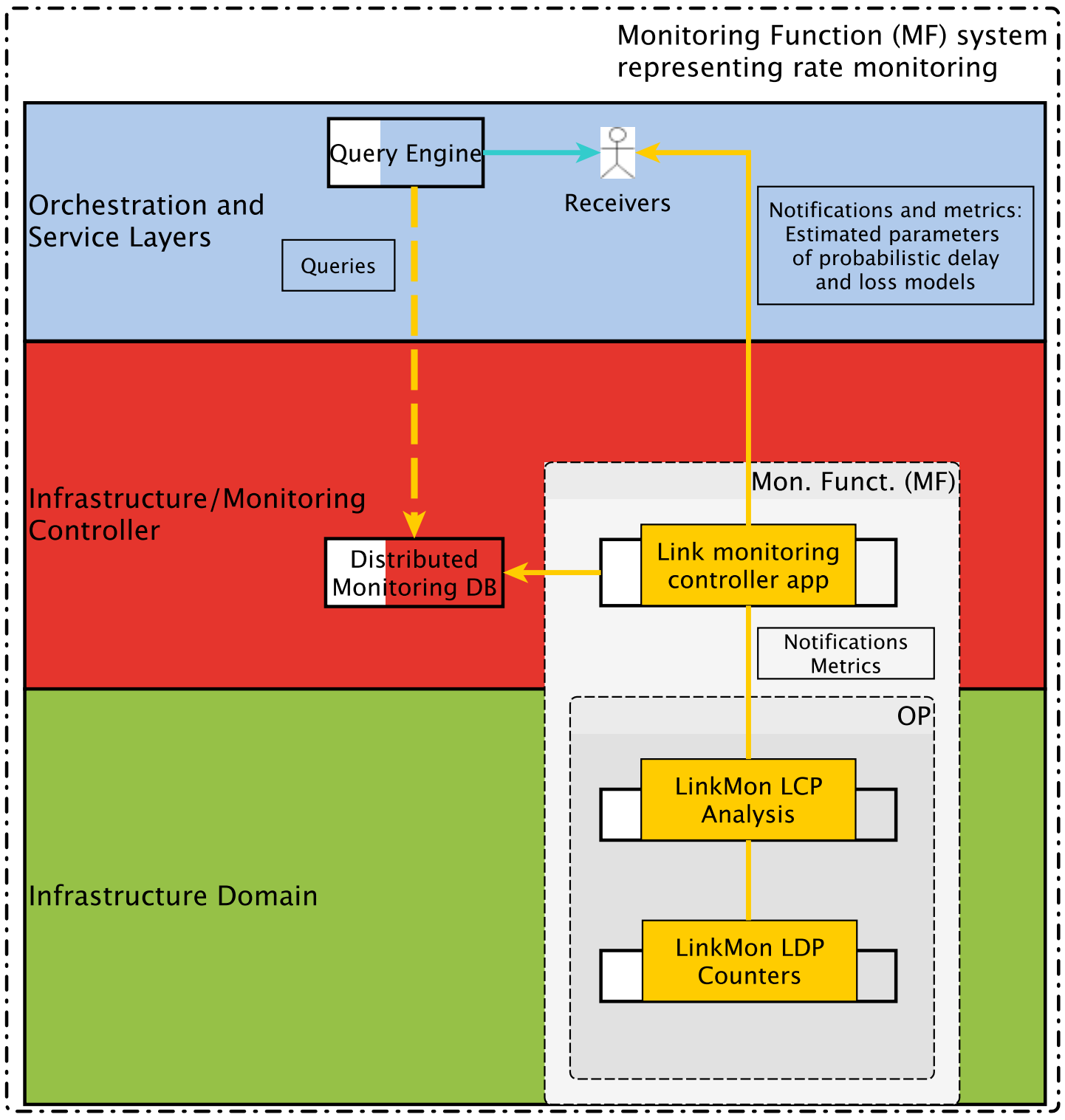

*Figure 5.22: The link monitoring function mapping to the infrastructure layer of the functional architecture.*

All the nodes included in the MF are managed by a link monitoring MF controller app, which manages configuration (received from the rest of the architecture over the Ca-Co reference point) of each type of node (sender, receiver, or intermediate) as well as forwarding of notifications (change detection, or when a measurement is finished in line with required precision) and parameter estimates of individual links. The control app is running in the infrastructure controller layer, either on a UN that is included on the monitored path, or on an auxiliary node. The exact placement and instantiation of one or several MFs is part of the decomposition and mapping operations in the Orchestration layer based on available resource information (e.g. which node has capability of running the MF, and current resource availability for running such a function). Similarly to other tools, the monitoring information (Figure 5.22) is pushed on the DoubleDecker (Section 5.1) and forwarded to specified recipients (e.g. via the MEASURE language) - this may include storage in a DB accessible by e.g. the query language as described in Section 5.2 and 5.3, respectively.

**Universal node:** There are three types of OPs running in the nodes on the monitored path: the sender, that controls the measurement intensity, the receiver that performs the modelling of the one-way measurements, and intermediate nodes that forward the measurement and mule packet-pairs and perform necessary actions for accessing stored information and piggybacking in the sender and intermediate nodes.

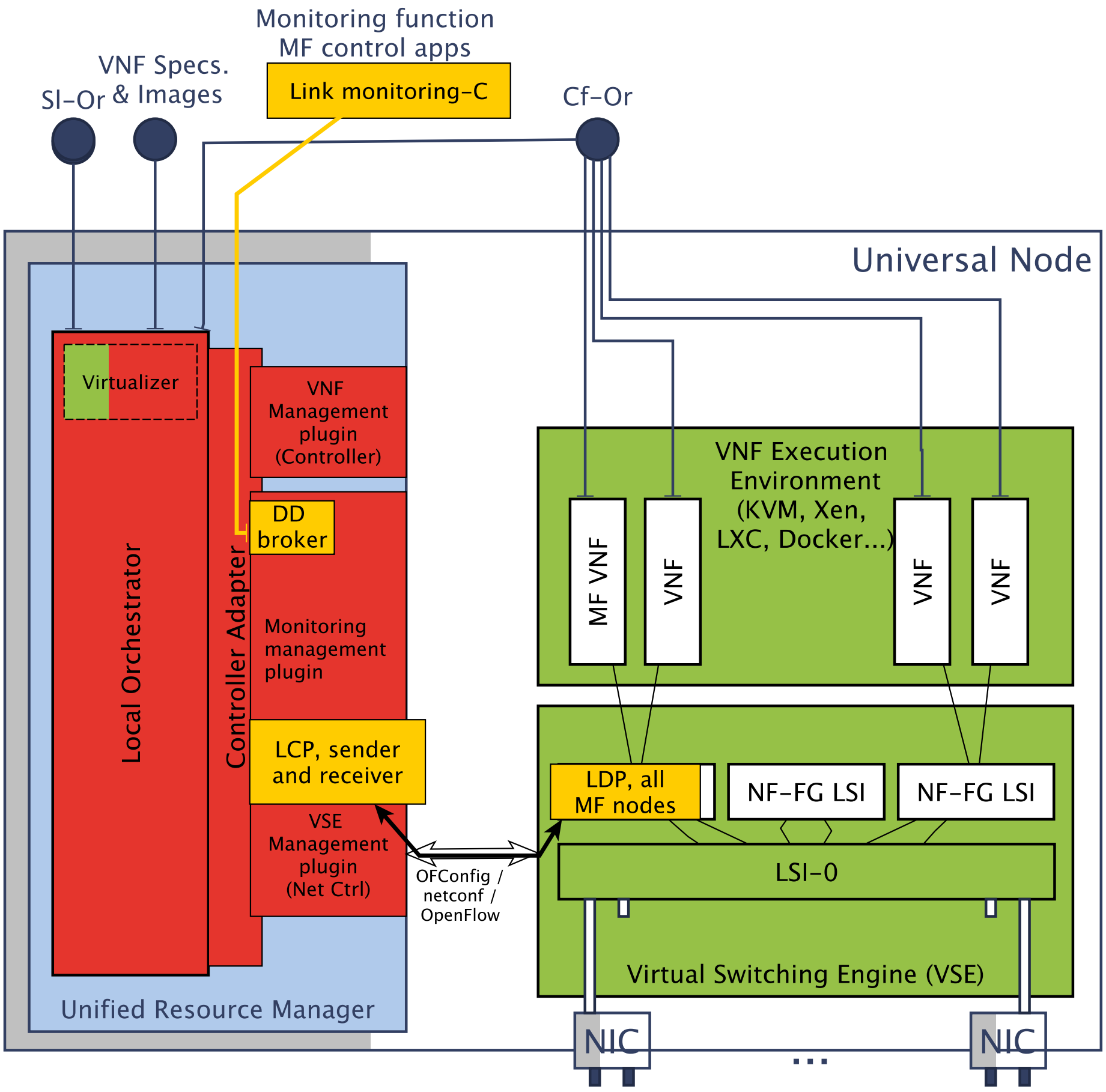

*Figure 5.23: OP components mapped to the UN for E2E link delay and loss monitoring. Note that the LCP component is only implemented for the sender and receiver nodes, whereas the LDP component is implemented for all nodes on the monitored path.*

The sender and receiver OPs both consist of active LCP and LDP components that may read statistics and handle measurements over the OpenFlow channel in the UN (Figure 5.23). The intermediate nodes implement the LDP part in terms of necessary actions and rules for storing timestamps and perform piggybacking [OpenFlow14][RST15], installed by the controller app (similar rules/actions are also installed in the receiver and sender).

Configuration updates for the sender node are received by the local orchestrator on the UN, as well as notifications and parameter estimates from the receiver, are sent over DoubleDecker (see Section 3.3). Figure 5.23 illustrates the mapping of the monitoring function in the UN.

### 5.6.3 Support to SP-DevOps processes and objectives addressed

The delay and loss monitoring tool is another instance of probabilistic in-network monitoring methods addressing RC1 as described in D4.1. In the longer term, this tool is aimed for detecting drift or changes in the observed link delay or loss based on the described detection approach, thereby partially addressing RC9 in terms of troubleshooting support. Addressing both these research challenges is part of achieving O4.2 and O4.3. Specifically, this tool contributes to increased network observability as intermediate link metrics can be derived from successive end-to-end measurements without explicitly setting up individual measurements sessions for each link on a monitored path. The approach relates to several requirements in D4.1: NL1 (NL1.2-1.5), NL2 (NL2.1), NL4 (NL4.1-4.8), OL2, OL3 (OL3.1-3.2) and OL4(4.1-4.3).

Hence, this tool is designed to support both observability and troubleshooting processes, in terms of

- providing statistics on the observed delay and loss on a monitored path represented in compact parameter form, supporting resource management and deployment processes;
- can be used to report individual link statistics for the purpose of dynamic resource management and troubleshooting
- triggering local actions for handling locally detected performance degradations on one or several links used by service;
- troubleshooting processes at higher levels for analyzing and correlating input from one or several MFs across the network in case of more severe failures or performance degradations.

### 5.6.4 Evaluation and individual prototyping results

The approach has been evaluated in terms of parameter estimation accuracy for delay and loss in relation to how the measurements are obtained in an assumed SDN setting – initial simulation results were presented in [RST15]. The simulation setup (implemented within the ns-3 framework) comprises a monitored path of six nodes (i.e. linear topology) connected via 1Gbps links, and a traffic generator creating UDP flows according to statistical characteristics observed in real world Internet traces [CAIDA] and previous work [MKA03]. The measurement intensity, represented by fraction $\alpha$ of the traffic, can be configured as a trade-off between accuracy and overhead cost. Fraction $\alpha$ reflects the importance of obtaining measurements and is varied in the simulation in steps of $0.05$ up to $0.5$ , and is used for randomly selecting, managing, and forwarding packet-pairs consisting of measurement and mule packets. Other parameter settings worth mentioning are (*i*) a probabilistic error model causing packet loss of a per-link-fraction of $0.04$ packets out of the total traffic on average, (*i*) packet inter-arrival times (which determine the intervals of packets reaching the sender node) drawn from a log-normal distribution, (*iii*) packet sizes drawn from the empirical CDF created from the data obtained in [RST15], and (*iv*) link delay following a Gamma distribution (see e.g. [MKA03]).

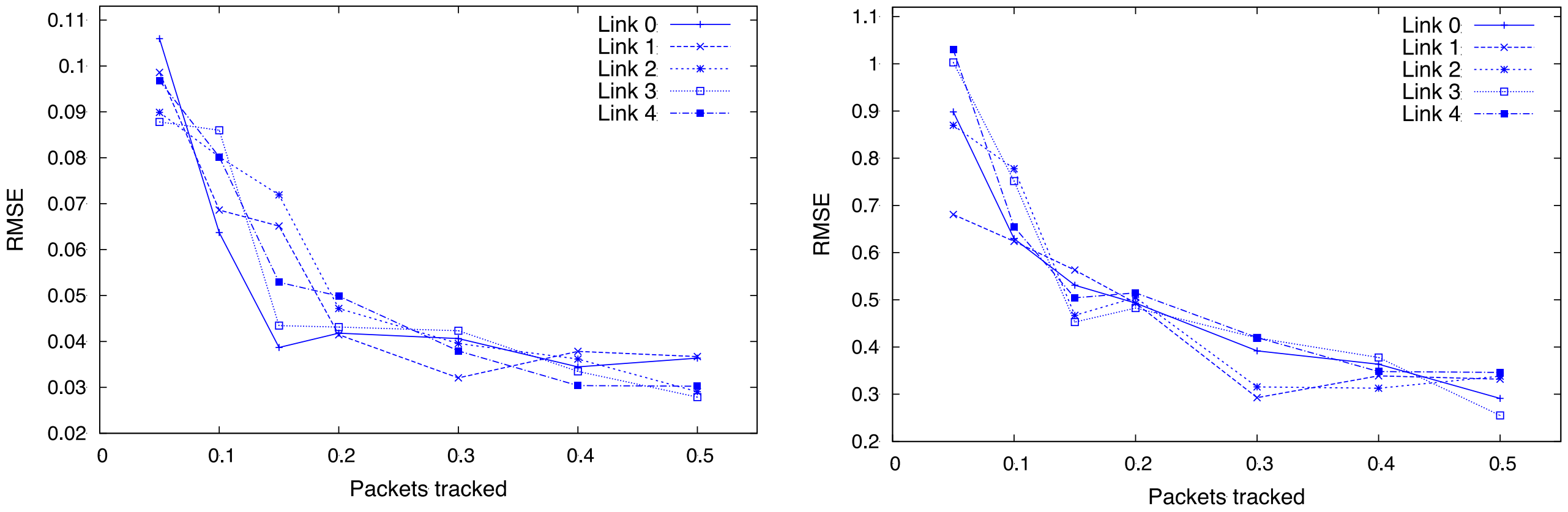

*Figure 5.24: Error of link delay estimation for different α values (fraction of observed packets) ranging from 0.05 to 0.5, with estimated mean (left) and variance (right).*

As indicator for the accuracy of the estimation models the Root-Mean-Squared Error (RMSE) between ground truth and estimated values has been calculated over 20 simulation runs conducted for each setting. The accuracy of the estimated mean and variance for link delay (Figure 5.24) improves continuously with increasing fraction of tracked packets for all individual links.

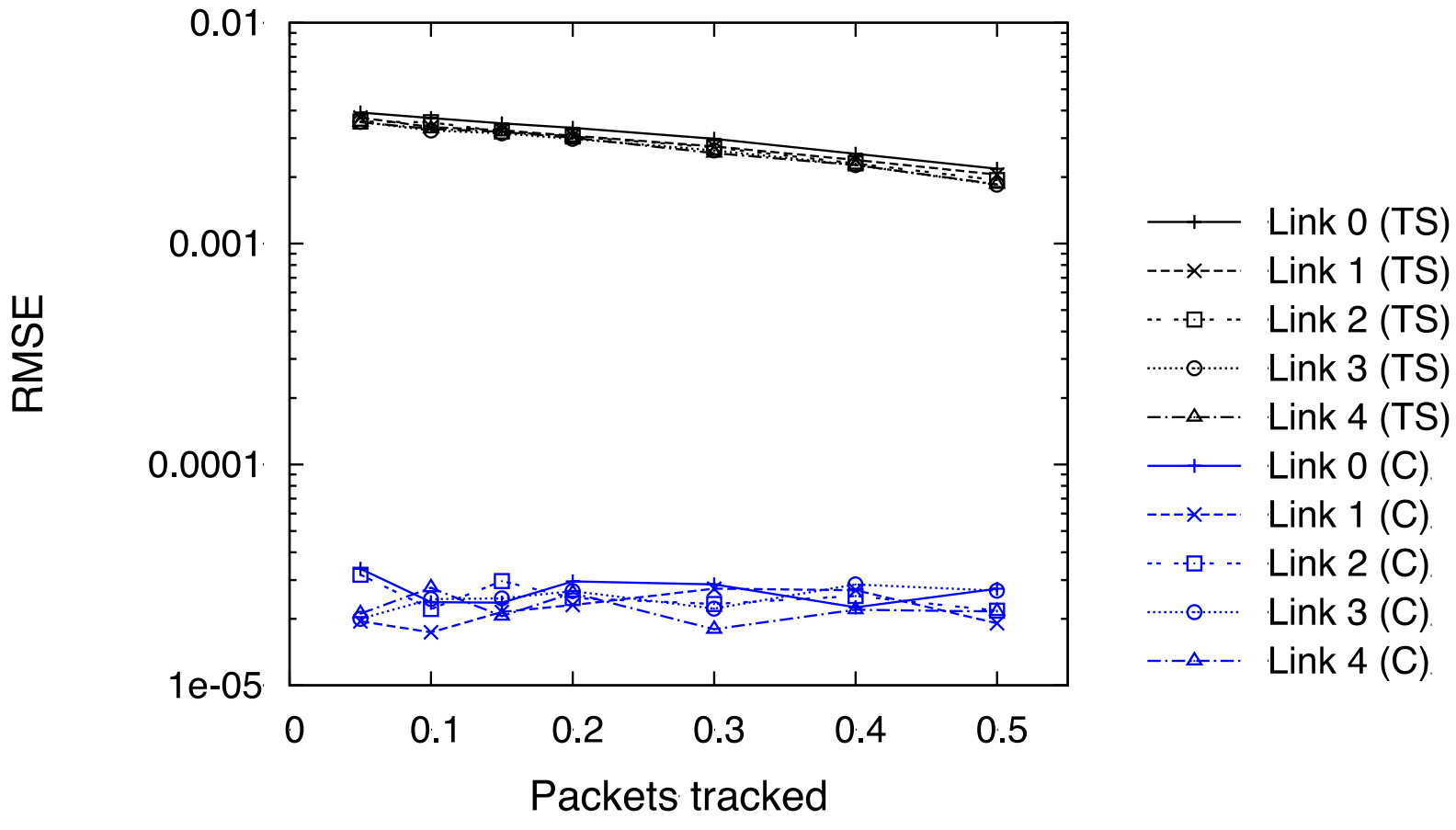

*Figure 5.25: Error of packet loss estimation for different α values for timestamp-based (abbreviated as TS) and counter-based (abbreviated as C) loss model.*

Note that the accuracy strongly depends on the number of samples, which in turn depends on the number of generated packets and the simulation time [RST15]. Here, an acceptable RMSE is achieved when tracking at least $30\%$ of the packets (out of about $108000$ packets forwarded in total), whereas in practice it is expected that a tracking fraction of less than $10\%$ is sufficient at normal or high traffic volume.

Moreover, Figure 5.25 shows the RMSE between the loss model and the ground truth when employing the information-consistency or counter-based measurement strategies (see Section 5.6.1). The result indicates that the counter-based method is superior in accuracy when piggybacking counter information between the end-points at any point in time during the monitoring session. However, exploiting already existing information carried by the mule packet for consistency assessment (here based on the number of collected timestamps along the path) can yield an acceptable RMSE as well, without the need for piggybacking and processing of packet counter values for loss estimation.

### 5.6.5 Next steps

So far, the development of our method has been focused on supporting the increased observability process in a scalable manner. Theoretically, in-network link monitoring can scenario reduce measurement overhead from $O(n^2)$ to $O(n)$ [RST15] compared to centralized approaches exploiting SDN controller messages (e.g. [CYU14]). We will refine the approach with respect to modelling robustness and measurement strategy under varying network conditions. The method has been tested under strict assumptions of the network behavior, with the advantage of eliminated need for signalling and thereby reduced processing overhead in involved nodes. However, for increased robustness under more realistic conditions, signalling via manipulation of packet headers will be considered as part of future work. Further, we will develop the method towards supporting automated monitoring and troubleshooting workflows.

## 5.7 Efficient Packet Loss Estimate - EPLE

EPLE is a monitoring function that estimates packet loss for aggregated flows in OpenFlow switches in a manner that is efficient from both dataplane (no probe packets) and control plane (piggyback on OpenFlow messages and relieve the controller from analyzing each individual flow contained within an aggregated descriptor) perspectives.

### 5.7.1 Technical description

As outlined in D4.1 [D41] and [WJO14], our work was inspired by FlowSense [CYU13] for low-cost monitoring and DevoFlow [JMO13] in terms of using an automated devolving mechanism. Flows described with wildcard rules typically do not expire automatically, and therefore the FlowSense method of using the OpenFlow OFPT_FLOW_REMOVED message that is automatically generated at flow expiration cannot be used for such flows. The validity of packet loss calculations made by directly comparing per-flow packet counters is negatively affected by grouping flows using wildcard rules in two ways: (1) There is no guarantee that wild-carded flow definitions at one Forwarding Element (FE) correspond to the same definition on other FEs, i.e. wild-carded flow definitions do not reflect the same set aggregated microflows. This can be caused by differing flow definitions or use of microflows at intermediate FEs, diverting parts of the original wildcarded flow to other network paths. In such cases, comparison of per-flow counters is meaningless. (2). Receiving counters asynchronously at the end of the flow-lifetime (i.e. an implicit push mechanisms, with counters piggy-packed by the flow-removal notification message from forwarding elements to the SDN controller as proposed by FlowSense) is an elegant way, since it creates zero overhead in terms of measurement traffic, and at the same time it ensures known state of all packets belonging to a flow. Pull-based mechanisms applied during flow lifetime, on the other hand, create synchronization problems due to the unsure state of in-flight packets when comparing packet counters for packet-loss measurements. DevoFlow is a method for managing flows in high-performance networks. It introduces the automatic devolving of wildcard rules onto more specific single flow rules for every flow that matches a wildcard rule. DevoFlow addresses throughput and link utilization. A naïve extension of DevoFlow with FlowSense ideas would mean that the controller could be overwhelmed by the amount of OpenFlow OFPT_PACKET_IN and OFPT_FLOW_REMOVED messages generated by automatically devolved flows and long-lived flows will occupy the flow table without providing statistics for significant amounts of time.

EPLE addresses these shortcomings by:

- supporting aggregated flows
- providing a policy-steered selection of flows to be devolved, as well as a timer and policy-based mechanism for selecting which flows should be communicated to the controller as candidate flows
- using a policy-steered timer that forces expiration of a flow from the table (through setting the hard_timeout parameter in the OpenFlow ofp_flow_mod structure associated with the devolved flow), thus forcing the transmission of a OFPT_FLOW_REMOVED message even while a long-lived flow is still being active in the network.

The monitoring function has three major components:

- monitoring and control app: estimates packet loss based on packet counter data received from the nodes. This is the EPLE-C component in the mapping to Functional Architecture (Figure 5.26)
- extended OpenFlow protocol implementation supporting additional messages for provisioning specific types of flows and transmitting additional counter statistics when flows are expelled from the forwarding table due to an timeout. This is the EPLE-LCP component in the mapping to Functional Architecture (Figure 5.26)
- virtual switch support for specific functionality in the flow tables that allow automatic creation of individual packet flow rules based on incoming packets that match an aggregated rule installed by the controller. This is the EPLE-LDP component in the mapping to the Functional Architecture (Figure 5.26)

EPLE calculates a simple packet loss estimate based on plain reading of packet counters at the beginning and at the expiration of a devolved flow. In the default setting, the counters are read at the ingress and egress points of the flow, although intermediary measurement points can also be configured in case more detailed investigations are desirable. There may be opportunities to augment EPLE by replacing the simple estimate with more advanced statistical analysis based on the work reported in Section 5.5.

EPLE is implemented as an application on top of the Service Abstraction Layer in OpenDaylight, and requires a switch that supports a devolve capability in the flow table that is able to automatically create specific entries for flows based on flows matching a wildcarded OpenFlow descriptor.

**Design choices and practical considerations**

EPLE attempts to address scalability issues related to monitoring provider networks with millions of active flows that belong to potentially tens or hundreds of thousands of service instances deployed through NF-FGs. Active measurements that insert packet probes in the datapath introduce significant bandwidth consumption overhead if all these services need to be monitored at highly granular time intervals. By employing user-generated traffic as probes, EPLE eliminates this problem.

However, improved scalability on the dataplane user traffic comes at the expense of increased traffic on the control plane towards the management systems of the infrastructure provider. Here, EPLE makes use of OpenFlow OFPT_FLOW_REMOVED messages which provide lower protocol overhead than SNMP notifications and are significantly less bandwidth and resource consuming than transporting the information using a request-response protocol. Naturally, the EPLE-related signalling would need to be accounted for when dimensioning the controller,

### 5.7.2 Architectural mapping

**Functional architecture:** In terms of mapping towards the UNIFY Functional Architecture, EPLE is a typical monitoring function that operates primarily at the Infrastructure layer. When deployed as an infrastructure layer monitoring function, EPLE needs to receive information from a Monitoring Controller with respect to which aggregated flows have to be monitored, with the option to enable it to monitor by default all the aggregated flows installed in a VSE. Each NF-FG in the Infrastructure Domain could contain a description of the aggregated flows that are configured on the ports contained in the graph, and this information could be transmitted to the Monitoring Controller. Data collected by EPLE is transferred to the Distributed Monitoring DB via the DoubleDecker Messaging system and queried from there via the Query Language. The access control is thus taken care of by the query language that collects data based on a particular NF-FG. The description of the aggregated flows from the NF-FG, along with the NF-FG identifier, is expected to be used as an automatically generated key for retrieving the data using the Query Language.

Since it is possible that EPLE monitors only certain flows, it is possible to deploy one EPLE instance for one NF-FG if so desired. In this scenario, the EPLE-C part would be deployed with the NF-FG, with a restriction that the execution environment needs to contain a controller and switch agent that support EPLE-LCP and EPLE-LDP capabilities. Multiple EPLE-C instances could thus co-exist and estimate packet losses on their own infrastructure slice defined through the respective NF-FG. Results from each of such instances are sent to the data store through the messaging system and made available through the query language. In this scenario, the operator of the infrastructure will be unaware of the packet losses measured by the EPLE-C instances. This deployment could be used as a way to validate the performance of the infrastructure from the service layer, provided that cooperation is offered by the infrastructure in terms of measurement capabilities in line with EPLE-LDP.

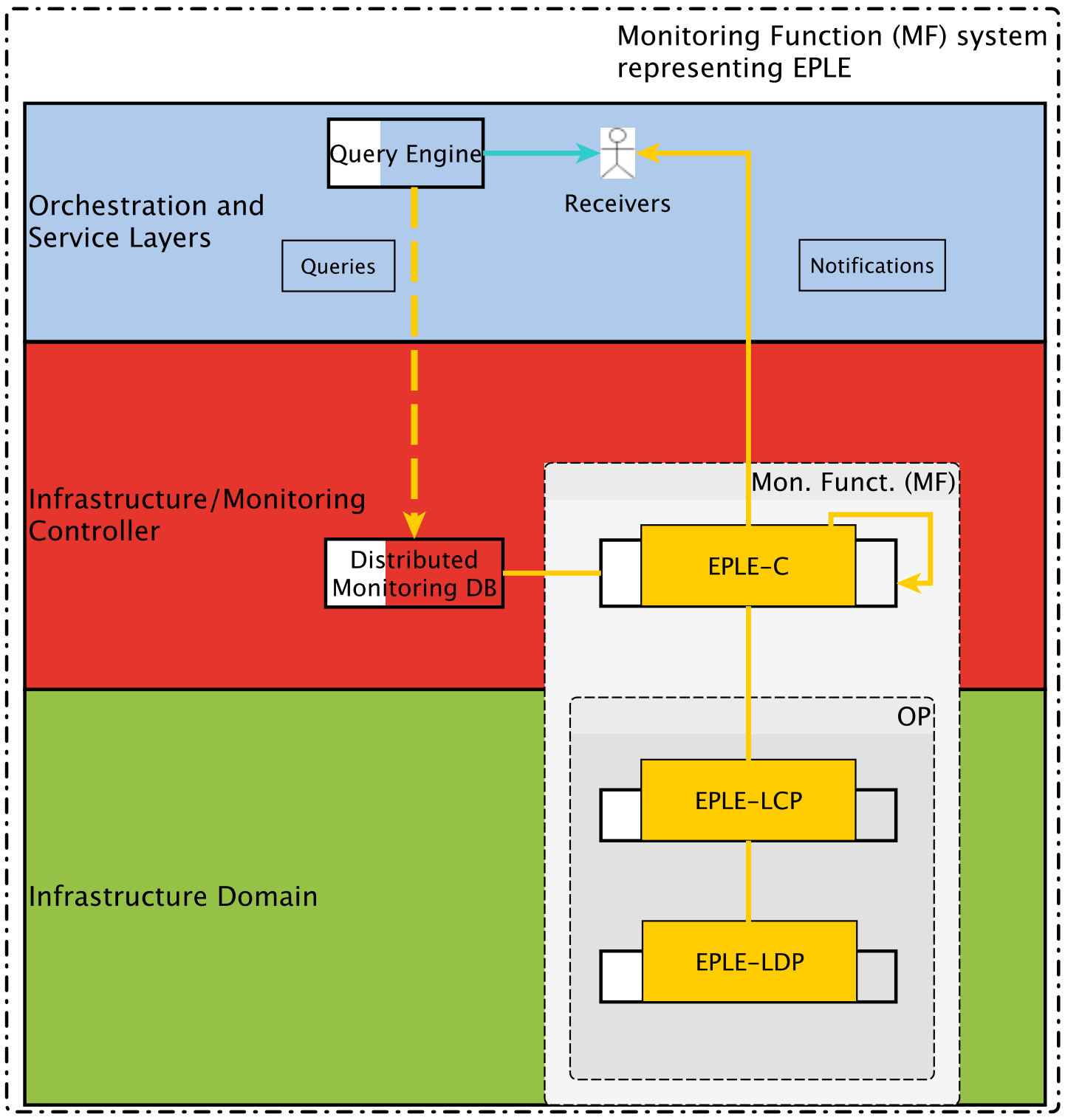

*Figure 5.26: EPLE mapping to UNIFY Functional Architecture as infrastructure monitoring function*

**Universal Node:** We present two mappings between EPLE functionality and the UNIFY infrastructure layer. The first option, depicted in Figure 5.27 (a) shows EPLE executing as a Monitoring Function in an SDN controller and communicating with the EPLE Devolve Table (EDT) via the OpenFlow protocol that is supported by the FE (OF switch). This option is the more general approach, valid in SDN domains with a logically centralized control plane, realized as one (or more federated) controller instance(s). Usually, in this case the ports monitored by the EPLE instance are located on separate FEs, but they could also be within the same element but with strong isolation requirements between them.

The second option, depicted in Figure 5.27 (b), presents EPLE mapped onto a UN. In this case, EPLE is realized as a node-external EPLE-C and a node-local control plane (LCP) executing as part of the Monitoring plugin, utilizing the VSE Management plugins (i.e. OpenFlow controller) to access the EDT in the corresponding logical switch instance (LSI). The LCP would then make measured loss information available on the DoubleDecker messaging system (DoubleDecker broker). The MF control app (EPLE-C) runs on a monitoring controller for coordination of several OPs (i.e. LCP/LDP pairs) running on several different UNs. The execution environment of the monitoring controller would typically either be a dedicated, external hardware or a VNF in a UNs.

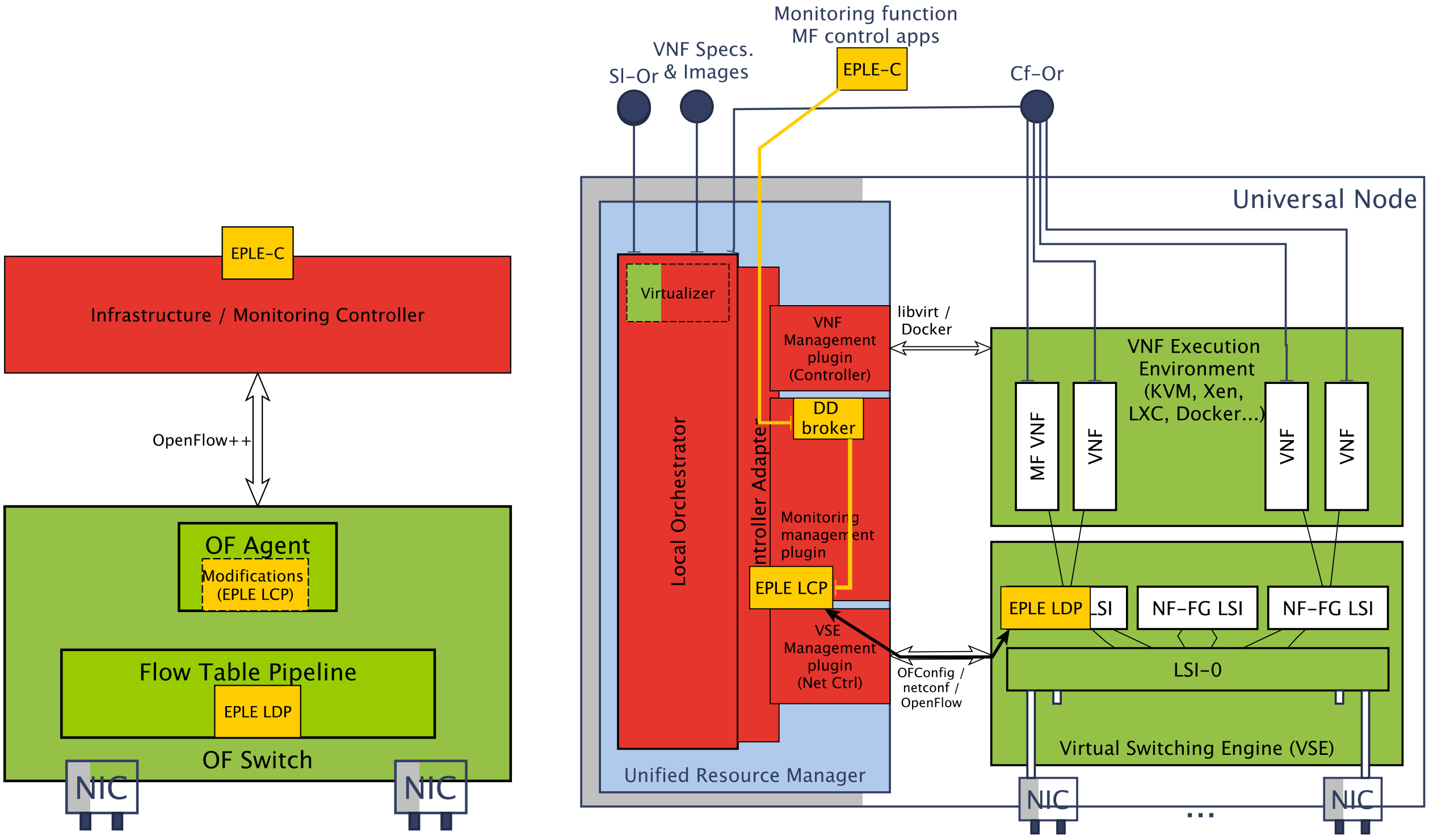

(a) (b)

*Figure 5.27: In (a): mapping of EPLE on a switch. In (b): mapping of EPLE on the UN.*

The exact distributions of control functions between EPLE-C and LCP in the UN case is quite flexible and could range from a void EPLE-C on one extreme (i.e. all control functionality placed on the LCP) to a void LCP on the other extreme (i.e. all control functionally placed in the EPLE-C, with the LCP reduced to an adapter function connecting the node internal OF channel and DoubleDecker).

### 5.7.3 Support to SP-DevOps processes and objectives addressed

EPLE focuses on addressing RC3: Low-overhead performance monitoring for SDN (due to zero overhead added on the dataplane and limited additional overhead on the control plane) and RC2: Scalable observability data transport and processing (the method addresses flows described with wildcarded aggregate descriptors which are typical for transport networks and scales well with respect to the overhead introduced both on the flow table and on the control connections) challenges outlined in D4.1. From the original WP4 objectives, O4.3 Develop scalable service monitoring approaches, adapted to software-defined networks is supported through the fact that EPLE extends OpenFlow, a typical software-defined network control protocol and switch specification. O4.2 Define conditional observability points located on UNs is supported through the placement of the EDT on the VSE within the UN, as well as the programmability features included in the method in terms of choosing which wildcarded flows to monitor, programming the flow table for the devolve action and the expiration timers, as well as choosing through policies which types of flows should be devolved.

With respect to the requirements outlined in D4.1, EPLE addresses OR1 by reducing the amount of traffic introduced in the network for measuring packet loss and NL3 by allowing instantiation and removal its Observability Points at the same time with the aggregate flow being observed. When deployed in an UN, it also supports NL1, NL2 and NL4 as well as OL2 and OL4.

EPLE supports directly the SP-DevOps Observability Process by providing a scalable way for estimating packet losses on dynamic service chains. Through the planned implementation of interfaces that allow for automated configuration (both through the orchestration-related interfaces and on-demand) via the MEASURE annotations and publishing the estimated values on the messaging system, EPLE supports also the SP-DevOps Troubleshooting and VNF Development Support processes.

### 5.7.4 Evaluation and individual prototyping results

At the time of writing this text, implementation is ongoing and therefore no evaluation data is available. OpenDaylight Hydrogen and Open vSwitch are being extended to support EPLE-C and EPLE-LDP components, respectively. We expect to include EPLE evaluation data in the final WP4 deliverable.

### 5.7.5 Next steps

Apart from finalizing the implementation of EPLE and evaluating it with a publicly available traffic trace, the following steps are planned as part of the development:

- integration with the DoubleDecker messaging system API
- integration with the recursive monitoring language
- integration with MEASURE

## 5.8 Automatic Test Packet Generation Tool - AutoTPG

The AutoTPG tool is an OpenFlow based tool and can be used to verify the data plane functionality of OpenFlow switches/routers (i.e., correct delivery of packets through FlowTables). The existing automatic test packet generation tool (ATPG) [HZG12] verifies only one of the packet-headers that matches correctly with the wildcarded flow or not. The matching of all the other packet-headers remains untested. In comparison with the existing automatic test packet generation tool (ATPG), AutoTPG verifies that all the packet headers match correctly with aggregated flows.

### 5.8.1 Technical description

An OpenFlow [OpenFlow13] switch/router contains FlowTables. An entry in a FlowTable contains: (1) a Flow Match Header or matching header part, which defines a flow, (2) Actions, which define how packets should be processed or forwarded (i.e., forward to an output port or to a different FlowTable) and (3) some additional fields such as priority, statistics and cookie identifier. The AutoTPG tool verifies the Flow Match part of the Flow Entries (i.e., correct delivery of data packets through FlowTables or Flow Entries). This may be hard to verify in the case when matching of packets with a Flow Entry only affects some of the packets [HZG12]. For example, when a flow matches a packet header that it should not match, or a flow does not match a packet header that it should. These issues may occur due to incorrect translation of some of specific packet-headers, incorrect translation of a flow or incorrect flow matching in hardware or software. For the verification of the matching-header part, our mechanism uses a header field such as EtherType (or VLAN ID) for differentiating test packets from data packets and assumes that this header field (e.g., EtherType) is wildcarded in the matching-header part of Flow Entries.

Our mechanism performs three steps for verification: (1) flow duplication, (2) test packet generation, and (3) matching errors identification. In the flow duplication step, the mechanism duplicates the Flow Entries from a FlowTable to another FlowTable. In the test packet generation step, the mechanism generates and transmits all the test packets that can match with the flow (e.g., aggregated flow) of duplicated Flow Entries. In the matching errors identification step, the mechanism calculates the matching errors either by reading the counters (statistics) of the duplicated Flow Entries (we call this method as the binary search method, as binary search is used to find matching errors) or by comparing the sent and received test packets (we call this method as packet reception method). In the binary search method, the action of all the duplicated Flow Entries is DROP (i.e., drop all the matched test packets)and in the packet reception method, the action of all the duplicated Flow Entries is CONTROLLER (i.e., send the matched test packets to the controller). The advantage of the binary search method is that it reduces the upstream bandwidth of the controller compared to the packet-reception method. However, the disadvantage is that it takes more time to find matching errors.

The flow duplication step can be performed by the controller, as it only needs to insert additional Flow Entries in the switches for verification. Verification using the controller can increase the bandwidth requirements between the controller and switches (for example, in an out-of-band network). Therefore, we propose that the steps (test packet generation and matching-error identification) that increase the bandwidth requirement of the controller significantly can be performed by a virtual machine using an in-band network. In this case, the bandwidth of the data plane network can be used for verification. We propose that the VM should establish an OpenFlow session with the switches in order to perform the verification activity. In this proposal, the VM works like an additional controller in the network. This is proposed because of the following two capabilities of a controller application: (1) The capability to generate test packets with an ingress port as a matching field and (2) the capability to verify two or more switches at the same time. The main cost of the tool is the bandwidth requirements between the controller and the switches  for out-of-band networks and also the bandwidth requirement between switches for in-band networks. Additionally, in case the AutoTPG is placed in a VM, there is also the additional cost on having a VM in the network.

### Design choices and practical considerations

An existing tool closely related with our work is ATPG [HZG12]. It performs the verification of the data plane functionality for error conditions (e.g., incorrect firewall rules, software, hardware, and performance errors) by transmitting real test packets. Debugging by transmitting test packets is important because finding all the errors is difficult by just analyzing the configuration of networks. However, ATPG verifies only one of the packet headers matching an aggregated Flow Entry (i.e., containing wildcards in the matching header part) for matching issues, whereas all the other packet headers remain untested.  This is an issue because there can be two kinds of matching issues  in switches: (1) matching issue due to incorrect OpenFlow configuration and (2) bugs in OpenFlow switch implementation. Firstly, incorrect OpenFlow configuration may occur if wrong Entries (e.g., wrong matching-header part or priority number due to a bug in the controller or due to addition of manual entries) are installed in the FlowTable and therefore, either no match or incorrect match for some are found.  Secondly, bugs in OpenFlow switch implementation can be: (1) matching-header specific i.e., if Flow Entries contains the same matching-header part, the matching errors will occur always or will occur after a race condition happened in the switch (may be due to software or hardware bugs of a switch), (2) Flow Entry specific i.e., it is specific to the Flow Entry which was installed in a FlowTable (may be due to a software or hardware bug). Using our mechanism, we can find errors related OpenFlow configurations and bugs related matching-header specific implementation of OpenFlow switches. As the errors related Flow Entries-specific are specific to the original Entries installed in the switches, verification of Flow Entries in our tool cannot find these errors. For finding these errors, test packets are needed to be transmitted through the Flow Entries used for forwarding data packets.  Future work will be focused on finding the errors specific to a Flow Entry.

### 5.8.2 Architectural mapping

**Functional Architecture:** We can map AutoTPG into the UNIFY functional architecture as Monitoring Function (MF) control app (shown in Figure 5.28). Using AutoTPG as the Monitoring function control app, all the forwarding entries can be verified by building a module in the MF control app. The control app can perform all the three steps (Flow duplication, test packet generation, and matching error identification) of verification . The LDP unit in Figure 5.28 can make an out-of-band or in-band network with the controller app. In case of an in-band network, the control app has to establish OpenFlow sessions with all data plane unit through the other data plane unit (in-band path), as described in VM-induced emulation in the previous subsection.

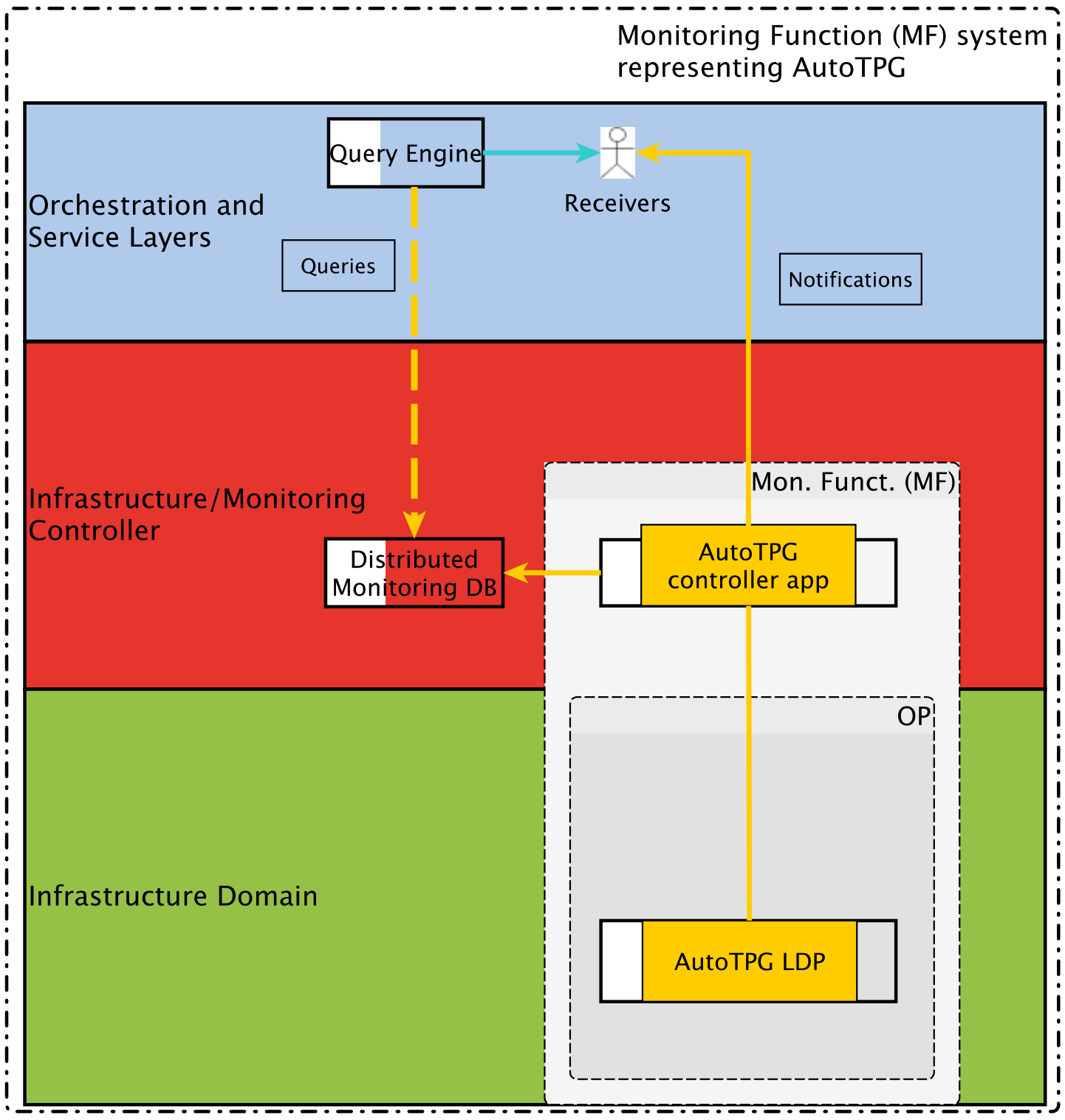

*Figure 5.28: Functional Architecture Mapping of AutoTPG*

**Universal node:** We present two mappings of AutoTPG into the UN node and both the mappings will be implemented on the UN node (Figure 5.29). The first option shows AutoTPG as a Monitoring Function (LCP) executing as part of the virtual switching Engine (VSE) Management plugins. In this option, the AutoTPG monitoring function makes an OpenFlow session with the logical switching instance (LSI). This option is valid when AutoTPG needs to verify all the forwarding entries of LSI.

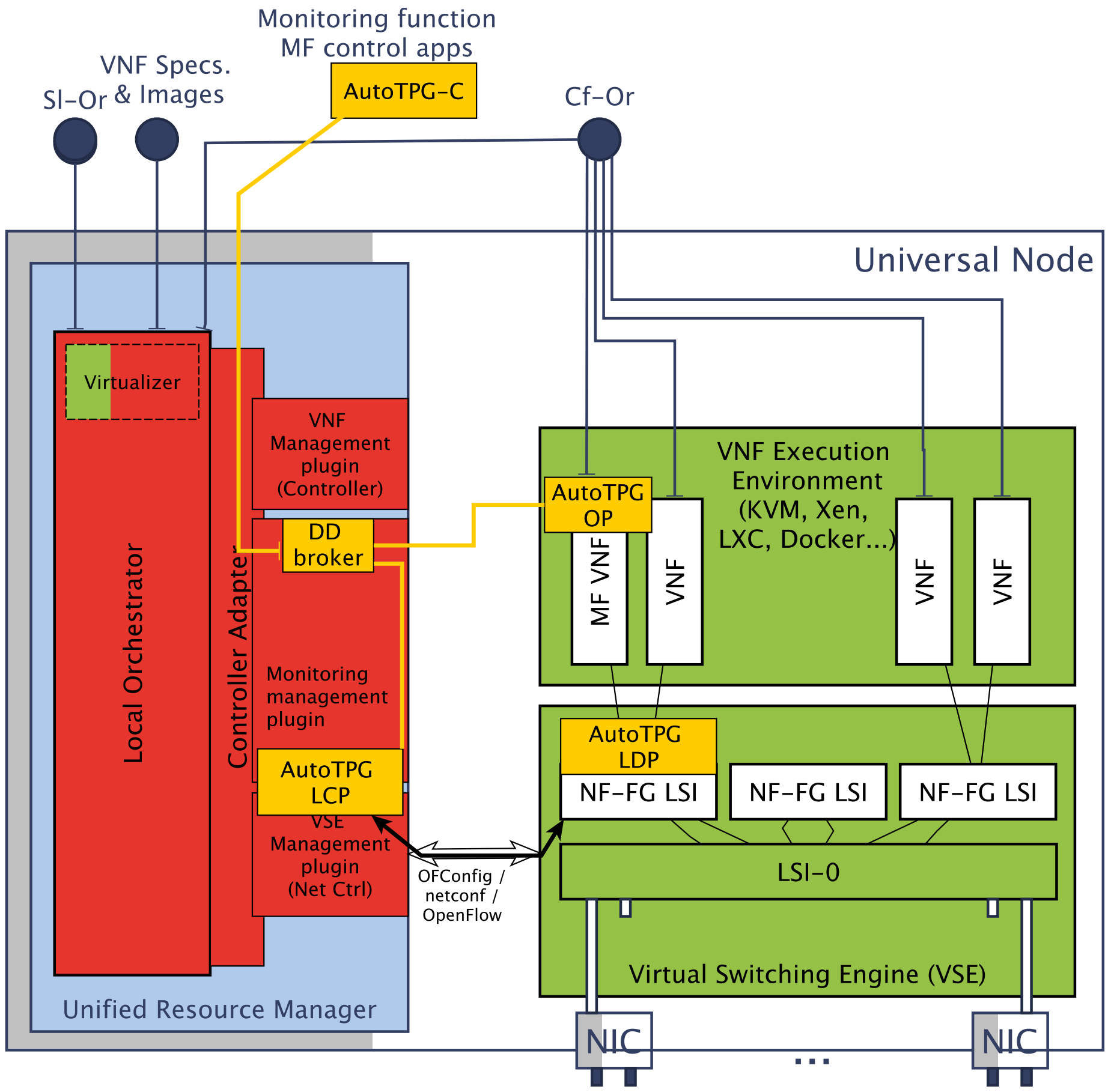

*Figure 5.29: AutoTPG mapping into the universal node*

In the second option, AutoTPG is located in one of the VNF as OP and therefore, it verifies all the entries of LDP through the in-band LDP path (i.e., through in-band verification explained in the previous subsection). In this verification, VNF has to establish an OpenFlow session with all the other VNFs through the NF-FG LSI, as shown in Figure 5.29.

### 5.8.3 Support to SP-DevOps processes and objectives addressed

The AutoTPG tool is an instance of run time verification (RC7) and troubleshooting with active measurement (RC10). Addressing both these research challenges is part of achieving O4.2 and O4.3. The approach relates to several requirements in D4.1: NL1 (NL1.2-1.5), NL2, and NL4 (NL4.2-4.6). The tool implements an observability process performing active measurements which are used for both verification and troubleshooting processes. The verification process is supported in terms of verification of all the Flow Entries present in the network. In the case of troubleshooting, the tool can be used to find the root cause of detected errors.

The tool supports both verification and troubleshooting processes. The verification process is supported in terms of verification of all the Flow Entries present in the network. In the case of troubleshooting, the tool can try to find the root cause of errors.

### 5.8.4 Evaluation and individual prototyping results

We performed emulations on the Fed4fire testbed facility at iMinds [FTD14]. Figure 5.30 shows an emulated pan-European topology that contains 16 OpenFlow switches connected with each other in a mesh fashion.

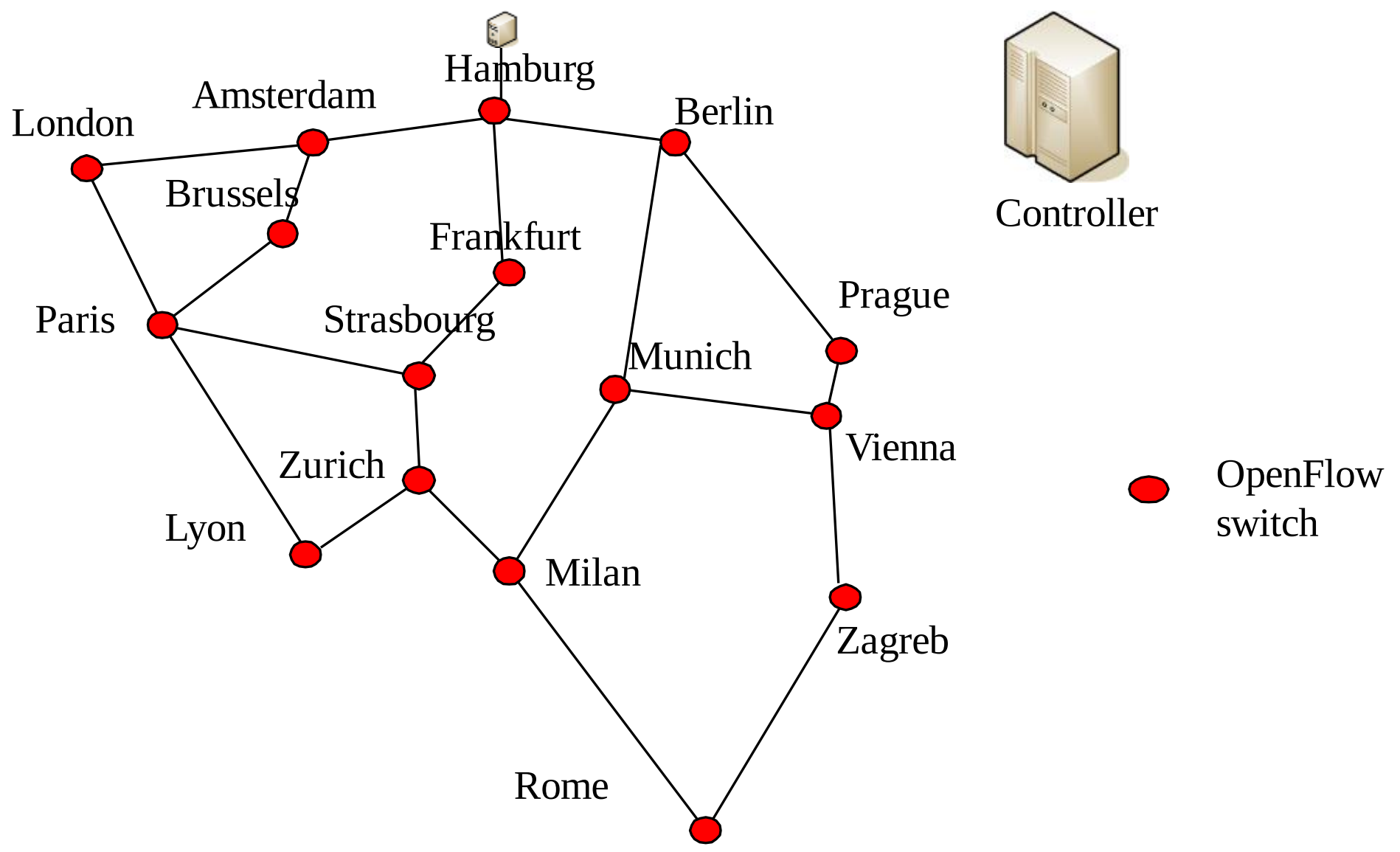

*Figure 5.30: Emulated Pan-European Topology.*

We perform two different experiments - (1) controller-induced verification, (2) VM-induced verification. In the controller-induced verification, the controller makes an out-of-band connection with switches (shown in Figure 5.30) and performs all the steps of verification and the Flow Match Header errors are detected either via the packet-reception or via the binary-search method. In the VM-induced verification, the controller triggers the creation of a VM (light weight VM, Linux container) and connects it with one of the switches in the network (in-band connection [SSH15]). For this experiment, we use the packet-reception method for calculating the matching errors in all Flow Entries in the network. For our emulation, we implemented the proposed verification mechanism in the Floodlight controller that uses OpenFlow version 1.3 [OpenFlow13] in its implementation. In addition, we used Open vSwitch for running OpenFlow in the switches of the emulated topology. In the experiments, software matching errors are emulated by translating packets of a flow to incorrect headers.

We present the results gathered by performing all the described experiments.

1) Controller-induced verification experiment

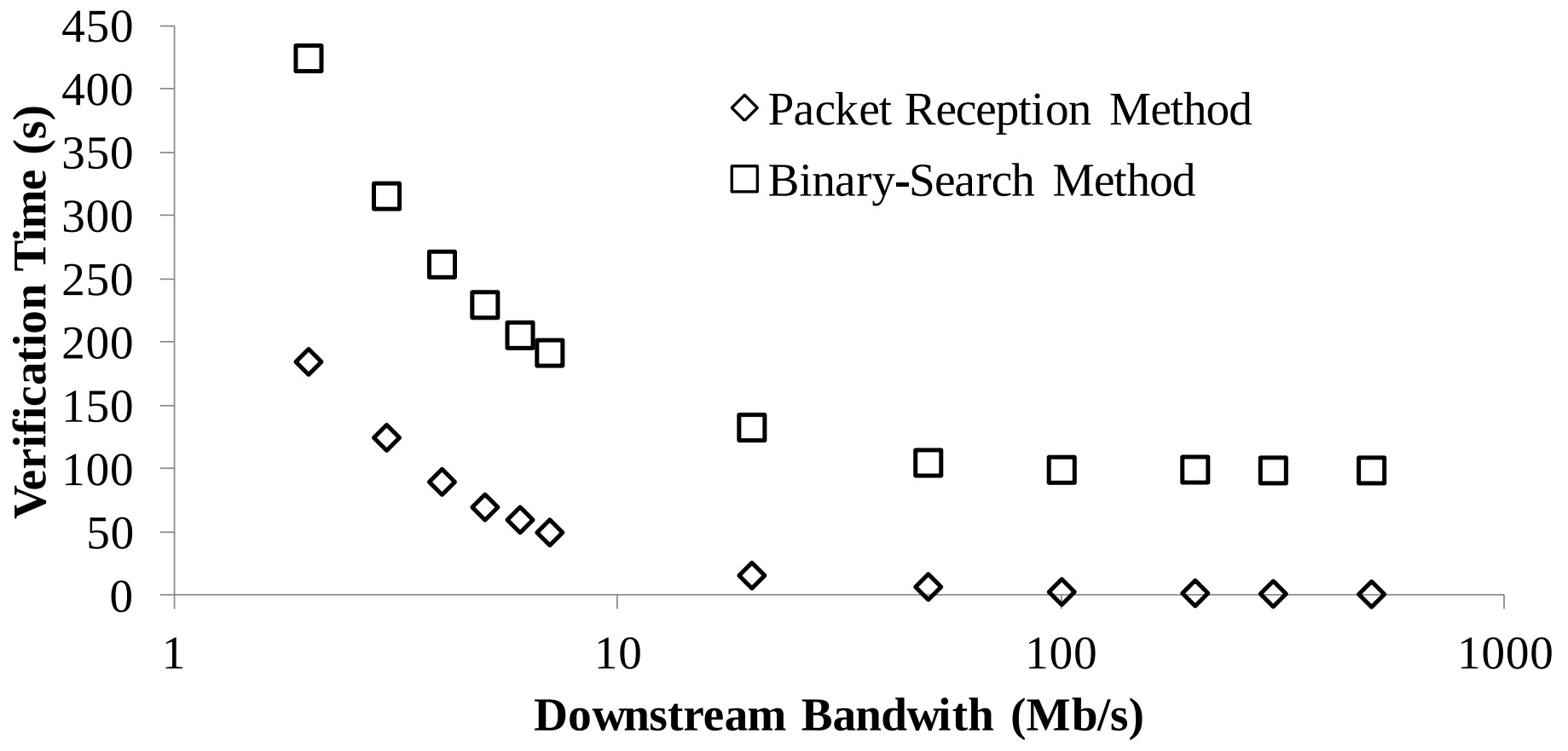

*Figure 5.31: Verification time using (controller out-of-band network scenario)*

In this experiment, we vary the bandwidth between the controller and switches for verification and the verification time is calculated. The controller in our implementation sets the rate of the test packets according to the bandwidth available between the controller and switches. As expected, Figure 5.31 shows that the verification time of the Flow Entries decreases with the increase in the bandwidth reserved for verification. Figure 5.31 also shows that the verification time in the binary-search method is longer than the verification time in the packet-reception method. This is because the binary-search method sends a large number of test packets for verification through binary search iterations, leading to increase in the verification time. In addition, as the counter update interval is 2 seconds in our emulation, it further increases the verification time in the binary-search method.

2) VM-induced verification experiment

For this experiment, a VM is connected at Hamburg and the verification time is calculated with respect to the bandwidth reserved in each switch link for verification. In this experiment, the bandwidth is limited in each switch link and the VM makes an in-band connection with switches in the network. As the paths for verification for some of switches is through other switches in the network, the switches may have to wait for verification until the intermediate switches in the path perform verification. This leads to increase in the verification time in the VM-induced verification experiment as compared to the controller verification experiment in which the controller makes an out-of-band connection with the switches.

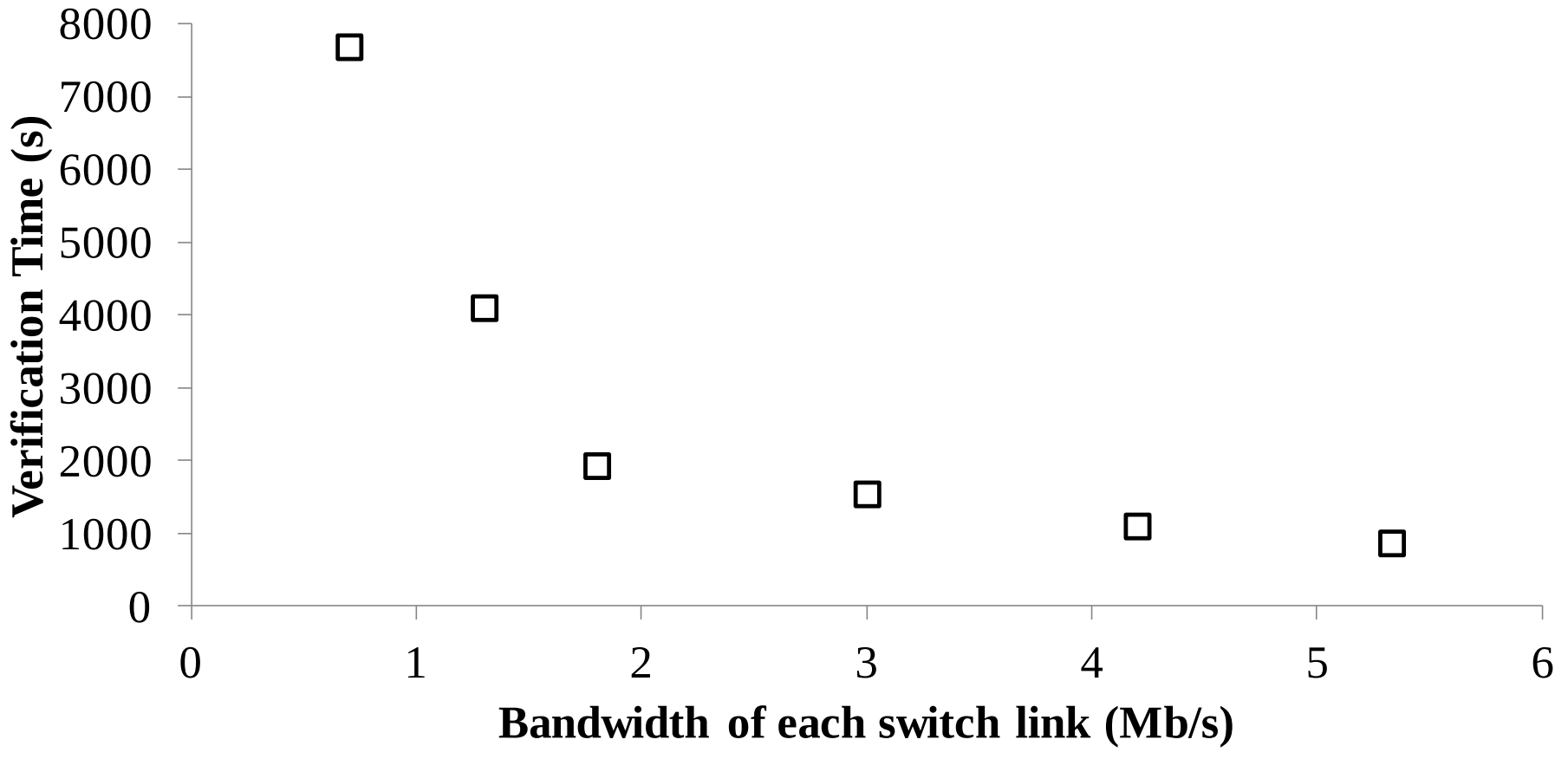

*Figure 5.32: Limited bandwidth scenario (VM in-band control scenario)*

Figure 5.32 also shows that if the bandwidth in each switch link for verification increases, verification can be performed in limited time.

### 5.8.5 Next steps

Currently, AutoTPG just detects that there are matching errors present in a switch or not. The next step is to find or establish a valid Flow Entry if errors are present in a switch and re-route traffic using newly established path. The challenge in the implementation is to verify the new path before traffic is re-routed to this path. The next plan is also to integrate AutoTPG in the ESCAPE framework [D22][D32] and using DoubleDecker messaging system.

## 5.9 IPTV Service Quality Monitor

Today, the consumption of multimedia content is responsible for the major fraction of the total Internet traffic, therefore assessing the quality of the media content delivery is becoming crucial to both network carriers and content providers. Moreover, for a complex service creation platform such as UNIFY, the flexible placement of such a monitoring tool is an especially emphasized requirement. The IPTV service quality monitor is a monitoring function that is designed for measuring and verifying the quality level of a specific service (i.e., IPTV).

### 5.9.1 Technical description

The IPTV service quality monitor aims at assessing the Mean Opinion Score (MOS) of the quality of experience (QoE) given on an IPTV service based on the standardized model, namely ITU P.1201.2 [ITU1202]. The tool requires packet-level network traffic as an input and classifies IPTV traffic (i.e., MPEG-2 transport streams) from the overall traffic. The tool extracts required parameters such as video/audio codec, bitrates, timestamps, and GoP (Group of Pictures) structures and feeds them into the P.1201.2 model that estimates MOS of the multimedia stream. As a result of the assessment, the tool reports various quality indicators such as packet losses, startup delay, and audio/video/audiovisual quality in MOS.

While consuming multimedia content, user's quality experience may differ from what is reported by link (Section 5.5) or rate (Section 5.5) MFs mainly due to the different aspects of quality properties. For example, packet losses and jitters affect the quality of video keenly during the scene change, while the impact may not be significant when the motion in the video is static. Thus, a detecting mechanism of scene changes plays an important role in the multimedia quality assessment. An interesting quality indicator of the IPTV service that the tool reports is the duration of performance degradation during the multimedia playback. The IPTV service quality monitor takes such multimedia-specific quality properties into consideration for measuring the quality of user's experience.

### 5.9.2 Architectural mapping

**Functional architecture:** The IPTV quality monitor is currently at the beginning phase of the development. However, it is clear that the actual assessment components (LCP and LDP) of the tool need to locate on the infrastructure layer where the network traffic can be directly fed to the monitor.

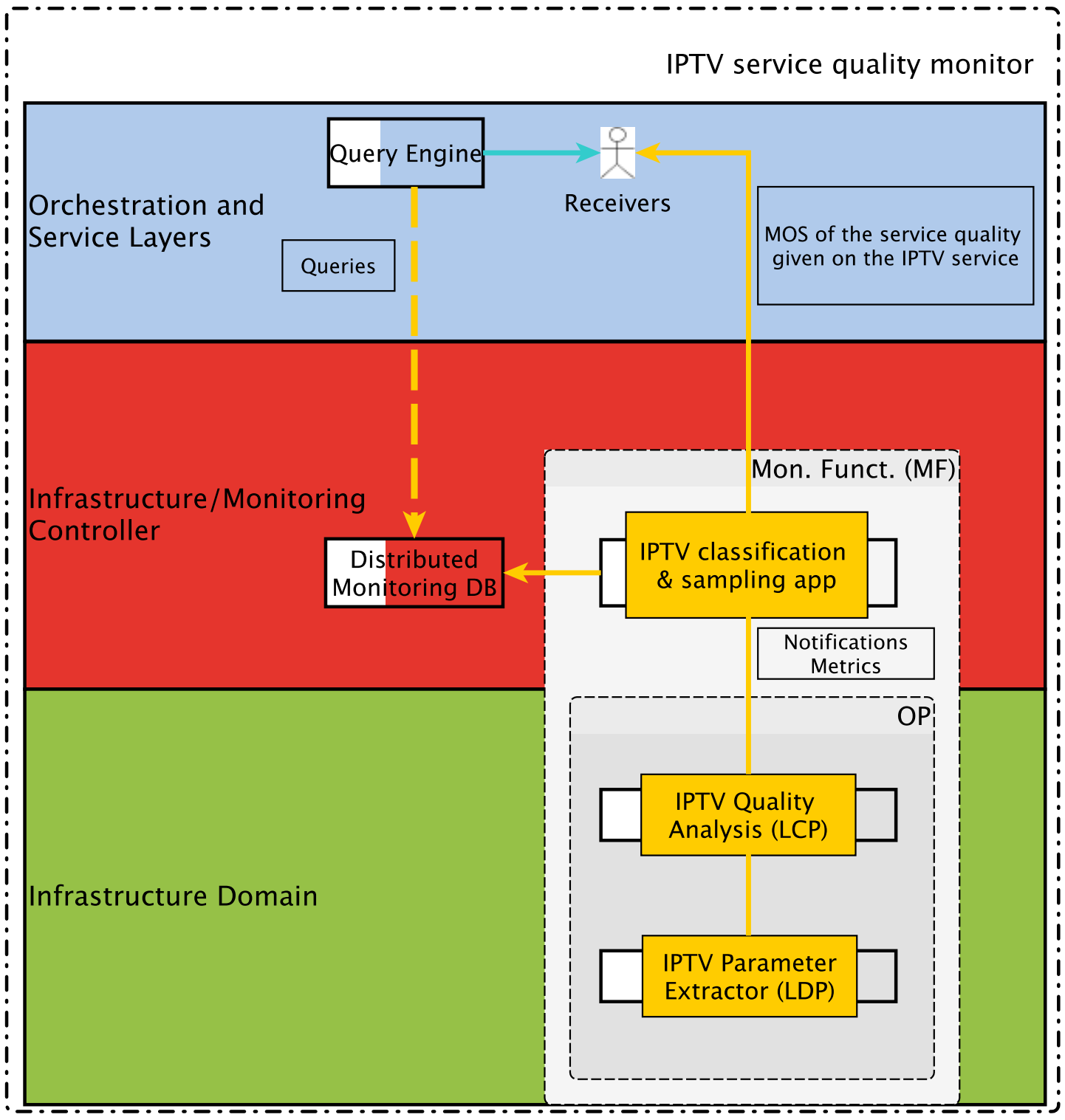

*Figure 5.33: Potential mapping strategies of the IPTV service quality monitor into the functional.*

The role of the IPTV service quality monitor controller (i.e., IPTV classification & sampling app in Figure 5.33) is yet to be fully discovered, but the three of the main roles are the configuration of the traffic classification rule, management of the sampling rate, and the notification to users. For handling these tasks, this application should be placed on the infrastructure/monitoring controller layer. Figure 5.33 illustrates the potential mapping of IPTV service quality monitoring components to the UNIFY functional architecture.

**Universal Node:** In Figure 5.34, the mapping of the IPTV service quality monitor into a UN may follow the similar concept as the one presented by the link monitoring tool (Section 5.6) and EPLE (Section 5.7).

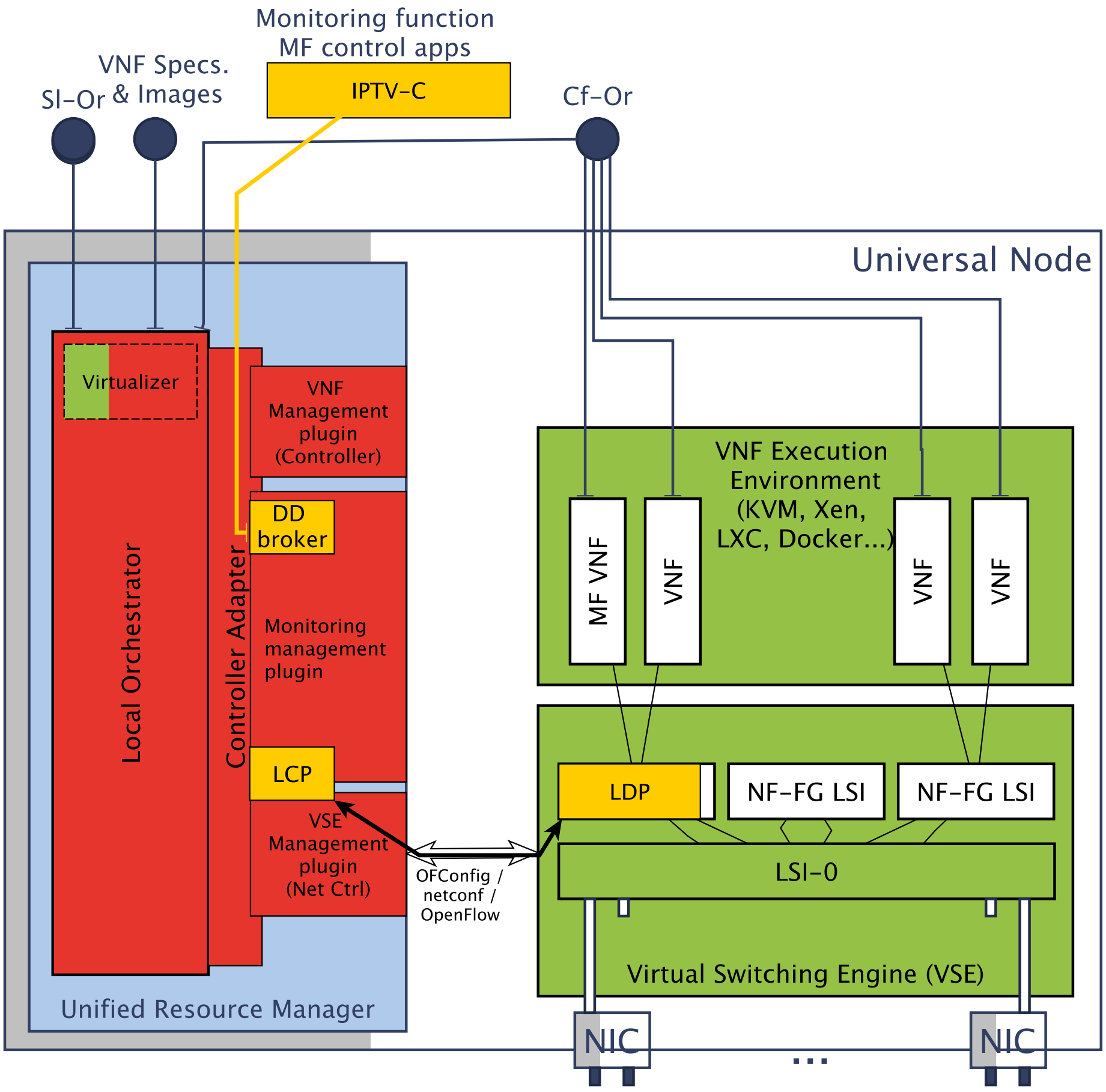

*Figure 5.34: Potential mapping of the IPTV service quality monitor into UN.*

Besides calculating the service quality (based on ITU P.1202.2) of IPTV traffic, translating traffic classification policy delivered from the controller application into the OpenFlow rule is one of the main tasks that LCP carries out.

Similarly, LCP implements a flow sampling function that probabilistically selects IPTV streams for the quality assessment with the rate configured by the controller application. The main task of LDP in IPTV service quality monitor is to extract the desired parameters from filtered multimedia streams.

### 5.9.3 Support to SP-DevOps processes and objectives addressed

The IPTV service quality monitor is passively measuring the quality of multimedia content delivery and verifies whether or not the level of its quality fulfils the requirement of the service. Therefore, this tool mainly supports the observability process of the SP-DevOps and it deals with O4.2 and O4.3 by addressing RC1. Optionally, the SLA verification can be covered by this tool from the perspective of operators which is partially related to RC9. For this, the IPTV service quality monitor specifies several requirements such as NL1, NL2, NL3, NL4 (4.4, 4.5, 4.7, and 4.8), Although the tool will not be publically available due to its closed license, the demonstration of this tool within the scope of the UNIFY framework is expected to be a good use-case example of UNIFY.

### 5.9.4 Evaluation and individual prototyping results

There is no evaluation result performed with the IPTV service quality monitor since it still needs to be ported to a virtualized system environment. However, the evaluation is planned to be carried out when it is successfully implemented on the test-bed. The evaluation will mainly focus on discovering the characteristics of IPTV traffic as well as scalability properties of the tool.

### 5.9.5 Next steps

The tool is currently implemented for specific content providers in a specific hardware. Therefore, the future implementation effort needs to go into the virtualization of the function and the integration of the tool into the prototype platform (i.e., ESCAPE [D22][D32]).

## 5.10 Service Model Verification

### 5.10.1 Technical description

The verification process aims to check certain properties related to the UNIFY service models (e.g., network node reachability, forwarding loop and black hole absence) so as to reduce the risk of critical and erroneous situations at deploy time. In order to achieve this goal, the verification process is based on formal methods, i.e. mathematically founded methods that can be used to prove that the involved models (SG, NF-FG, OpenFlow rules) fulfil certain properties. Such formal techniques should allow the verification process to be completed in a reasonable amount of time and with fair processing resources (e.g. CPU, memory and so on). Given these requirements, we decided to explore SMT based techniques since they seem to be more promising than traditional model checking methods, which usually suffer memory scalability problems.

In addition, the verification process should be able to handle real networks that contain active network functions, implemented as VNFs, which dynamically change the traffic forwarding path according to local algorithms and an internal state based on traffic history. Examples of those active functions (or middleboxes) are NATs, load balancers, packet marking modules, intrusion detection systems and more. In order to perform formal verification on a service model that includes such functions, we need to represent the internal state of a middlebox or an abstract version of it, in order to perform the verification in a reasonable time and also to make this process scalable even in case of SG/NF-FG changes. The abstraction level to be used in the active VNF models is the most critical issue: finding out an appropriate level of detail so that a significant analysis can be done in reasonable time is the main challenge we are addressing.

#### Design choices and practical considerations

The verification process exploits the kernel of a verification engine recently proposed at UC Berkeley [APA14], which follows the same approach that we designed for our tool. In order to achieve high performance, the verification engine from UC Berkeley exploits Z3, a state of the art SAT solver, and also addresses network scenarios with multiple active network functions connected together to form a complex network graph. Z3 is able to solve reachability problems thanks to the translation of these problems into SAT problems. The model of all the involved middleboxes and the overall network behavior are represented as a set of first order logic formulas and completed with other formulas that express the properties to be verified,for example reachability properties between two nodes in the network,in such a way that the satisfiability of the formulas implies the truth of the specified properties.

As an example of how network function models can be expressed, we present a simplified version of an antispam network function. Here, we simplify and abstract the email protocol by assuming each client interested in receiving a new message addressed to him, sends a POP3 REQUEST to the mail server in order to retrieve the message content. The server, in turn, replies with a POP3 RESPONSE which contains a special field (named *email_from*) representing the message sender. In order to be able to introduce our model within the verification tool, we need to express our function behavior by means of logic constraints and implications. As evident from the first formula in Equation 5, our function rejects any message containing a black listed email address. On the other hand, Equation 6 is needed in order to state that a POP3 REQUEST message is forwarded only after having received it in a previous time instant.

$$(send(antispam, n_0, p_0, t_0) \wedge p_0.protocol = POP3_RESPONSE) \rightarrow \,!\, isInBlackList(p_0.email_from) \wedge \exists(n_1, t_1) \mid (t_1 < t_0 \wedge recv(n_1, nat, p_0, t_1)) \qquad \forall n_0, p_0, t_0 \tag{5}$$

$$(send(antispam, n_0, p_0, t_0) \wedge p_0.protocol = POP3_REQUEST) \rightarrow \exists(n_1, t_1) \mid (t_1 < t_0 \wedge recv(n_1, nat, p_0, t_1)) \qquad \forall n_0, p_0, t_0 \tag{6}$$

.

The formal verification task is split into multiple sub-tasks, so that the whole process is simpler and faster. More precisely, at SG/NF-FG deploy time, or when the graphs undergo modifications in response to higher level events (e.g. administration events or user requests), the VNF chains belonging to the graph are calculated and then for each of them a formal model is generated. Each single chain contains the models of each VNF in the chain itself. These models are expressed by means of Python classes that generate all the logic formulas that describe the models and feed them to Z3 through the Z3 Python API. Since we are using abstract models of the real middleboxes, we assume that these models are correctly defined. The properties to be verified are also expressed by means of logic formulas and given to Z3 in the same way. Each property is such that it holds if and only if the generated Boolean formulas are satisfable. Z3 processes the whole set of logic formulas, including the VNF chain model and the properties to be verified, by checking its satisfiability.

The UC Berkeley verification tool kernel had some limitations that have been overcome by extending the way VNFs are modelled. Specifically, new VNF models have been added for the purposes of including network functions that modify the in-line traffic header. In this way, the set of available services for a UNIFY user was enriched with functionalities like, for example, web caching and anti-spamming. Moreover, the generation of the overall model from a NF-FG or from OF rules is another part that will need to be implemented, as it is not part of the verification kernel provided by UC Berkeley.

### 5.10.2 Architectural mapping

The verification process involves the entire UNIFY Functional Architecture, with four different modules included in the three architecture layers, i.e., the Service Layer, the Orchestration Layer, and the Infrastructure Layer (Figure 5.35).

At the Service Layer, the verification activity is implemented by applying graph-theoretic algorithms to the SG received by the Service Management Sublayer (Figure 5.35). This task includes the verification of the user defined policies. After this stage, we move to the Orchestration layer, where the NF-FG is generated. At this point, the Resource Orchestration Sublayer will trigger the verification process sending the generated NF-FG to the verification engine. From this graph, the verification activity is performed by verifying network properties considering only the input topology.

Concerning the verification of the OpenFlow rule correctness, we plan to perform another verification stage in the Infrastructure Layer  (Figure 5.35), more precisely in the Controllers. For this purpose we plan to deploy VeriFlow [AKH12], one of the state-of-the-art OpenFlow rule verification tools.

In the Infrastructure layer, we can also verify the correctness of some network properties (e.g., reachability property) in the NG-FG. This NF-FG verification differs from the one in the upper-layer (i.e., Orchestration layer) as it consider both the NF-FG topology and also the configuration of the middleboxes contained within the graph. Hence the NF-FG verification at this level follows these steps:

- Find the VNF chains belonging to the NF-FG
- Create the chain model exploiting the Z3 API
- Perform the verification task, also considering the possible policies defined by the users (enforcing a given traffic flow to traverse a desired waypoint is an example of such policies)

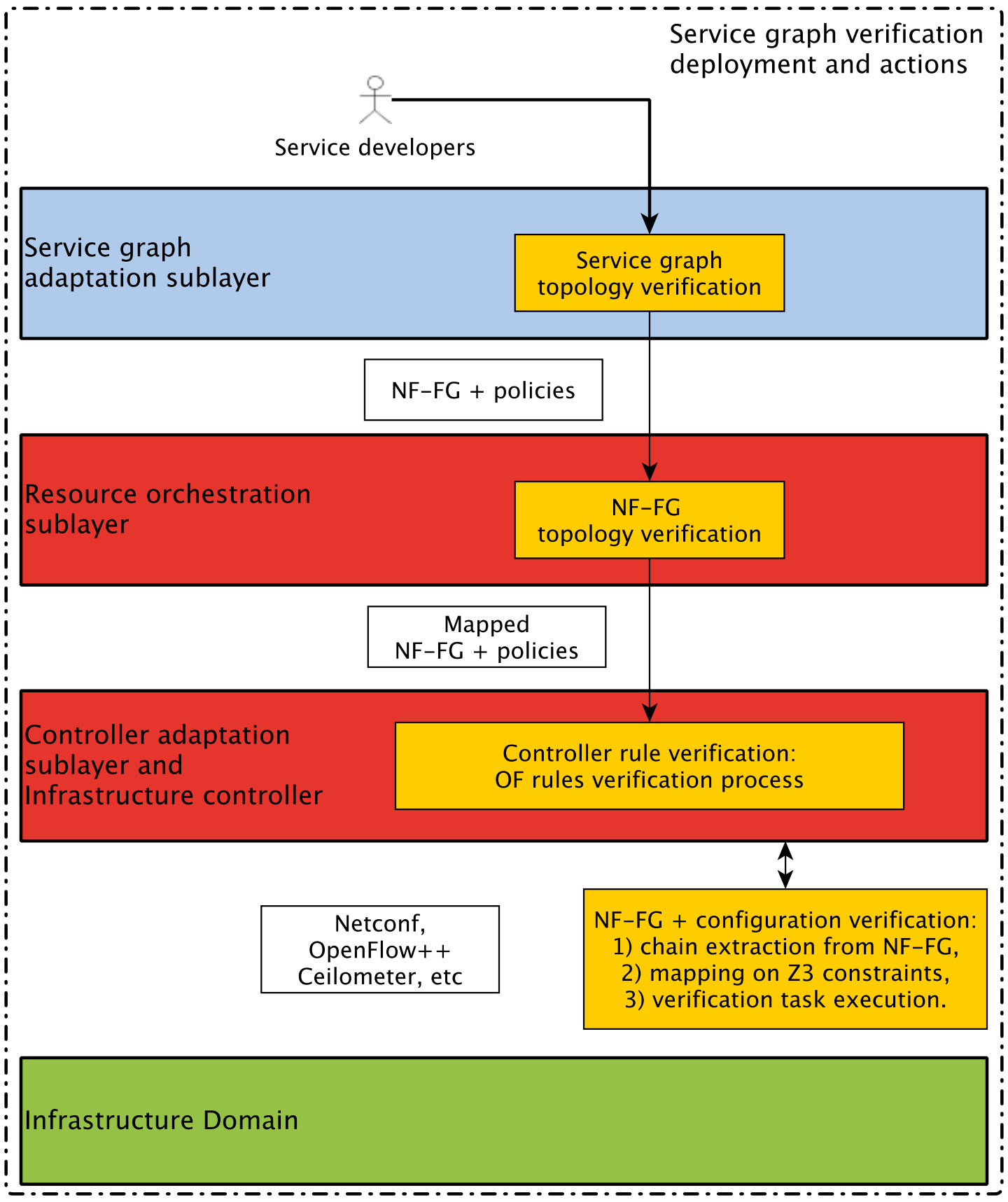

*Figure 5.35: Mapping of the verification tool to the UNIFY Functional Architecture.*

In addition, the verification module could use the U-Sl interface to get both the SG and the policies defined by an UNIFY user for his SG. Instead a Sl-Or interface should be used by the verification module to get the NF-FG and its policies. In addition to these interfaces, there will be some internal API at each architecture layer, in order to trigger the verification process (e.g., verification Request message) and to send back the verification result (e.g., verification Reply message).

### 5.10.3 Support to SP-DevOps processes and objectives addressed

The verification engine is an instance of the functional verification of dynamic SGs defined in the RC6. The engine also achieves the O4.6 objectives related to this research challenge (defined in D4.1), which consists in enabling the possibility to verify service chains within the limit of one development cycle. This also implies that requirements OL3 OL5 OL6 must be covered.

The verification engine is able to support the verification process, checking certain properties on the abstract service models (SG, NF-FG, OpenFlow rules). It provides this specific support:

* Checking the existence of a path that connects two nodes in the topology of the SG, at the Service Layer.
* Solving reachability problems in the NF-FG, taking into account certain user-defined policies related to the NF-FG, at the Orchestration Layer.
* Verifying the consistency of network configurations expressed in the form of SDN OpenFlow rules, at the Orchestration Layer.

### 5.10.4 Evaluation and individual prototyping results

A preliminary version of the verification engine based on the UC Berkeley tool kernel and Z3 has been implemented. In order to evaluate the extended verification engine, we consider the NF-FG (shown in Figure 5.36) as a use case. In our reference graph, four end-hosts (two clients and two servers) can generate either HTTP or POP3 and also SMTP traffic, which is processed by different middleboxes when traversing the graph. In this graph, the forwarding is configured such that the web traffic is forwarded to the web cache, while the email traffic (both POP3 and SMTP) is routed to an anti spam function. Moreover, some of those network functions process traffic based on the received packets and each of them may require a different configuration. In our use-case, we have considered:

* a NAT that must be configured in order to know which hosts belong to the private network (as the web cache) and which IP address must be used as masquerading address;
* an ACL firewall must be provided with a set of Access Control List (ACL) entries that specify which couples of nodes (e.g., the web client and server) are authorized to exchange traffic;
* a web cache, which does not require a particular configuration. It only needs to know which hosts belong to the private network.

A first step towards the NF-FG verification is the VNF chains extraction. In our use case, two chains are extracted from the NF-FG (Figure 5.36):

* the first chain (Chain A) is composed by a web cache, a NAT and an ACL firewall;
* the second chain (Chain B), instead, has an anti-spam function, a NAT and an ACL firewall.

Several tests have been run over these two chains in order to verify reachability properties between the graph nodes. Specifically the performed tests differ for the anti-spam and firewall configuration and for the traffic directions (from client to server and vice versa). Moreover, we have used a workstation with 32GB of memory and anIntel i7-3770 processor running an Ubuntu 14.04.01 (kernel 3.13.0-24-generic for x86 64 architecture).

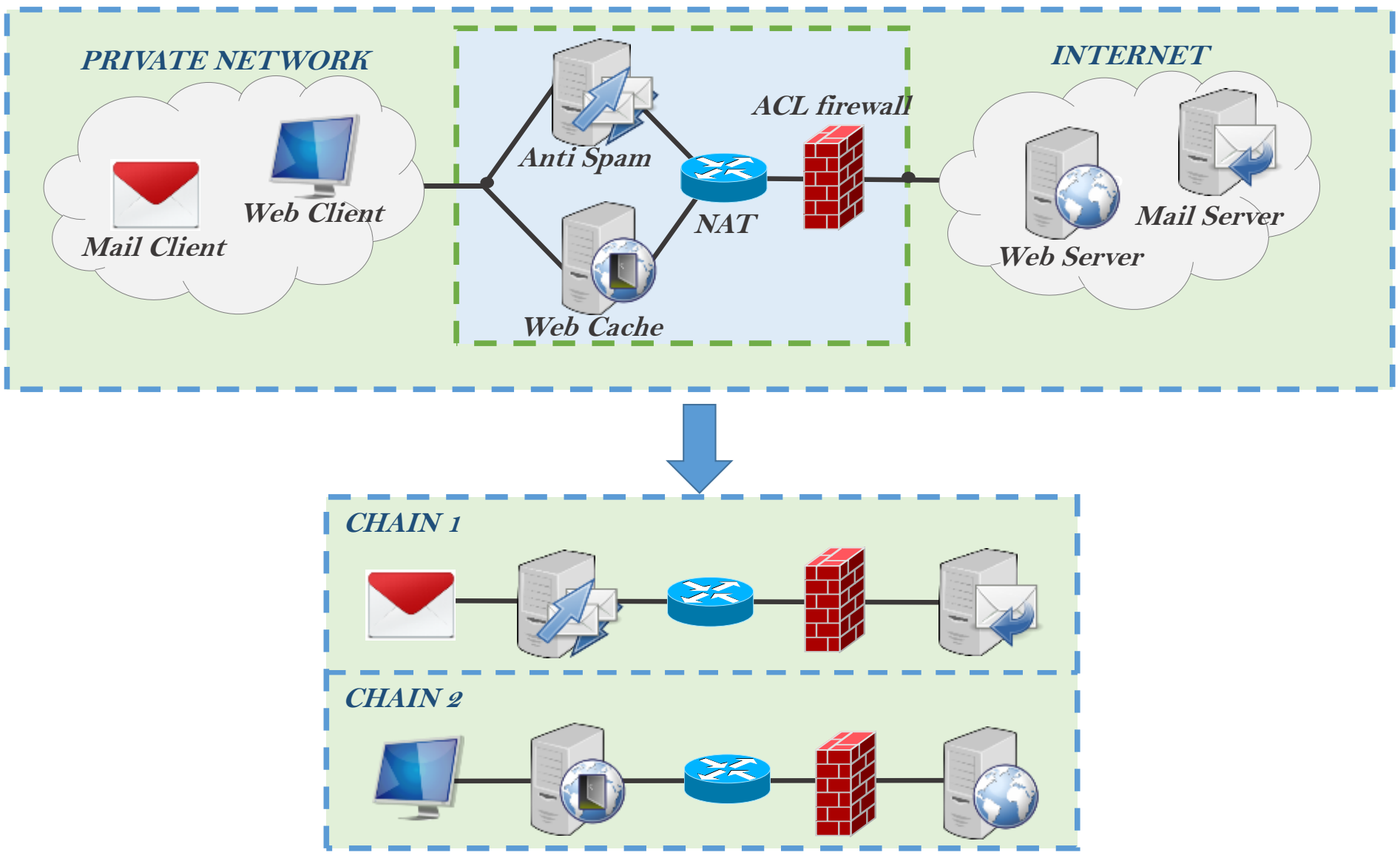

*Figure 5.36: NF-FG use case.*

On the first chain (Chain A), only the ACL firewall can be configured, hence we setup:

- a reachability test from the web client to the server and vice versa, where no ACL entry is set on the firewall;
- another reachability test is such that the firewall drops all packets exchanged between the client and server and vice-versa.

The reachability problem from the web-client to the server is satisfied as expected (the stripped bar in Figure 5.37). It is worth noting that the unsatisfiability of the problem in the opposite direction (the plain coloured bar in Figure 5.37) is due to the fact that client and server can exchange traffic only if the connection is initiated by the client. In the second test, in both traffic direction the reachability problems are not satisfied (Figure 5.37) because of the firewall VNF configuration.

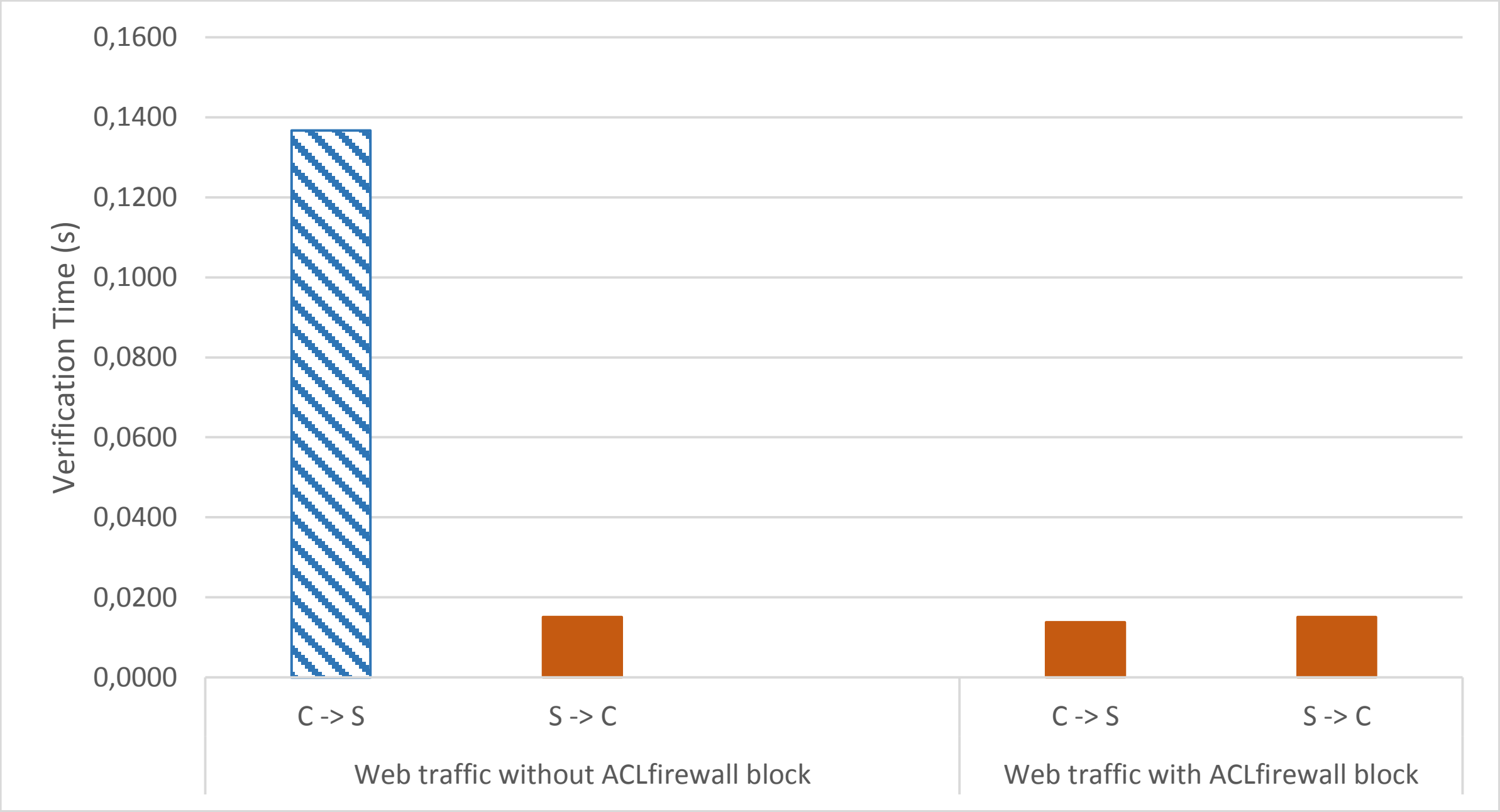

*Figure 5.37: Reachability test performed on a WebCache-NAT-Firewall chain.*

Instead the Chain B is tested in three scenarios, obtained by changing the firewall and anti-spam configurations as follows:

- the first test is performed without any function configured to drop the received traffic;
- in the second case, the firewall drops the traffic between the mail client and server;
- the last test is such that the anti spam is configured to drop all the emails sent by the mail client, while the traffic originated by the server is allowed.

In all of these tests (Figure 5.38), the verification problem is satisfiable in case of traffic sent by the mail client, while the reachability property is not verified for the traffic sent by the mail server for the abovementioned reasons.

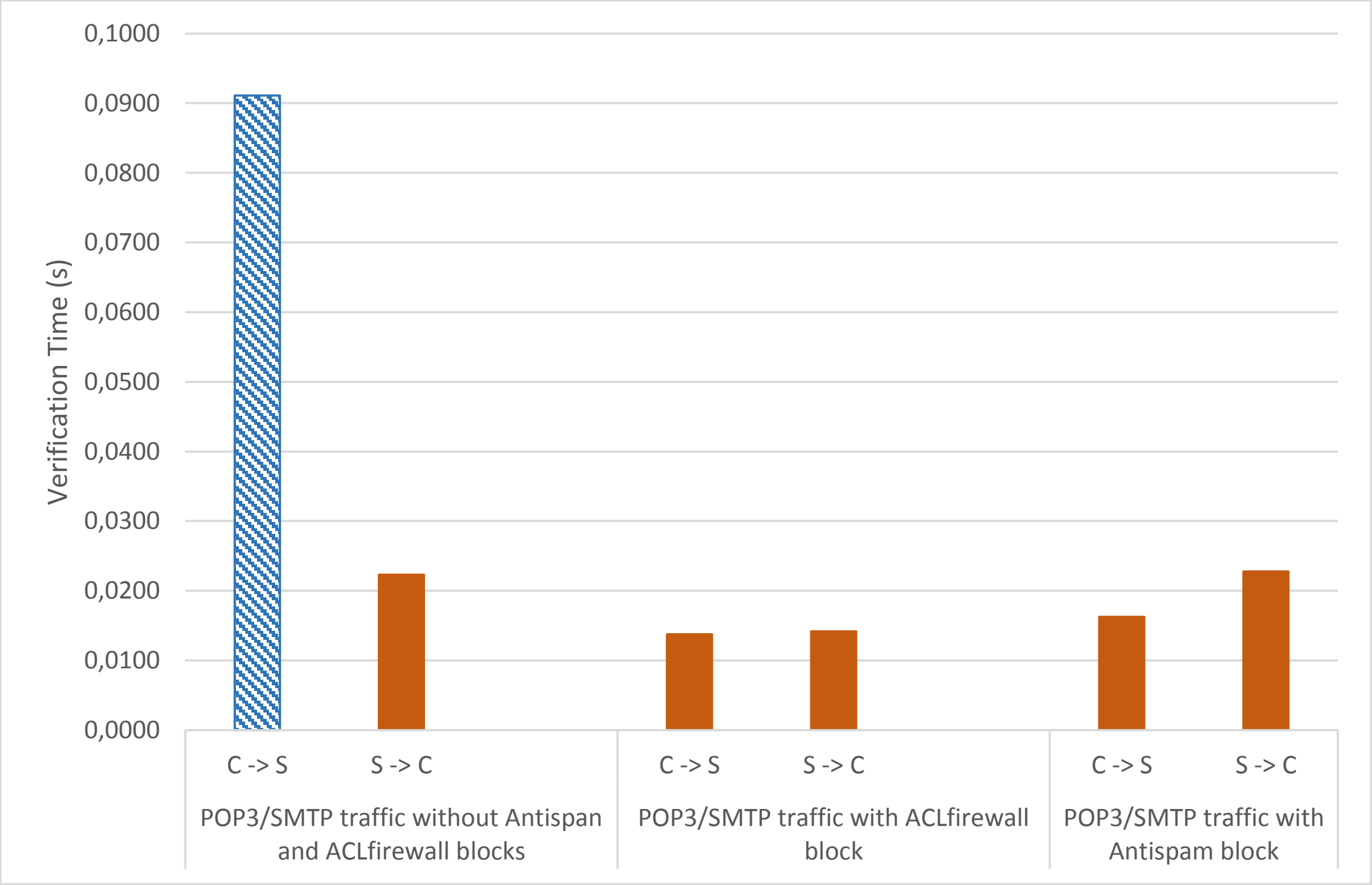

*Figure 5.38: Reachability test performed on an anti-spam-NAT-Firewall chain.*

As showed by our preliminary results, we have obtained promising performance in all the tests performed. In fact both diagrams show that the verification engine is able to solve a reachability problem in a reasonable time of about 200ms at worst, while the verification time is generally less than 50 ms.

### 5.10.5 Next steps

The next steps involve the investigation of new methods for:

- optimizing the incremental computation of the VNF chains starting from the NF-FG;
- designing of network property specifications in order to allow different UNIFY layers to drive the verification process (e.g. the user may want to verify a reachability property between the nodes A and B in the graph);
- enriching the current verification tool engine in terms of VNF models in order to be able to verify more complex SG/NF-FG use cases;
- concerning the prototyping activity, integrating the verification modules in the UNIFY architecture by considering the available platforms (e.g. ESCAPE [D22][D32]). For what concern the SG/NF-FG information retrieval, the verification prototype will be able to interact with a database optimized for storing and managing graphs, for example Neo4j [NEO4J].

## 5.11 Run-Time Verification of NF-FGs

The verification tool is an OpenFlow based tool and can be used to verify the ordered traversal of service chains as per the application-traffic. We plan to extend this tool to verify paths or trajectories taken by packets in the data plane as well.

### 5.11.1 Technical description

All kinds of application-traffic like HTTP, SMTP, VOIP etc have their own service chains or Service paths. For instance, HTTP has its service chain in the order Firewall>> IDS >> Proxy and VOIP has Firewall >> IDS. As shown in the below diagram in Figure 5.39:

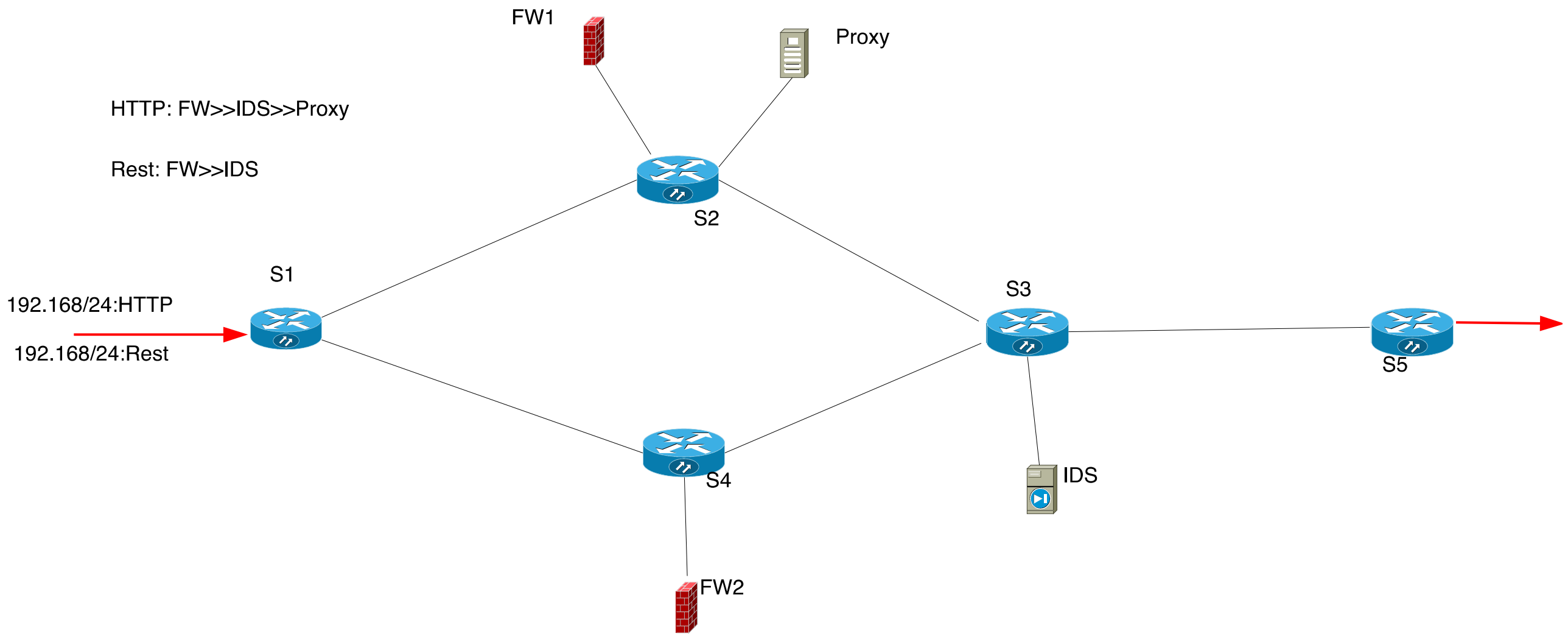

*Figure 5.39: Overview of a service chain*

An OpenFlow switch contains the Flow tables which contain rules pushed by the controller. These rules contain the data path traversal according to the order of service chains. Therefore, its very important for the Service Providers to ensure that there is correct traversal of service chains in the right order as well. Due to any issue, the packet from any particular application-traffic might violate the expected path. Sometimes, there are whiteholes or a switch that broadcasts the packets wrongly.

As of now, the tagging is done using VLAN ID or MPLS tag in the packets by using REST API. For every, service or middlebox, there is one unique pre-decided tag associated with it which is predefined and specific to a certain service. We plan to use MPLS which maintains a stack of values in order to preserve service chaining. We are planning to extend functionality to tag in the controller in the form of Controller Application as well. We randomly pick a packet to tag it at middleboxes and switches as well [SNA14][PTA14][NDU01][PKA12][HZG12]. Finally, at the end of service path or network domain, the collector gathers the tagged packets to do the analysis of the tags, in order to determine the service chain as well as the order in which the service chain was traversed. More specifically, in REST we define a criterion for e.g. HTTP port 80, followed by sending a REST command to the controller to tag every packet as specified. A packet matching the criterion may be tagged as VLAN 50, if it goes to the outport to any service or middlebox. This is achieved by installing a flow entry in the switch or middlebox - for instance, "Action : Set VLAN ID 50". The sampling rate considered here is 1 per 1000 packets. By the end of the service path, the collector perform the analysis of the gathered and tagged packets.

## Design choices and practical considerations

This verification tool will be advantageous to verify the service path/service chain of the packet on the data plane. The closely related tool with our work is Compiling-Path Queries [SNA14]. It performs the verification of the data plane paths taken by the packets using Pyretic and maintaining automaton states on the packet. However, our verification tool verifies the service path/service chain of the packet belonging to any kind of application-traffic. This is an issue because there can be : (1) white-holes in the paths like switches broadcasting packets and (2) bugs in OpenFlow switch implementation. These conditions might influence the packet to take the wrong path and not the expected path. For finding these errors, the verification tool is needed for detection. Future work will be focused on verifying the entire trajectory (Figure 5.39) and not just the service chains.

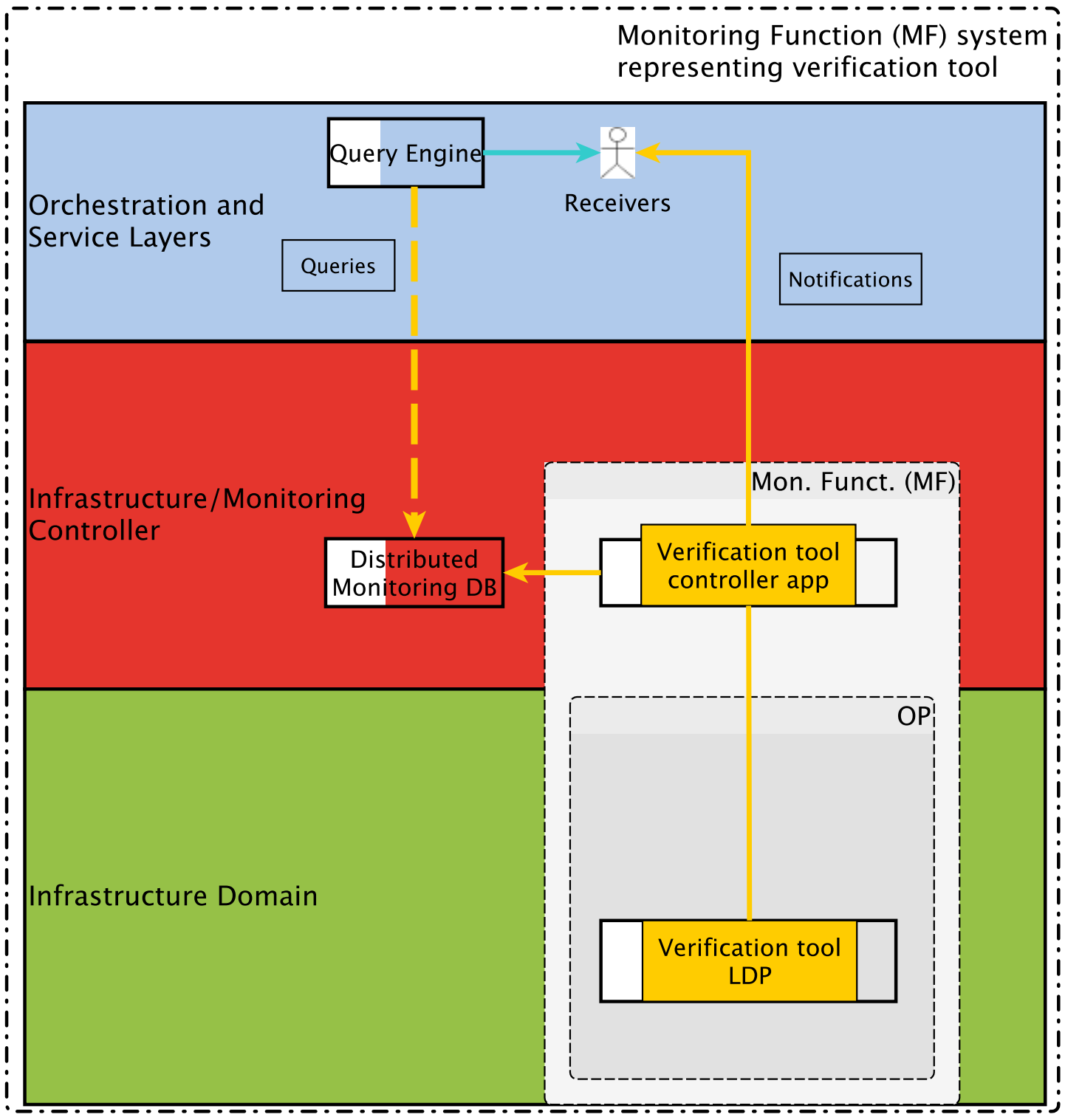

*Figure 5.40: Functional Architecture Mapping of the verification tool.*

## 5.11.2 Architectural mapping

**Functional architecture:** We can map Verification tool into the UNIFY functional architecture as MF control app (shown in Figure 5.40). Using this tool as the Monitoring function control app, all the service chains traversed but the packet can be verified by building a module in the MF control app.

**Universal node:** We present two mappings of the verification tool into the UN node and both the mappings will be implemented on the UN node. The first option (most preferred) depicted in Figure 5.41, shows the verification tool as a monitoring function (LCP) executing as part of the virtual switching engine management plugins. In the second option, the verification tool is located in one of the VNF as OP and therefore, it verifies the LDP. In this verification, VNF has to establish an OpenFlow session with all the other VNFs through the NF-FG LSI, as shown in Figure 5.41.

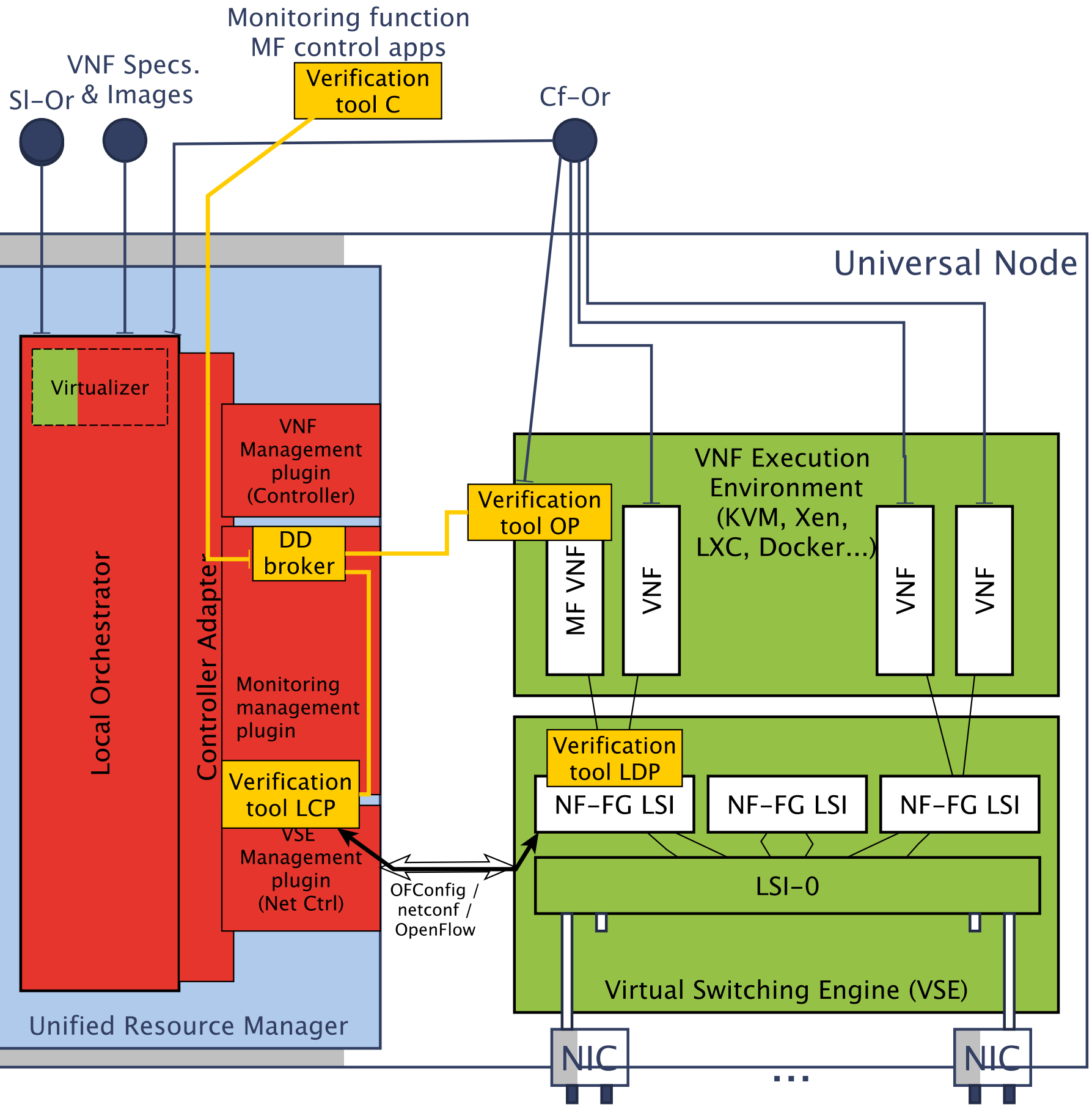

*Figure 5.41: The verification tool mapping into the universal node*

### 5.11.3 Support to SP-DevOps processes and objectives addressed

This verification tool addresses run time verification (RC7) and troubleshooting (RC10) research challenges. Addressing both research challenges is part of achieving O4.2 and O4.3. The approach relates to several requirements in D4.1: NL1 (NL1.2-1.5), NL2, and NL4 (NL4.2-4.6).

Based on tagging of randomly selected packets, the tool implements an observability process that in turn supports both verification and troubleshooting processes. The run-time service graph verification process is based on the analysis of tagged packets obtained via deployed observability points. The analysis allows for identifying SLA violations, which can be used for triggering a troubleshooting process in order to correct the problem.

### 5.11.4 Evaluation and individual prototyping results

We have preliminary evaluation of our novel prototype, through the RESTful API fired through the POSTMAN UI (REST Client) of Open Daylight Controller [ODL]. Given this setup, tagging works successfully on a set of 6 switches. Next step is to integrate this tagging mechanism through SDN controller application which links to controller through the northbound interface of the UN or switch.

### 5.11.5 Next steps

Currently, the run-time verification tool is focused on verifying service graphs. In addition, we plan to extend the approach to include verification of the trajectory taken by packets and identification of DoS attacks.

## 5.12 Network Watchpoints

The Watchpoint tool is a standalone debugging tool that helps to define network watchpoints for certain datapath or control traffic events (such as policy violations or OpenFlow rule matching) that trigger different actions (packet drops, notifications, etc). Our watchpoint tool operates in the spirit of ndb/NetSight [NHA14], but it contains more features and does not duplicate and collect every data plane packet transmitted through OpenFlow-compliant switches. The tool has two working modes. In the first, proxy mode, it intercepts OpenFlow control traffic. In the second, data path mode, it adds extra OpenFlow rules to the flow tables of switches. The technical details of the tool have already been reported in Section 5.2 of M4.1 and presented at the Y1 review as part of an integrated demonstration [FNE15] (Section 7).

### Design choices and practical considerations

In line with the goals of flexibility, extensibility and reusability of the UNIFY Architecture, the watchpoint tool is based on a modular design consisting of four main components [M41].

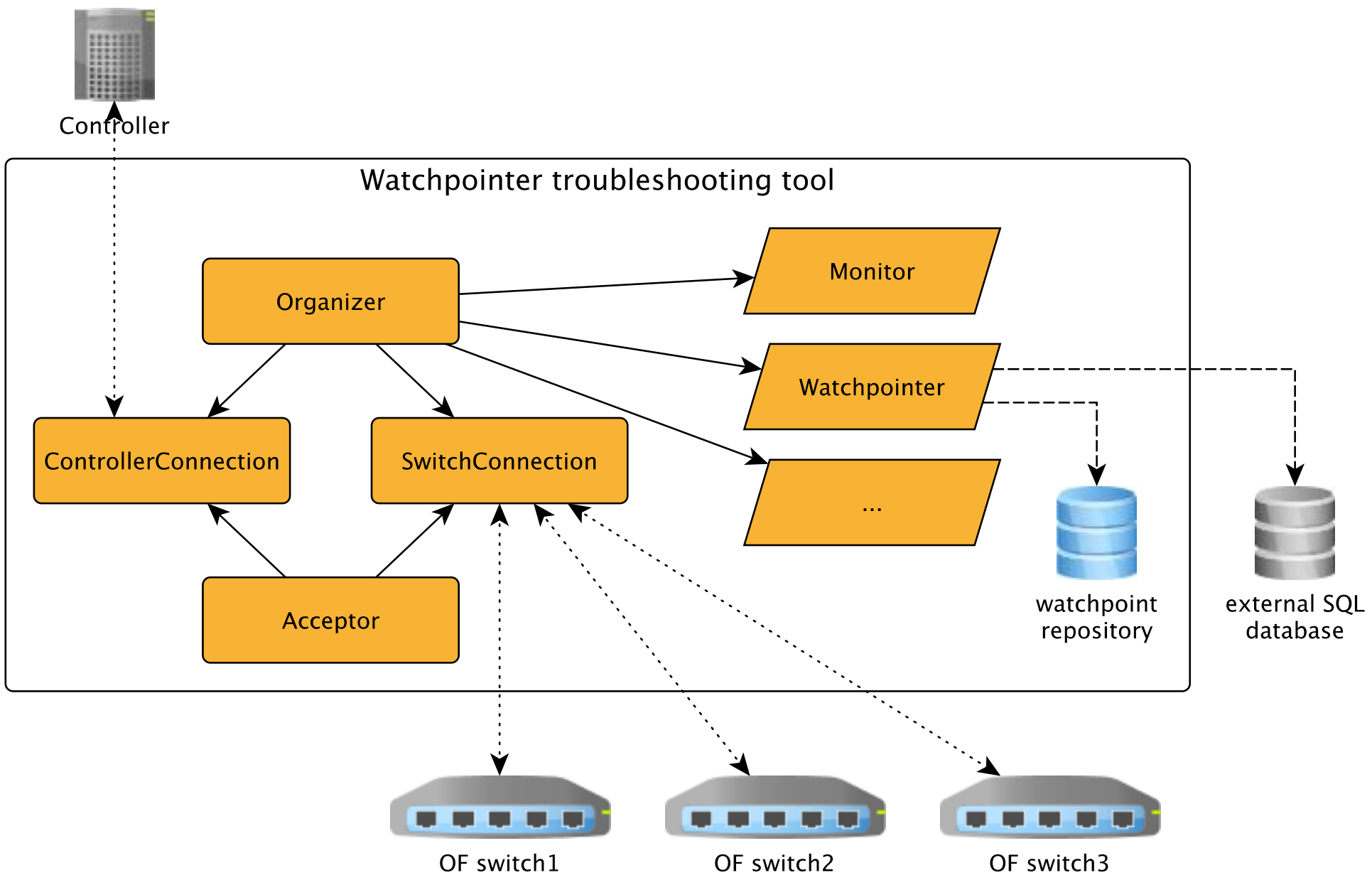

*Figure 5.42: Overview of the Watchpoint modules.*

In Figure 5.42, the Acceptor sub-module is responsible for the basic management of the control channel and handles the join intentions from an SDN element. The SwitchConnection and ControllerConnection sub-modules handle the corresponding OpenFlow connection, which is organized and controlled by the Organizer sub-module. This Organizer module is also responsible for managing application-specific troubleshooting modules.

### 5.12.1 Architectural mapping

Regardless of the operational mode of the tool, the users have to define watchpoints, i.e., a traffic pattern to search for and actions that are executed when matching traffic is found. The main question for a tool like this in conjunction with the UNIFY framework is who will define the watchpoints. In M4.1 [M41] we outlined two use-cases, summarized in the following.

In the first use case, the VNF developers explicitly add the Watchpoint tool to their service graphs; as a consequence they will be able to use the tool to monitor the behavior of their running NF-FG only. That is, if a developer wants to add policy enforcement to the NF-FG, it is necessary to include the Watchpoint tool as a VNF to the NF-FG.

In the second use case, the Watchpoint tool is part of the UNIFY infrastructure and user's policy definition is somehow transformed into watchpoints, which are installed in a Watchpoint tool running as part of the infrastructure. There is more than one possibility to transfer these policy definitions to configuration commands:

- The OSS module can collect policy definitions and configure the Watchpoint tool as shown in Figure 5.43. This approach nicely extends the resource policy functionalities described in Section 3.2.8 of D2.2 [D22].

- The policy definitions can be added to the service graph as annotation. This service graph, and later the NF-FG, is then transformed into a more detailed one by each layer of the architecture. Figure 5.44 and 5.45 depict this approach. The policies at the Service layer are expressed in the MEASURE language or other annotations of the NF-FG. At the Orchestration layer, the MEASURE annotation is decomposed and/or the NF-FG is extended with MFs. Finally at the Infrastructure layer, the Controller configures the Watchpoint tool according to the policies received in MEASURE description.

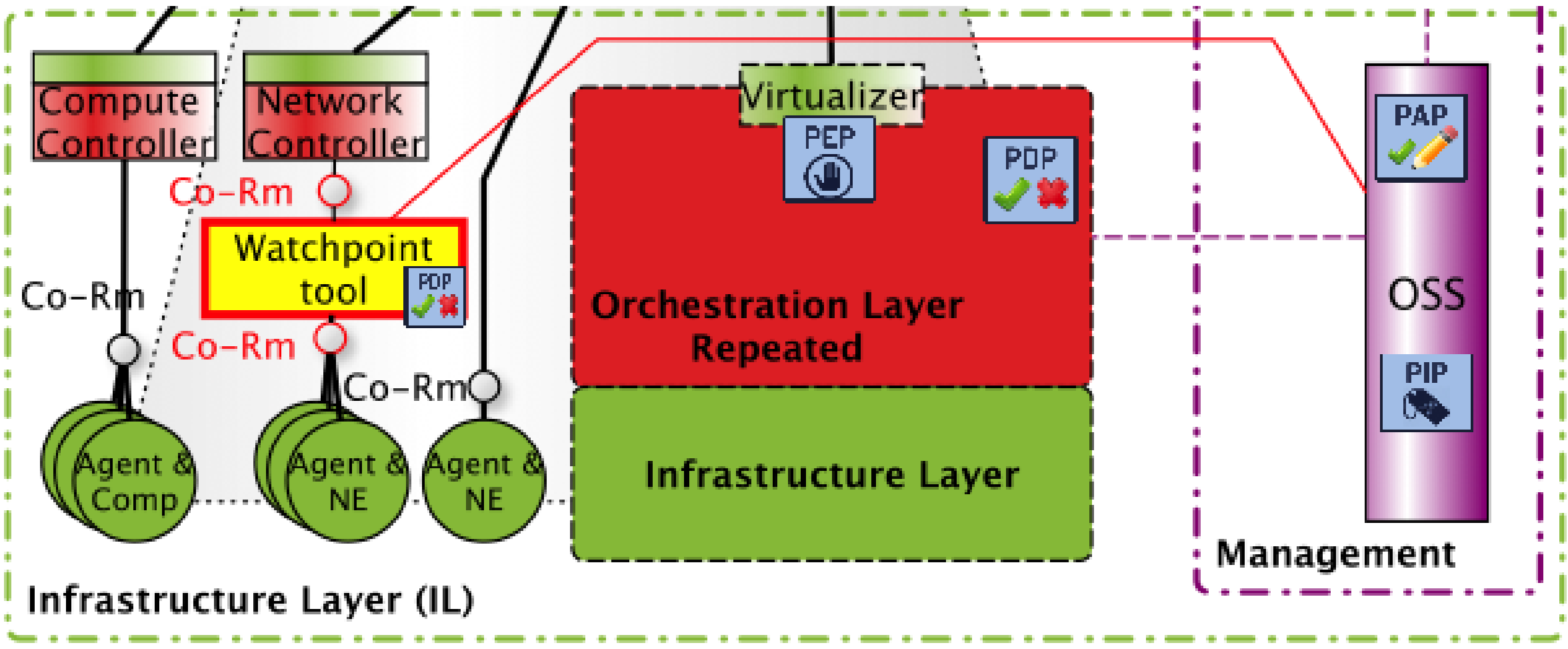

*Figure 5.43: Watchpoint tool and other policy functionalities in the UNIFY architecture (The figure is a modified version of the infrastructure layer part of Figure 12 of D2.2 [D22])*

Currently, there is one unresolved issue with the second approach: we assume that the Co-Rm interface between the network manager and the network element is OpenFlow, because the Watchpoint tool intercepts the OpenFlow control channel. However, this interface then is not capable of configuring the tool itself (i.e. it is not possible to install watchpoint rules with the standard OpenFlow protocol). To overcome this issue the UNIFY architecture can be extended with an additional interface, or the OpenFlow protocol itself can be extended with vendor specific (experimental) extension. Finding the exact method for configuration remains as a future work.

In case of policy violation, the Watchpoint tool can use the alarm/notification channels of the UNIFY architecture, or it can rely on the same communication channels that are used for collecting measurements.

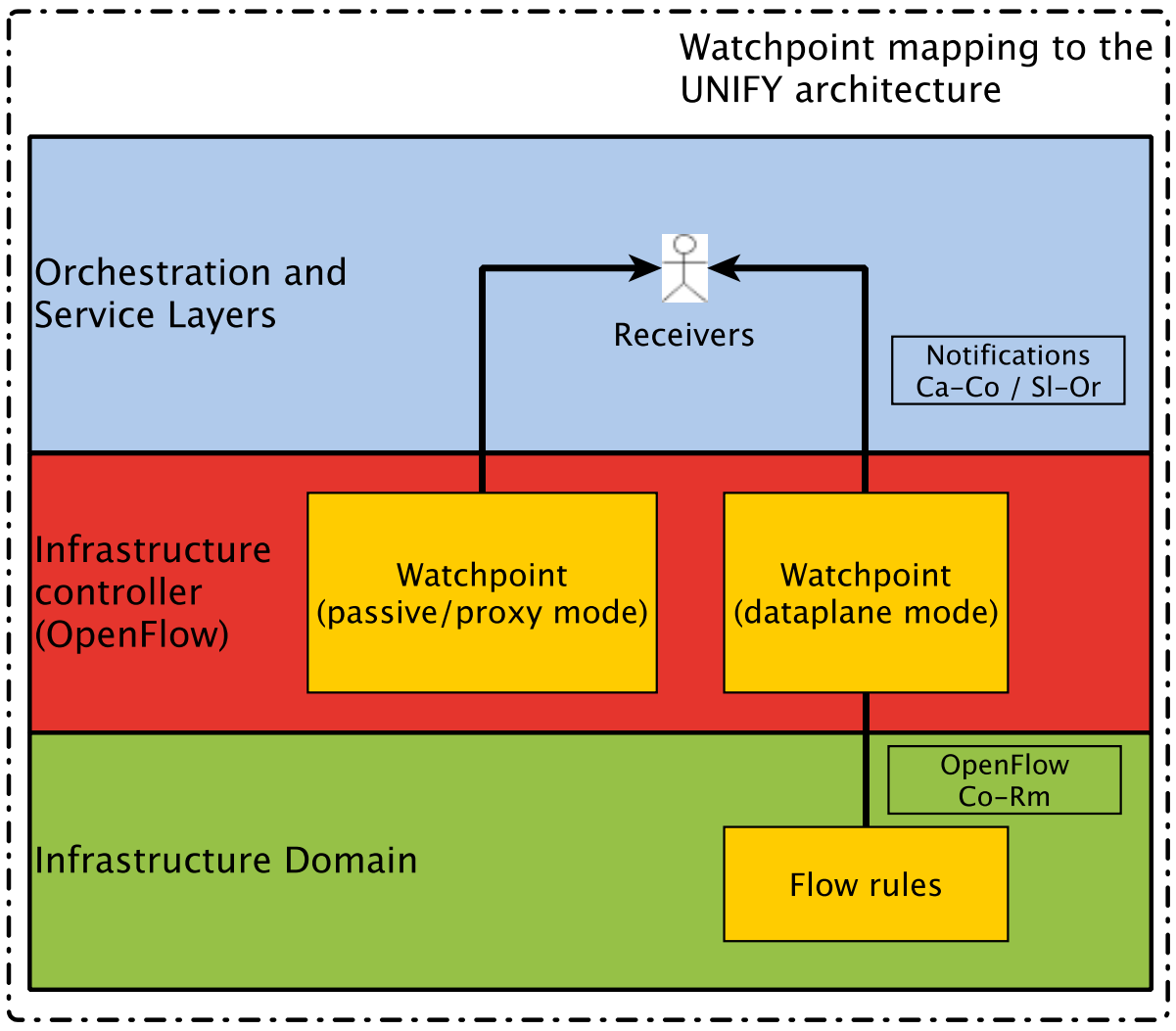

*Figure 5.44: Watchpoint tool in the functional architecture when it is part of the infrastructure*

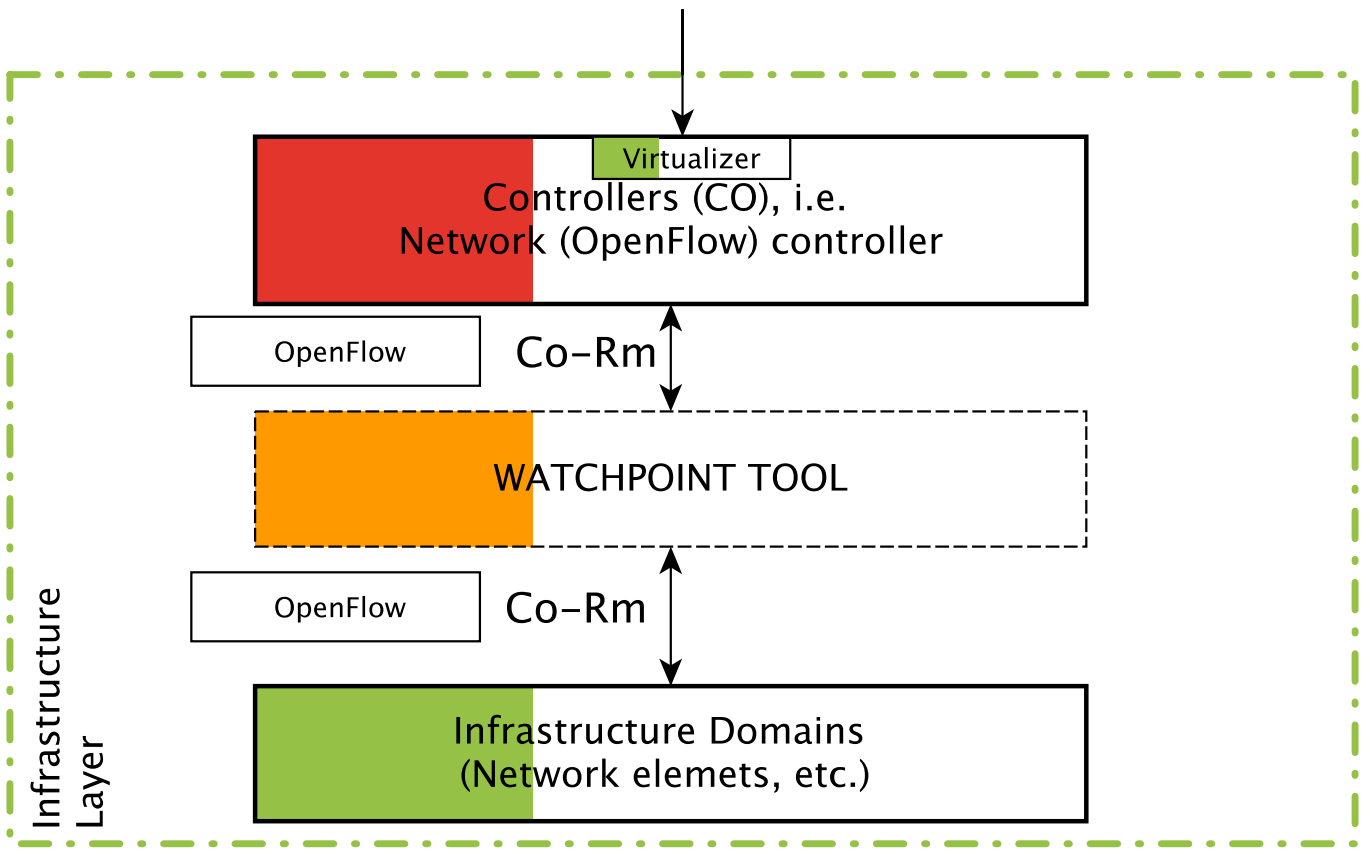

*Figure 5.45: Watchpoint tool in the Infrastructure layer when it is part of the infrastructure.*

### 5.12.2 Support to SP-DevOps processes and objectives addressed

The Network Watchpoint mechanism is useful as a standalone tool, but trying to use it in the UNIFY framework helps to refine the architecture (if it is part of the infrastructure) and helps to understand how a VNF developer can use it as a debugging tool (if it is part of a service graph). Therefore it addresses the same research challenges as the Epoxide tool (Section 5.13): **RC8** (Automated troubleshooting workflows) and **RC11** (VNF development support). Addressing both of these research challenges is part of achieving objective **O4.5** (Enable automatic definition of workflows for verification and activation tests for dynamic service chains). The approach relates to the following high level requirements in D4.1: **OR2** (The proposed SP-DevOps methods and tools must be in harmony with existing operational processes), **OL1** (The UNIFY architecture must support capabilities to develop and test components).

The Watchpoint tool can support the VNF Developer process if it is part of a service graph by allowing a developer to directly define watchpoints. But it can also support the troubleshooting process if it is part of the infrastructure by sending alarms in case of policy violations.

### 5.12.3 Evaluation and individual prototyping results

The prototype version of the Watchpoint tool is written in Java and based on FlowVisor. Watchpoint relies on persistent storage module of FlowVisors and extends its message handling and parsing modules extensively. The prototype is a standalone tool [M41], but it can also send alerts and notification through the DoubleDecker messaging system as we demonstrated it during the Y1 review (see Section 7) and at IM'15 [FNE15] as well.

### 5.12.4 Next steps

If the Watchpoint tool is used to intercept OpenFlow control traffic, currently it only processes packet_in messages looking for matching watchpoints. It could be extended to process packet_out messages as well. Next steps may also include tighter integration with a prototype version of the UNIFY framework (e.g., with ESCAPE [D22][D32]).

## 5.13 Epoxide: a Multi-Component Troubleshooting Tool

Due to the heterogeneous and distributed nature of computer networks, the detection of misconfigurations and software/hardware failures is frequently reported to be notoriously non-trivial. The advent of SDN complicates the situation even more, since besides troubleshooting, the problem of finding software bugs in controller/switch/VNF implementations also has to be solved. Today a wealth of general and SDN-specific troubleshooting tools is available which are usually tailored to identify network-related errors and bugs of a particular nature. Based on our paper [IPE15] this section highlights a troubleshooting framework, which can assemble many of these tools in a single platform and enables flexible combinations of them. In practice, network operators and SDN developers manually execute similar tasks in various ways to see what is going on in the network, for example by combining ping, traceroute and tcpdump (or more complex tools). Users can write wrappers around existing tools if the tool has a command line interface; or its standard input or output can be redirected; or it can somehow be automated, for example, through a remote procedure call interface. Creating wrappers requires writing a handful of simple functions in *elisp* programming language. Our framework can simplify their work by consolidating the available troubleshooting tools in a flexible, automated, reproducible and re-usable manner.

### 5.13.1 Technical description

Instead of writing a complex troubleshooting software that fully incorporates the functionalities of all the available troubleshooting tools (such as ping, traceroute, tcpdump, or OFRewind, VeriFlow, etc), we designed Epoxide to be a lightweight framework, which can flexibly combine these tools to achieve specific troubleshooting tasks tailored to the network under examination. Our aim is to allow creating combinations of the existing (and future) special-purpose tools more effective, using a troubleshooting framework, while keeping as much flexibility as we can to give the engineers more freedom.

The combinations of the existing tools can be properly defined in the form of troubleshooting graphs. A troubleshooting graph contains nodes as representatives of the tools themselves or additional convenience functions, connected by directed links indicating the flow of data between these nodes. Such graphs can represent practical patterns a troubleshooter may use to identify networking bugs. For an illustrative example, let us take a look at the SDN network depicted in Figure 5.46. The network contains four switches s1 . . . s4, two hosts h1, h2 and a VNF. We also have a management network using an additional switch, which all above nodes and the OpenFlow controller are connected. Now suppose that we implement a load balancing mechanism by marking the packets going from h1 to h2 at the VNF and by installing flow entries at s1, which decode the marks and forward the packets to s2 or s4 accordingly. Let us assume that there is a bug in the VNF, which marks the packets in so they use only the path through s2. For testing the system, a troubleshooter may want to do the following steps: (i) run an iperf from h1 to h2 (this reports that the connection is up but slow), (ii) get the flow entries from s1 to see if there is a problem in the switch (she sees that the rules seem to be OK, but the counters indicate that there are no packets through s4) and (iii) inspect the behavior of the VNF by attaching *gdb* to the VNF process running on the host, where she finally finds the bug.

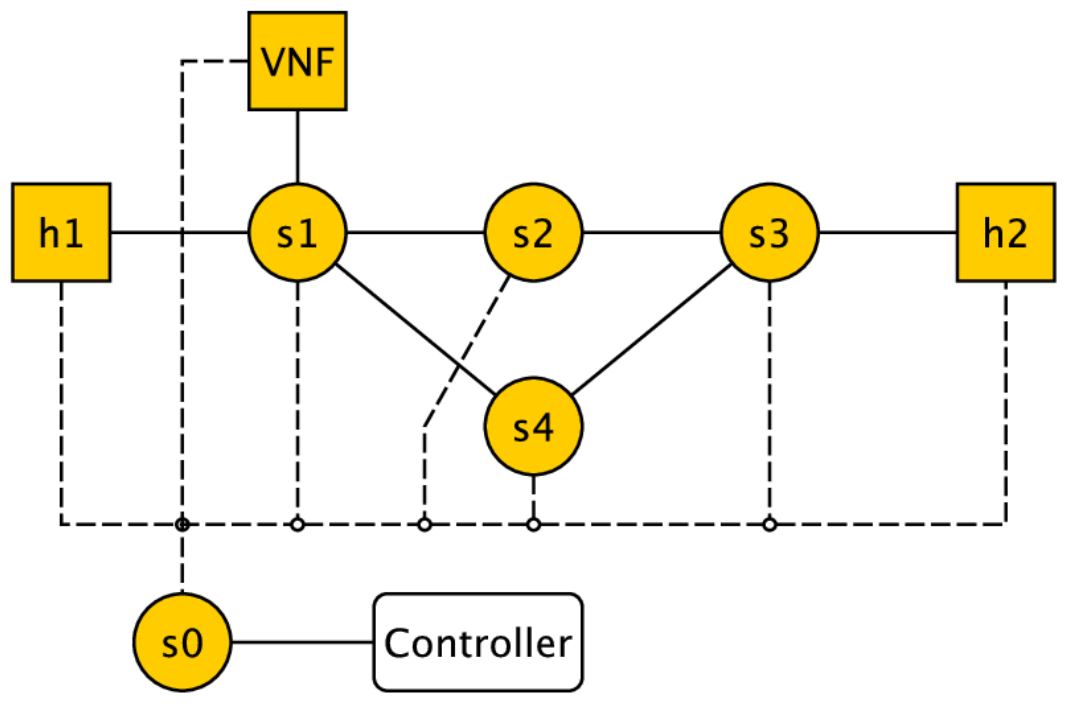

*Figure 5.46: An example SDN scenario.*

Instead of manually executing these three troubleshooting processes, the troubleshooter can use Epoxide to define three simple troubleshooting graphs (Figure 5.47), one for testing the network by injecting a flow, one for retrieving a particular piece of information from the flow table of a switch and finally a graph for determining the PID of a given VNF process and attaching *gdb*. Currently, the troubleshooter has to define these troubleshooting graphs either in an abstract, high-level domain specific language, or in an interactive manner. Epoxide then executes troubleshooting graph allowing the troubleshooter to examine not just the final result, but the intermediate steps as well.

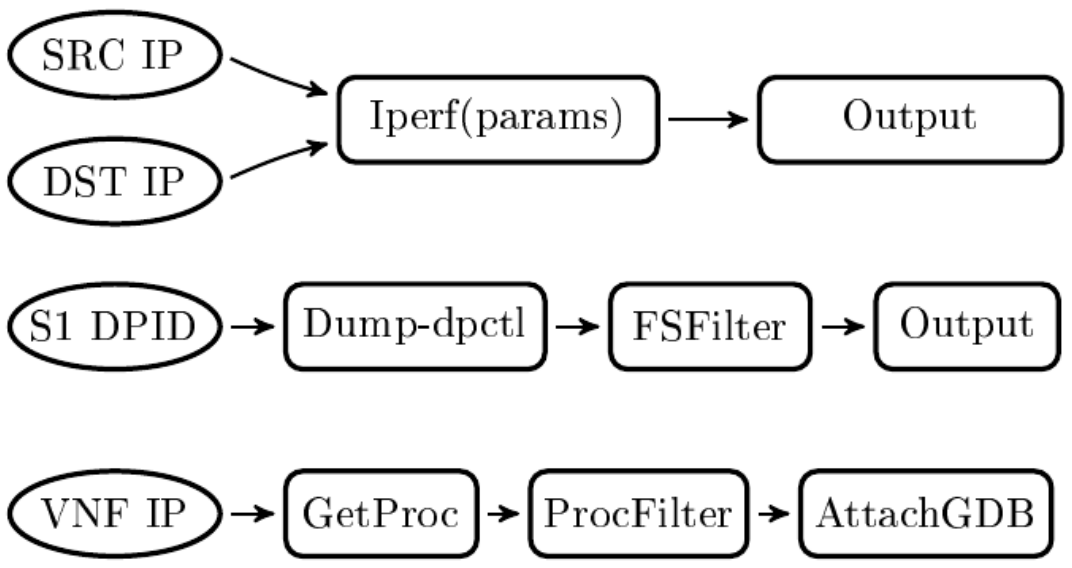

*Figure 5.47: Simple troubleshooting graphs for the scenario in Figure 5.46.*

Building around the concept of these troubleshooting graphs we have designed a modular framework called Epoxide which can describe and interpret such graphs and notify the troubleshooter about the outcome in a convenient way. Epoxide has the following properties (see [IPE15] for details):

- It defines and implements a common interface for different troubleshooting tools.
- It enables flexible assembling of the nodes to implement practical troubleshooting configurations represented by troubleshooting graphs.
- It provides an intuitive, human readable language and the corresponding interpreter to represent and execute configurations given by the troubleshooting graphs.
- It enables the developer to observe the operation step-by-step, but also provide a convenient alerting mechanism and representation of the outcomes of her configuration.
- It enables intelligent navigation between the invoked processes of a running configuration.
- It provides easy extendibility.
- It is relatively lightweight and fast, because Emacs is less demanding in terms of required CPU, memory, and storage resources compared to other modern integrated development environments like Eclipse and Visual Studio.

### Design choices and practical considerations

We have implemented a prototype of the modular troubleshooting framework on top of the *GNU Emacs* text editor with its extension language *Emacs Lisp*. Similarly to the Unix, where "everything is a file", a core concept of Emacs is the *buffer*, which can display and allow editing textual information. A buffer can hold different kinds of texts like files of source code, outputs of external programs called *sub-processes*, manual pages, other help information, web pages, email messages, etc.

Even interactive shells can be started in a buffer, where the user input is sent to the shell sub-process and the shell output is appended to the buffer. Running shells in buffers instead of terminal windows (e.g. xterm) is beneficial because of many reasons. First, the user interface makes possible to easily and efficiently switch between and work with an extremely large number of buffers. Second, information exchange between buffers is not limited to the clipboard - for example, the feature of completing partially written words considers candidates from every buffer. Third, the editing capabilities of Emacs are more advanced than the command line interface of shells.

We created an even scheduler, because Emacs does not currently support coordinated communication between buffers (i.e. sub-processes of the tools) according to a predefined pattern to implement our troubleshooting graphs. We designed our prototype to map the nodes and also the links of the troubleshooting graph to Emacs buffers. With the help of this event scheduler, Epoxide keeps tracks of modified buffers. As an immediate consequence, users can take advantage of the familiar core features of Emacs. For example, they can easily navigate among links and nodes, and by selecting a buffer they can observe or modify the state of a node or data of a link.

Somewhat similarly to the troubleshooting graph description of Epoxide, Business Process Modeling Language (BPLM) [AAR03] is used for defining complex relations among distributed computing entities. Additionally, configuration tools like Ansible [Ansible], Puppet [Puppet], Chef [Chef] also wraps available elementary tools to automate workflows.

### 5.13.2 Architectural mapping

Theoretically, Epoxide can be used to troubleshoot any kind of SDN system and the UNIFY framework is an SDN system, hence Epoxide with proper wrapper nodes could be used to debug / troubleshoot an implementation of the UNIFY framework, e.g., in the ESCAPE [D22][D32]. However, in D4.1 we decided that we focused on two DevOps aspects: the DevOps of service creation (i.e. creation of service graphs from available VNF components) and the DevOps of individual VNFs (i.e. developing a new VNF). Therefore, we have been investigating how Epoxide as a general purpose tool can fit into the UNIFY architecture. That is whether the U-Sl and the Sl-Or interfaces allow a VNF Developer to run Epoxide to troubleshoot a running NF-FG (Figure 5.48).

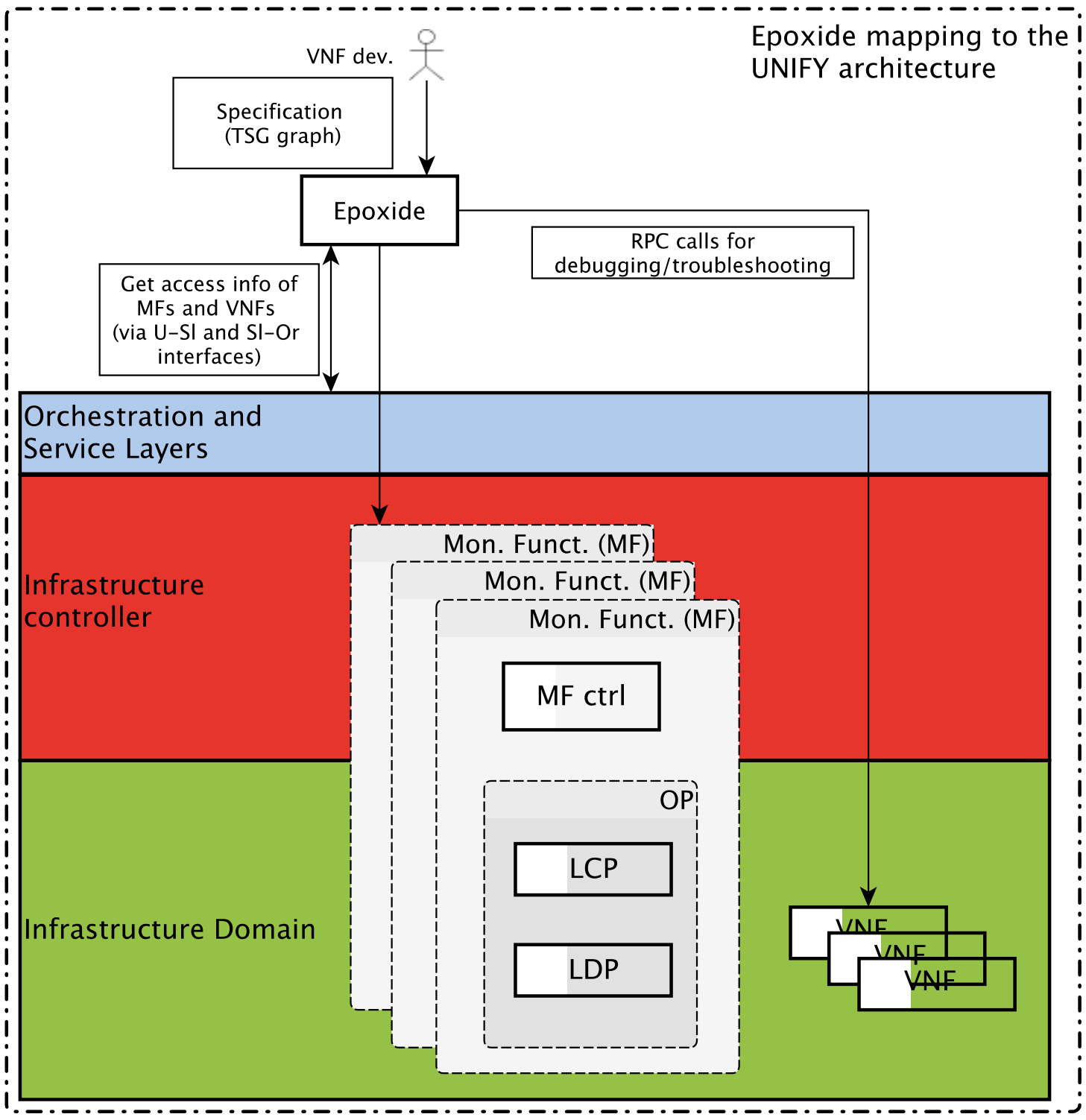

*Figure 5.48: Mapping Expoide to the UNIFY architecture.*

Let us assume that a VNF developer wants to develop a modified version of an existing VNF called NF2. In a typical scenario, this NF2 cannot operate in isolation, i.e., it has to be a part of an NF-FG (Service Graph #1, Figure 5.49). If the developer wants to test a new version of the VNF (NF2') with operational, live traffic, then it would be good if the VNF Developer could use the Sl-Or interface to ask for a modified service graph depicted as Service Graph #2 in Figure 5.49. The traffic duplication functionality (DUP1 and DUP2) could be a general purpose network function. However, a comparator network function (CMP) that compares some aspects of the output traffic should be probably application specific.

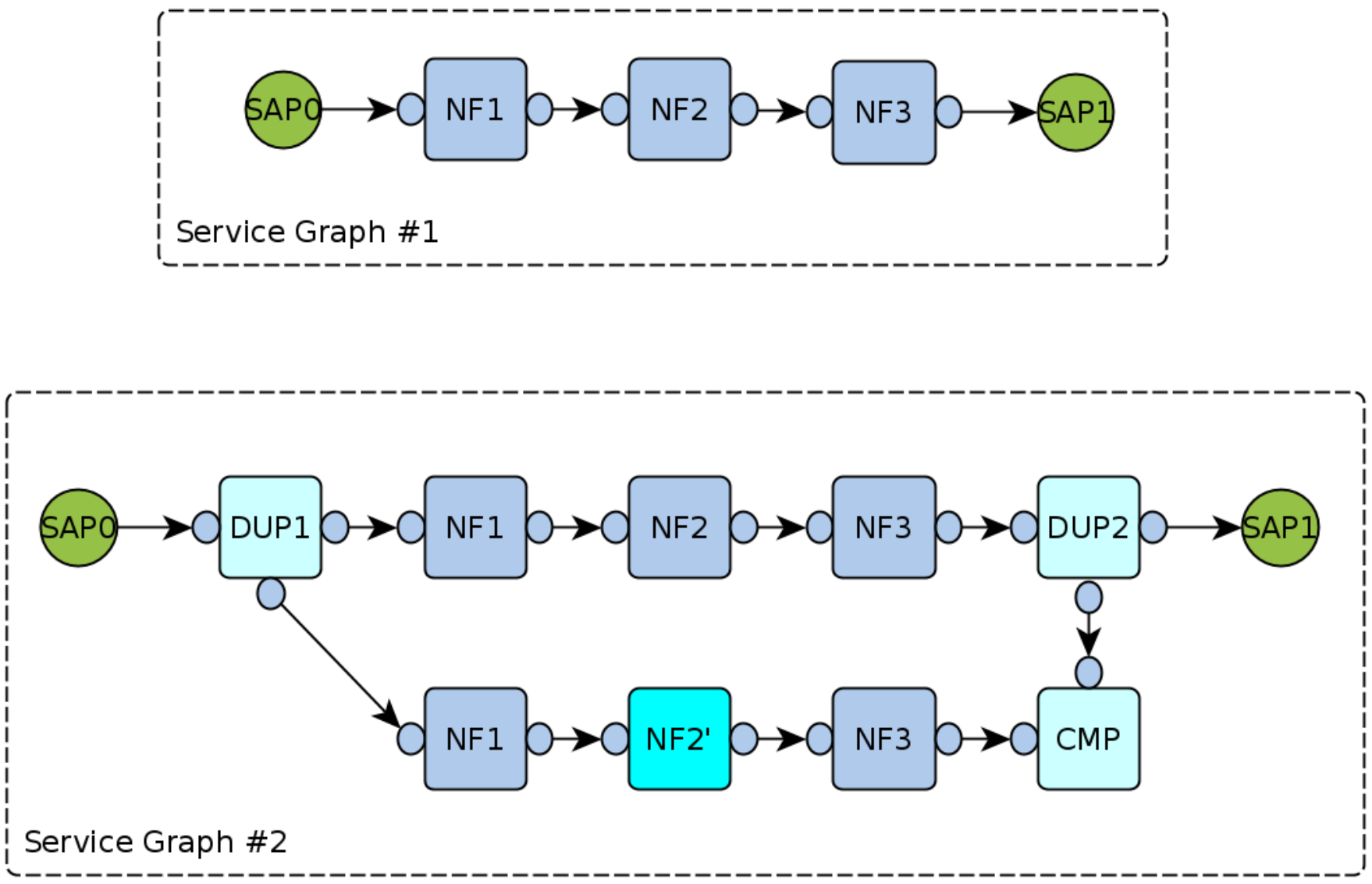

*Figure 5.49: Modification of a Service Graph in order to test a network function with operational traffic.*

If the VNF developer wants to debug NF2', then she should have access to the execution environment to be able to attach a debugger to her network function. Once again there is currently no predefined interface call to conveniently set up *ssh* or other management access to the VNF of a NF-FG, however the service graph can be further extended to have a management network as shown in Figure 5.50.

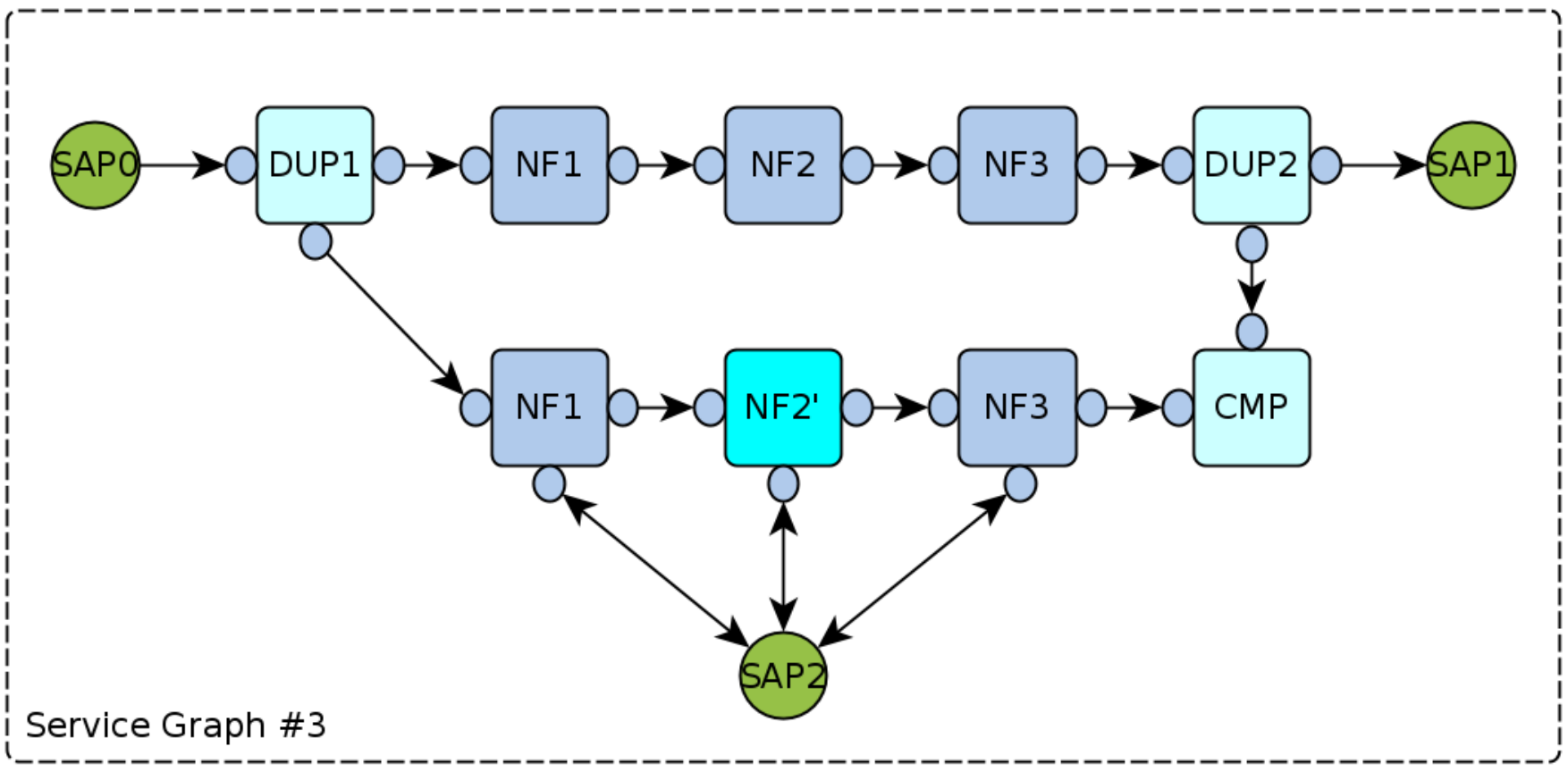

*Figure 5.50: Service Graph extended with a management network.*

### 5.13.3 Support to SP-DevOps processes and objectives addressed

The Epoxide tool addresses the following research challenges of D4.1: **RC8** (Automated troubleshooting workflows) and **RC11** (VNF development support). Addressing both of these research challenges is part of achieving objective **O4.5** (Enable automatic definition of workflows for verification and activation tests for dynamic service chains). The approach relates to the following high-level requirements in D4.1: **OR2** (The proposed SP-DevOps methods and tools must be in harmony with existing operational processes), and **OL1** (The UNIFY architecture must support capabilities to develop and test components).

The VNF Developer process described in D4.1 has many aspects. It facilitates the VNF developers with general observability, verification and troubleshooting capabilities. However, there are three additional sub-processes that Epoxide relies on. If a VNF Developer would like to use Epoxide, she will need to use the following sub-processes. Or more generally, if we want to integrate a tool like Epoxide with the UNIFY framework, we need to rely on the following sub-processes and they should be accessible via the U-Sl and the Sl-Or interfaces:

- Adding a new VNF to the production environment.
- Modifying an already deployed service graph with a new or updated VNF
- Attaching VNF to software IDE.

However, even without explicit support for attaching a VNF to software IDE, Epoxide could be used by "manually" extending the service graphs as Figure 5.50 demonstrates.

### 5.13.4 Evaluation and individual prototyping results

In an illustrative example to show the benefits of Epoxide, we executed two troubleshooting sessions of finding bugs in the setup of Figure 5.46. First, we started the elementary tools in "xterm" windows. The second time, we defined and executed the troubleshooting graphs of Figure 5.47. Although we have not tried to scientifically quantify the effectiveness of the user interfaces of the two approaches, we can definitively claim that the user can more easily lost its way among her large number of xterm windows. On the other hand, defining troubleshooting graphs might require learning Epoxide a bit, whereas script writing in, for example, bash can be regarded as common knowledge that current system administrators / network engineers all possess. Troubleshooting graph is easier to share and enhance, and it is more straightforward to turn them into a completely automated workflow.

### 5.13.5 Next steps

The current version of Epoxide provides only a framework and a limited set of illustrative elements. At this point in time we have wrapper classes for ping, iperf, ofctl, gdb, classes for consolidating basic POX, OpenDaylight and Floodlight functionality, plus some basic filtering classes. Writing more practical node classes (e.g., wrappers for a wide range of existing troubleshooting tools) and providing a core set of troubleshooting graphs and their definitions in the high-level language of Epoxide are clear directions for future works. Additionally, we are turning our attention to integrating NF-FG troubleshooting tools developed in the UNIFY project. Moreover, we plan to set up a public repository in order to invite the networking community to share its troubleshooting know-how by contributing node implementations and troubleshooting graphs.

Another interesting future direction, which we plan to address in Task 4.3, is the implementation of the automated generation of Epoxide configurations. We envision a specialized constraint logic solver that can generate large parts of the configurations by requesting only the essential information from the developer. Such solver could speed up the definition of valid configuration files and let the developer focus on her primary work.

# 6 Case study

## 6.1 Introduction

Today's Telecommunications networks rely on a large number of specialized appliances or middle/boxes, executing network function (from ISO OSI Layer 4 to Layer 7). Diffusion of more and more advanced terminals (e.g., smartphones) for wireless video streaming services are expand even more the range of middle-box applications. If from one side middle-boxes offer improved security (e.g., firewalls and intrusion detection systems), improved performance (e.g., proxies) and reduced bandwidth costs (e.g., WAN optimizers), on the other side middle-boxes imply high (CAPEX and OPEX) costs; in fact, not only middle-boxes are complex and specialized processing nodes, but they implies variations in management tools across vendors' solutions, and the need to address policy interactions between these appliance and other network infrastructure. Then it is argued that middle-boxes are playing a key role in current Telecommunications networks: systems and methods for simplifying deployment and management of middle-boxes would have important impacts in terms of optimising CAPEX and OPEX.

Starting from this overall motivation, this section is addressing the use-case Elastic Network Function (which also has been reported in Deliverable D2.1 [D21a]), specifically referring to the provisioning of middle-boxes functions (such as firewall) with the capability of scaling in real time so to dynamically adapt to the traffic or overall requirements changes. In addition to that, the use-case considers that the middle-box – which is a generic Elastic Network Function - could be instantiated by a service graph including different functions and/or a number of instances of the same function.

According to D2.1 Section 3.4.1 [D21a], the scalability depends on the application and can be separated in three potential options:

1. Scalable Network Function which scales up/down by itself in the resource space pre-assigned by management actions;
2. Semi-transparent scalable Service Chain which scales in/out by an action of a supervising management element adding/removing elements to the service chain;
3. Non-transparent scalable Service Chain which scales in/out by actions of a control/management entity in the service chain itself.

## 6.2 Elastic Firewall

In this sub-section, the analysis is focussed on a specific middle-box, i.e., Network Function, which is the Elastic Firewall. The Elastic Firewall can be modelled as a non-transparent scalable Service Chain which is managed by an Elastic Firewall Control App inside the running application (i.e. the third option mentioned above). How an Elastic Firewall implementation could look like and which architectural components it is communicating with is shown in Figure 6.1 below:

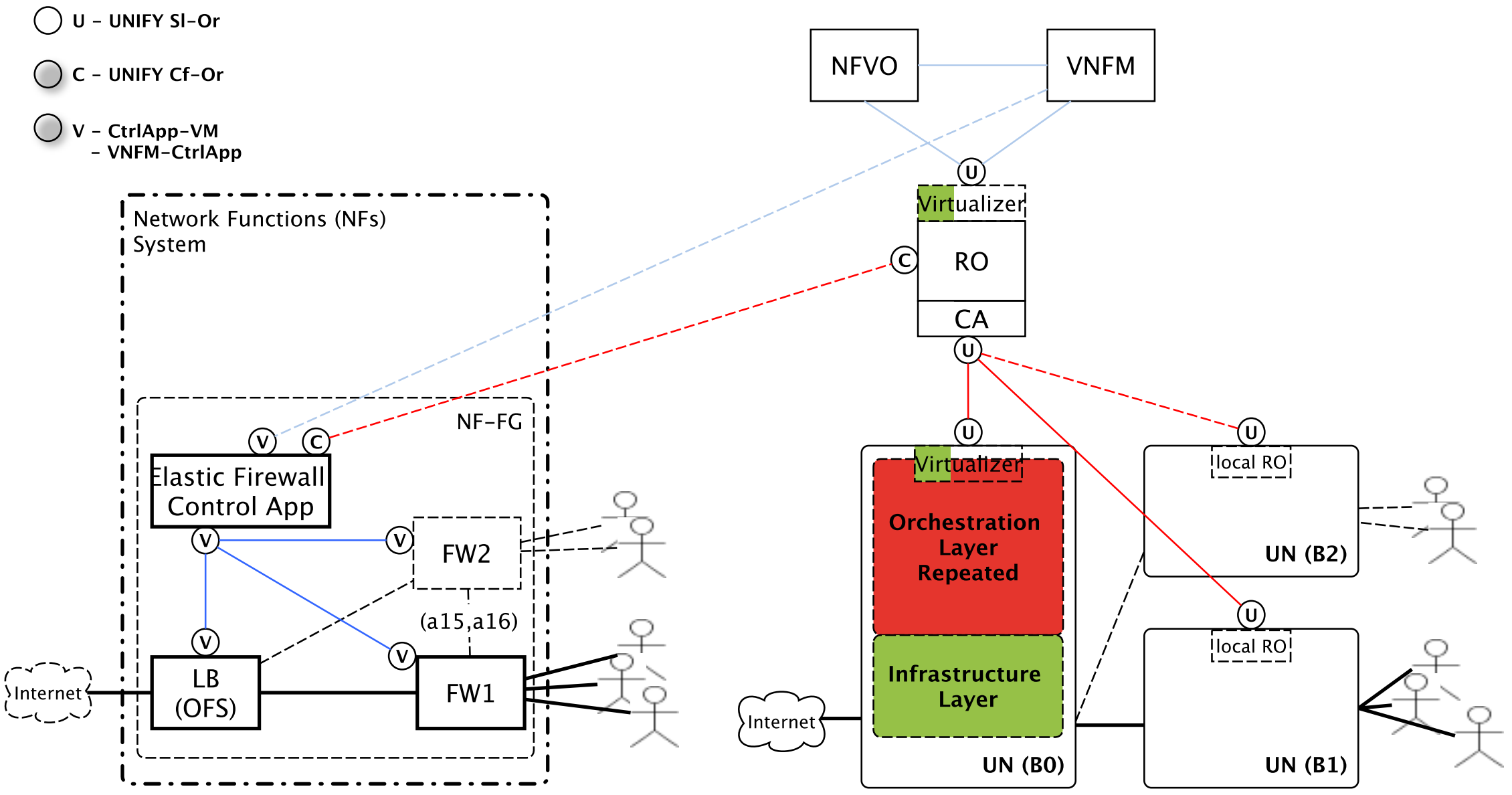

*Figure 6.1: Elastic Firewall example and relation to UNIFY architecture - the figure is from Fig. 4.2 in the D2.1 amendment [D21b].*

The NF-FG shown in Figure 6.1 (and the D2.1 amendment [D21b]) consists of:

- Elastic Firewall Control App designed for the Elastic Firewall control providing the external interfaces to the RO (Cf-Or) and to the VNFM
- Load balancer (LB), a data plane element implemented as OpenFlow switch (OFS) which receives forwarding rules from the Elastic Firewall Control App and forwards traffic accordingly to the stateful Firewall elements
- Stateful Firewall elements which perform the firewalling as well as supporting the migration of state in-between the firewall elements based on the Elastic Firewall Control App requests.

The right side of the picture shows a simplified version of the UNIFY Architecture focused on the use case relevant elements only:

- the service layer consists of the two management elements NFVO and VNFM;
- the orchestration layer consists of a resource orchestrator with virtualizer and controller adaptation (CA);
- the infrastructure layer consist of the resources with three UNs (B0, B1, B2) connected with each other, the customers as well as the Internet. In addition each UN has its own (repeated) orchestration layer in each element (applied recursion in the architecture).

## 6.3 DevOps Support by SP-DevOps Processes

To understand the role of DevOps in a telecommunication service use case like the elastic firewall, we list the main phases of a service lifecycle and recap how UNIFY SP-DevOps processes can support them:

- Plan, Software Development (Code, Build, Test), Release and Update
  - Most stages in this phase are not covered in UNIFY SP-DevOps and this deliverable. **VNF development**, however, is partly supported by *SP-DevOps VNF Development tools* (e.g. by EPOXIDE, the multi-component debugging tool – see Section 5.13)
- Test and Verification
  - **Verification** is not limited to the code validation but it includes the assurance that the service configuration is compliant with the service definition, e.g., quality and performance indicators. Supported by *SP-DevOps Verification tools* (Sections 5.10-5.12.
- Operations
  - **Observability** can be defined as the property that provides visibility on the status of both physical and virtual components of the Elastic Firewall at the relevant time scale. Observability is supported by *SP-DevOps Observability tools* (Section 5.5-5.9), and will require:
    - placement of the monitoring functions/capabilities;
    - programmability features of the monitoring functions/capabilities;
    - prediction capabilities of instabilities;
    - triggering reactions and adjustments like scale in/out of elastic network functions
  - **Troubleshooting** concerns identifying the cause of potential failures, and could be triggered manually or automatically. *SP-DevOps Troubleshooting tools* include almost all proposed tools in this deliverable (see Sections 5.5-5.13). In the remainder of the project duration (Task 4.3), we will focus on automation of troubleshooting workflows in more detail.

## 6.4 Example of SP-DevOps Tools used for Elastic Firewall Fulfilment and Assurance

For a more concrete understanding of how SP-DevOps process and tools will aid operators, we describe two scenarios related to the elastic firewall, and point out how and when SP-DevOps can play a role. The first scenario deals with the deployment of an elastic firewall, while the second scenario addresses the operations phase, including dynamic scaling based on real-time monitoring data.

### 6.4.1 Deployment of an elastic firewall

This scenario describes the deployment steps of an elastic firewall and corresponding monitoring functions. The deployment process is supported by SP-DevOps Verification, to allow identification of problems early in the service lifecycle. This can significantly reduce times and costs spent on debugging and troubleshooting processes, which are especially complex for telecommunications services due to the high spatial distribution and the lower levels of redundancy in an operator environment.

The SP-DevOps tools included in this scenario are the general support functions DoubleDecker messaging system (Section 5.1), the MEASURE description language (Section 5.2), network performance monitoring functions for rate-, delay- and loss monitoring (Sections 5.5 – 5.7), the service model and configuration verification tools (Section 5.10); as well as the watchpointer tool (Section 5.12).

A deployment scenario includes the following steps (SP-DevOps involvement pointed out in **bold**):

1. The service with its components (i.e. an elastic firewall) is defined as a Service Graph (SG) together with KPIs/KQIs in the Service Layer by a service developer.
2. **The SG is verified by the SP-DevOps Service model verification module.** If the verification fails, the SG is returned to the service developer for refinement.
3. An NF-FG is defined and generated by the service layer and forwarded to the lower Orchestration layer. The NF-FG in this example represents the elastic firewall consisting of a control app, a load balancer and a first FW instance. **The appropriate KQIs/KPIs are translated into resource requirements and MEASURE code for service assurance.**
4. **The NF-FG is verified by the SP-DevOps NF-FG verification module**. If the verification fails, the NF-FG is returned to the service layer, where it can either be redefined based on the SG, or the SG can be returned to the service developer for refinement.
5. The NF-FG, including the associated MEASURE code, is split into the scopes of several controllers (UN (B0) and UN (B1) - **requirements (configuration/scope/actions) for requested monitoring functions are translated and expressed via MEASURE**.
6. Before deploying/installing the mapped NF-FG on the physical topology, **the network configuration (expressed in OpenFlow), is verified by control rule and configuration verification modules. Additionally, the watchpointer tool can be used for further policy enforment or debugging purposes**.
7. The performance monitoring functions, required to follow up on KPIs and to fulfil the requested monitoring, are **deployed and set up automatically according to MEASURE expressions**. The deployment of Observability Points is done with a controller southbound interface (e.g. OpenFlow, etc.)
8. The monitoring functions and their OPs operate in the infrastructure according to the configuration (including notification thresholds, etc.) expressed in MEASURE. **Monitoring data is collected via the DoubleDecker bus, which is also responsible for transmitting automatic notifications generated by measurement aggregation points in case of performance degradations.**

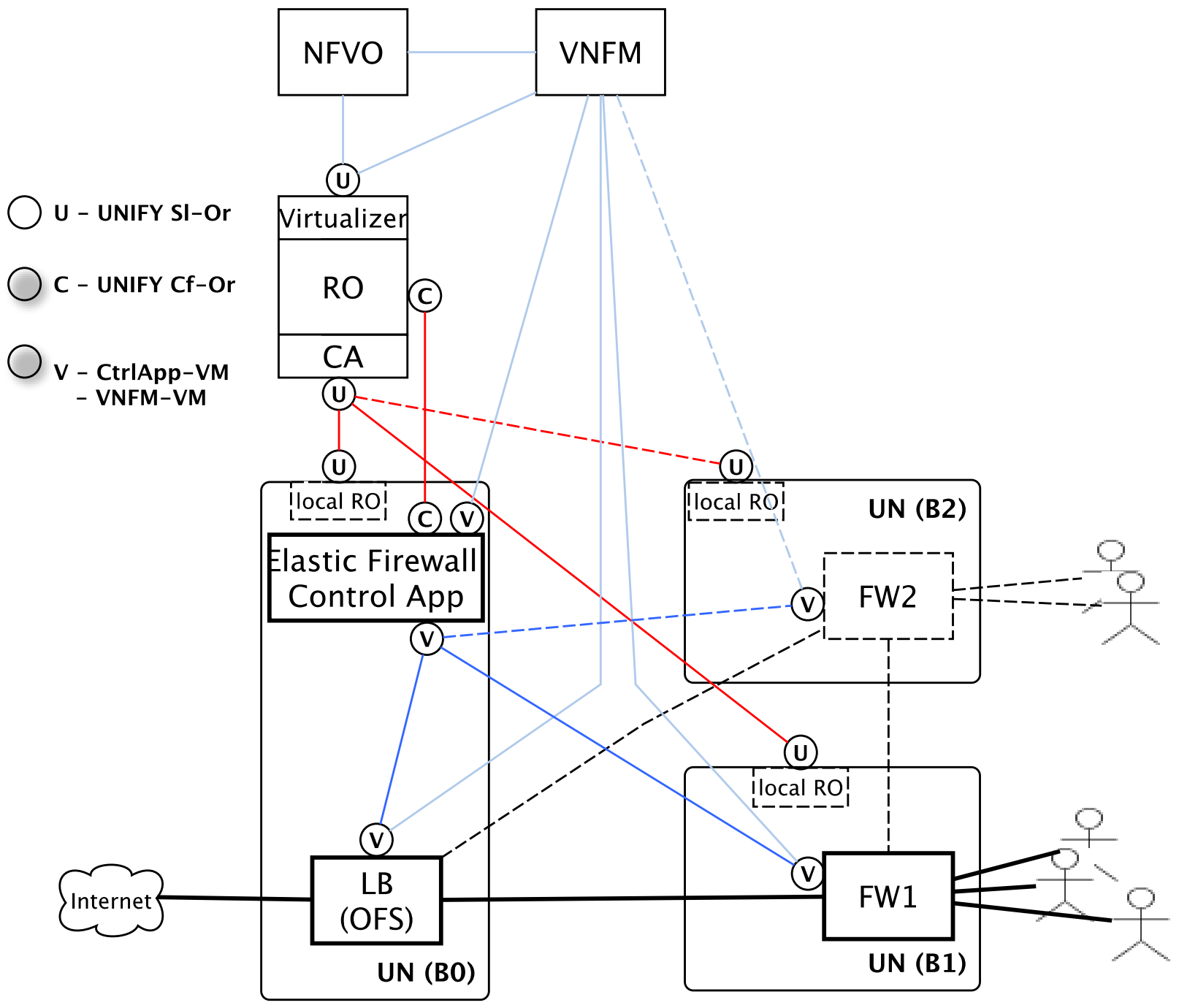

*Figure 6.2: Elastic Firewall deployed as a non-transparent scalable Service Chain with own control app.*

### 6.4.2 Dynamic scaling of an elastic firewall

This scenario describes the steps of scaling out an elastic firewall based on SP-DevOps observability tools. The scenario considers the deployment option 3 of a non-transparent scalable service chain which scales in/out by actions of a control entity in the service chain itself (i.e. the elastic firewall control app). The reference scenario given after deployment is shown in Figure 6.2.

The SP-DevOps tools included in this scenario are the general support functions DoubleDecker messaging system (Section 5.1), the MEASURE description language (Section 5.2), the recursive query engine and language (Section 5.3), network performance monitoring functions for rate-, delay- and loss monitoring (Sections 5.5 – 5.7), as well as the configuration verification tool (Section 5.10).

A dynamic scaling scenario includes the following steps (SP-DevOps involvement pointed out in **bold**):

1. The Elastic FW control app is deployed within the VNF EE in UN (B0). On the data plane, a load balancing OpenFlow switch (LB) is deployed in the VSE in UN (B0), and a first firewall instance (FW1) is deployed on another UN (B1). As initial setup, all traffic to the elastic firewall is forwarded to FW1 by the LB.
2. The network performance monitoring functions (MFs) are automatically deployed in UN1 (B1) and UN (B1) and are **operating in line with the configuration expressed by MEASURE. The Elastic firewall control app initiates a subscription on the DoubleDecker messaging system for the topic "performance degratations at FW1".**

3. **The MFs send monitoring information via the DoubleDecker messaging system towards a monitoring DB** on in the resource orchestration layer for purposes of resource utilization updating and SLA reporting. Monitoring information include link rate, delay, and loss estimates.
4. **A service developer/operator can follow up and report on SLAs by querying the monitoring DB using the recursive query engine in the Service layer,** Queries can be kept on a high level of abstraction (e.g. for one elastic firewall), without knowledge of the actual deployment (with load-balancer, multiple FW instances and overlay network).
5. The traffic rate is increasing towards FW1 - **the risk of overload is detected by the rate monitoring function. The warning is published on the DoubleDecker "performance degratations at FW1" topic, the overload warning triggers a scale-out event** at the Elastic firewall control app.
6. The Elastic firewall control app requests additional resources by augmenting the NF-FG and the annotated MEASURE expressions, sending it to the Resource orchestrator for instantiation via the Cf-Or interface.
7. **The resource orchestrator RO queries the monitoring DB** for up-to-date resource utilization status to find suitable infrastructure resources for an additional firewall instance with sufficient network resources for the overlay (satisfying service KPIs).
8. The Resource orchestrator decides to use free resources on UN (B2). Before deploying/installing the updated NF-FG with UN (B2), **the NF-FG and related network configuration is verified by control rule and configuration verification modules.**
9. The Resource orchestrator pushes the updated NF-FG and MEASURE annotation on the local resource orchestrator in UN B2 to instantiate a new Firewall instance (FW2) and overlay links.
10. Additional network performance monitoring functions (MFs) are automatically deployed in UN (B2) and are **operating in line with the configuration expressed by MEASURE**.
11. The Resource orchestrator updates the NF-FG for UN (B0) and UN (B1) to update the overlay. Also MEASURE expressions for UN (B0) and UN (B1) are updated in order to re-configure the performance monitoring functions (e.g. new link between LB and FW2; new thresholds for LB and FW1; etc.). **The updated NF-FGs and related network configuration is verified by control rule and configuration verification modules.**
12. The state migration is controlled by the elastic firewall control app according to the procedures described in Section 5.2 of the D2.1 amendment [D21b]:
    a. The Elastic firewall control application instructs firewall FW2 to buffer any incoming traffic related to state s1.
    b. Buffering command is acknowledged.
    c. The Elastic firewall control application configures the load-balancer LB to redirect traffic related to state s1 to firewall FW2
    d. Load-balancer update is acknowledged.
    e. The Elastic firewall control application instructs firewall FW1 to send state s1 to firewall FW2 through a direct connection.
    f. Firewall FW1 initiates state transfer to firewall FW2.

g. State transfer of state s1 is acknowledged from firewall FW2 to FW1.
h. State transfer of state s1 is acknowledged from firewall FW1 to Elastic firewall control application.
i. The Elastic firewall control application instructs firewall FW2 to process any buffered traffic for state s1 and stop buffering once the buffer has been emptied.
j. Firewalls FW2 acknowledges buffer processing and stop commands.
k. Steps a. to j. are repeated until all the desired state has been transferred.

## 6.5 Concluding Remarks

Operations processes (e.g., e-TOM) carried out by current OSS/BSS of Telecommunications networks are rather static and time consuming: for example, most of configurations are still done by human operators; the services provisioning time is much longer (weeks or months) than the OTTs (minutes or hours).

The evolution of Telecommunications and ICT infrastructures towards the UNIFY production environment will definitely request faster and more flexible way for implementing said Operations processes. The lessons learnt in this use-case are showing that DevOps can be considered as one of the best candidate for pursuing this evolution of Operations processes.

In general, DevOps tools already used today in Data Centers, such as [Chef], [Puppet], and [Ansible], have to be properly extended to support network nodes, systems and products. This is another lesson learnt in this use-case, when trying to apply the developed tools. In fact, the goal, at the end of the day, will be to automate configuration and provisioning of end-to-end services, as part of an orchestrated delivery across multiple nodes and systems including physical or virtual compute, storage, and network resources.

Moreover the DevOps tools criteria of success depend on how closely they will match with the Service Providers' (SP) specific roles (e.g., Smart Connectivity Provider, Business or Service Enabler, etc). In this sense, it would be strategic, for the evolution towards UNIFY production environments, developing a sort of catalog with a number of DevOps tools, each of them described by its own "certificate".

Needless to say that a big "cultural" gap has also to be filled, for educating SPs in the use of DevOps: in fact, the required IT-oriented mindset is very different from the established one in current SPs, where IT and Network domains are still very separated in the Operations, and the level of automation is very low.

# 7 Integrated Prototyping

Since the previous deliverable, D4.1, WP4 presented one integrated demonstration (described in Annex 4) and one stand-alone demonstration (showcasing the status of AutoTPG, Section 5.8) at the first year's review of the UNIFY project. Most of the efforts in Task 4.4 (Evaluation and prototyping) went to creating and organizing these demonstrations. Since then, the conclusions of this development process helped us to find viable workflow of integrating individual prototypes. The loose coupling provided by the SP-DevOps support functions (in particular the DoubleDecker messaging system) was the key for a fast and convenient integration.

Additionally, we have started to work on the second phase of integration. WP4 started to work on the development of an integrated prototype under the name of SP-DevOps Pro, which focuses on the Service Provider DevOps aspects of the UNIFY architecture. A status update on SP-DevOps Pro is given in Section 7.1. The following Section 7.2 summarizes the successive integration approach of SP-DevOps Pro into the project-wide prototype called IntPro, common to all UNIFY work packages.

## 7.1 Integration within WP4

Currently, WP4 has one integrated prototype combining several individual prototypes. This prototype was demonstrated at the Y1 review and at the IFIP/IEEE International Symposium on Integrated Network Management [FNE15]. In the demonstrator, the operation of a simple *Load balancer* network function was supported by four SP-DevOps functions and tools: the DoubleDecker messaging system; configuration tool using a rudimentary version of MEASURE; the rate monitoring tool as a probabilistic congestion detector; and the Network Watchpoint mechanism. The demonstration also presents potential roles and use-cases for the tools. It exemplifies two perspectives of network observability and the interaction between selected monitoring, debugging, and control messaging components of the UNIFY architecture. It shows how detection of link congestion triggers automated load balancing, and how the information about the observed state and associated actions are presented to the VNF developer and operator. The information forwarded can be used either for supporting continuous service development or for supporting operator management actions. The demo also illustrates how relations and communication between components and levels of an overall architecture could be supported.

Annex 4 contains the shortened version of the paper describing the demo showcased at the symposium. It includes several lessons learnt during the integration process. For instance, the DoubleDecker messaging system was identified as key for coupling of several component, facilitating rapid and smooth integration. The demo also gave some insights about the implementation choices for MFs, i.e. whether they are part of the infrastructure of instantiated as a VNF within the VNF EE.

Based on those lessons next phase ( phase 2), of the agile integration will follow a bottom-up approach. The communication method will be built around the DoubleDecker messaging system. Each individual tool prototype will first be hooked into the ESCAPE framework [D22][D32] by means of the SP-DevOps support functions, and then in the following phase some of the ESCAPE-embedded versions will be integrated into a SP-DevOps Pro scenario.

Figure 7.1 shows how the tools considered for phase 2 (DoubleDecker, MEASURE, Rate Monitoring, EPLE, AutoTPG, Deploy-time Verficiation, Watchpoint tool, Epoxide,) are mapped to the functional architecture. These tools collectively cover functionalities in almost all functional blocks. Notable exception is the OSS, but during phase 1 we had already showed possible integrations with OSS tools (OpenNMS, OpenTSDB, see Annex 4).

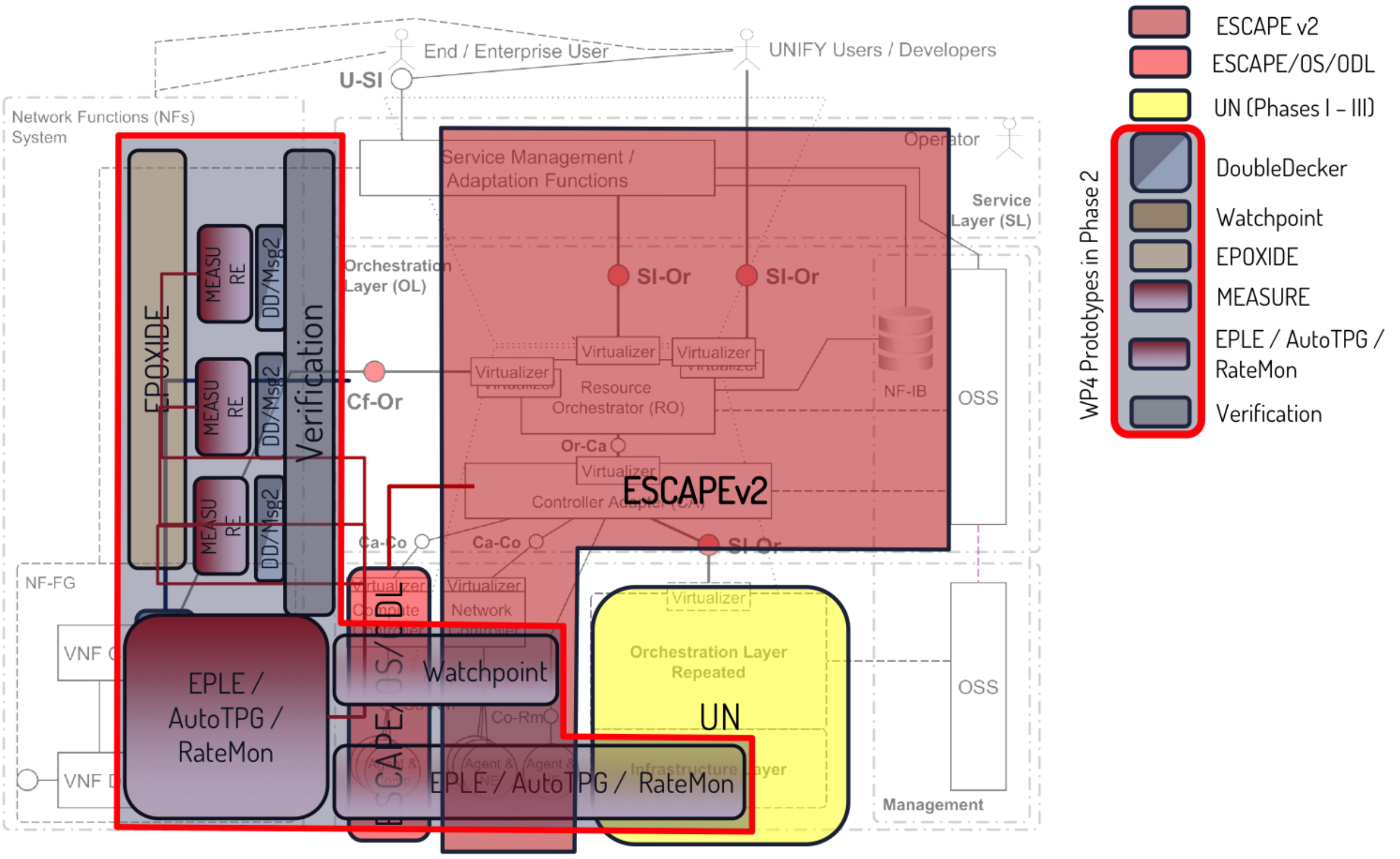

*Figure 7.1: Functionalities of some of the individual SP-DevOps tool prototypes mapped to the UNIFY architecture.*

Besides these integration efforts, WP4 is planning to publicly offer a set of individual prototypes under the aegis of the SP-DevOps Toolkit (Section 4.2)

## 7.2 Towards IntPro — Project-Wide Integration

Beside the work on SP-DevOps Pro with the integration of the individual prototypes into an overarching showcase scenario, the integration with the other activities in UNIFY is under consideration. The concept is called Integrated

Prototype, IntPro, and the idea is to show the overall concept of UNIFY (to be reported in the planned D2.4 "Integration of prototypes"). The approach is to initially collect the prototype information from the project, understand and define the overall picture and finally, propose an integration plan. As a first step of the WP2-WP4 cooperation, we prepared answers to a WP2 questionnaire about our prototyping activities. On the one hand, the questionnaires are going to be discussed in WP2 in order to define a common scenario, identify software and hardware requirements and dependencies, etc. On the other hand, the questionnaires are also useful for finding synergies within WP4 as well because they help us to (i) understand the partners' prototyping results, activities, and future plans in order to foster additional co-operations, (ii) facilitate the ongoing discussions on the mapping of the tools to the functional blocks of the UNIFY architecture (Figure 7.1) and the implementation/realization of such a mapping in practice. Due to space constrains these descriptions about the prototypes are omitted from this deliverable and can be found in the Section 3 and Section 4 of the internal milestone report M4.2 [M42]. So far it can be concluded that the ESCAPEv2 demonstrator [D32] will be central component representing the service and orchestration layer elements including support for controllers.

Due to internal prioritization of resources, WP5 will not implement node-local monitoring functions. That means that special requirements of WP4 like high frequency counters might not be available on the prototype implementation of the UN. As one solution, WP5 does not foresee problems for the case when the Monitoring Functions implemented as VNFs. This, however, is likely to come with an extra performance penalty, which makes some of the SP-DevOps tools presented in Section 5 unsuitable for integration with the UN prototype (see e.g. Section 5.5-5.7). As a second solution, WP4 and WP5 are currently discussing possible alternatives of how to embed MFs and their OPs in the current UN architecture by the means of the monitoring management plugin, MMP, as described in Section 3.3.

## 7.3 Concluding Remarks

The ESCAPEv2 [D32] framework allows for integrating the SP-DevOps support functions (in particular DoubleDecker and MEASURE) and therefore provides support for other elements of phase 2 of SP-DevOps-Pro integration. The other important integrating element will be the NF-FG itself, which is documented in detail in D3.1 [D31]. The specification allows to implement the Sl-Or interface and by doing so to recursively attach other orchestrators or UN prototypes. By the time of writing, the integration with the ESCAPEv2 is under development as part of the work in T4.4.

# 8 Conclusions

## 8.1 Summary and Results

In this deliverable, we have reported on the implementation of the SP-DevOps concept, encompassing 13 support functions and tools that technically support the four SP-DevOps processes concerning observability, troubleshooting, verification and VNF development. The presented SP-DevOps functions and tools enable:

- low-latency and scalable communication between SP-DevOps tools and other functional components (Section 5.1);
- specification and parsing of NF-FGs for monitoring and information retrieval within the scope of an NF-FG (Section 5.2 and 5.3);
- support of OAM legacy tools in programmable frameworks (Section 5.4);
- scalable and adaptive high rate monitoring for increased observability capable of reducing the measurement rate by up to 75% compared to fixed rate monitoring (Section 5.5);
- efficient link and loss monitoring of flow aggregates, capable of significantly reducing the measurement overhead from $O(n^2)$ to $O(n)$ (Section 5.6) as well as the load on the controller (Section 5.7);
- run-time verification of aggregated flows for ensuring correct delivery of data packets and troubleshooting of incorrect Flow Entries (Section 5.8 and 5.11);
- deploy-time verification of NF-FG reachability and resolution to detected reachability problems in less than 200ms (Section 5.10);
- novel opportunities to monitor control traffic events (such as policy violations or rule matching) on the OpenFlow control plane (Section 5.12), as well as tools for automated troubleshooting support (Section 5.13).

Based on the technically more matured SP-DevOps concept, the integration of the SP-DevOps components and processes in the functional architecture as well as on the UN has been updated.

The role of SP-DevOps functions and tools with respect to orchestration has been described across the architecture, layer-by-layer, in terms of relevant actions during deployment and information dissemination of monitoring data during the operational phase of a service. The set of SP-DevOps metrics defined in this deliverable provides a number of important KPIs that are relevant for evaluating different aspects of the service lifecycle, starting from development and deployment to operations.

The programmability aspects researched in UNIFY are supported by a number of functions developed in both WP3 and WP4, covering communication, configuration, support to legacy monitoring tools, as well as information dissemination (Section 5.1-5.4). These functions are essential for facilitating SP-DevOps processes in an efficient and scalable manner, and are mapped across all the layers of the functional architecture down to the infrastructure layer and the UN.

Support functions and tools are described in detail together with their architectural mapping, giving a wider understanding of the SP-DevOps concept as a whole, and how SP-DevOps tools can be used for supporting orchestration and programmability in the UNIFY framework. The concept is further exemplified in a case study for deployment and scaling of an Elastic Firewall. As a result of the technical development in WP4 and the SP-DevOps concept, an open source toolkit involving a subset of the support functions and tools is planned for a first release before the end of 2015.

## 8.2 Gap Analysis

**WP2-WP4 relevant topics:** Based on joint WP2-WP4 discussions and the work reported in D2.2 and M4.2, the functional architecture and the mapping of the SP-DevOps concept have been refined since the publication of D4.1. Input to the updated functional architecture has been provided by WP4 mainly from the perspective of the more matured SP-DevOps observability and verification processes (as most of the SP-DevOps tools perform monitoring or verification). Tools developed in WP4 support SP-DevOps processes across all layers of the architecture, with a strong presence in the Infrastructure layer as many of the tools to a large degree implement observability functionality for the purpose of service and infrastructure monitoring. Integration of existing monitoring, troubleshooting, verification and VNF development support tools with the other parts of the project in a joint integrated prototype, carried out in Task 2.3 and Task 4.4. From individual prototypes and joint demos already presented by WP4, the steps towards a project-wide prototype includes migration to the more UNIFY architecture compliant ESCAPEv2 [D32], which is ongoing work in T4.4. Further steps towards a project-wide prototype will include discussions on the definition and implementation of relevant interfaces supporting SP-DevOps tools and functions.

**WP3-WP4 relevant topics:** Since D4.1 the joint work between WP3 and WP4 has involved discussions on how SP-DevOps observability and troubleshooting processes support resource orchestration, as well as the role of SP-DevOps during service deployment with respect to decomposition and verification. In this deliverable, we have outlined the role of SP-DevOps in these respects in Section 3, which is subject for further refinement as part of the work towards automated workflows in T4.3. Moreover, we have exemplified in a case study in Section 6 how SP-DevOps processes support deployment and dynamic scaling for the Elastic Firewall scenario [D21b]. In the deployment scenario we assume that SP-DevOps monitoring functions are deployed by the use of the MEASURE language. As defined in M4.1, the MF concept admits different data access for different roles (Operator, Service/VNF Developer) and types of monitoring (infrastructure or service level). Currently WP3, to a large extent, defines these orchestration processes based on existing OAM tools and the assumption of static network environments. However, a stronger integration between WP3 and WP4 regarding the use of SP-DevOps monitoring functions for pre-deployment and resource management (e.g. bootstrapping and resource discovery) is subject for further discussions.

Since the Milestone M4.2, very recent developments in WP3 involved discussions on changing the service model from the NF-FG representation of service graphs to an aggregated form of service representation per tenant. In all WP4 deliverables and milestones (including this deliverable) the SP-DevOps concept and orchestration support have been described with the notion of NF-FGs as defined in D3.1. The implications of the alternative NF-FG representation for service deployment, decomposition and operations with respect to the SP-DevOps processes (as well as the integrated prototyping) will be followed-up in the next deliverables.

**WP4-WP5 relevant topics:** Deployment of SP-DevOps tools on a UN based Infrastructure layer has since the initial description in M4.1 been further discussed between WP4 and WP5, resulting in an updated UN architecture that allows for instantiating MFs and OPs locally on the nodes with access to various data sources (e.g. OF counters). Part of this work also involved mapping of the DoubleDecker messaging system enabling communication between SP-DevOps components and functional blocks of the architecture. In Section 3 of this deliverable, we introduced the notion of the MMP, which is part of the local infrastructure controller layer in the UN and enables deployment of MFs. There is still an open question regarding where and how an MF control app can be deployed and whether it is running inside the control layer of the UN or on a dedicated network entity. This issue, as well as the implementation and interfaces of the MMP, are part of the work towards a project-wide integrated prototype - discussions between WP4 and WP5 are expected for answering these questions.

The majority of the SP-DevOps tools implement observability functionality related to the networking performance aspect, involving network measurements and in some instances node-local processing on the UNs, but not monitoring of compute and storage performance. Currently, the resource management interface of the unified resource manager in the UN reports CPU and memory usage via the Sl-Or interface. Further discussions are needed for strengthening the use of SP-DevOps tools that conceptually can be used for monitoring and reporting of compute and storage parameters in addition to network performance. How SP-DevOps processes can be integrated for accessing and reporting storage and compute performance parameters for monitoring and troubleshooting is therefore subject for further discussions, and is relevant to the definition of automated workflows in T4.3.

## 8.3 Next Steps and Future work

This deliverable concludes the work in T4.2 focused on SP-DevOps tools and support functions. Remaining tasks will be focused on the definition of automated workflows with respect to monitoring, troubleshooting and verification processes. Initial definitions of these processes were outlined in D2.1 [D21a] and D4.1 [D41], but are subject to further refinement given the technically more matured SP-DevOps concept presented in this deliverable. Several prototyping activities within T4.4 currently includes migration of existing prototypes to ESCAPEv2 [D32] and are to a high degree based on the results in the first year, as an intermediate step towards integration in the project-wide prototype. The open issues highlighted in Section 8.2 will be addressed as part of the work in both tasks, and will be reported in the upcoming deliverables D4.3 and D4.4.

# Annex 1 SP-DevOps metrics

In the following tables , different sets of SP-DevOps metrics (Section 3.5) are shown in Tables A1.1-A1.3. The SP-DevOps metrics are categorized per Operator, VNF Developer and Service Developer role, covering technical, process-related, and cost-related aspects.

| No. | Name | Importance for stakeholder | Importance for internal evaluations | Easiness to measure |
|---|---|---|---|---|
| Telecom Operator | | | | |
| Technical metrics | | | | |
| 1 | Service performance (RTT, delay in msec, jitter) | high | high | high |
| 2 | Throughput (in Mbps/Kbps) | high | high | high |
| 3 | Congestion (number of dropped packets) | high | high | high |
| 4 | Energy consumption | high | low | low |
| 5 | Availability (number/duration/frequency of outages) | high | medium | medium |
| 6 | Reaction to increasing/excessive load (time in msec, from congestion event, e.g. packet drop, to service recovery) | high | high | high |
| 7 | Infrastructure overhead | medium | high | medium |
| Process-related metrics (concerning Business or Network operation) | | | | |
| 8 | Service creation (time) | medium | high | medium |
| 9 | Service creation success rate (%) | high | high | medium |
| 10 | Service failure rate (%) | high | high | medium |
| 11 | Maintenance (number/duration/frequency of actions) | high | medium | medium |
| 12 | Time to repair (in hours/mins) | high | high | medium |
| 13 | Orchestration overhead (CPU, storage, | medium | medium | high |
| | bandwidth necessary to perform scalable orchestration) | | | |
| 14 | Hardware resources (routers, switches, etc.) | low | low | high |
| 15 | Software resources (including VNFs) | high | medium | high |
| 16 | Network resources (capacity, slices, etc.) | high | medium | high |
| 17 | Number of probes/robots for e2e QoS monitoring (provide global level of QoS) | low | low | high |
| 18 | Manual analysis for e2e QoS (provides detailed level of QoS) | high | high | medium |
| 19 | Number of on-site interventions | low | low | High |
| 20 | Number of unused alarms (used for roo cause analysis) | low | low | medium |
| 21 | Number of configuration mistakes (for incidents resolution that impacts other customers) | high | medium | medium |
| 22 | Delay between order and delivery (includes service creation time) | medium | medium | medium |
| 23 | Trouble reaction delay | high | high | medium |
| 24 | Time for pre-deployment verification (from development and test-bed trial to pilot and production network) | low | low | high |
| Cost-related metrics | | | | |
| 25 | Cost of service outage (monetary potential based on SLAs) | high | medium | high |
| 26 | Labour | high | medium | medium |
| 27 | OPEX (per subscriber, service minute, site, etc.) | high | medium | low |
| 28 | Effort to maintain inventory | high | low | medium |

*Table A1.1: Overview of SP-DevOps metrics related to the Operator role.*

| No. | Name | Importance for stakeholder | Importance for internal evaluations | Easiness to measure |
|---|---|---|---|---|
| VNF Developer | | | | |
| Technical metrics | | | | |
| 1 | Service performance (RTT, delay in msec, jitter) | high | high | high |
| 2 | Throughput (in Mbps/Kbps) | high | high | high |
| 3 | Congestion (number of dropped packets) | high | high | high |
| 4 | Energy consumption (Kwh) | medium | medium | medium |
| 5 | Availability (number/duration/frequency of outages) | high | medium | medium |
| 6 | Level of virtualization/access (i.e. in what extend the developer can modify the testing environment) | high | high | high |
| Process-related metrics | | | | |
| 7 | Time to fix (i.e. time between discovering a bug and deploying s solution or workaround) | high | high | medium |
| 8 | Length of compile-deploy-modify cycle (different than the software development life cycle) | high | low | medium |
| 9 | Length of software development life cycle | high | low | medium |
| Cost-related metrics | | | | |
| 10 | Required skills (user-friendly architecture) | medium | medium | medium |
| 11 | Idle cost (vs. shutdown-start cost) | high | medium | medium |

*Table A1.2: Overview of SP-DevOps metrics related to the VNF Developer role.*

| No. | Name | Importance for stakeholder | Importance for internal evaluations | Easiness to measure |
|---|---|---|---|---|
| Service Developer | | | | |
| Technical metrics | | | | |
| 1 | Number/volume of defects | high | high | low |
| 2 | Continuous runtime | high | high | medium |
| Process-related metrics | | | | |
| 3 | Decomposition and deployment processing time from the moment of a ready NF-FG definition until the service is up and running | high | high | medium |
| 4 | Development success rate (or the other side of the coin: percentage of failed deployments) | high | high | medium |
| 5 | Deployment frequency (per week/month) | high | medium/low? | medium |
| 6 | Deployment cycle (steps/phases/time) | high | medium | medium |
| 7 | Time per release | high | high | medium |
| 8 | Mean time to recover | high | medium | medium |
| 9 | Rate of number of automated operations to total number of operations | medium | high | medium |
| Cost-related metrics | | | | |
| 10 | Cost per release | medium | medium | low |
| 11 | Cost of outages (monetary, potentially based on SLAs) | high | medium | low |
| 12 | Cost of necessary resources (storage, computation) | high | medium | medium |

*Table A1.3: Overview of SP-DevOps metrics related to the Service Developer role.*

# Annex 2 Example of DoubleDecker Deployment in a Docker Environment

To illustrate how the DoubleDecker messaging system can be used we here introduce an example deployment, connecting a broker to a higher layer broker and providing connectivity to some clients. In this example the broker and clients are all implemented as containers in a Docker environment, a popular platform for deploying containers.

The basic primitive in Docker is the *image,* which is similar to any virtual machine image, and contains a file system, and initial configuration. Instantiating an image creates an isolated environment called a *container,* in which one or more processes run. Containerized processes in the container are obtained in a restricted view of their environment, a so called namespace. A namespace can restrict the processes view of for example process IDs, the file system structure, available network interfaces, and SysV IPC mechanisms, allowing multiple processes, which would normally interfere with each other, to run independently. By presenting the processes in a container with restricted namespace views, a container can act similarly to a virtual machine but with significantly lower overhead. It is however less isolated compared to its virtual machine counterpart, as multiple containers share important resources (such as the kernel).

In Docker different kinds of relationships between containers can be easily configured (and set if necessary when a container is instantiated). Container relationships can be defined using *links, exposed ports, and volumes. Links* and *exports ports* refer to network connectivity, and linking container B to container A will cause container B to have a number of environment variables automatically set. These variables will then contain the IP address of container A and the port numbers A have chosen to expose. *Volumes* can be used to share files and directories between containers, for example container A may generate data files in the directory it perceives as /output that should be processed by container B. When starting B, we can specify that all data volumes from container A should be mapped to B, such that processes running in container B will then be able to access files mounted in /input on B. Volumes can be used to share files not only between containers but also between containers and between the node hosting them. Processes running in containers can be configured by for example mounting a directory containing configuration files when starting the container, e.g. for preparing a directory with configuration files and mounting it as /etc on the container. A more flexible way of providing small configuration values is by using environmental variables supplied when starting the container - these environmental variables will then be available for all processes running in the container.

To deploy a broker we start by creating a Docker image (called "unify/broker") with a broker included, and configure it to automatically start waiting for connections on UNIX socket "/ipc/broker.ipc", as well as on TCP port 5555. We can also set it to try to connect to the TCP/IP port defined in environmental variable BROKER_ PORT, if it exists. This way we can start the container with or without that environmental variable to determine if it should be a standalone broker or part of a hierarchy.  Similarly for clients, we create a Docker image (called "unify/client") with the client included, and configure it to automatically try to connect to a broker on UNIX socket "/ipc/broker.ipc" if it exists, or otherwise try to connect to "BROKER_PORT".

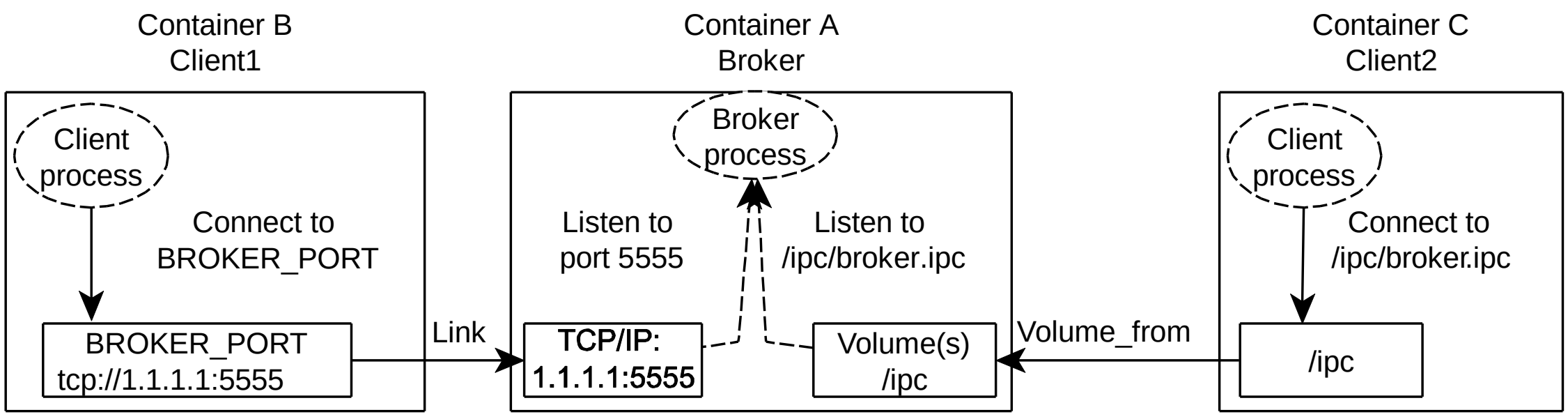

*Figure A2.1: Example of connecting clients using links (left) and shared volumes (right).*

Now, to start a broker and client and connect them we can 1) start the broker, 2) get its IP address, and 3) start client with BROKER_ PORT set to include the IP we found.

```
1.    $ docker run -name broker unify/broker
2.    $ IP=$(docker inspect --format '{{ .NetworkSettings.IPAddress }}' broker)
3.    $ docker run -name client1 -e BROKER_PORT=tcp://$IP:5555 ponsko/client
```

This process is however a bit cumbersome - to automatically set the BROKER_PORT environment variable in the client we can use the Docker link concept, which automatically creates this variable if we link the client to the broker. The new process becomes 1) start broker, 2) start client with link to broker (illustrated in the left of Figure A2.1).

```
1.    $ docker run -name broker unify/broker
2.    $ docker run -name client1 --link broker:broker unify/client
```

Now client1 is automatically connected to the TCP port the broker is listening to, where it registers and can start sending messages. However, since these are running on the same host using TCP causes unnecessary overhead - this we can avoid by using the UNIX socket instead. For automation we can 1) start the broker with the "/ipc" volume defined, and 2) import the "/ipc" volume to the client when we start it (illustrated in the right of Figure A2.1).

```
3.    $ docker run -name broker -volume /ipc unify/broker
4.    $ docker run -name client1 --volumes-from=broker unify/client
```

Now the "/ipc" directory is mounted in the client1 container where it can connect to the listening broker.  If we want to run multiple separate brokers (e.g. to not allow certain clients to communicate), brokers can be started with different names, the same methods still work. Finally, if we want to connect the broker to an external broker, e.g. in a higher layer, we simply provide the BROKER_ PORT variable when starting the broker.

# Annex 3 Definition of the Recursive Query Language

The query language (Section 5.3) is based on Datalog which is a declarative logic programming language. A Datalog program consists of a set of declarative rules and a query. A rule has the form

$$h <= p_1, p_2,..., p_n.$$

which can be defined as "$p_1$ and $p_2$ and ... and $p_n$ implies q". "h" is the head of the rule, and "$p_1$, $p_2$, ..., $p_n$" is a list of literals that constitutes the body of the rule. Literals "$p(x_1, ..., x_i, ..., x_n)$" are either predicates applied to arguments "$x_i$" (variables and constants), or function symbols applied to arguments. The program is said to be recursive if a cycle exists through the predicates, i.e., predicate appearing both in the head and body of the same rule. The order in which the rules are presented in a program is semantically irrelevant. The commas separating the predicates in a rule are logical conjuncts (AND); the order in which predicates appear in a rule body also has no semantic significance. The names of predicates, function symbols and constants begin with a lower-case letter, while variable names begin with an upper-case letter.

A variable appearing in the head is called distinguished variable while a variable appearing in the body is called non-distinguished variable. The head is true of the distinguished variables if there exist values of the non-distinguished variables that make all sub goals of the body true.

In every rule, each variable stands for the same value. Thus, variables can be considered as placeholders for values. Possible values are those that occur as constants in some rule/fact of the program itself.

In the program, a query is of the form "$query(m, y_1, ..., y_n)$", in which "query" is a predicate contains arguments "m" and "$y_i$". "m" represents the monitoring function to be queried, e.g., end to end delay, average CPU usage, and etc. "$y_i$" is the arguments for the query function. The meaning of a query given a set of Datalog rules and facts is the set of all facts of query() that are given or can be inferred using the rules in the program.

The predicates can be divided into two categories: extensional database predicates (EDB predicates), which contains ground facts, meaning it only has constant arguments; and intensional database predicates (IDB predicates), which correspond to derived facts computed by Datalog rules.

In order to perform recursive monitoring query, the resource graph in NF-FGs need be transformed and represented as Datalog ground facts which are used by the rules in the program. The following keywords can be defined to represent the NF-FG graph into Datalog facts, which are then used in the query scripts:

- sub(x, y) which represents 'y' is an element of the directly descend sub-layer of 'x';
- link(x, y) which represents that there is a direct link between elements 'x' and 'y';
- node(z) which represents an node in NF-FG.

It is to be noted, more keywords can be defined in order to describe more properties of NF-FG.

In addition, a set of functions call are to be defined for the monitoring query. The function call will start with "fn_" in the syntax and may include 'boolean' predicates, arithmetic computations and some other simple operation. The function calls can be provided by the query engine or developers.

The following BNF describes the syntax of the Datalog based language.

```
<program>          ::=   <statement>*
<statement>        ::=   <rule> | < fact>
<rule>             ::=   [<rule-identifier>] <head> <= <body>
<fact>             ::=   [<fact-identifier>]<clause> |
                         <fact_predicate>(<terms>)
<head>             ::=   <clause>
<body>             ::=   <clause>
<clause>           ::=   <atom> | <atom>, <clause>
<atom>             ::=   <predicate> ( <terms> )
<predicate>        ::=   <lowercase-letter><string>
<fact_predicate>   ::=   ("sub" | "node" | "link")( <terms> )
<terms>            ::=   <term> |
                         <term>, <terms>
<term>             ::=   <VARIABLE> |
                         <constant>
<constant>         ::=   <lowercase-letter><string>
<VARIABLE>         ::=   <Uppercase-letter><string>
<fact-identifier>  ::=   "F"<integer>
<rule-identifier>  ::=   "R"<integer>
```

# Annex 4 Integrated WP4 Prototype

We have implemented and integrated four SP-DevOps tools that are useful in their own right, but showcase also an integrated scenario [FNE15], where they form the basis for a more complete SP-DevOps Toolkit (see also Section 4 and 7). The demo consists of several interacting components developed in WP4 (and partially WP3):

- DoubleDecker Messaging system, based on ZeroMQ [ZeroMQ] (Section 5.1);
- MEASURE configuration tool (Section 5.2);
- Congestion detector based on the rate monitoring tool (Section 5.5)
- Network watchpoint intercepting OF control messages (Section 5.12)

The demonstration of the integrated prototype also presents potential roles and use-cases for the tools. It exemplifies two perspectives of network observability and the interaction between selected monitoring, debugging, and control messaging components of the UNIFY architecture. It shows how detection of link congestion triggers automated load balancing, and how the information about the observed state and associated actions are presented to the VNF developer and operator. The information forwarded can be used either for supporting continuous service development or for supporting operator management actions. The demo also illustrates how relations and communication between components and levels of an overall architecture could be supported.

**Observability aspects**

In a service chaining architecture like UNIFY, monitoring, troubleshooting, and messaging components can serve several distinct purposes. They can be part of the infrastructure layer of the programmability framework and observe physical resources, e.g., link or node status; or they can be part of a service chain and instantiated as VNFs for observing and controlling service-level properties, e.g., SLA and resource usage monitoring, or as input to autonomous reconfiguration.

Moreover, the very same component can be placed in the infrastructure layer or the service chain based on users' intent. For instance, the scalable monitoring component should be part of the infrastructure if it used for link congestion detection, but it should be instantiated as a VNF if it measures the traffic intensity for one particular service chain.

Finally, the users of the monitoring information can be VNF developers or service operators. For example, a VNF developer can use the Watchpoint tool to detect bugs in her code while testing a new VNF implementation, but later when the VNF is deployed in production the Watchpoint tool can be used to detect configuration errors and notify an operator or an Operations Support System. Our prototype implementation is flexible enough to support all these aspects.

## Demonstration setup

In the demo scenario as shown in Figure A4.1, a Tier 1 ISP is providing network transit for three smaller ISPs, over its backbone SDN network. To forward the transit traffic the Tier 1 ISP has two links available, one low-latency, reliable link and a secondary link with higher latency and low reliability. Due to its SLA the traffic of the smaller ISP should never be routed on the secondary link whereas the other two ISPs should use the first link unless there is a significant risk of congestion, in which case the secondary link is preferred. The Tier 1 ISP additionally wants to keep its monitoring and messaging overhead low and avoid transferring data to central points in its network unless necessary.

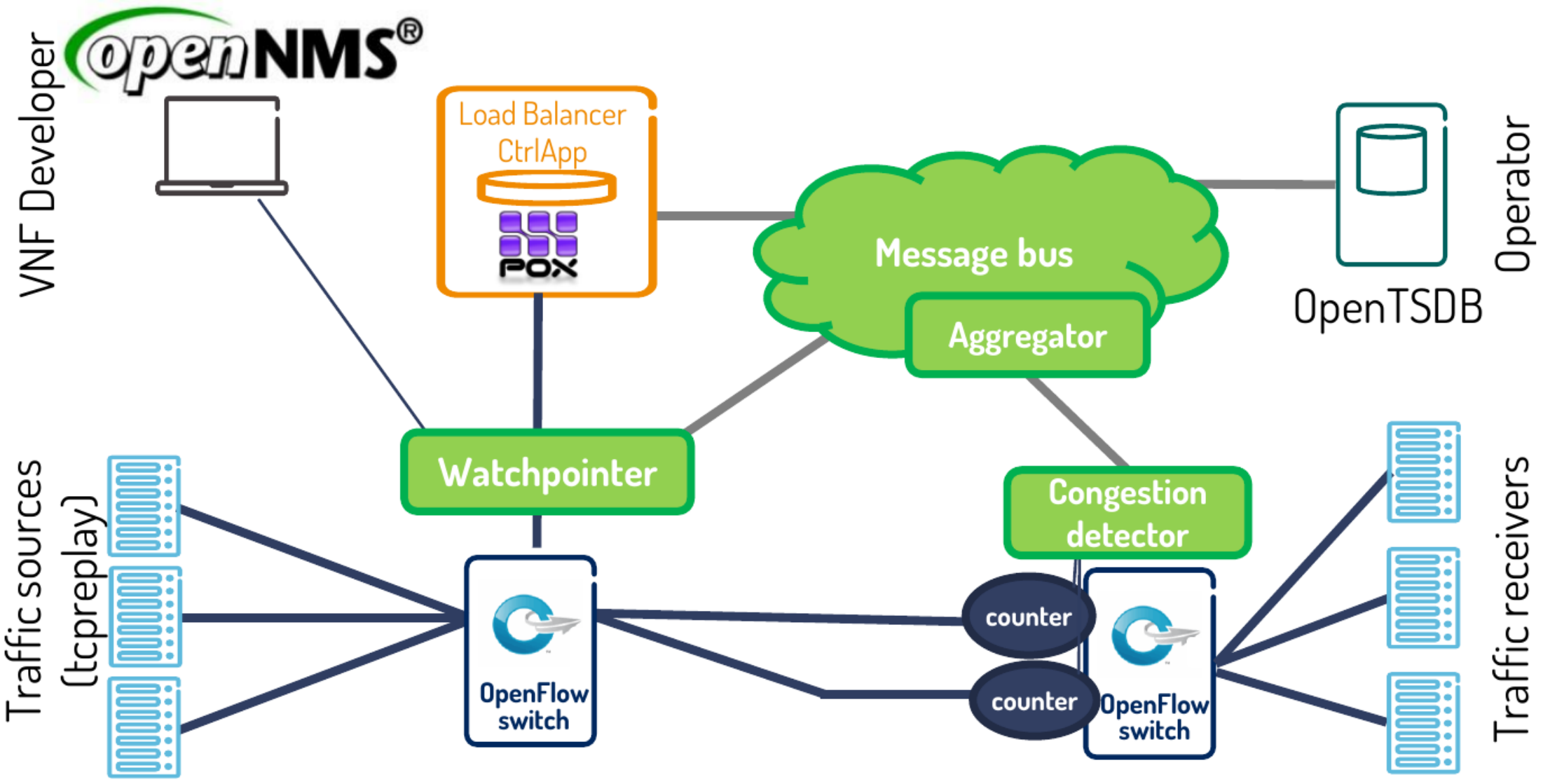

*Figure A4.1. The demo setup containing the four DevOps tools. Here, the configuration tool is responsible for initialising the Aggregator component.*

To take link congestion risk into account, the Tier 1 ISP deploys rate monitors on the links, which calculate an estimate every second and send it to their local aggregation point. The local aggregation point has been dynamically programmed in the setup phase to apply hysteresis to the overload risk estimates and compare the results to a threshold, which if breached will cause a signal to be sent to the load balancing application on the SDN network controller. Additionally, the estimates are also sent to an OpenTSDB [OpenTSDB] instance. To ensure that policies are not violated the ISP has deployed a watchpoint beneath the SDN controller, which monitors all OF control traffic and stops any control message which would cause a violation of the policies. To keep network overhead low; monitoring results, triggers, and configuration data are sent over a hierarchical distributed messaging system which forwards messages locally if possible.

The demo shows three data flows arriving in sequence to the Tier 1 network, from the different ISP, which are allocated to the primary or secondary links depending on the current overload risk. The first flow, arriving to an empty network, is placed on the primary link. The second flow is placed on the secondary link if the primary link is above a one percent risk of overload. The third flow should always be placed on the first link (due to the SLA), but due to a bug in the load balancing software causing it to ignore this policy it will be assigned to the secondary link if the overload risk on the primary exceeds one percent. This bug however will be caught by the watchpoint which will block the traffic assignment and raise an alarm in the NMS.

The demo is running in a virtualized environment based on Mininet (mininet.org) in combination with POX (www.noxrepo.org/pox ) which provides the SDN network and load balancing functionality. Other open source projects used are OpenTSDB (opentsdb.net), OpenNMS (opennms.org), tcpreplay (tcpreplay.synfin.net), and ZeroMQ (zeromq.org) for the messaging system. The congestion detector, DoubleDecker, and the aggregation component are written in Python.

### Lessons learned

Creating the demonstration was useful for us, because we learnt the following lessons that help us the current and forthcoming prototyping / integration / demonstration activities.

- The messaging system provides a scalable messaging subsystem, for example, to collect monitoring data, but it is also a general communication mechanism. Being a general mechanism, it has additionally turned out to be a key to loosely couple the components, and therefore it was the main raison of rapid and successful integration.
- The some components can either be part of the infrastructure or instantiated as a VNF. For example, the Congestion Detector can be an OP in the infrastructure layer of the UN or if the performance requirements allow it, it can be VNF measuring the traffic of just one NF-FG.
- Presenting a demo on a single laptop is possible, but precise measurement is problematic in case moderate traffic load in the presented scenario.

Rate monitoring has to have access to high frequency traffic counters. In order to port this demo to the Universal Node, WP5 and WP4 need to find a way to provide this measurement information to the Rate monitoring module.